\documentclass{USC-Thesis}
\usepackage{graphicx}
\usepackage{subfigure}
\usepackage[subfigure]{tocloft}

\usepackage{amsmath}
\usepackage{amssymb}
\usepackage{amsthm}
\usepackage{mathrsfs}
\usepackage{wrapfig}
\usepackage{boxedminipage}
\usepackage{epsfig}
\usepackage{setspace}

\usepackage{rotating}
\usepackage{hyperref}

\DeclareGraphicsExtensions{eps,.pdf}

\newcommand{\be}{\begin{equation}}
\newcommand{\ee}{\end{equation}}
\newcommand{\ber}{\begin{eqnarray}}
\newcommand{\eer}{\end{eqnarray}}

\newcommand{\Dbar}{\ensuremath D \kern-0.63em/}
\newcommand{\delbar}{\ensuremath \partial \kern-0.63em/}

\def\cL{{{\cal L}}}
\def\cZ {{{\cal Z}}}
\def\cF{{{\cal F}}}
\def\cS{{{\cal S}}}
\def\cO{{{\cal O}}}
\def\cJ{{{\cal J}}}
\def\cE{{{\cal E}}}

\def\lsim{\mathrel{\rlap{\lower4pt\hbox{\hskip1pt$\sim$}}
    \raise1pt\hbox{$<$}}}                
\def\gsim{\mathrel{\rlap{\lower4pt\hbox{\hskip1pt$\sim$}}
    \raise1pt\hbox{$>$}}}                

\DeclareSymbolFont{AMSb}{U}{msb}{m}{n}
\DeclareMathSymbol{\IN}{\mathbin}{AMSb}{"4E}
\DeclareMathSymbol{\IZ}{\mathbin}{AMSb}{"5A}
\DeclareMathSymbol{\IR}{\mathbin}{AMSb}{"52}
\DeclareMathSymbol{\Q}{\mathbin}{AMSb}{"51}
\DeclareMathSymbol{\II}{\mathbin}{AMSb}{"49}
\DeclareMathSymbol{\IC}{\mathbin}{AMSb}{"43}
\DeclareMathSymbol{\IP}{\mathbin}{AMSb}{"50}
\DeclareMathSymbol{\IH}{\mathbin}{AMSb}{"48}
\DeclareMathSymbol\IA{\mathalpha}{AMSb}{"41}
\DeclareMathSymbol\IS{\mathalpha}{AMSb}{"53}

\makeatletter
\let\l@tableOLD \l@table
\renewcommand{\l@table}{\vspace{\baselineskip}\l@tableOLD}
\makeatother

\makeatletter
\let\l@figureOLD \l@figure
\renewcommand{\l@figure}{\vspace{\baselineskip}\l@figureOLD}
\makeatother

\begin{document}

 \title{EXTERNAL FIELDS AND THE DYNAMICS OF FUNDAMENTAL FLAVOURS IN HOLOGRAPHIC DUALS OF LARGE N GAUGE THEORIES}

 \author{Arnab Kundu}
 \major{Physics}
 \month{August}
 \year{2010}
 \maketitle

\chapter*{Dedication}

\begin{center}

{  \it To my parents: }

\end{center}

\begin{center}

{ \it Anup Kumar Kundu and Purabi Kundu,}

\end{center}

\begin{center}

{\it  to whom I owe it all.}

\end{center}

\addcontentsline{toc}{chapter}{Dedication}



\topmatter{Acknowledgments}

First, I would like to thank my supervisor Clifford V. Johnson for his guidance and supervision throughout the entire course of my graduate years at USC. He has always been wonderfully supportive and encouraging of new ideas and independence of thought, which has shaped me become a mature researcher over the years.

I am deeply grateful to other senior members of the Theoretical Physics group at USC for numerous interactions: in the form of course-work, light informal chats and serious collaborations. Several courses offered by Itzhak Bars, Clifford Johnson, Dennis Nemeschansky, Hubert Saleur, Nicholas Warner and Paolo Zanardi have taught me not just a number of advanced topics, but also an interesting blend of humor, rigor and clarity. These memories will be cherished for years to come.

Among the senior members, a special note of acknowledgement goes to Krzysztof Pilch and Nicholas Warner. I continue to enjoy a great experience of collaborating with Krzysztof and Nick in recent times. The intoxicating sense of humor, work ethics and the supersonic speed have deeply influenced my overall outlook.

The life of a graduate student is never complete without mentioning his fellow ones. I have been blessed to have Tameem Albash and Veselin Filev as my collaborators as well. A big part of this thesis is based on the work done in collaboration with them, which was initially pioneered by Veselin. They were invaluable supports to my baby-steps in the world of research. I am also grateful to Nikolay Bobev for the recent collaborations, which has broadened my horizon in magnitudes. Besides direct collaboration, I deeply acknowledge numerous engaging conversations with Tameem Albash, Nikolay Bobev, Holden Chen, Veselin Filev, Ramakrishnan Iyer, Guillaume Quelin, Aditya Raghavan, Koushik Roychoudhury about physics, life and Los Angeles.

I am also fortunate to have Faizal Ahmed, Susmita Basak, Ramakrishnan Iyer, Mithun Kumar Mitra, Sayak Mukherjee, Dhritiman Nandan, Aditya Raghavan, Amrita Rajagopalan, Koushik Roychoudhury for sharing some great experiences with jamming, photography, camping, traveling, rock concerts and jazz clubs.

I most humbly acknowledge the contribution of my family in India: my parents to whom this thesis is dedicated to and my sister. Without their sacrifice, unconditional support and love none of these would be possible. Credits are due to the taxpayers in India for supporting a large part of my education there.

A special note of thanks goes to Stephan Haas for his readiness and ability to solve any bureaucratic issue associated with graduate studies within a matter of seconds.

Thanks are also due to Francis Bonahon, Werner Dappen and Stephan Haas for serving in my qualifier and my dissertation committee.

And last but not the least, I express my deepest gratitude to Kasturi for her support, encouragement and love that kept me sane through the trying times and magnified the good ones.

\tableofcontents
\updatechaptername

\listoftables

\listoffigures

\chapter*{Abstract}
\addcontentsline{toc}{chapter}{Abstract}

Using the gauge-gravity duality we study strongly coupled dynamics of fundamental flavours in large $N_c$ gauge theories in a constant external field. We primarily focus on the effects of an external magnetic field. We use two holographic models realized in the Type IIB and Type IIA supergravity and present a comparative case study. In both these models, by studying the dynamics of probe branes, we explicitly demonstrate and discuss the magnetically induced chiral symmetry breaking effect (``magnetic catalysis") in the flavour sector. We also study the associated thermodynamics and the meson spectrum and realize {\it e.g.} Zeeman splitting, stability enhancement of the mesons in the presence of an external magnetic field {\it etc}. By studying the quasinormal modes of the probe brane fluctuation in the hydrodynamic limit we also obtain an analytic dispersion relation in the presence of a magnetic field in the Type IIA model. This dispersion relation consists of a propagating sound mode in the otherwise diffusive channel and is sourced by the quantum anomaly of the global U(1) current. We briefly discuss the effects of an external electric field and observe that the flavour bound states dissociate for sufficiently high electric fields and an electric current is induced.

\mainmatter

%
%
%
%
%
%
%
%
%

\chapter*{Chapter 1:  \hspace{1pt} Introduction}

\addcontentsline{toc}{chapter}{Chapter 1:\hspace{0.15cm}
Introduction}

\section*{1.1 \hspace{2pt} Motivation}
\addcontentsline{toc}{section}{1.1 \hspace{0.15cm} Motivation}

It is well-known that elementary particles interact by four fundamental interactions: gravitation, electromagnetism, the weak interaction and the strong interaction. The framework of quantum field theory has been extremely successful in describing many aspects of the last three interactions within remarkable experimental precision. There are however numerous aspects which remain poorly understood. One such long-standing puzzle comes from the strong interaction. Conventional quantum field theory methods are mainly based on (but not limited to) perturbative techniques and are therefore generally inadequate to address various aspects of strong coupling dynamics.

The quantum field theory describing the strong interaction is called Quantum Chromodynamics (QCD). The elementary degrees of freedom of QCD are massive spin-$1/2$ particles called ``quarks" and massless spin-$1$ particles called ``gluons". There are several different quarks (named ``up", ``down", ``charm", ``strange", ``top" and ``bottom"). However depending on their masses the quarks can be broadly divided into two categories: the light quarks (including the up, down and the strange) and the heavy quarks (including the charm, bottom and the top). As a first approximation we can set the masses of the light quarks to zero and the masses of the heavy quarks to infinity. In this approximation the only parameter in QCD is the dimensionless coupling constant.

This coupling constant is a function of the energy scale, which is encoded in the QCD beta function. The QCD beta function was calculated a long time ago in refs. \cite{Gross:1973id, Politzer:1973fx} and it was shown that it is an asymptotically free theory. This means that the coupling constant becomes small as the energy scale increases (in the UV) and becomes large when the energy scale becomes small (in the IR). This notion of UV or IR energy scale in QCD is decided in terms of the only relevant scale $\Lambda_{\rm QCD}$ generated by dimensional transmutation. Physically, $\Lambda_{\rm QCD}$ (around $200$ million electron volt (MeV)) corresponds to the energy scale where the QCD coupling constant becomes of order one, hence the theory becomes non-perturbative.

Thus perturbative field theory methods based on expansion in the small coupling constant are applicable only in the energy regime much larger than $\Lambda_{\rm QCD}$. Indeed a lot of efforts have been made to address aspects of QCD within such perturbative regime, {\it e.g.} in ref.~\cite{Rajagopal:2000wf} aspects of dense quark matter have been addressed in a perturbative framework, refs.~\cite{Kajantie:2002wa, Vuorinen:2003fs, DiRenzo:2006nh} have perturbatively computed thermodynamic pressure of hot QCD. There is a vast literature on similar aspects.

Although such perturbative approach has led us to remarkable insights about phases of QCD, this is only a small part of what we would want to understand. It is of vital theoretical interest to build up useful and efficient non-perturbative techniques to address strong coupling dynamics in general. QCD is one such strongly coupled theory which posses (at least) two remarkable features in the low energy regime: confinement and chiral symmetry breaking. Both these effects have been observed in experiments, however their dynamical origin is not well-understood. In a confining theory, the elementary degrees of freedom exist only in the form of  bound states. In QCD, we can only observe the quark bound states such as mesons or baryons (collectively known as the hadrons) in the low energy regime, but not free quarks. Chiral symmetry breaking, on the other hand,  occurs {\it via} the formation of a condensate consisting of fermions of opposite chiralities. The understanding of both these features remains an open problem till today.

There exist powerful non-perturbative techniques. Lattice QCD is one such example. In this framework, spacetime is represented by discrete lattice points and quantum fields are represented by degrees of freedom living on each lattice site. Such a discrete formulation of spacetime provides a natural cut-off for the energy scale given by the inverse lattice spacing. The QCD path integral, and therefore any observable, can in principle be computed by numerical simulations on a computer.

Although lattice QCD is a very useful computational tool, it may not shed any light on the conceptual understanding. Moreover, the lattice computations have technical limitations of their own. Specifically, this framework is not (yet) equipped well-enough to address non-equilibrium time dependent dynamics. These computations are also limited to vanishing chemical potential due to the so called ``sign problem". However various methods have been developed over the past decade to circumvent the sign problem; we refer to ref.~\cite{deForcrand:2010ys} for a recent review. Moreover, lattice QCD remains a very useful tool in studying thermal properties of QCD at vanishing chemical potential; for a recent review see {\it e.g.} ref.~\cite{DeTar:2009ef}. Some pedagogical introduction on lattice QCD can be found in {\it e.g.} refs.~\cite{creutz, MM, DD}.

The effort towards understanding the strong coupling dynamics of QCD is not purely theoretically motivated. In recent years, the Relativistic Heavy Ion Collision (RHIC) experiments at Brookhaven National Laboratory have probed the strong coupling regime of QCD and thus pose a challenge for a theoretical understanding of the data. The quark-gluon plasma (QGP) produced at RHIC is at the energy scale roughly same as the $\Lambda_{\rm QCD}$, thus it is expected that the dynamics is presumably strongly coupled. This intuition has further been supported by the fact that RHIC data is best described or fitted in the framework of ideal fluid dynamics\cite{Shuryak:2003xe, Shuryak:2004cy}.

Fluid dynamics\footnote{The standard pedagogical textbook for this is {\it e.g.} ref.~\cite{Landau:1987fl}.} can be viewed as an effective description of a system when the mean free path of the constituent microscopic degrees of freedom is much smaller than any characteristic length scale of the system. Thus, by definition it implies the presence of strong coupling in the microscopic description of the corresponding system. For the case of RHIC, some direct evidences of strong coupling further exist and can be measured in {\it e.g.}, the large elliptic flow, jet quenching {\it etc.} For a pedagogical introduction of heavy ion physics and discussions on related phenomena see {\it e.g.} ref.~\cite{Heinz:2004qz}. The tools of lattice QCD are not very helpful to understand the RHIC results since many of the observed phenomena are time dependent non-equilibrium dynamics.

Surprisingly string theory provides an unique toolkit to study strongly coupled aspects of QCD, at least in some qualitative sense. This is rooted on the idea of holography or gauge-gravity duality. At present, the holographic principle is a conjecture. The progress in understanding of black hole entropy and its dependence on its area rather than its volume led to the key idea\cite{'tHooft:1993gx, Susskind:1994vu, Bousso:2002ju}. The precise statement of holography is as follows: any theory of quantum gravity in a given background is entirely equivalent to a non-gravitational theory describing the degrees of freedom living at the boundary of the spacetime and {\it vice versa}. This is also referred as the gauge-gravity duality. Furthermore in the known examples of this duality the non-gravitational theory turns out to be a certain class of conventional quantum field theories. String theory provides us with a natural and well-defined framework where most of such examples of gauge-gravity duality can be understood and used. However, much of the effective and working knowledge of the gauge-gravity duality can be learnt with minimal involvement of string theory. We refer the reader to ref.~\cite{McGreevy:2009xe} for a recent review from this point of view. We will take a more conventional route and briefly review how this duality emerges in string theory.

In ref.~\cite{'t Hooft:1973jz} 't Hooft made the crucial step in recognizing the rank of the non-Abelian gauge group as an expansion parameter for the non-Abelian gauge theory. Namely, he proposed to consider an ${\rm SU}(N_c)$ Yang-Mills theory in the limit $N_c \to\infty$ keeping the 't Hooft coupling $\lambda = g_{\rm YM}^2 N_c$ constant. In this so called large $N_c$ limit many simplifications happen. It can be shown that the partition function of the large $N_c$ Yang-Mills theory is dominated by the planar diagrams only and the next corrections occur at the order of $1/N_c^2$. Viewed as an approximation scheme for QCD (where $N_c = 3$), this gives a result correct up to about $10\%$ error margin and thus not completely irrelevant. Furthermore, the $1/N_c$ expansion of the Yang-Mills partition function takes an identical form of the genus expansion of the string partition function, thus suggesting a possible connection between string theory and large $N_c$ Yang-Mills theory.

It was realized that string theory naturally contains various extended objects called ``branes". A certain kind of such ``branes" called the D-branes was originally realized in ref.~\cite{Polchinski:1995mt}; however now it is well-known that there are several kinds of branes in string theory. We refer to the book in ref.~\cite{Johnson:2003gi} for an extensive discussion on the properties of various kinds of branes. Here we will focus entirely on the D-branes. These extended objects are the places where an open string can end. Such an extended object with $p$ spatial dimensions is called a D$p$-brane. It was further realized that these D$p$-branes arise as solutions of supergravity which is obtained as the low energy limit of string theory. Now the crucial step was to realize that the low energy dynamics of $N_c$ coincident D$p$-branes can be equivalently described by the gravitational background of the D$p$-brane and an ${\rm SU}(N_c)$ Yang-Mills theory in $(p+1)$-dimensions living at the boundary of the D$p$-brane background.

String theory provides a large class of such examples. The existence of D$p$-branes in both Type IIA and Type IIB supergravity (obtained as the low energy limits of Type IIA and Type IIB string theory respectively) provides a rich arena to explore and realize numerous examples of gauge-gravity duality. The most popular example of this is the duality between ${\cal N} = 4$ Super Yang-Mills (SYM) theory in $(3+1)$-dimensions and the gravitational background of anti de-Sitter (AdS) space in $5$-dimensions.

However none of the gauge theories mentioned above resemble QCD in precise details. It is still an open problem to construct the precise gravitational dual to QCD. The idea of using, {\it e.g.} $\mathcal{N} = 4$ SYM to understand features of QCD is therefore completely based on the concept of universality. This idea was initially explored in refs.~\cite{Chamblin:1999tk, Chamblin:1999hg} to learn about the phase diagram of QCD using the dual gravitational background. See also refs.~\cite{Gubser:1998jb, Cvetic:1999ne, Cvetic:1999rb} for discussions on related physics. For an early review exploring this idea we refer to the Chapters 18 and 19 of ref.~\cite{Johnson:2003gi}. A more recent and less technical review can be found in ref.~\cite{Gubser:2009md}.

The notion of universality class is quite old in statistical mechanical systems. Phase transitions in physical systems are described by critical exponents which capture the behavior of physical quantities near the phase transition. Various physical systems have been observed to have the same critical exponents based on which many macroscopic systems can be grouped into one set of universality class. We refer the reader to ref.~\cite{Jean:2002} for a more detailed discussion of the physics of phase transitions and the notion of universality class. However, here we will consider a much relaxed notion of universality. We will merely concentrate on any statement (qualitative or quantitative) that holds universally for such large $N_c$ gauge theories.

That such universal statements exist is further supported by the following observation. At zero temperature, the physics of QCD differs considerably from the physics of $\mathcal{N} = 4$ SYM (or any such large $N_c$ gauge theory for which we know the gravitational dual). Firstly, the degrees of freedom do not match. Moreover QCD is confining, with a discrete spectrum and running coupling constant whereas $\mathcal{N} = 4$ SYM does not share any of these features. However at finite temperature both these descriptions consist of deconfined strongly coupled plasma of gluons and various matter fields. Furthermore as it has been emphasized before, the best model describing the dynamics of quark gluon plasma (QGP) at RHIC is fluid dynamics which is an effective long-wavelength description and does not necessarily rely on the microscopic details.

Interestingly RHIC provides us with a remarkable quantitative evidence of such possible universality. Recent investigations, {\it e.g.} ref.~\cite{Song:2007ux} seem to suggest that the ratio of viscosity to entropy density for the QGP plasma $\eta / s\sim \hbar/ ( 4\pi k_B)$, where $\hbar$ is the Planck constant and $k_B$ is the Boltzmann constant. Although presently we do not have any QCD calculation for this ratio, using AdS/CFT refs.~\cite{Policastro:2001yc, Policastro:2002se, Kovtun:2003wp, Kovtun:2004de} calculated the ratio to be $\eta / s = \hbar / ( 4\pi k_B)$. It was further argued in refs.~\cite{Kovtun:2004de, Buchel:2003tz, Buchel:2004qq} that the ratio $\eta / s = \hbar / ( 4\pi k_B)$ is universal for any gauge theory with a classical gravity dual. This remarkable qualitative and quantitative (in terms of orders of magnitude) resemblance to the observed data at RHIC is suggestive of an universality we alluded to.

Let us briefly comment on a further interesting observation. Generically, the finite temperature strong coupling physics seems to have universal properties which can be obtained using the methods of gauge-gravity duality and ideal fluid dynamics. Thus a natural question arises: whether there is a connection between the dual gravitational and the fluid dynamic descriptions. During the past few years this possible connection has been analyzed in detail and it has been concluded that there is a precise map between a class of inhomogeneous, dynamical black hole solutions in asymptotically AdS spacetime and arbitrary fluid flows in the strongly coupled field theory living at the boundary. This is now known as the fluid-gravity correspondence\cite{Bhattacharyya:2008jc}. We refer to the ref.~\cite{Rangamani:2009xk} for an extensive review and further references on this topic.

Thus we can boldly step forward in using techniques of gauge-gravity duality to further understand the strong coupling dynamics of QCD in some universal sense. However we will not attempt to compare any of our results with any experimental data, at most we will merely comment or compare with the results and expectations from QCD whereever necessary. Our approach should thus be viewed as an attempt to use the tool of the AdS/CFT correspondence (or gauge-gravity duality) to learn about strong coupling dynamics of large $N_c$ non-abelian gauge theories in as much generality as possible.

\section*{1.2 \hspace{2pt} Gauge-gravity duality from string theory}
\addcontentsline{toc}{section}{1.2 \hspace{0.15cm} Gauge-gravity duality from string theory}

String theory is a vast subject and we will just refer the standard pedagogical textbooks~\cite{Green:1987sp, Green:1987mn, Polchinski:1998rq, Polchinski:1998rq2, Johnson:2003gi} for interested readers. Here we will briefly review some basic facts about string theory. The basic degrees of freedom of string theory are one dimensional objects called strings. The fundamental parameter of the theory is the characteristic length scale $\ell_s = \sqrt{\alpha'}$, which sets the string tension $T_s = 1/ (2\pi\alpha')$. The interaction strength between the strings is given by the dimensionless string coupling constant denoted by $g_s$. Strings can be both open and closed. Open strings are one-dimensional line segments with a boundary and two end points. On the other hand, closed strings are loops with no boundary or end points.

Based on the presence of supersymmetry in the underlying action, string theory can be divided into two broad categories: the bosonic string theory and the superstring theory. There are various kinds of superstring theories, which go under the technical names of Type I, Type IIA, Type IIB and heterotic string theories {\it etc}. These apparently different theories are actually related by certain duality transformations and are connected to what is known as M-theory.

Although presently we lack a complete understanding of string theory (or M-theory), the low energy limit of this theory is rather well-studied. In the low energy limit, string theory/M-theory reduces to ten/eleven dimensional supergravity. It is in the framework of supergravity where many aspects of the gauge-gravity duality have been discovered and used. Note however that many aspects ({\it e.g.} later when we consider branes and fundamental strings with gauge fields on them) of the duality depends on the full string theory and not just its low energy limit. Nonetheless, all our computations in the later chapters are performed within the supergravity approximation. In this thesis we will discuss two different (the Type IIB and the Type IIA) supergravity frameworks in particular.

\subsection*{1.2.1 \hspace{2pt} The Type IIB theory: low energy dynamics on D3-branes}
\addcontentsline{toc}{subsection}{1.2.1 \hspace{0.15cm} The Type IIB theory: low energy dynamics on D3-branes}

In Type IIB theory, we can have D$1$, D$3$, D$5$ and D$7$ (stable) branes. Here we will restrict ourselves to the low energy limit of D3-branes only.

To lead to the conjecture, let us start with a stack of $N_c$ parallel D3-branes. In this set-up there are two kinds of excitations: the open strings that start and end on the D3-branes and the closed strings that propagate through the ten dimensional bulk. At low energies ({\it i.e.} energies much smaller than $\ell_s^{-1}$), we can integrate out the massive modes to obtain an effective action given by $S = S_{\rm bulk} + S_{\rm brane} + S_{\rm int}$. The bulk action $S_{\rm bulk}$ is the action for Type IIB supergravity up to corrections in $\alpha' = \sqrt{\ell_s}$ (which appear as higher curvature corrections to the Type IIB supergravity action). On the other hand, the massless spectrum of open strings starting and ending on the D3-branes constitute the vector supermultiplet of $\mathcal{N} = 4$ SYM in $(3+1)$-dimensions with ${\rm U}(N_c)$ gauge group. The presence of the gauge group can be understood as follows. The massless open string spectrum ending on a stack of $N_c$ parallel D3-branes is labeled by an index known as the Chan-Paton factor. In case of oriented strings, it can be shown that these indices transform in the adjoint representation of an ${\rm U}(N_c)$ gauge group. This gives the brane action $S_{\rm brane}$. There are also interactions between the bulk modes and the brane modes which is denoted by $S_{\rm int}$. In the limit $\alpha'\to 0$, these interactions are suppressed reflecting that gravity becomes free at large distance scales. Thus in the low energy limit, we get a complete decoupling between the open and the closed string modes. Let us now discuss this decoupling in some details.

\subsubsection*{1.2.1a \hspace{2pt} The decoupling limit and the correspondence}
\addcontentsline{toc}{subsubsection}{1.2.1a \hspace{0.15cm} The decoupling limit and the correspondence}

In supergravity D-branes are massive charged objects sourcing the supergravity fields. To illustrate the decoupling limit, let us start from the extremal D3-brane background in flat space obtained as a solution of Type IIB supergravity equations of motion. Following the notation of ref.~\cite{Johnson:2003gi} this background is given by
\begin{eqnarray}\label{eqt: d3full}
&& ds^2 = H_3^{-1/2} \eta_{\mu\nu} dx^\mu dx^\nu + H_3^{1/2} dx^i dx^i\ , \nonumber\\
&& C_{(4)} = H_3^{-1} g_s^{-1} dx^0 \wedge \ldots \wedge dx^3\ , \nonumber\\
&& e^{\Phi} = g_s\ , \qquad H_3 = 1 + \frac{R^4}{r^4}\ , \qquad R^4 = 4\pi g_s N_c \alpha'^2\ , 
\end{eqnarray}
where $\mu = 0, \ldots 3$ and $i = 4, \ldots 9$ and $\eta_{\mu\nu}$ is the flat Lorentz metric; $r$ is the radial coordinate and $H_3$ is a function harmonic in the six directions transverse to the D3-branes. $C_4$ is the four-form RR potential sourced by the D3-brane and $N_c$ is an integer measuring the flux quanta of the RR field strength. This supergravity background has a horizon at $r = 0$ and the boundary is located at $r = \infty$. From the point of view of a boundary observer, the low energy Type IIB string modes near the horizon will be infinitely redshifted by the factor $H_3^{-1/4}$ and thus these low energy excitations near the horizon should describe the low energy massless degrees of freedom seen by the asymptotic observer. On the other hand, the low energy massless excitations of Type IIB string theory is described by the Type IIB supergravity as seen by the boundary observer.

Now any bulk excitation in the full background interacts with the near-horizon excitations, which is captured by the absorption cross-section of graviton scattered by the near-horizon regime, $\sigma \sim \omega^3 R^8$ as obtained in refs.~\cite{Gubser:1997yh,Klebanov:1997kc}, where $\omega$ denotes the energy of the graviton. However, in the low energy limit (denoted by $\omega \to 0$), $\sigma \to 0$ since the bulk excitations have a wavelength much larger than the gravitational size of the brane. Thus it is appropriate to conclude that bulk excitations decouple from the near-horizon excitations in the low energy limit.

Now we will precisely define what we mean by the ``near-horizon" limit of the background in (\ref{eqt: d3full}). We achieve the near-horizon limit of (\ref{eqt: d3full}) by sending $\alpha' \to 0$ and keeping $u = r/\alpha'$ fixed. In leading order in $\alpha'$ we get the following background\cite{Maldacena:1997re}
\begin{eqnarray}\label{eqt: d3ads}
&& ds^2 = \frac{u^2}{R^2} \eta_{\mu\nu} dx^{\mu} dx^{\nu} + R^2 \frac{du^2}{u^2} + R^2 d\Omega_5^2\ , \nonumber\\
&& C_{4} = \frac{R^4}{g_s u^4} dx^0 \wedge \ldots \wedge dx^3\ , \nonumber\\
&& e^\Phi = g_s\ .
\end{eqnarray}
The metric in (\ref{eqt: d3ads}) is the metric of AdS$_5\times S^5$ background. Thus we learn that the near-horizon geometry of D3-branes is identical to an anti-de Sitter space.

Therefore we arrive at two physical descriptions for the low energy physics of a stack of $N_c$ D3-branes: the $\mathcal{N} = 4$ SYM theory and the Type IIB string theory in AdS$_5\times S^5$ background. The essence of Maldacena's AdS/CFT correspondence is to conjecture that these two descriptions are entirely equivalent. Let us now formally state the correspondence.

{\it Type IIB string theory on AdS$_5\times S^5$ background is equivalent to $\mathcal{N} = 4$ SYM theory in $(3+1)$-dimensions.}

One hint in favor of this correspondence comes from the non-trivial matching of the global symmetry group in the gauge theory with the isometry group of the geometry. Let us therefore discuss this in some more details. We will not explicitly prove or demonstrate any of the following assertions, for our purposes it would suffice to just review them.

$\mathcal{N} =4$ SYM is a maximally supersymmetric gauge theory with ${\rm SU}(N_c)$ gauge group. The degrees of freedom are contained in the gauge multiplet of this theory which consists of gauge fields, four left-handed Weyl fermions and six real scalar fields. All these fields transform in the adjoint representation of the ${\rm SU}(N_c)$ gauge group. This theory has only one gauge coupling constant usually denoted by $g_{\rm YM}$. A remarkable feature of this theory is the exact conformal invariance implying that the coupling constant $g_{\rm YM}$ does not run at all. The global symmetry group of this theory is the conformal group in $(3+1)$-dimensions, which is given by ${\rm SO}(4,2)$. $\mathcal{N} =4$ SYM also has sixteen real supercharges which can be rotated into each other by an ${\rm SU}(4)_{\rm R}$ group called the R-symmetry group. Thus the global symmetry group is given by ${\rm SO}(4,2)\times {\rm SU}(4)_{\rm R}$. For future reference, let us also note that ${\rm SU}(4)$ is locally isomorphic to ${\rm SO}(6)$, {\it i.e.} ${\rm SU}(4)\simeq {\rm SO}(6)$.

Now we focus on the geometric side. Anti-de Sitter spaces are defined as spaces with constant negative curvature. These spaces are obtained as a solution to Einstein's equations with a negative cosmological constant. The $(4+1)$-dimensional anti-de Sitter space can be represented as the following hyperboloid in $\mathbb{R}^6$
\begin{eqnarray}\label{eqt: adshyper}
X_0^2 + X_5^2 - \sum_{i=1}^4 X_i^2 = R^2 \ , 
\end{eqnarray}
where $R$ is the radius of curvature of the hyperboloid. It is evident from this description that AdS$_5$ space has ${\rm SO}(4,2)$ isometry acting as the rotation group in $(4+2)$-dimensions. A simple way to parametrize the solution of eqn. (\ref{eqt: adshyper}) is to set
\begin{eqnarray}
X_0 &=& \frac{R^2}{2 u} \left(1 + \frac{u^2}{R^4} \left(R^2 + \vec{x}^2 - t^2 \right)\right)\ , \quad X_5 =  \frac{u t}{R}\ , \nonumber\\
X^4 &=& \frac{R^2}{2 u} \left(1 - \frac{u^2}{R^4} \left(R^2 - \vec{x}^2 + t^2 \right)\right)\ , \quad X^i =  \frac{u x^i}{R}\ , \quad i = 1,2,3
\end{eqnarray}
and $u>0$. These coordinates are called the Poincar\'{e} patch which cover only half of the hyperboloid. There exists also a global patch which cover the entire hyperboloid, but we refrain from discussing it since it will not be relevant for out purposes. For interested readers, we refer to the review in ref.~\cite{Aharony:1999ti}. In the Poincar\'{e} patch, the induced metric on the AdS$_5$ space is given by
\begin{eqnarray}\label{eqt: ads}
ds^2 = \frac{u^2}{R^2} \left( -dt ^2 + d\vec{x}^2 \right) + R^2 \frac{du^2}{u^2} \ ,
\end{eqnarray}
which is what we have met in eqn.~(\ref{eqt: d3ads}). In eqn.~(\ref{eqt: d3ads}) we obtained a background in a product form AdS$_5\times S^5$. The isometry group of $S^5$ is ${\rm SO}(6)$ corresponding to the rotation group. Thus the geometric background has an isometry group ${\rm SO}(4,2)\times {\rm SO}(6)$, which is exactly identical to the global symmetry group of the $\mathcal{N} = 4$ SYM theory. The relation between the gauge coupling and the string coupling is simply given by $g_s = g_{\rm YM}^2$. For further details we refer the review~\cite{Aharony:1999ti}.

We have, so far, stated the strong form of the correspondence. However, it is not known how to quantize string theory on the curved AdS$_5$ background. Thus the strong form of the conjecture is hard to confirm. A weaker form of the conjecture can be based on the 't Hooft limit with a fixed 't Hooft coupling defined by $\lambda = g_{\rm YM}^2 N_c$, while taking the limit $N_c\to \infty$. In this large $N_c$ expansion the planar $\mathcal{N} = 4$ SYM is conjectured to be equivalent to Type IIB string theory perturbatively expanded in powers of the string coupling $g_s = \lambda / N_c$. A further weaker form of the conjecture is in the large 't Hooft coupling limit, {\it i.e.} $\lambda\gg 1$. Using the definition of $R$ from eqn.~(\ref{eqt: d3full}), we can conclude that in this limit $R^4/ \alpha'^2 \gg1$, meaning that the curvature scale is much larger than the string scale, where classical supergravity is a valid description. Thus the weakest form of the conjecture states that classical Type IIB supergravity is dual to the $\mathcal{N} = 4$ SYM in the strong 't Hooft coupling limit. In this thesis we will use the weakest form of the conjecture and use it to learn strong coupling aspects in a purely classical (super) gravity set-up.

\subsubsection*{1.2.1b \hspace{2pt} The dictionary of the correspondence}
\addcontentsline{toc}{subsubsection}{1.2.1b \hspace{0.15cm} The dictionary of the correspondence}

Let us now review the precise map between the gauge invariant operators in $\mathcal{N} = 4$ SYM and the supergravity fields. To demonstrate this let us consider a simple massive scalar field $\phi$ propagating in the background AdS$_5$ space. The equation of motion for the scalar is given by the massive Klein-Gordon equation 
\begin{equation}
\nabla^2 \phi - m^2\phi = 0\ ,
\end{equation}
which has two independent asymptotic solutions given by
\begin{eqnarray}
\phi (u) \sim u^{\Delta - 4} \phi_0 + u^{-\Delta} \langle \mathcal{O}_\phi\rangle \ , \quad \Delta = 2 + \sqrt{4 + m^2 R^2}\ ,
\end{eqnarray}
where $\Delta$ is the conformal dimension of the field. The first term with coefficient $\phi_0$ is the non-normalizable solution and the second term with the coefficient $\langle \mathcal{O}_\phi\rangle$ is the normalizable solution. It was suggested in ref.~\cite{Witten:1998qj} that the non-normalizable mode should be identified with the source and the normalizable mode should be identified with the expectation value corresponding to the source. Furthermore it was suggested in refs.~\cite{Witten:1998qj, Gubser:1998bc} that the precise form of the dictionary is given by equating the generating functional of the conformal field theory with the partition function of classical supergravity 
\begin{eqnarray}\label{eqt: weakcorr}
\left<{\rm exp}\left(\int d^4 x \phi_0 \mathcal{O}_\phi\right)\right>_{\rm CFT} = \mathcal{Z}_{\rm sugra} \left[\phi_0\right] = {\rm exp} \left(i S_{\rm sugra} \left[\phi_0\right] \right)\ ,
\end{eqnarray}
where $\phi_0$ is the boundary value of the bulk scalar field. The proposal of refs.~\cite{Witten:1998qj, Gubser:1998bc} holds true for any bulk supergravity field irrespective of whether it is a scalar, vector or tensor. This proposal can therefore be viewed as the operational definition of the AdS/CFT correspondence in its weak form. Furthermore, it is noteworthy that this proposal can be generally extended to any AdS$_{d+1}$/CFT$_d$ correspondence.

However the precise correspondence in (\ref{eqt: weakcorr}) does not give a prescription of how to determine the dual bulk field corresponding to a given operator in the gauge theory and {\it vice versa}. Usually this identification is uniquely fixed based on the scaling dimension and the symmetry property of the operator or the dual bulk field. Another caveat with this precise statement is that on the face of it both the field theory generating functional and the on-shell supergravity action are divergent quantities. Thus we need to introduce a consistent regularization and renormalization scheme. The divergence in the supergravity action comes from the infinite volume of AdS space. Thus while evaluating the supergravity action integral we introduce a cut-off at some $u = {\rm const}$ hypersurface and add appropriate counterterms on the hypersurface to cancel the divergences. This is formally known as the ``holographic renormalization" scheme and we refer the review~\cite{Skenderis:2002wp} for further details on this.

It is interesting to note that the regularization scheme we just described is very similar to the familiar regularization scheme in field theory if we identify the radial direction $u$ of the AdS background to be the renormalization scale of the dual field theory. This identification is indeed true and can be understood simply as follows. If we scale $x_{\mu}\to \alpha x_{\mu}$, where $\alpha$ is some real positive number; then for the AdS-metric in (\ref{eqt: ads}) to be invariant under the scaling transformation we must require $u\to \alpha^{-1} u$, which is how the energy must transform. In the AdS background, $u\to 0$ is the region deep inside and $u\to\infty$ corresponds to the region near the boundary. Motivated from the above observation we will sometimes refer to these as ``the IR" and ``the UV" respectively.

\subsubsection*{1.2.1c \hspace{2pt} The correspondence at finite temperature}
\addcontentsline{toc}{subsubsection}{1.2.1c \hspace{0.15cm} The correspondence at finite temperature}

Let us review the key aspects of the correspondence at finite temperature. As we will see later, for the most of the cases we will be interested in finite temperature physics. The notion of a temperature can be introduced very easily in any quantum field theory. This is simply achieved by Euclideanizing the time-coordinate by an Wick rotation $t\to - i t_{\rm E}$ and compactifying this Euclidean direction on a circle with period $\beta = 1/T$, where $T$ is the corresponding temperature. The frequency modes corresponding to this compact direction naturally becomes discrete, which are known as the ``Matsubara frequencies". This is known as the ``imaginary time formalism". It is to be noted that in this formalism, once temperature is introduced we loose the notion of time and therefore we cannot describe any time-dependent phenomenon.

The dual gravity background of finite temperature $\mathcal{N} = 4$ SYM is given by the AdS$_5$-Schwarzschild$\times S^5$ background, where the Hawking temperature and the entropy of the black hole is identified with the temperature and entropy of the field theory\cite{Gubser:1996de, Witten:1998qj}. In view of the precise map in (\ref{eqt: weakcorr}), the generating functional of the quantum field theory becomes the thermal partition function. Thus we get that the free energy of the field theory is given by the Euclidean on-shell supergravity action
\begin{eqnarray}
F = T S_{\rm sugra} \left[\phi_0\right]\ .
\end{eqnarray}
Therefore the thermodynamics of $\mathcal{N} = 4$ SYM can be calculated from the Euclidean gravity action. Finite temperature thermodynamic quantities such as the free energy or the entropy density, thus computed, scales as $N_c^2$ as observed in ref.~\cite{Gubser:1996de}. This factor of $N_c^2$ is understood as counting the degrees of freedom for the ${\rm SU}(N_c)$ gauge theory. Since all the degrees of freedom contribute to thermodynamic observables, we define this as the ``deconfined" phase of the $\mathcal{N} = 4$ SYM theory (as opposed to the ``confined" phase where degrees of freedom exist only in the form of bound states and thus provide a factor of $N_c^0$ in front of thermodynamic observables). All our analysis in the subsequent chapters will mostly be restricted to this ``deconfined" phase only.

As clear from the formalism itself, this is not just limited to the finite temperature $\mathcal{N} = 4$ SYM.

\subsection*{1.2.2 \hspace{2pt} The Type IIB theory: introducing flavours}
\addcontentsline{toc}{subsection}{1.2.2 \hspace{0.15cm} The Type IIB set-up: introducing flavours}

As we have already reviewed, $\mathcal{N} = 4$ SYM theory contains only adjoint matter sector whereas QCD contains both adjoint (the gluons) and fundamental matters (the quarks). We will focus in particular to understand the dynamics of these fundamental flavours and thus we need to introduce the flavour fields in the gravity set-up.

So far we have focussed on the low energy dynamics of D3-branes only. In this case, open strings that start on the D3-branes must end on the D3-branes itself. These two end-points of the open string correspond to point charge in the fundamental and anti-fundamental representation of the ${\rm SU}(N_c)$ gauge group. Together they transform under the adjoint representation. It is therefore clear that in order to introduce fundamental flavour degrees of freedom we need to introduce open strings with only one end on the D3-brane.

The simplest way in which this is achieved is by considering an additional stack of $N_f$ D7-branes\cite{Karch:2002sh}. The relative orientation of the $N_c$ D3-branes and $N_f$ D7-branes is given by the following array.
\begin{table}[h]
\begin{center}
\begin{tabular}{|c|c|c|c|c|c|c|c|c|c|c|}
\hline
 &0&1&2&3&4&5&6&7&8&9\\\hline
 D3&--&--&--&--&$\bullet$&$\bullet$&$\bullet$&$\bullet$&$\bullet$&$\bullet$\\\hline
 D7&--&--&--&--&--&--&--&--&$\bullet$&$\bullet$\\\hline
 
\end{tabular}
\end{center}
\caption{ The brane construction including fundamental flavours. }
\label{default}
\end{table}
Here $``-"$ represents that the brane is extended along the corresponding direction whereas $``\bullet "$ represents that the brane is point-like in the corresponding direction. It is to be noted that the D3-branes' directions are aligned with four of the D7-branes' directions.

The low energy dynamics of the above configuration is now given by open strings that start and end on the D3-branes, open strings that start and end on the D7-branes and finally open strings that start on the D3/D7-branes and end on the D7/D3-branes. The low energy dynamics of the open strings that start and end on the D3-branes give the $\mathcal{N} = 4$ SYM as before. The D7-branes are extended in the non-compact direction transverse to the D3-branes and thus have infinite volume. Therefore the strings that start and end on the D7-branes yield a global symmetry as opposed to a gauge symmetry since the gauge coupling vanishes.

The strings that are stretched between the D3 and the D7-branes give rise to the so called $\mathcal{N} = 2$ hypermultiplet. The $\mathcal{N} = 2$ hypermultiplet consists of two complex scalars and two Weyl spinors of opposite chirality. One scalar and Weyl fermion transform under the fundamental of ${\rm SU}(N_c)$ and the other pair transform under the conjugate representation. In an abuse of language, we will call that the hypermultiplet transforms under the fundamental representation. We will not discuss the explicit Lagrangian of this theory in details.

A few words about the R-symmetry group are in order. As seen from the explicit orientation of the D7-branes the full ${\rm SO}(6)_{\rm R}$ symmetry is broken down to an ${\rm SO}(4)_{\rm R}$  symmetry corresponding to the group rotation in the $\{4,5,6,7\}$-directions and an ${\rm SO}(2)_{\rm R}$ symmetry corresponding to the group rotation in the $\{8,9\}$-directions.

This ${\rm SO}(2)_{\rm R}\simeq {\rm U}(1)_{\rm R}$ is identified as the chiral symmetry under which the two Weyl fermions are oppositely charged. Note that the D3 and the D7-branes can be separated in the $\{8,9\}$-plane in which case the strings stretched between these branes will have a finite length and finite energy. This corresponds to giving mass to the hypermultiplet fields, which explicitly breaks the ${\rm U}(1)_{\rm R}$ symmetry. This is analogous to the chiral symmetry breaking in QCD when the quarks become massive. Note that in principle we can give different mass to each flavour sector by separating each of the $N_f$ different D7-branes from the D3-branes by different amounts in the $\{8,9\}$-plane. However, we will content ourselves with the simple case when all the $N_f$ D7-branes are coincident.

In what follows we will restrict ourselves to the so called ``probe limit" which means that we will take the large $N_c$ limit keeping $N_f$ fixed such that $N_c\gg N_f$. In this limit, the quantum corrections leading to the ${\rm U}(1)_{\rm R}$ anomaly are suppressed\cite{Witten:1979vv}, thus the ${\rm U}(1)_{\rm R}$ symmetry becomes an honest global symmetry.

This probe limit has another interesting consequence. In the case when the mass of the hypermultiplet is zero and there is no background temperature, the theory is still conformal. It can be shown that introducing $N_f$ flavours modifies the beta function of this theory by adding a positive contribution at the order $N_f /N_c$ and is thus negligible in the probe limit\cite{Kirsch:2005uy}. However in the parametric regime $N_f\sim N_c$, the coupling of the theory increases with increasing energy and finally acquires a Landau pole where the gauge coupling constant becomes infinitely strong. This behaviour is completely different from QCD and requires an UV completion.

The gravity dual picture is rather simple in this probe approximation. We will replace the near-horizon geometry of the D3-branes by the AdS$_5\times S^5$ background (or with the AdS-Schwarszchild background in the finite temperature case) and treat the D7-branes as probes in this gravity background which only ``sense" the background geometry but do not backreact on it. This is similar to the ``quenched approximation" used in the Lattice QCD studies to analyze the effect of a test quark in the background of gluons. In this probe limit, the action of the probe brane is given by the Dirac-Born-Infeld (DBI) action. This action will be our main tool and we will discuss some general aspects of the DBI action in a later section.

\subsection*{1.2.3 \hspace{2pt} The Type IIA theory: low energy dynamics on D4-branes}
\addcontentsline{toc}{subsection}{1.2.3 \hspace{0.15cm} The Type IIA theory: low energy dynamics on D4-branes}

Immediately after Maldacena's AdS/CFT correspondence, the low energy dynamics of a general D$p$-brane was investigated in ref.~\cite{Itzhaki:1998dd} to realize more examples of gauge-gravity duality. Here we will not review the details of Type IIA string theory or Type IIA supergravity. However, we will just recall that in Type IIA supergravity D$p$-branes exist only with even values for $p$. In particular we will focus on the low energy dynamics of D4-branes.

As clear from the discussions in the previous sections, the gauge theory representing the low energy dynamics of a stack of $N_c$ D4-branes is $(4+1)$-dimensional. It was illustrated in ref.~\cite{Itzhaki:1998dd} that the decoupling limit is now taken by sending the string scale $\alpha'\to 0$ and the asymptotic string coupling $g_s \to\infty$ while keeping the $(4+1)$-dimensional gauge coupling $g_5 = (2\pi)^2 g_s \sqrt{\alpha'}$ fixed. The $(4+1)$-dimensional ${\rm SU}(N_c)$ gauge theory in this case is not a conformal theory. Moreover this is not a renormalizable theory and thus requires an UV completion at high energies. This UV completion is given by the $(5+1)$-dimensional $\mathcal{N} = (2,0)$ superconformal theory defined on the worldvolume of $M5$-branes compactified on a circle\footnote{M-branes are objects like D-branes in M-theory. For an extensive list of the examples of gauge-gravity duality arising from various D-branes and M-branes we refer to the classic review~\cite{Aharony:1999ti}. }. However, we will not elaborate on this issue since it will not be relevant for our subsequent discussions and refer the interested reader to ref.~\cite{Itzhaki:1998dd} for further discussions on this.

\subsubsection*{1.2.3a \hspace{2pt} The decoupling limit}
\addcontentsline{toc}{subsubsection}{1.2.3a \hspace{0.15cm} The decoupling limit}

As we have motivated ourselves to learn aspects of QCD (or at least $(3+1)$-dimensional gauge theory) using the gauge-gravity duality, we will restrict ourselves to the construction described by Witten in ref.~\cite{Witten:1998zw}. Let us discuss this set-up in some detail. Witten proposed to consider the near-horizon limit of a stack of $N_c$ D4-branes compactified on a spatial circle (an $S^1$) with a spin structure on the $S^1$ that breaks supersymmetry (achieved by imposing an anti-periodic boundary condition for fermions in the adjoint of the ${\rm SU}(N_c)$ gauge theory). The near-horizon gravity background is given by
\begin{eqnarray}\label{eqt: lowTd4}
&& ds^2 = \left(\frac{u}{R}\right)^{3/2}\left(-dt^2+dx_idx^i+f(u)(dx^4)^2\right)+\left(\frac{u}{R}\right)^{-3/2}\left(\frac{du^2}{f(u)}+u^2d\Omega_4^2\right)\ ,   \nonumber\\
&&   e^{\Phi}=g_s\left(\frac{u}{R}\right)^{3/4}\ , \quad F_{(4)}=\frac{2\pi N_c}{V_4}\epsilon_4\ ,\quad f(u)=1-\left(\frac{U_{\rm KK}}{u}\right)^3\ ,
\end{eqnarray}
where $t$ and $x^i, i = 1,2,3$ are the four gauge theory directions and $x^4$ is the compact $S^1$. The radial variable is denoted by $u$ and $U_{\rm KK}$ is a constant. The ${\rm SO}(5)$ invariant line element is denoted by $d\Omega_4^2$, $\epsilon_{(4)}$ denotes the corresponding volume form and $V_4$ is the volume of an unit $4$-sphere. Also, $\Phi$ is the
dilaton and $F_{(4)}$ is the RR four-form field strength. Since $x^4$ is a compact direction, to avoid a
conical singularity in the $\{x^4,u\}$ plane one should make periodic
identification:
\begin{equation}
\delta x^4=\frac{4\pi}{3}\left(\frac{R^3}{U_{\rm KK}}\right)^{1/2}=2\pi R_4\ .
\end{equation}
This endows the background with a smooth cigar geometry in the $\{x^4,u\}$ plane. When $U_{\rm KK} = 0$, there is a naked singularity at $u = 0$ with no restriction on the period of the $x^4$ direction.

The radius of curvature of the background is given by
\begin{equation}
R^3=\pi g_sN_c \, \ell_s^3\ .
\end{equation}
The gauge coupling of the five-dimensional field theory is given by $g_5 = (2\pi)^2 g_s \sqrt{\alpha'}$ as alluded to above. The corresponding five-dimensional 't Hooft coupling is given by
\begin{eqnarray}
\lambda_5 = g_5^2\, N_c\ .
\end{eqnarray}
Since the five-dimensional gauge coupling and the 't Hooft coupling are dimensionful quantities, we find a power law running of the dimensionless effective coupling given by\cite{Itzhaki:1998dd} $g_{\rm eff}^2 = g_5 ^2 \, N_c u / (2\pi\alpha')$. There is a tower of Kaluza-Klein (KK) modes arising from the compactification along the $x^4$ direction, which sets the typical glueball mass scale as obtained by studying excitations around the background in (\ref{eqt: lowTd4}): $M_{\rm KK} = 1/R_4$. Below this scale, the effective low energy gauge coupling is given by $g_4^2 = g_5^2 / \left(2\pi R_4\right)$ which defines a $(3+1)$-dimensional 't Hooft coupling $\lambda_4 = g_4^2 \, N_c$.

Let us now comment on the regime of validity of supergravity approximation. In order for this to be true, we impose that the spacetime curvature must be much smaller than the string length scale. The maximum value of the Ricci scalar of the background in (\ref{eqt: lowTd4}) goes as $\left(U_{\rm KK} R^3\right)^{-1/2}$, hence we impose the condition
\begin{equation}
\frac{\left(U_{\rm KK} R^3\right)^{1/2}}{\ell_s^2} \gg 1 \implies \quad g_4^2 \, N_c \gg 1\quad \implies \frac{\lambda_5}{R_4} \gg 1\ ,
\end{equation}
which corresponds to the large $(3+1)$-dimensional  't Hooft coupling limit.

Note that the dilaton is not constant in this background and string loop effects are not suppressed for arbitrary large values of $u$. However, the critical value of $u$ at which the dilaton is order one can be obtained to be
\begin{eqnarray}
u_{\rm cutoff} \sim \frac{N_c^{4/3} \ell_s}{\lambda_5 R_4}\ ,
\end{eqnarray}
which does not play any important role in the large $N_c$ limit (with fixed $\lambda_5$).

It was shown in ref.~\cite{Witten:1998zw} that the theory develops a mass gap. In the regime $\lambda_5\ll R_4$, the scale of the mass gap is exponentially suppressed compared to $1/R_4$. Therefore the gauge theory is effectively $(3+1)$-dimensional below the KK-scale set by $1/R_4$. However in the strongly coupled regime $\lambda_5\gg R_4$, the dual gauge theory develops a mass gap of the order $1/R_4$ and therefore cannot be separated from the KK scale. Thus the gravity calculation (which is valid at the strong coupling regime) corresponds to a $(4+1)$-dimensional gauge theory which is non-renormalizable. The presence of the cut-off in the radial coordinate is a reflection of this fact. The scheme of holographic renormalization in this set-up is therefore somewhat subtle. The systematics of the holographic renormalization of such non-conformal systems (including the example of the near-horizon D4-brane geometry) has been worked out recently in ref.~\cite{Kanitscheider:2008kd}.

A few words about the dual gauge theory are in order. Let us briefly mention the various degrees of freedom that constitute the low energy description of the D4-branes below the KK-scale. The massless degrees of freedom that arise from the open strings that start and end on the D4-branes are fermions, gauge fields and six real scalars transforming in the adjoint representation of the ${\rm SU}(N_c)$ gauge group. The component of the gauge field along the compact direction $x^4$ also appears to be an adjoint scalar as seen from the effective $(3+1)$-dimensional perspective. Since there is no supersymmetry, these scalars obtain mass via one-loop corrections and thus decouple from the low energy physics.

The dictionary of this correspondence works in a similar manner as we discussed earlier. Any bulk supergravity field will have two types of modes near the boundary (which is now placed at some cut-off radial value of $u$): the non-normalizable and the normalizable, which are identified with the source and the vev respectively. The identification of a bulk field with an appropriate boundary gauge theory operator is again based on its symmetry properties and engineering dimension.

\subsubsection*{1.2.3b \hspace{2pt} The finite temperature background}
\addcontentsline{toc}{subsubsection}{1.2.3b \hspace{0.15cm} The finite temperature background}

We can discuss finite temperature aspect of the correspondence by Euclideanizing the time coordinate $t\to - i t_{\rm E}$ in the gravity background (\ref{eqt: lowTd4}) and compactifying along $t_{\rm E}$. Now we have two directions which are compact. In this
case, the $x^4$ circle smoothly shrinks away at $u=U_{\rm KK}$ but the $t_{\rm E}$ circle is
fixed. One can also construct a finite temperature version by
interchanging the role of the $t$ and $x^4$ circles so that now time
circle shrinks away at some value $u=U_T$ but the $x^4$ circle remains
fixed. It turns out that these are the only known Euclidean continuations
of the background in equation~(\ref{eqt: lowTd4}) with the right
asymptote. This second solution is given by (in Euclidean signature)
\begin{eqnarray}\label{eqt: highTd4}
&& ds^2=\left(\frac{u}{R}\right)^{3/2}\left(dx_idx^i + f(u)d t_{\rm E}^2+\left(dx^4\right)^2\right)+\left(\frac{u}{R}\right)^{-3/2}\left(\frac{du^2}{f(u)}+u^2d\Omega_4^2\right)\ ,\nonumber\\
&& t_{\rm E} = t_{\rm E} +\frac{4\pi R^{3/2}}{3U_T^{1/2}}\ ,\quad T=\frac{1}{\beta}=\left(\frac{4\pi R^{3/2}}{3U_T^{1/2}}\right)^{-1}\ ,\quad f(u)=1-\left(\frac{U_T}{u}\right)^3\ ,
\end{eqnarray}
where $U_T$ is a new constant and $T$ is the temperature.

Therefore at finite temperature we have two candidate gravity backgrounds: the Euclidean version of (\ref{eqt: lowTd4}) and the background in (\ref{eqt: highTd4}) and hence we have to resort to energetics to determine the favoured background. This can be done by comparing the on-shell actions for the two gravity backgrounds, which in the light of our earlier discussions on finite temperature gauge-gravity dictionary corresponds to different phases of the dual gauge theory. It can be shown\cite{Witten:1998zw} that at temperature $T_d = 1/ (2\pi R_4)$, there is a first order phase transition between these two backgrounds. At low temperatures $T_d < 1/ (2\pi R_4)$ the Euclidean version of (\ref{eqt: lowTd4}) is energetically favoured, whereas at high temperature $T_d > 1/ (2\pi R_4)$ the background in (\ref{eqt: highTd4}) is favoured. Geometrically the energetics of this transition is fairly intuitive: it is a competition between the two circles along $t$ and $x^4$ and the smaller circles prefers to shrink.

The gauge theory interpretation of this transition is fairly straightforward. A comparison of the appropriately renormalized free energies in these two phases shows that in the low temperature phase the free energy scales as $N_c^0$ whereas in the high temperature phase it scales as $N_c^2$. This is, according to our broad characterization, a confinement/deconfinement transition. Thus this model, as suggested by Witten in ref.~\cite{Witten:1998zw}, is a holographic model dual to a non-supersymmetric gauge theory which undergoes a confinement/deconfinement transition, which is similar to QCD.

\subsection*{1.2.4 \hspace{2pt} The Type IIA theory: introducing flavours}
\addcontentsline{toc}{subsection}{1.2.4 \hspace{0.15cm} The Type IIA theory: introducing flavours}

We will now discuss introducing flavour degrees of freedom in this model following refs.~\cite{Sakai:2004cn, Sakai:2005yt} which is known as the Sakai-Sugimoto model. We will call this the Witten--Sakai-Sugimoto model. Our approach will be very similar to the Type IIB construction discussed before. In this case  flavour degrees of freedom are introduced by intersecting the $N_c$ D4-branes by $N_f$ $\overline{\rm D8}$-branes at $x^4=-\frac{L}{2}$ and $N_f$ D8-branes at $x^4=\frac{L}{2}$ with the obvious constraint $L \le \pi R_4$. This configuration can be represented by the following array.
\begin{table}[h]
\begin{center}
\begin{tabular}{|c|c|c|c|c|c|c|c|c|c|c|}
\hline
 &0&1&2&3&4&5&6&7&8&9\\\hline
 D4 &--&--&--&--&--&$\bullet$&$\bullet$&$\bullet$&$\bullet$&$\bullet$\\\hline
 D8,  $\overline{\rm D8}$ &--&--&--&--&$\bullet$&--&--&--&--&-- \\\hline
 
\end{tabular}
\end{center}
\caption{ The brane construction including fundamental flavours. }
\label{default}
\end{table}

From the open strings with one end attached to the D4-branes and the other end on the D8 or the $\overline{\rm D8}$-brane we obtain $N_f$ fundamental fermions of opposite chirality. Since the D4-branes are intersected by the D8/$\overline{\rm D8}$-branes these fermions are necessarily massless. As seen from the D4-brane worldvolume perspective the flavour branes introduce a global flavour symmetry group ${\rm U}(N_f)_L\times {\rm U}(N_f)_R$, which is exactly analogous to the flavour symmetry group in massless QCD. Henceforth we will call this the chiral symmetry.

Since the brane construction here involves both branes and anti-branes, we may worry about the stability of the construction. Usually a brane--anti-brane pair is unstable since oppositely charged branes annihilate each other, which is manifest as the presence of a negative mass squared tachyon in the spectrum of the open string stretching between the brane--anti-brane pair. The mass of this ``tachyon" in this case is explicitly given by
\begin{equation}
m^2 = \left(\frac{L}{2\pi\alpha'}\right)^2 - \frac{1}{2\alpha'}\ ,
\end{equation}
which is positive definite for $L> \pi \sqrt{2\alpha'} $. Therefore as long as the brane--anti-brane pair is separated by a length much larger than the string length scale, this system should be stable. We will again treat the flavour branes as probes in the background of D4-branes.

\subsection*{1.2.5 \hspace{2pt} The Type IIA theory: an instructive limit}
\addcontentsline{toc}{subsection}{1.2.5 \hspace{0.15cm} The Type IIA theory: an instructive limit}

Before proceeding further, we will comment on a particularly instructive limit of the previous model. This has been realized recently in ref.~\cite{Antonyan:2006vw} and further generalized in refs.~\cite{Gao:2006up, Antonyan:2006qy, Antonyan:2006pg}. Let us consider sending the radius of the compact direction $x^4$ to infinity making it a flat direction. The flavour brane--anti-brane pair are aligned (separated by a distance $L$ in the $x^4$-direction) in exactly the same way as illustrated in table 2. We will focus on the low energy dynamics from the point of view of a $(3+1)$-dimensional observer living in the intersection region. We will concentrate in the limit $g_s\to 0$, $N_c\to \infty$ with $g_s N_c$ and $N_f$ held fixed.

In this special limit ($R_4\to\infty$) we have three dimensionful parameters: the brane--anti-brane separation $L$, the five-dimensional 't Hooft coupling $\lambda_5$ and the string length $\ell_s$. The chiral fermions are located at the $(3+1)$-dimensional intersections but they interact via a $(4+1)$-dimensional gauge field. This interaction strength is represented by the dimensionless ratio $\lambda_5/ L$. There is another dimensionless ratio given by $\lambda_5/\ell_s$ which is relevant in this model. In the regime of the following hierarchy of scales
\begin{eqnarray}
\lambda_5 \ll \ell_s \ll L
\end{eqnarray}
we can neglect all stringy effects. We can also neglect all gauge field dynamics on the worldvolume of the D4 and the D8-branes. The only relevant interaction is the exchange of (five-dimensional) gluons between the chiral fermions located at the brane-intersections. In this regime the theory is weakly coupled with a natural UV cut-off given by $\ell_s$. In the regime $\ell_s \ll \lambda_5 \ll L$, the gauge theory is still weakly coupled, but the natural UV cut-off is given by $\lambda_5$. Finally in the regime $\ell_s  \ll L \ll \lambda_5$ the gauge theory becomes strongly coupled. In this limit we can reliably use the supergravity description (as demonstrated in ref.~\cite{Itzhaki:1998dd}) to address the strong coupling dynamics.

It was further shown in ref.~\cite{Antonyan:2006vw} that in the weak coupling regime the effective description of the gauge theory is given by a Nambu-Jona-Lasinio (NJL) model\cite{Nambu:1961tp} with a non-local four fermi interaction between the chiral fermions. The corresponding Lagrangian was obtained by writing down an effective action for a Weyl fermion located at the $(3+1)$-dimensional intersection and a $(4+1)$-dimensional ${\rm SU}(N_c)$ gauge field. After integrating out the gauge field, one arrives at the non-local NJL model with a non-local four fermi interaction term. This non-locality arises from the $(4+1)$-dimensional gauge field exchange between the chiral fermions. We will not elaborate more on this topic here, but refer the interested reader to ref.~\cite{Antonyan:2006vw}.

The NJL model was first proposed in ref.~\cite{Nambu:1961tp}. This model is an useful effective low energy description of QCD as well. It has been successfully used in several phenomenological studies of hadrons and therefore provide an interesting avenue to understand many aspects of QCD. For a review of some such applications see {\it e.g.} ref.~\cite{Klevansky:1992qe}. We will not use this model explicitly to provide quantitative estimates or comparisons on the topics we will study using the gauge-gravity duality, however, we will comment on the qualitative aspects where necessary. It is very exciting that the gravity dual of a certain kind of NJL model can be realized in a precise sense.

\section*{1.3 \hspace{2pt} The probe brane action: the tool of the game}
\addcontentsline{toc}{section}{1.3 \hspace{0.15cm} The probe brane action: the tool of the game}

Our general strategy will be to study the dynamics of a probe brane (of various dimensions) in an appropriate background. In subsequent chapters we will make heavy use of the probe action, so let us review some general aspects of this action. Here we will restrict ourselves only to the bosonic part of the probe action.

The bosonic action consists of two parts: the Dirac-Born-Infeld (DBI) action and the Wess-Zumino (WZ) term. In supergravity we have two kinds of fields to which a D-brane can couple to, namely the Neveu-Schwarz (NS) fields (containing the metric, dilaton and the B-field) and the Ramond-Ramond (RR) fields (containing the various potentials). The DBI action describes the coupling of the D-brane with the background NS fields and the WZ action describes its coupling to the background RR fields.

The DBI action was formulated by Born and Infeld\cite{Born:1934ab} to avoid various singularities of Maxwell's theory of electrodynamics and later studied in more detail by Dirac\cite{Dirac:1960ab}. This action was obtained by requiring Lorentz and gauge invariance and also the condition that for small electromagnetic fields it should reduce to the familiar Maxwell action. The DBI action is also naturally equipped to provide a cut-off for the maximum electric field. This fact will play an interesting role in the later chapters.

The DBI action was derived explicitly from string theory in refs.~\cite{Fradkin:1985qd,Leigh:1989jq, Tseytlin:1996it}. Furthermore, the fermionic part of the DBI action has been analyzed in refs.~\cite{Marolf:2003ye, Marolf:2003vf, Martucci:2005rb}; however we will not need it. The bosonic part of the DBI action for a D$p$-brane is given by
\begin{eqnarray}\label{eqt: dbi0}
S_{\rm DBI} = - \mu_p \int_{\mathcal{M}_{p+1}} d^{p+1}\xi  e^{-\Phi} \sqrt{-{\rm det} \left( P\left[G_{ab} + B_{ab} \right] + (2\pi\alpha') F_{ab}\right)}\ ,
\end{eqnarray}
where $P\left[G_{ab} + B_{ab} \right]$ is the pull-back of the background metric and the B-field, $F_{ab}$ is the field strength corresponding to the gauge field $A_a$ defined on the worldvolume of the probe brane. The constant $\mu_p$ is related to the tension of the D-brane and is given by\cite{Polchinski:1995mt}, $\mu_p = \left(2 \pi\right)^{-p} \alpha'^{-(p+1)/ 2}$.

The Wess-Zumino action describing the coupling of the D-brane to the background RR potentials (denoted by $C_{(p+1)}$) is a topological term. This action in complete generality is given by (see {\it e.g.} ref.~\cite{Johnson:2003gi} for a review)
\begin{eqnarray}\label{eqt: wz0}
S_{\rm WZ} = \mu_{p} \int_{\mathcal{M}_{p+1}} \left[\sum_{p} C_{(p+1)}\right] \wedge {\rm Tr} \, e^{2\pi\alpha' F + B}\ .
\end{eqnarray}
For most of the cases that we will consider in later chapters this term will not play any role in determining the brane's classical profile. It will, however, play an important role in the study of the fluctuation spectrum around this classical profile of the brane.

\section*{1.4 \hspace{2pt} Outline}
\addcontentsline{toc}{section}{1.4 \hspace{0.15cm} Outline}

This thesis is organized as follows. We begin with a brief review of chiral symmetry breaking in QCD in chapter 2. As an effective model for QCD, we also review the so called NJL model and the pattern of chiral symmetry breaking within this model. Our focus is to obtain a qualitative understanding of the physics of chiral symmetry breaking under the influence of several external parameters such as temperature or an external electromagnetic field in the NJL model.

Then we start discussing these aspects in the holographic set-up. In chapter 3 we first discuss introducing fundamental flavour in finite temperature $\mathcal{N} = 4$ SYM theory. At finite temperature the flavour sector undergoes a first order meson melting transition. Now we introduce an external constant magnetic field and discuss its effect on the phase diagram of the fundamental flavours in finite temperature $\mathcal{N} = 4$ SYM theory. We find that temperature and magnetic field are two competing parameters which give rise to an interesting phase diagram. We further realize that an external magnetic field can induce spontaneous breaking of chiral symmetry which was first observed in ref.~\cite{Filev:2007gb} in the zero temperature set-up. We begin by
establishing the non--trivial phase structure that results from finite
temperature.  We observe, for example, that above the critical value
of the field that generates a chiral condensate spontaneously, the
meson melting transition disappears, leaving only a discrete spectrum
of mesons at any temperature.  We also compute several thermodynamic
 properties of the  plasma.

Following the same logical path, we investigate the effect of an external electric field in chapter~4. At zero temperature, we observe that the electric field induces a phase transition associated with the dissociation of the mesons into their constituent quarks. This is an analogue of an insulator--metal transition, since the system goes from being an insulator with zero current (in the applied field) to a conductor with free charge carriers (the quarks). At finite temperature this phenomenon persists, and the dissociation transition becomes subsumed into the more familiar meson melting transition. Here, the dissociation phenomenon reduces the critical melting temperature.

We change gears in chapters~5 and 6 and work in a different model. In chapter~5 using the Witten--Sakai-Sugimoto model we study the effect of an external
magnetic field on the dynamics of fundamental flavours in both the
confined and deconfined phases of a large $N_c$ gauge theory. We find
that an external magnetic field promotes chiral symmetry breaking, consistent
with the ``magnetic catalysis'' observed in the field theory
literature, and seen in the Type IIB set-up in chapter~3. The
external field increases the separation between the deconfinement
temperature and the chiral symmetry restoring temperature. In the
deconfined phase we investigate the temperature-magnetic field phase
diagram and observe, for example, there exists a maximum critical
temperature (at which symmetry is restored) for very large magnetic
field. We find that this and certain other phenomena persist for the
Witten--Sakai-Sugimoto type models with probe branes of diverse dimensions.
We comment briefly on the dynamics in the presence of an external
electric field.

In chapter~6 we continue our study of the dynamics of the flavour sector of the
Witten--Sakai-Sugimoto model in the presence of an external magnetic field,
uncovering several features of the meson spectrum at high and low
temperatures. We employ both analytical and numerical methods to study
the coupled non--linear equations that result from the gravity dual. Finally in chapter~7 we conclude with open problems and future directions.


 \chapter*{Chapter 2:  \hspace{1pt} Review of Chiral Symmetry Breaking}

\addcontentsline{toc}{chapter}{Chapter 2:\hspace{0.15cm}
Review of Chiral Symmetry Breaking}

In this chapter we briefly review the phenomenon of chiral symmetry breaking in QCD and the NJL model which is an effective model of QCD. We also discuss the effect of an external parameter (such as temperature and an external electromagnetic field) on the chiral symmetry breaking in the latter model. Our discussions here are entirely based on the more detailed reviews in ref.~\cite{Schafer:2005ff, Gusynin:1995nb}.

\section*{2.1 \hspace{2pt} QCD and Symmetries}
\addcontentsline{toc}{section}{2.1 \hspace{0.15cm} QCD and Symmetries}

We begin with a brief review of the symmetries of QCD. We will closely follow the notations and discussions in ref.~\cite{Schafer:2005ff}. The fundamental degrees of freedom of QCD are the fermionic quark fields (denoted by $\psi_i ^ a$) and the non-abelian gauge fields called gluons (denoted by $A_\mu^a$). We denote flavour indices by the letter $i$ and color indices by the letter $a$ whereas $\mu$ labels the spacetime indices. The quarks transform under the fundamental representation of the ${\rm SU}(N_c)$ gauge group (where $N_c = 3$) and the gluons transform under the adjoint representation. The QCD Lagrangian is given by
\begin{eqnarray}
&& \cL _{\rm QCD} =  \overline{\psi} \left( i  \Dbar - m \right) \psi - \frac{1}{4} F_{\mu\nu}^a F^{\mu\nu a} \ , \\
&& {\rm where} \quad \Dbar = \gamma^\mu \left( \partial_\mu - i g A_\mu^a t^a \right) \ ,  \quad F_{\mu\nu}^a = \partial_\mu A_\nu^a - \partial_\nu A_\mu^a + g f^{abc} A_\mu^b A_\nu^c \ ,
\end{eqnarray}
where $g$ is the gauge coupling, $t^a$ are the generators of the ${\rm SU}(3)$ gauge group and $f^{abc}$ are the structure constants. In writing the fermionic term in the Lagrangian we have implicitly assumed a summation over the flavour indices.

To begin with, the QCD Lagrangian is invariant under a local ${\rm SU}(3)$ gauge symmetry
\begin{eqnarray}
\psi(x) \to U(x) \psi(x) \ , \quad A_\mu^a t^a (x) \to U(x) A_\mu^a t^a U^\dagger(x) + i U(x) \partial_\mu U^\dagger(x) \ .
\end{eqnarray}
If the masses of the three lightest quarks are the same then the Lagrangian is invariant under the following global flavour rotations
\begin{eqnarray}
\psi_i = V_{ij} \psi_j \ , \quad V \in {\rm SU}(3) \ ,
\end{eqnarray}
which is known as the flavour isospin symmetry. Furthermore if the masses are equal to zero, then this global symmetry is enlarged. Let us define the chiral quark fields as
\begin{eqnarray}\label{eqt: chiralferm}
\psi_{L, R} = \frac{1}{2} \left( 1 \pm \gamma_5 \right) \psi \ ,
\end{eqnarray}
and rewrite the fermionic part of the QCD Lagrangian as
\begin{eqnarray}
\cL _{\rm fermion} = \overline{\psi}_{L} i \Dbar \psi_{L} + \overline{\psi}_{L}  M \psi_{R} + \overline{\psi}_{R} i \Dbar \psi_{R} + \overline{\psi}_{R}  M \psi_{L} \ ,
\end{eqnarray}
where $M$ is the diagonal mass matrix. Clearly if $M = 0 $, the Lagrangian is invariant under the following independent flavour transformations
\begin{eqnarray}
\psi_{{L}, i} \to V_{{L}, ij} \psi_{{L}, j} \ , \quad \psi_{{R}, i} \to V_{{R}, ij} \psi_{{R}, j} \quad {\rm where} \quad \{V_{L}, V_{R}\} \in {\rm SU}(3)_L\times {\rm SU}(3)_R \ .
\end{eqnarray}
This is known as the chiral symmetry. In the real world the quark masses are not vanishing. However, for the three lightest quarks the masses are negligible compared to the scale $\Lambda_{\rm QCD}$ and therefore it is an approximate symmetry of the theory.

Lastly, the QCD Lagrangian has the following two global ${\rm U}(1)$ symmetries as well
\begin{eqnarray}
&& {\rm U}(1)_{\rm B} : \quad \psi_{L, R} \to e^{i \phi} \psi_{ L, R} \ , \\
&& {\rm U}(1)_{\rm A} : \quad \psi_{ L} \to e^{i \alpha} \psi_{ L} \ , \quad \psi_{ R} \to e^{- i \alpha} \psi_{ R} \ ,
\end{eqnarray}
where $\phi$ and $\alpha$ are constant angles. The ${\rm U}(1)_{\rm B}$ is an exact symmetry for massive quarks as well. On the other hand, the axial ${\rm U}(1)_{\rm A}$ is exact for the massless quarks only at the classical level and is broken by quantum anomalies. This can be represented by writing the divergence of the axial current 
\begin{eqnarray}
\partial^\mu j_\mu^5 = \frac{N_f g^2}{16 \pi^2} F_{\mu\nu}^a \tilde{F}^{\mu\nu a} \ , \quad \tilde{F}^{\mu\nu a} = \frac{1}{2} \epsilon^{\mu\nu\alpha\beta} F_{\alpha\beta}^a \ .
\end{eqnarray}

At zero temperature and density, the QCD vacuum breaks chiral symmetry {\it via} the formation of a chiral condensate $\langle \overline \psi \psi \rangle$. The symmetry group breaks down to ${\rm SU}(3)_L \times {\rm SU}(3)_R \to {\rm {\rm {\rm SU}}}(3)_V$ and an octet of approximately massless pseudoscalar Goldstone bosons appear. Chiral symmetry restricts the interactions of these Goldstone modes which can be obtained from the low energy effective chiral Lagrangian. The low energy effective chiral Lagrangian is given by
\begin{eqnarray}
\cL_{\rm chiral} = \frac{f_\pi^2}{4} {\rm Tr} \left[ \partial_\mu \Sigma \partial^\mu \Sigma^\dagger \right] + \left[ B {\rm Tr} \left( M \Sigma^\dagger \right) + {\rm h.c.} \right] + \ldots \ ,
\end{eqnarray}
where $f_\pi$ is the pion decay constant, $\Sigma$ collectively denotes the Goldstone modes, $M$ is the quark mass matrix and $B$ is a constant. The Goldstone modes associated with the chiral symmetry breaking satisfies the Gell-Mann--Oakes--Renner (GMOR) relationship\cite{GellMann:1968rz}
\begin{eqnarray}
m_\pi^2 \propto \frac{m}{f_\pi^2} \langle \overline \psi \psi \rangle \ ,
\end{eqnarray}
where $m$ is the bare quark mass.

At finite temperature, QCD undergoes the deconfinement and the chiral phase transition. However, close to this critical temperature the essential dynamics remain non-perturbative. We will now turn to an effective model of QCD known as the Nambu--Jona-Lasinio (NJL) model. It turns out to be a useful toy model for understanding the physics of chiral symmetry breaking among other features of QCD. We will discuss the effect of external parameters on chiral symmetry breaking within this model.

\section*{2.2 \hspace{2pt}The Nambu--Jona-Lasinio model} 
\addcontentsline{toc}{section}{2.2 \hspace{0.15cm} The Nambu--Jona-Lasinio model}

This model was originally proposed in ref.~\cite{Nambu:1961tp}. The Lagrangian of the NJL model is given by
\begin{eqnarray}
\cL_{\rm NJL} = \overline{\psi} i \delbar \psi + G \left[\left(\overline{\psi} \psi \right)^2 + \left(\overline{\psi} i \gamma^5 \psi \right)^2 \right]  \ ,
\end{eqnarray}
where $G$ is the coupling corresponding to the four-Fermi interaction. For simplicity we will focus only on the single flavour case. Due to the presence of the four-Fermi terms, this is a non-renormalizable theory and therefore requires a UV cut-off. In writing the four-Fermi interaction terms we have also assumed a summation over the color indices. 

This bare Lagrangian is invariant under an ${\rm U}(1)_L\times {\rm U}(1)_R$ symmetry (which from now onwards we refer as the chiral symmetry)
\begin{eqnarray}
\psi_{L,R} \to {\rm exp} \left( \frac{i}{2} \theta_{L,R}\right) \psi \ , \quad \overline{\psi}_{L,R} \to {\rm exp} \left( - \frac{i}{2} \theta_{L,R}\right) \overline{\psi} \ ,
\end{eqnarray}
where $\psi_{L,R}$ is defined as in equation (\ref{eqt: chiralferm}). Clearly the NJL model has an obvious generalization for $N_f$ flavours.

There are two simple cases to consider: when $\theta_L = \theta_R = \theta_V$, the chiral symmetry group is the ${\rm U}(1)_V$ group; the other choice is $\theta_L = - \theta_R = \theta_A$ for which the invariance group is the axial ${\rm U}(1)_A$ group.

Now we want to couple the NJL model with an external electromagnetic field. The corresponding Lagrangian is then given by
\begin{eqnarray}
\cL_{\rm NJL} = \overline{\psi} i \Dbar \psi + G \left[\left(\overline{\psi} \psi \right)^2 + \left(\overline{\psi} i \gamma^5 \psi \right)^2 \right] \ , \quad D_\mu = \partial_\mu - i e A_\mu \ .
\end{eqnarray}
Note that we can rewrite this Lagrangian by introducing auxiliary fields (denoted by $\sigma$ and $\pi$)
\begin{eqnarray}\label{eqt: lagnjl}
&& \cL_{\rm NJL} = \overline{\psi} i \Dbar \psi - \overline{\psi} \left(\sigma + i \gamma^5 \pi \right) \psi - \frac{1}{4G} \left(\sigma^2 + \pi^2 \right) \ , \\
&& {\rm where} \quad \sigma = - 2 G \left(\overline{\psi} \psi\right) \ , \quad \pi = - 2 G \left(\overline{\psi} i \gamma^5 \psi \right) \ .
\end{eqnarray}
To analyze the effect of the external field on the chiral symmetry we have to study what happens to the dynamically generated mass associated with the chiral symmetry breaking. One way to achieve this is to compute the one-loop Coleman-Weinberg effective potential\cite{Coleman:1973jx}. We will not review the calculation of this potential, but point the interested reader to the relevant reference, {\it e.g.} ref.~\cite{Gusynin:1995nb}. A pedagogical introduction to Coleman-Weinberg potential can be found in {\it e.g.} ref.~\cite{Peskin:1995ab}. The effective potential for the NJL model, denoted by $V_{\rm eff}$, depends on the ${\rm U}(1)_L\times {\rm U}(1)_R$ invariant combination $\rho^2 = \sigma^2 + \pi^2$. Therefore without any loss of generality, we consider a configuration with $\pi = 0$. We will now review the effective potential for this model under various external parameters, such as temperature and constant electromagnetic field.

\subsection*{2.2.1 \hspace{2pt} The effective potential at finite temperature}
\addcontentsline{toc}{subsection}{2.2.1 \hspace{0.15cm} The effective potential at finite temperature}

Let us first review the properties of the ground state by analyzing the effective potential at finite temperature. The effective potential at finite temperature is given by (see {\it e.g.} ref.~\cite{Schwarz:1999dj})
\begin{equation}
V_{\rm eff} (\rho) = \frac{\rho^2}{4 G} - 2 N_c \int_0^\Lambda \frac{d^3 p}{(2\pi)^3}\left[ \sqrt{p^2 + \rho^2} + 2 T \log\left(1 + {\rm exp}\left(- \frac{1}{T} \sqrt{p^2 + \rho^2}\right)\right)\right] \ , 
\end{equation}
where $T$ is the temperature.

The true vacuum of the system is given by $dV_{\rm eff}/d\rho = 0$. This is popularly known as the Gap equation. At $T = 0$, the Gap equation gives\cite{Gusynin:1995nb}
\begin{eqnarray}\label{eqt: gapt0}
\rho \Lambda^2 \left( \frac{1}{g} - 1 \right) = - \rho^3 \log\left(\frac{\Lambda}{\rho}\right) + \cO(1/ \Lambda^2) \ , \quad g = \frac{N_c G \Lambda^2}{\pi^2} \ .
\end{eqnarray}
Clearly in order for the above equation to have a real non-trivial solution one should have $g > g_c = 1$. Also, any solution of the Gap equation (\ref{eqt: gapt0}), $\rho = m_{\rm dyn}$ basically gives the dynamically generated quark mass (from the definition of $\sigma$ in the Lagrangian in (\ref{eqt: lagnjl})) associated with the chiral symmetry breaking. In this case, the chiral symmetry is spontaneously broken if $g>g_c$\cite{Gusynin:1995nb}.

At finite temperature the resulting Gap equation is given by
\begin{eqnarray}\label{eqt: gapt1}
\frac{\rho}{2 G} - \frac{N_c \rho}{\pi^2} \int_0^{\Lambda} \frac{p^2 dp}{\sqrt{p^2 + \rho^2}} \left(1 - 2 {\rm exp} \left(-\frac{1}{T} \sqrt{p^2 + \rho^2}\right)\right) = 0 \ .
\end{eqnarray}
We will not explore this equation in detail, but we will try to estimate the effect of introducing a temperature. Specifically we want to estimate how the coupling constant must be tuned if we introduce an infinitesimally small temperature keeping the value of $\rho$ fixed.

First let us note that a solution of equation (\ref{eqt: gapt0}), $\rho\sim \cO(\Lambda)$. Therefore for small temperature, the exponential factor in the integrand in (\ref{eqt: gapt1}) can be well approximated by a small positive constant, which we denote by $\epsilon$. Further we assume that the change in the coupling is given by $g = g_{(0)} + \delta g$, where $g_{(0)}$ is the coupling at zero temperature. In this limit the Gap equation becomes
\begin{eqnarray}
&& \rho \Lambda^2 \left(\frac{1}{g} - (1-\epsilon) \right) = - \rho^3 (1 - \epsilon) \log\left(\frac{\Lambda}{\rho}\right) \nonumber\\
&& \implies \quad  \delta g = \epsilon g_{(0)} > 0 \ ,
\end{eqnarray}
where we have used the fact that $\rho$ satisfies the Gap equation at vanishing temperature. This suggests that in order to have a chiral symmetry breaking the four Fermi coupling needs to be stronger in finite temperature background. In other words, finite temperature tends to restore chiral symmetry. Physically this is intuitive to understand. Raising temperature will presumably melt the quark bound states and break the quark condensate, therefore tend to restore the chiral symmetry. For more detailed analysis we refer the reader to the review in ref.~\cite{Klevansky:1992qe}.

\subsection*{2.2.2 \hspace{2pt} The effective potential in a magnetic field}
\addcontentsline{toc}{subsection}{2.2.2 \hspace{0.15cm} The effective potential in a magnetic field}

In the presence of an external constant magnetic field the effective potential is given by\cite{Gusynin:1995nb}
\begin{eqnarray}\label{eqt: effb}
V_{\rm eff}(\rho) = \frac{\rho^2}{4 G} + \frac{N_c}{8\pi^2} \int_{1/\Lambda}^\infty \frac{ds}{s^2} e^{- s \rho^2} e H \coth\left(e H s\right) \ .
\end{eqnarray}
The corresponding Gap equation in this case is given by\cite{Gusynin:1995nb}
\begin{eqnarray}\label{eqt: gapb}
\rho \Lambda^2 \left(\frac{1}{g} - 1 \right) & = & - \rho^3 \log\left[\frac{\left(\Lambda \ell\right)^2}{2}\right] + \gamma \rho^3 + \frac{\rho}{\ell^2} \log\left[\frac{\left(\rho \ell\right)^2}{4\pi}\right] \nonumber\\
& + & \frac{2\rho}{\ell^2} \log\left[\Gamma \left(\frac{\rho^2 \ell^2}{2}\right)\right] + \cO(1/\Lambda) \ ,
\end{eqnarray}
where $\gamma$ is the Euler constant and $\ell^2 = (e H)^{-1} $ is the magnetic length.

The magnetic field modifies the Gap equation in a rather interesting way. To see this, let us assume that $\rho = m_{\rm dyn}$ is a solution of (\ref{eqt: gapb}) and $\rho = m_{\rm dyn}^{(0)}$ is a solution of (\ref{eqt: gapt0}). Further, let us work in the small magnetic field limit, {\it i.e.} $m_{\rm dyn}^{(0)} \gg \ell$. Then from (\ref{eqt: gapb}) we can obtain\cite{Gusynin:1995nb}
\begin{eqnarray} \label{eqt: mdynb}
m_{\rm dyn} ^2 \approx \left( m_{\rm dyn}^{(0)}\right)^2 \left[ 1 + \frac{\left(e H\right)^2}{3 \left( m_{\rm dyn}^{(0)}\right)^4 \log \left(\Lambda/ m_{\rm dyn}^{(0)} \right)^2}\right] \ ,
\end{eqnarray}
which means that the magnetic field enhances the breaking of the chiral symmetry.

Interestingly, the Gap equation in the presence of a magnetic field admits a solution even for\cite{Gusynin:1995nb} $g\ll g_c$ given by
\begin{eqnarray}
m_{\rm dyn}^2 \sim e H {\rm exp} \left( - \frac{\Lambda^2 (1 - g)}{ e H g}\right) \ .
\end{eqnarray}
It is therefore clear that any non-zero magnetic field in this model induces a spontaneous chiral symmetry breaking for arbitrarily small values of the coupling. This feature of a magnetic field inducing spontaneous breaking of chiral symmetry, however, is rather universal and commonly known as the magnetic catalysis in chiral symmetry breaking. In section 2.3, we will review the basic mechanism behind this catalysis.

\subsection*{2.2.3 \hspace{2pt} The effective potential in an electric field}
\addcontentsline{toc}{subsection}{2.2.3 \hspace{0.15cm} The effective potential in an electric field}

At zero temperature, the unbroken Lorentz invariance particularly simplifies the expression for the effective potential. Specifically, we can obtain the effective potential in a constant electric field from the potential in a constant magnetic field just by sending $H \to i E$, and {\it vice versa}, in {\it e.g.} equation (\ref{eqt: effb}).

The consequences are simple to observe. Now, the dynamically generated mass $m_{\rm dyn} < m_{\rm dyn}^{(0)}$ by virtue of equation (\ref{eqt: mdynb}). Hence we conclude that an electric field tends to restore the chiral symmetry. Physically this is expected, since an external electric field will tend to dissociate the quarks bound states into its constituents.

\section*{2.3 \hspace{2pt} Fermions in a constant magnetic field}
\addcontentsline{toc}{section}{2.3 \hspace{0.15cm} Fermions in a constant magnetic field}

In this section we will review the basic properties of a relativistic fermion in a constant magnetic field in $(3+1)$-dimensions. We will closely follow the discussions in ref.~\cite{Gusynin:1995nb}. The corresponding Lagrangian is simply given by
\begin{eqnarray}
\cL_{\rm fermion} = \overline{\psi} \left( i  \Dbar - m \right) \psi \ , \quad D_\mu = \partial_\mu - i e A_\mu \ .
\end{eqnarray}
In the so called Landau gauge, the gauge potential corresponding to a constant external magnetic field (pointing in the $x^3$-direction) is given by
\begin{eqnarray}
A_\mu = - \delta_{\mu1} H x^2 \ . 
\end{eqnarray}
The energy spectrum of the fermions is given by the Landau levels
\begin{eqnarray}
E_n \left(k_3 \right) = \pm \sqrt{m^2 + k_3^2  + 2 |e H| n} \ , \quad {\rm where} \quad n \in \mathbb{Z}_+ \cup \{0\} \ .
\end{eqnarray}
Clearly, $n = 0$ corresponds to the lowest Landau level (henceforth abbreviated as LLL). The LLL has degeneracy and the number of states is given by
\begin{eqnarray}
dN_0 = V \frac{|e H|}{2 \pi} \frac{dk_3}{2\pi} \ ,
\end{eqnarray}
where $V$ is the spatial $3$-volume.

Let us first argue that the LLL plays a special role in the infrared dynamics. Clearly the energy of the LLL is given by $\sqrt{m^2 + k_3^2}$ whereas the energy for the higher Landau levels goes as $\sqrt{|e H|}$. It is therefore obvious that the infrared dynamics in a strong magnetic field, {\it i.e.} when $\sqrt{|e H|} \gg m, k_3$, is insensitive to the higher Landau levels.

We want to demonstrate that the dynamics of the LLL plays a crucial role in catalyzing the chiral symmetry breaking. To do so, we consider the fermion propagator in a background constant magnetic field computed in ref.~\cite{Schwinger:1951nm}
\begin{eqnarray}
S \left(x, y\right) = {\rm exp} \left[ \frac{ie}{2} \left(x - y \right)^\mu A_\mu \left(x + y \right)\right] \tilde{S} \left(x - y\right) \ , 
\end{eqnarray}
where
\begin{eqnarray}
\tilde{S} (k) & = & \int_0^\infty ds {\rm exp} \left[ i s \left( k_0^2 - k_3^2 - k_{\perp}^2 \frac{\tan\left(e H s\right)}{e H s} - m^2 \right)\right] \nonumber\\
&& \left[\left(k_0 \gamma^0 - k_3 \gamma^3 + m \right) \left(1 + \gamma^1 \gamma^2 \tan\left( e H s\right)\right) - k_{\perp} \gamma_{\perp} \left( 1 + \tan^2\left( e H s \right)\right)\right] \ , \nonumber\\
\end{eqnarray}
where $k_\perp= \{k_1 , k_2\}$ and $\gamma_\perp = \{\gamma_1, \gamma_2 \}$. Now using expression for the generating functional for Laguerre polynomials (denoted by $L_n^\alpha$)
\begin{eqnarray}
(1 - z)^{-\alpha - 1} {\rm exp} \left(\frac{x z}{z - 1}\right) = \sum_{n = 0}^\infty L_n^\alpha(x) z^n \ ,
\end{eqnarray}
we can decompose the fermionic propagator over the Landau poles
\begin{eqnarray}
\tilde{S} (k) = i {\rm exp} \left( - \frac{k_\perp^2}{|e H|}\right) \sum_{n=0}^\infty (-1)^n \frac{D_n \left(e H, k\right)}{k_0^2 - k_3^2 - m^2 - 2 |e H| n} = \sum_{n=0}^\infty \tilde{S}^{(n)} (k) \ ,
\end{eqnarray}
where
\begin{eqnarray}
&& D_n \left(e H, k\right)  =  \left(k_0 \gamma^0 - k_3 \gamma^3 + m\right) \left[ \left(1 - i \gamma^1\gamma^2 {\rm sign} \left(e H\right)\right) L_n^0 \left(\frac{2 k_\perp^2}{|e H|}\right)\right. \nonumber\\
 && \left. -  \left(1 + i \gamma^1\gamma^2 {\rm sign} \left(e H\right)\right) L_{n-1}^0 \left(\frac{2 k_\perp^2}{|e H|}\right) \right] + 4 \left(k_1\gamma^1 + k_2 \gamma^2 \right) L_{n-1}^1 \left(\frac{2 k_\perp^2}{|e H|}\right) \ . \nonumber\\
\end{eqnarray}
The contribution coming from the LLL is then given by
\begin{eqnarray}
\tilde{S}^{(0)} (k) = i {\rm exp} \left( - \frac{k_\perp^2}{|e H|}\right) \frac{k_0 \gamma^0 - k_3 \gamma^3 + m}{k_0^2 - k_3^2 - m^2} \left(1 - i \gamma^1\gamma^2 {\rm sign} \left(e H\right)\right) \ .
\end{eqnarray}
It is now clear from the LLL contribution to the propagator that in the infrared regime, {\it i.e.} when $k_\perp^2 \ll |e B|$, $\tilde{S}^{(0)}$ is independent of $k_\perp$ completely and the dynamics is effectively $(1+1)$-dimensional. The physical reason behind this is simple: the motion of a charged particle is constrained in the plane perpendicular to the magnetic field.

It is well-known that infrared dynamics is stronger in low dimensional field theories\footnote{We thank V. Miransky for a discussion on this point.}. This can be qualitatively understood as follows: the measure in the momentum space goes as $k^3 dk$ in $d = 3$ and $k dk$ in $d = 1$. Clearly in the infrared, {\it i.e.} for small $k$, $k^3$ suppresses the dynamics much stronger than $k$. Therefore it is expected that the infrared dynamics in $(1+1)$-dimensional theory will remain much strongly coupled as compared to the infrared dynamics of a $(3+1)$-dimensional theory. This therefore suggests an universal mechanism through which an external magnetic field facilitates the spontaneous breaking of chiral symmetry by generating a dynamical quark mass. In the forthcoming chapters we will further observe that a similar universal mass generation mechanism in the presence of a magnetic field exists within the specific holographic models that we consider in this thesis.


\chapter*{Chapter 3: \hspace{1pt} Flavoured $\mathcal{N} = 4$ Yang-Mills theory at finite temperature and an external magnetic field}
\addcontentsline{toc}{chapter}{Chapter 3:\hspace{0.15cm}
Flavoured $\mathcal{N} = 4$ Yang-Mills theory at finite temperature and an external magnetic field}

\section*{3.1 \hspace{2pt} Introductory remarks}
\addcontentsline{toc}{section}{3.1 \hspace{0.15cm} Introductory remarks}

We begin with a holographic analysis of thermal properties of large $N_c$ flavoured Yang--Mills theory. We will then consider the effect of both temperature and an external magnetic field in later sections. Our holographic set-up is the so called Type IIB framework reviewed in the first chapter. The presented material about the thermal properties is based on a close collaboration with Tameem Albash, Veselin Filev and Clifford Johnson\cite{Albash:2006ew}. Similar results were obtained in ref.~\cite{Mateos:2006nu} and also in ref.~\cite{Karch:2006bv}. The main result, however, was already known before in ref.~\cite{Kirsch:2006he}.

We study the geometry of AdS$_5$--Schwarzschild $\times S^5$ which is
the decoupled/near--horizon geometry of $N_c$ D3--branes, where $N_c$
is large and set by the (small, for reliability) curvature of the
geometry. The physics of closed Type~IIB string theory in this
background is dual to the physics of ${\cal N}=4$ supersymmetric
${\rm SU}(N_c)$ gauge theory in four dimensions, with the supersymmetry broken by being at finite
temperature\cite{Witten:1998zw}. The temperature is set by the horizon
radius of the Schwarzschild black hole, as we will recall below.

We introduce a D7--brane probe into the background. Four of the
brane's eight world--volume directions are parallel with those of the
D3--branes, and three of them wrap an $S^3\subset S^5$. The remaining
direction lies in the radial direction of the asymptotically AdS$_5$
geometry.

Such a D3--D7 configuration controls the physics of
the ${\rm SU}(N_c)$ gauge theory with a dynamical quark in the fundamental
representation\cite{Karch:2002sh}.  The configuration (at zero temperature)
preserves ${\cal N}=2$ supersymmetry in $D=4$, and the quark is part
of a hypermultiplet.  Generically, we will be studying the physics at finite
temperature, so supersymmetry will play no explicit role here.

We are studying the D7--brane as a probe only, corresponding to taking the $N_c \gg N_f$ limit, and therefore there is no backreaction on the background geometry. This is roughly analogous
to the quenched approximation in lattice QCD.  The quark mass and other flavour physics--such as the
vacuum expectation value (vev) of a condensate and the  spectrum of mesons
that can be constructed from the quarks--are all physics which are
therefore invisible in the background geometry. We will learn nothing
new from the background; our study is of the response of the probe D7--branes to the background, and this is where the new physics emerges from.

We carefully study the physics of the probe itself as
it moves in the background geometry. The coordinates of the probe in
the background are fields in an effective D7--brane world--volume
theory, and the geometry of the background enters as couplings controlling the dynamics of
those fields. One such coupling in the effective model represents the
local separation, $L(u)$, of the D7--brane probe from the D3--branes,
where $u$ is the radial AdS$_5$--Schwarzschild coordinate.

In fact, the asymptotic value of the separation between the D3--branes and
D7--brane for large $u$ yields the bare quark mass $m_q$ and the
condensate vacuum expectation value (vev) $ \langle \bar{\psi} \psi
\rangle$ as follows \cite{Kruczenski:2003uq,Polchinski:2000uf}:
\begin{equation} \label{eqt: L}
\lim_{u \rightarrow \infty} L(u) = m + \frac{c}{u^2} + \dots\ ,
\end{equation}
where $m = 2 \pi \alpha' m_q$ and $-c = \langle \bar{\psi} \psi \rangle / (8 \pi^3 \alpha' N_f \tau_{\mathrm{7}})$, in this chapter we will set $N_f = 1$. The fundamental string tension is defined as $T=1/(2\pi\alpha^\prime)$ here, and the D7--brane tension is $\tau_{\mathrm{7}}=(2\pi)^{-7}(\alpha^\prime)^{-4}$. The
zero temperature behavior of the D7--branes in the geometry is
simple. The D7--brane world--volume actually {\it vanishes} at finite~$u$, corresponding to the part of the brane wrapped on the $S^3$
shrinking to zero size.  The location in $u$ where this vanishing
happens encodes the mass of the quark, or equivalently, the separation
of the probe from the D3--branes.  In addition, in the zero temperature background, the only value of $c$ allowed is zero, meaning no condensate is allowed to form, as is expected from supersymmetry.

The finite temperature physics introduces an important new feature.
As is standard \cite{Gibbons:1979xm}, finite temperature is studied by
Euclideanizing the geometry and identifying the temperature with the
period of the time coordinate.  The horizon of the background geometry
is the place where that $S^1$ shrinks to zero size. The D7--brane
 is also wrapped on this $S^1$, so it can vanish at
the horizon, if it has not vanished due to the shrinking of the $S^3$.
For large quark mass compared to the temperature (horizon size), the
$S^3$ shrinking will occur at some finite $u>u_{\rm{H}} $, and the physics
will be similar to the zero temperature situation. However, for small quark
mass, the world--volume will vanish due to the shrinking of
the $S^1$ corresponding to the D7--branes going into the horizon. This
is new physics of the flavour sector.

The authors of ref.~\cite{Babington:2003vm} explored some of the
physics of this situation (the dependence of the condensate and of the meson mass on the bare quark mass), and predicted that a phase transition
should occur when the topology of the probe D7--brane changes.  However, they were not able to explicitly see this transition because of poor data resolution in the transition region, coming from using UV boundary conditions on the scalar fields on the D7--brane world--volume.  The
origin of this phase transition, as we shall see, is as follows: The
generic behavior of an allowed solution for $L(u)$ as in
equation~(\ref{eqt: L}), is not enough to determine whether the
behavior corresponds to an $S^3$--vanishing D7--brane or an
$S^1$--vanishing D7--brane. The choices of branch of solutions have
different values of $c$, generically. In other words, for a given
value of $m$ there can be more than one value of $c$. There are
therefore two or more candidate solutions potentially controlling the
physics. The actual physical solution is the one which has the lowest
value for the D7--brane's free energy. The key point is that, at a
certain value of the mass, the lowest energy solution may suddenly
come from a different branch, and, as the corresponding value of the
condensate changes discontinuously in moving between branches, we find
that the system therefore undergoes a first order phase transition.  On the gauge theory side, we can imagine a similar situation occurring; two different branches of solution are competing, and the lowest energy branch is always picked.  
We are able to uncover this physics by doing a careful numerical analysis of the equations of motion for the probe dynamics on the gravity side of the AdS/CFT correspondence, by using IR boundary conditions instead of UV boundary conditions.

\section*{3.2 \hspace{2pt} Holographic meson melting}
\addcontentsline{toc}{section}{3.2 \hspace{0.15cm} Holographic meson melting}

\subsection*{3.2.1 \hspace{2pt} General set up}
\addcontentsline{toc}{subsection}{3.2.1 \hspace{0.15cm} General set up}
We begin by reviewing the physics of the D7--brane probe in the
AdS$_5$--Schwarzschild background solution\cite{Babington:2003vm}.  The background metric is
given by:
\begin{eqnarray} \label{eqt: our metric}
&& ds^2 = - \frac{f(u)}{R^2} dt^2 + \frac{R^2}{f(u)}  du^2 + \frac{u^2}{R^2} d\vec{x}\cdot d\vec{x} + R^2 d\Omega_5^2 \ , \quad f(u) = u^2 - \frac{b^4}{u^2} \ , \nonumber\\
&& {\rm where} \quad d\Omega_5^2 = d\theta^2 + \cos^2\theta d\Omega_3^2 + \sin^2 \theta d \phi^2 \ , \nonumber\\
&& {\rm and} \quad d\Omega_3^2 = d\psi^2 + \cos^2 \psi d\beta^2 + \sin^2 \psi d \gamma^2 \ ,
\end{eqnarray}
where $\vec{x}$ is a three vector defining the three spacelike directions of the dual gauge theory, $u \in [0, \infty)$ is the radial coordinate and we are using standard polar coordinate on the $S^5$.  The quantity $R^2$ is given by:
\begin{eqnarray}
R^2 &=& \sqrt{4 \pi g_s N_c} \alpha' \nonumber \ ,
\end{eqnarray}
where $g_s$ is the string coupling (which, with the inverse string
tension $\alpha^\prime$ sets for example, Newton's constant).  The quantity $b$ is related to the mass of the black-hole, $b^2=8 G_5 m_\mathrm{b.h.} / (3 \pi)$.  The
temperature of the black hole can be extracted using the standard
Euclidean continuation and requiring regularity at the horizon.
Doing this in the metric given by equation~(\ref{eqt: our metric}), we
find that $\beta^{-1}=b/\pi R^2$.  Therefore, by picking the value of $b$, we are choosing at what temperature we are holding the theory.

Later we will use different coordinate system for convenience, so here we introduce the change of variables:
\begin{eqnarray} \label{eqt:changeofcoordinates}
&r^2& = \frac{1}{2}(u^2+\sqrt{u^4-b^4}) = \rho^2+L^2 \ ,\\ \mathrm{with}\quad
&\rho& = r\cos\theta\ , \,\, L = r\sin\theta\nonumber \ .
\end{eqnarray}
The expression for the metric now takes the form:
\begin{eqnarray}\label{eqt: metricchange}
ds^2/\alpha' & = & -\left(\frac{(4r^4-b^4)^2}{4r^2R^2(4r^4+b^4)}\right)dt^2+\frac{4r^4+b^4}{4R^2r^2}d\vec{x}^2 \nonumber\\
& + & \frac{R^2}{r^2}(d\rho^2+\rho^2d\Omega_{3}^2+dL^2+L^2d\phi^2)  \ . 
\end{eqnarray}
Note that with this variable change $r \to u$ as $u \to \infty$.

We choose to embed the D7--brane probe transverse to $\theta$ and $\phi$ {\it a la} the background in (\ref{eqt: our metric}).  In order to study the embeddings with the lowest value of the on--shell action (and hence the lowest free energy), we choose an ansatz of the form $\phi = 0$ and $\theta = \theta(u)$.  The asymptotic separation of the D3-- and D7--branes is given by $L(u) = u \sin \theta$.  Given this particular choice of embedding, the world--volume of the D7--brane is given by:
\begin{eqnarray}
 \sqrt{-g} &=& u^2 \cos^3 \theta(u) \sqrt{\det S^3} \sqrt{u^2 +\left(u^4-b^4 \right)\theta'(u)^2} \ ,
\end{eqnarray}
where $g$ is the determinant of the induced metric on the D7--brane given by the pull--back of the space--time metric $G_{\mu \nu}$.  We are interested in two particular cases.  First, there is the case where $u$ goes to~$b$,
which corresponds to the D7--brane probe falling into the event horizon.  In the Euclidean section, this case corresponds to the shrinking of the $S^1$ of periodic time.
we will refer to these solutions as the `black hole' embeddings. Second, there is the case where $\theta$ goes to~$\pi/2$, which corresponds to the shrinking of the $S^3$.  We will name these Minkowski embeddings.  It is this change in topology ($S^1$ versus $S^3$ shrinking) between the different solutions that will correspond to a phase transition.
The classical equation of motion for $\theta(u)$ is:
\begin{eqnarray} \label{eqt: eqt of motion}
\frac{d}{d u} \left( \frac{u^2\left(u^4-b^4 \right) \theta' \cos^3 \theta} {\sqrt{u^2 +\left(u^4-b^4 \right){\theta'}^2}} \right) + 3 u^2 \cos^2 \theta\sin \theta\sqrt{u^2 +\left(u^4-b^4 \right){\theta'}^2} &=& 0\ .
\end{eqnarray}
When $u$ goes to infinity and the background metric becomes asymptotically AdS$_5 \times S^5$, the equation of motion reduces to:
\begin{eqnarray}
\frac{d}{d u} \left(u^5 \theta'(u) \right) + 3 u^2 \theta(u) &=& 0\ ,
\end{eqnarray}
which has solution:
\begin{eqnarray}
\theta(u) &=& \frac{1}{u} \left(m + \frac{c}{u^2} \right) \ .
\end{eqnarray}
These two terms are exactly the non--normalizable and normalizable terms corresponding to a dimension 3 operator ($\bar{\psi} \psi$) in the dual field theory with source $m$ and vacuum expectation value (vev) $c$.

Next we solve equation~(\ref{eqt: eqt of motion}) numerically using a shooting technique. Shooting from infinity towards the horizon, physical solutions are those that have a finite value at the horizon. This will only be accomplished for a particular $m$ and $c$ value from equation
(\ref{eqt: L}) which would have to be delicately chosen by hand.
Therefore, if we instead start from the horizon with a finite
solution, it will shoot towards the physical asymptotic solutions we
desire. In order to be able to analyze the phase transition between
the black hole and Minkowski solutions, we shoot from the
horizon for the condensate solutions and from $\theta= \pi/2$ for the
Minkowski solutions. This technique avoids having to correctly
choose the boundary conditions at infinity, which is a sensitive
procedure, allowing us to have many more data points to analyze the
phase transition. We impose the boundary condition that, at
our starting point--the horizon, we have:
\begin{eqnarray} \label{eqt: bc}
\theta'(u) \big|_{S^1\to 0} &=&  \frac{3 b^2}{8}  \tan \theta(b) \nonumber \\
\theta'(u) \big|_{S^3 \to 0} &=& \infty\ .
\end{eqnarray}
We argue that this is the physical boundary condition to take; the first is simply a result of taking the limit of $u \to b$ in the equation of motion in equation~(\ref{eqt: eqt of motion}), whereas the second is a result of requiring no conical singularity as the $S^3$ shrinks to zero size \cite{Karch:2006bv}.

\begin{figure}[!ht]
\begin{center}
\includegraphics[angle=0,
width=11cm]{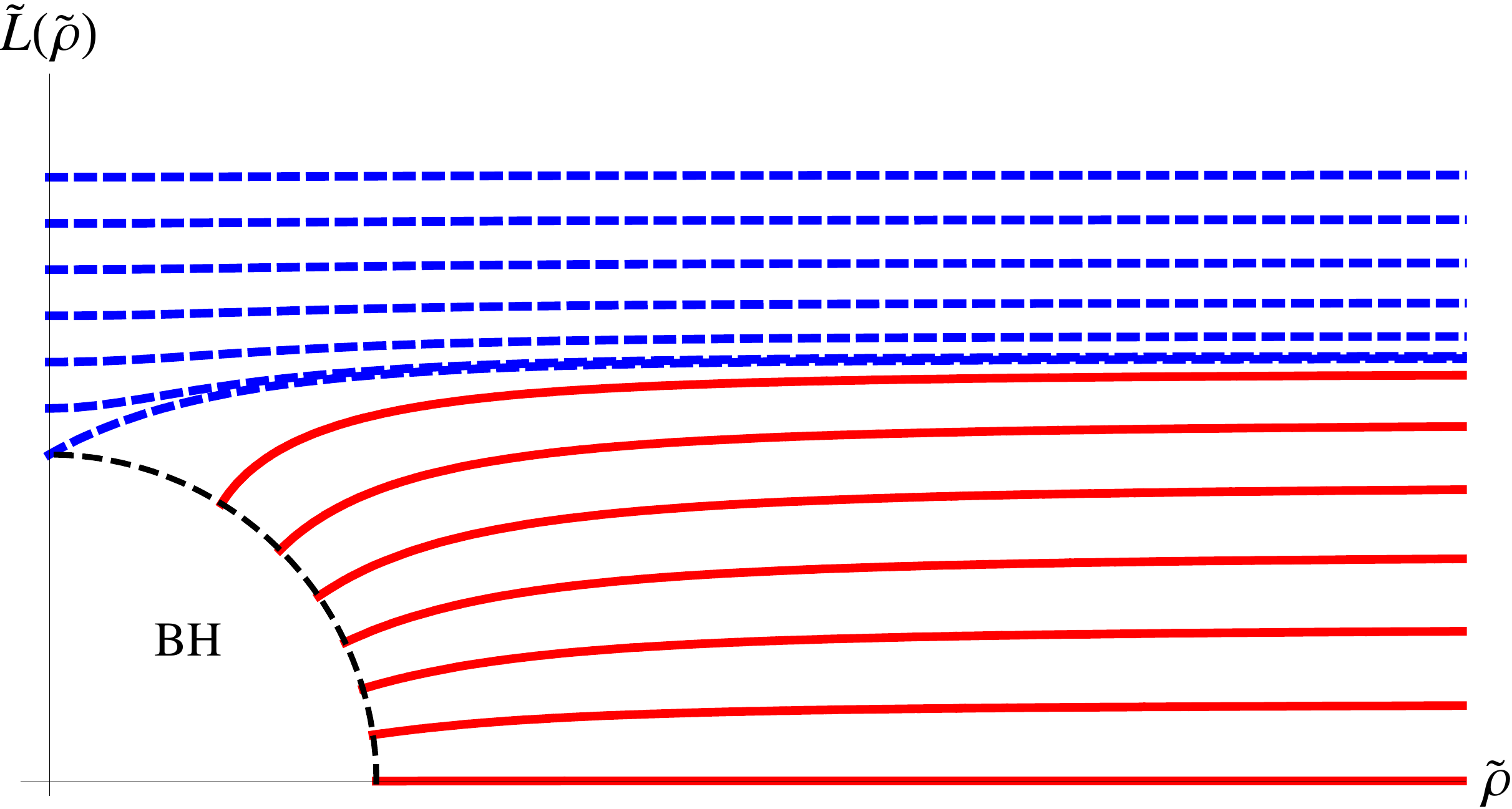}
\caption{\small Solutions for the D7--brane probe in the black hole background. }
\label{fig:solutionsinLandrho}
\end{center}
\end{figure}

For the sake of visualization, we will present the profile of the probe branes in $\{\rho, L\}$-plane. Therefore we define the following dimensionless quantities
\begin{eqnarray}
&& \tilde{\rho} = \frac{\rho}{b} \ , \quad \tilde{L} = \frac{L}{b} \ , \quad \tilde{m} = \frac{m}{b} \ , \\
&& \tilde{L} = \tilde{m} + \frac{\tilde{c}}{\tilde{\rho}^2} + \ldots \ .
\end{eqnarray}
In this coordinate system several D7--brane embedding solutions are shown in Figure~\ref{fig:solutionsinLandrho}.  The red (solid) lines correspond to Minkowski solutions, and the blue (dashed) lines correspond to black hole solutions.  From each of these solutions, we can extrapolate the bare quark mass and quark condensate vev.
%

\subsection*{3.2.2 \hspace{2pt} The first order phase transition}
\addcontentsline{toc}{subsection}{3.2.2 \hspace{0.15cm} The first order phase transition}

\begin{figure}[!ht]
\begin{center}
\subfigure[] {\includegraphics[angle=0,
width=0.45\textwidth]{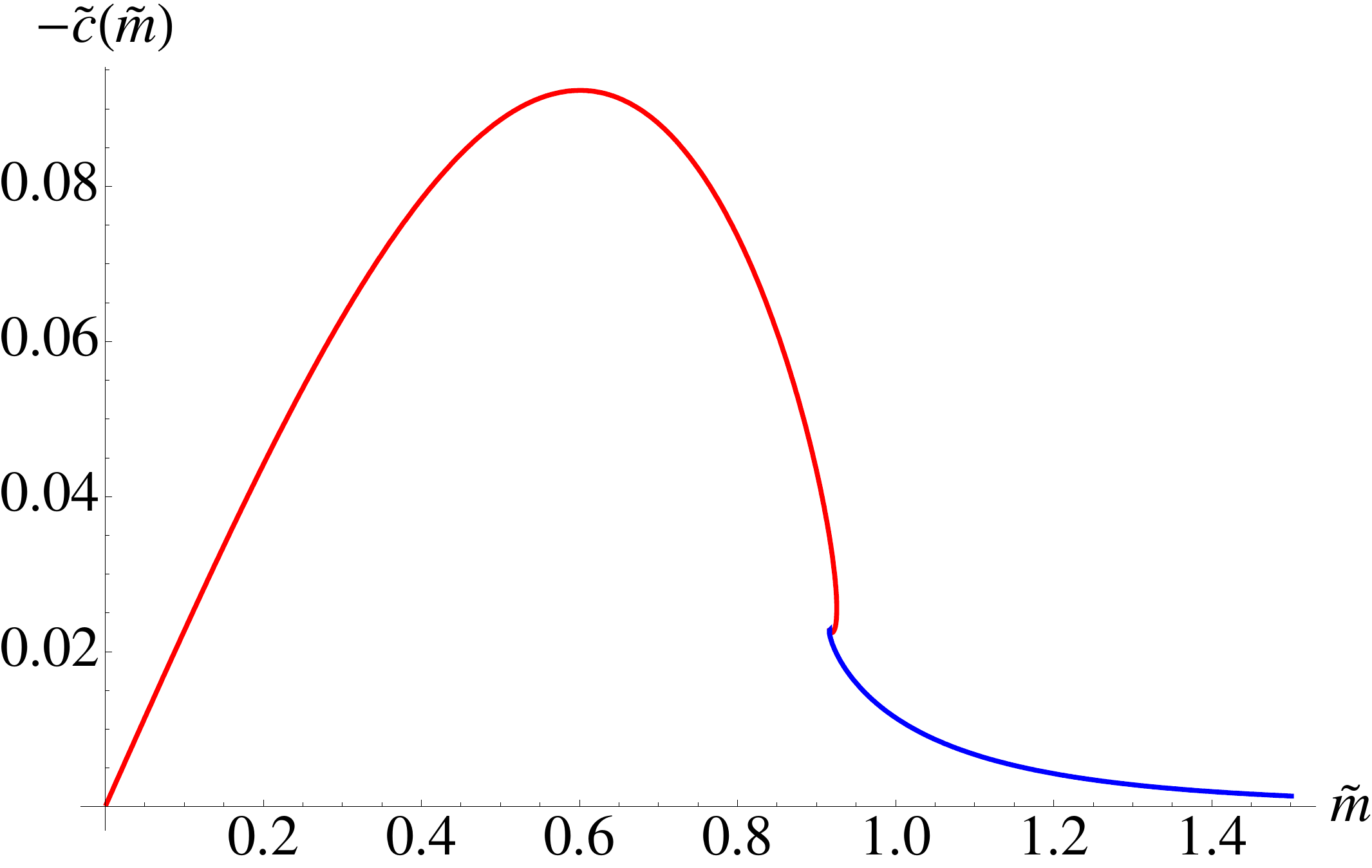} \label{fig: c vs m}}
\subfigure[] {\includegraphics[angle=0,
width=0.45\textwidth]{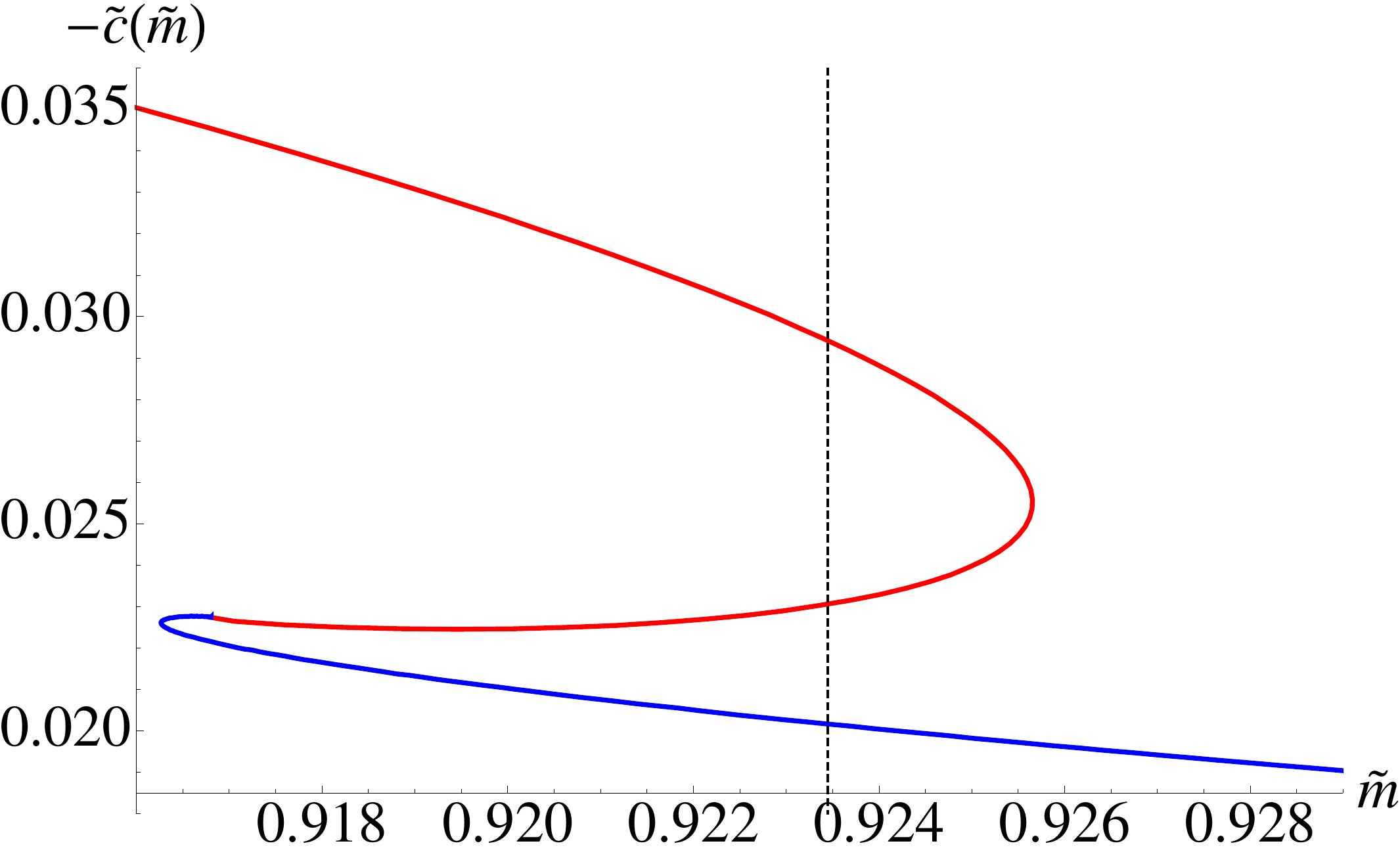} \label{fig: c vs m zoom}}
\caption{\small The condensate vev at the phase transition. See text for more details.}
\end{center}
\end{figure}

We plot the $c$ values as a function of $m$ in Figure~\ref{fig: c vs m}.  When enlarged, as shown in Figure~\ref{fig: c vs m zoom}, we find, as anticipated in the introductory remarks, the multi--valuedness in $c$ for a given $m$. Physics will choose just one answer for $c$. There is therefore the possibility of a transition from one branch to another as one changes $m$.

In order to determine exactly where the transition takes place, we have to calculate the free energy of the D7--brane.  In the semi-classical limit that we are considering, the free energy is given by the on--shell action times $\beta^{-1}$.  For our case, this is simply given by \cite{Kruczenski:2003uq}:
\begin{eqnarray}
\mathcal{F} &=& \beta^{-1} \tau_7 N_f \int d^4 x \ d\Omega_3 \ du \, \sqrt{-\det g}  \ ,
\end{eqnarray}
where here $N_f = 1$.  
Alternatively we can determine the phase transition point by considering an equivalent Maxwell construction. Since $c$ and $m$ are thermodynamically conjugate to each other, the free energy difference can be expressed as $d \mathcal{F} \sim - c \, dm$. Therefore the phase transition will happen exactly at the point where the areas under the curve separated by the vertical dashed line in Figure~\ref{fig: c vs m zoom} are equal. Thus we find a first order phase transition ---at $m\approx 0.92345$--- where the condensate's vev jumps discontinuously, which is shown with the dashed black line in figure~\ref{fig: c vs m zoom}.

This first order phase transition was studied in details in ref.~\cite{Hoyos:2006gb}, where the authors developed an appropriate formalism to study the meson spectrum of the theory before and after the phase transition. They were able to show that after the phase transition there are no bound meson states, but only quasi-normal excitations corresponding to a melting mesons. This is why the present understanding of the observed first order phase transition is that it corresponds to the confinement/deconfinement phase transition of the fundamental matter which in the case of consideration is the meson melting phase transition.

\section*{3.3 \hspace{2pt} Introducing an external magnetic field}
\addcontentsline{toc}{section}{3.3 \hspace{0.15cm} Introducing an external magnetic field}

We will investigate the effect of an external constant magnetic field in the dynamics of probe flavours in the thermal $\mathcal{N} = 4$ Yang-Mills theory. The material presented here is based on the work done in collaboration with Tameem Albash, Veselin Filev and Clifford Johnson\cite{Albash:2007bk}. Similar work was also done in ref.~\cite{Erdmenger:2007bn}. Ref.~\cite{Filev:2007gb} analyzed the dynamics of probe flavours in a background constant magnetic field at zero temperature, which is the premise of the finite temperature physics we will discuss in this chapter.

It was shown in ref.~\cite{Filev:2007gb} that the presence of an external magnetic field induces a spontaneous chiral symmetry breaking by generating a non-zero chiral condensate, where the chiral symmetry is an ${\rm U}(1)_{\rm R}$ symmetry as explained in chapter 1. However it was described in ref.~\cite{Babington:2003vm} that finite temperature melts the mesons and restores this symmetry at zero bare quark mass. Therefore we have two competing parameters and we will investigate their effects on the phase structure of the fundamental flavours.

We will consider similar embedding of the probe brane represented by $\phi = 0$, $ L \equiv L(\rho)$. An external constant magnetic field can be introduced in this set-up by exciting  an Abelian gauge field on the worldvolume of the probe brane\cite{Filev:2007gb}\footnote{It is easy to check {\it a posteriori} that this ansatz is consistent with the equations of motion and the Bianchi identity for the field strength of this gauge potential.}
\begin{equation}\label{eqt: bfield}
A_3 = H x^2 \ ,
\end{equation}
where $H$ is a real constant. Exciting a non-trivial gauge field on the worldvolume of the probe brane (resulting in a non-trivial value for $F_{ab}$) should not backreact on the background in the probe limit, however it will affect the probe brane. Also note that exciting a constant anti-symmetric field as in (\ref{eqt: bfield}) on the worldvolume of the probe brane corresponds to exciting a non-normalizable mode in the boundary gauge theory. It is fairly evident from the form of (\ref{eqt: bfield}) that this anti-symmetric field should be identified with an external constant magnetic field (the strength of which is denoted by $H$).

On the gauge theory perspective, this magnetic field therefore couples only to the flavour sector and does not affect the background adjoint fields. We will see that this magnetic field affects the finite temperature physics in many interesting ways.

To study the effects on the probe, let
us consider the general (Abelian) DBI action:
\begin{eqnarray}
S_{DBI}=- N_f  T_{D7} \int\limits_{{\cal M}_{8}}d^{8}\xi \ \mathrm{det} ^{1/2}(P[G_{ab}+B_{ab}]+2\pi\alpha' F_{ab})\ , \label{DBI1}
\end{eqnarray}
where $T_{D7}=\mu_7 / g_s = [(2\pi)^7\alpha'^4 g_s]^{-1}$ is the
D7--brane tension, $P[G_{ab}]$ and $P[B_{ab}]$ are the induced metric
and induced $B$--field on the D7--branes' world--volume, $F_{ab}$ is
the world--volume gauge field, and $N_f=1$ here. The resulting Lagrangian is:
\begin{equation}
{\cal L}=-\rho^3\left(1-\frac{b^8}{16 \left(\rho^2+L(\rho)^2\right)^4}\right) \left\{1+\frac{16 H^2 \left(\rho^2+L(\rho)^2\right)^2 R^4}{\left(b^4+4 \left(\rho^2+L(\rho)^2\right)^2\right)^2}\right\}^{\frac12}
   \sqrt{1+L'(\rho)^2} \ .
 \end{equation}
Note that at zero temperature the corresponding Lagrangian is given by
\begin{eqnarray}
{\cal L}=-\rho^3 \left\{1+\frac{ H^2 R^4}{ \left(\rho^2 + L(\rho)^2\right)^2 }\right\}^{\frac12} \sqrt{1+L'(\rho)^2} \ ,
\end{eqnarray}
which was analyzed in detail in ref.~\cite{Filev:2007gb}. 
For large $\rho \gg b$, the Lagrangian asymptotes to:
\begin{equation}
{\cal L}\approx-\rho^3\sqrt{1+L'(\rho)^2} \ ,
\end{equation}
which suggests the following asymptotic behavior for the embedding
function $L(\rho)$:
\begin{equation}
L(\rho)=m+\frac{c}{\rho^2}+\dots \ ,
\label{asymptote}
\end{equation}
where the parameters $m$ (the asymptotic separation of the D3-- and
D7--branes) and $c$ (the degree of transverse bending of the D7--brane
in the $(\rho,\phi)$ plane) are related to the bare quark mass
$m_{q}=m/2\pi\alpha'$ and the quark condensate
$\langle\bar\psi\psi\rangle\propto -c$ respectively
\cite{Kruczenski:2003uq}.

To proceed, it is convenient to define the following dimensionless
parameters:
 \begin{eqnarray}
 \tilde\rho&=&\frac{\rho}{b} \ , ~~~\eta=\frac{R^2}{b^2}H \ , \quad {\tilde m}=\frac{m}{b}\ ,\\
 \tilde L(\tilde\rho)&=&\frac{L(b\tilde\rho)}{b}=\tilde m+\frac{\tilde c}{\tilde \rho^2}+\dots \ .\nonumber
 \label{formula1}
 \end{eqnarray}
This leads to the Lagrangian:
\begin{equation}
\tilde{\cal L}=-\tilde\rho^3\left(1-\frac{1}{16 \left(\tilde\rho^2+\tilde L(\tilde\rho)^2\right)^4}\right) \left\{1+\frac{16\left(\tilde\rho^2+\tilde L(\tilde\rho)^2\right)^2 \eta^2}{\left(1+4 \left(\tilde\rho^2+\tilde L(\tilde\rho)^2\right)^2\right)^2}\right\}^{\frac12}\sqrt{1+\tilde L'(\tilde\rho)^2} \ .
\label{LagrangianM}
\end{equation}
For small values of $\eta$, the analysis of the second order,
non--linear differential equation for $\tilde L(\tilde\rho)$ derived
from equation (\ref{LagrangianM}) follows closely that performed in
refs.~\cite{Albash:2006ew,Babington:2003vm,Mateos:2006nu}. The
solutions split into two classes: the first class are solutions
corresponding to embeddings that wrap a shrinking $S^3$ in the $S^5$
part of the geometry and (when the $S^3$ vanishes) closes at some
finite radial distance $r$ above the black hole's horizon which is
located at $r=b/\sqrt{2}$.  These embeddings are referred to as
`Minkowski' embeddings.  The second class of solutions correspond to
embeddings falling into the black hole, since the $S^1$ of the
Euclidean section, on which the D7--branes are wrapped, shrinks away
there.  These embeddings are referred to as `black hole' embeddings.
As we have demonstrated in the previous section, this gives rise to a first order melting transition at some critical mass $m_{\rm cr}$. We will now 
show that the effect of the magnetic field is to
decrease this critical mass, and, at some critical magnitude of the
parameter $\eta_{\rm cr}$, the critical mass drops to zero.  For $\eta >
\eta_{\rm cr}$ the phase transition disappears, and only the Minkowski
embeddings are stable states in the dual gauge theory, possessing a
discrete spectrum of states corresponding to quarks and anti--quarks
bound into mesons. Furthermore, at zero bare quark mass, we have a
non--zero condensate and the chiral symmetry is spontaneously broken.

 \section*{3.4 \hspace{2pt} Properties of the solution}
\addcontentsline{toc}{section}{3.4 \hspace{0.15cm} Properties of the solution}

 \subsection*{3.4.1 \hspace{2pt} Exact results at large mass}
\addcontentsline{toc}{subsection}{3.4.1 \hspace{0.15cm} Exact results at large mass}

 It is instructive to first study the properties of the solution for
 $\tilde m \gg 1$. This approximation holds for finite temperature,
 weak magnetic field, and large bare quark mass $m$, or, equivalently,
 finite bare quark mass $m$, low temperature, and weak magnetic field.

 In order to analyze the case $\tilde m\gg 1$, let us write $\tilde
 L(\tilde{\rho})=\tilde a +\zeta(\tilde{\rho})$ for $\tilde{a} \gg 1$
 and linearize the equation of motion derived from equation
(\ref{LagrangianM}), while leaving only the first two leading terms in
 $(\rho^2+\tilde m^2)^{-1}$. The result is:
\begin{equation}
\partial_{\tilde\rho}(\tilde\rho^3\zeta')-\frac{2\eta^2}{(\tilde m^2+\tilde\rho^2)^3}\tilde m+\frac{2(\eta^2+1)^2-1}{2(\tilde m^2+\tilde\rho^2)^5}\tilde m+O(\zeta)=0 \ .
\label{bigm}
\end{equation}
 Ignoring the $O(\zeta)$ terms in equation (\ref{bigm}), the general solution takes the form:
 \begin{equation}
 \zeta(\tilde \rho)=-\frac{\eta^2}{4\rho^2(\tilde m^2+\tilde\rho^2)}\tilde m+\frac{2(\eta^2+1)^2-1}{96\tilde\rho^2(\tilde m^2+\tilde\rho^2)^3}\tilde m \ ,
 \label{weakL}
 \end{equation}
where we have taken $\zeta'(0) = \zeta(0) = 0$.  By studying the asymptotic behavior of this solution, we can extract the following:
\begin{eqnarray}
\tilde{m} &=& \tilde{a} - \frac{\eta^2}{4 \tilde{a}^3} + \frac{1+ 4 \eta^2 +2 \eta^4}{32 \tilde{a}^7} + O\left(\frac{1}{\tilde{a}^{7}} \right) \nonumber \ , \\
\tilde{c} &=& \frac{\eta^2}{4 \tilde{a}} - \frac{1+ 4\eta^2 +2 \eta^4}{96 \tilde{a}^5} + O\left(\frac{1}{\tilde{a}^7} \right)\ .
\end{eqnarray}
By inverting the expression for $\tilde{m}$, we can express $\tilde{c}$ in terms of $\tilde{m}$:
\begin{eqnarray}
\tilde{c} &=& \frac{\eta^2}{4 \tilde{m}} - \frac{1+ 4 \eta^2 + 8 \eta^4}{96 \tilde{m}^5} + O \left(\frac{1}{\tilde{m}^7} \right) \label{weakcond} \ .
\end{eqnarray}
Finally, after going back to dimensionful parameters, we can see that the theory has developed a fermionic  condensate:
\begin{equation}
\langle\bar\psi\psi\rangle\propto-c=-\frac{R^4}{4m}H^2+\frac{b^8 + 4 b^4 R^4 H^2 + 8 R^8 H^4}{96m^5} \ .
\label{weakc}
\end{equation}
The results of the above analysis can be trusted only for finite bare
quark mass and sufficiently low temperature and weak magnetic field.
As can be expected, the physically interesting properties of the
system should be described by the full non--linear equation of motion
of the D7--brane. To explore these we need to use numerical
techniques.

\subsection*{3.4.2 \hspace{2pt} Numerical analysis}
\addcontentsline{toc}{subsection}{3.4.2 \hspace{0.15cm} Numerical analysis}  

 We solve the differential equation derived from equation
 (\ref{LagrangianM}) numerically. It is convenient to
 use infrared initial conditions \cite{Albash:2006bs,Albash:2006ew}.
 For the Minkowski embeddings, based on symmetry arguments, the
 appropriate initial conditions are:
\begin{equation}
\tilde L(\tilde\rho)|_{\tilde\rho=0}=L_{\rm in}\ , \quad \tilde L'(\tilde\rho)|_{\tilde\rho=0}=0 \ .
\end{equation}
For the black hole embeddings, the following initial conditions:
\begin{equation}
\tilde L(\tilde\rho)|_{\rm e.h.}=\tilde L_{\rm in}\ , \quad \tilde L'(\tilde\rho)|_{\rm e.h.}=\left.\frac{\tilde L}{\tilde\rho}\right|_{\rm e.h.} \ ,
\end{equation}
ensure regularity of the solution at the event horizon. After solving
numerically for $\tilde L(\tilde\rho)$ for fixed value of the
parameter $\eta$, we expand the solution at some numerically large
$\tilde\rho_{\rm max}$, and, using equation (\ref{asymptote}), we generate
the plot of $-\tilde c$ vs $\tilde m$.

In section 3.2, we have discussed the nature of the phase transition and the associated Maxwell construction for the thermodynamic free energy. Here we will follow the same recipe in the presence of the additional parameter $\eta$ which controls the strength of the magnetic field.

\begin{figure}[!ht]
\begin{center}
\subfigure[] {\includegraphics[angle=0,
width=0.45\textwidth]{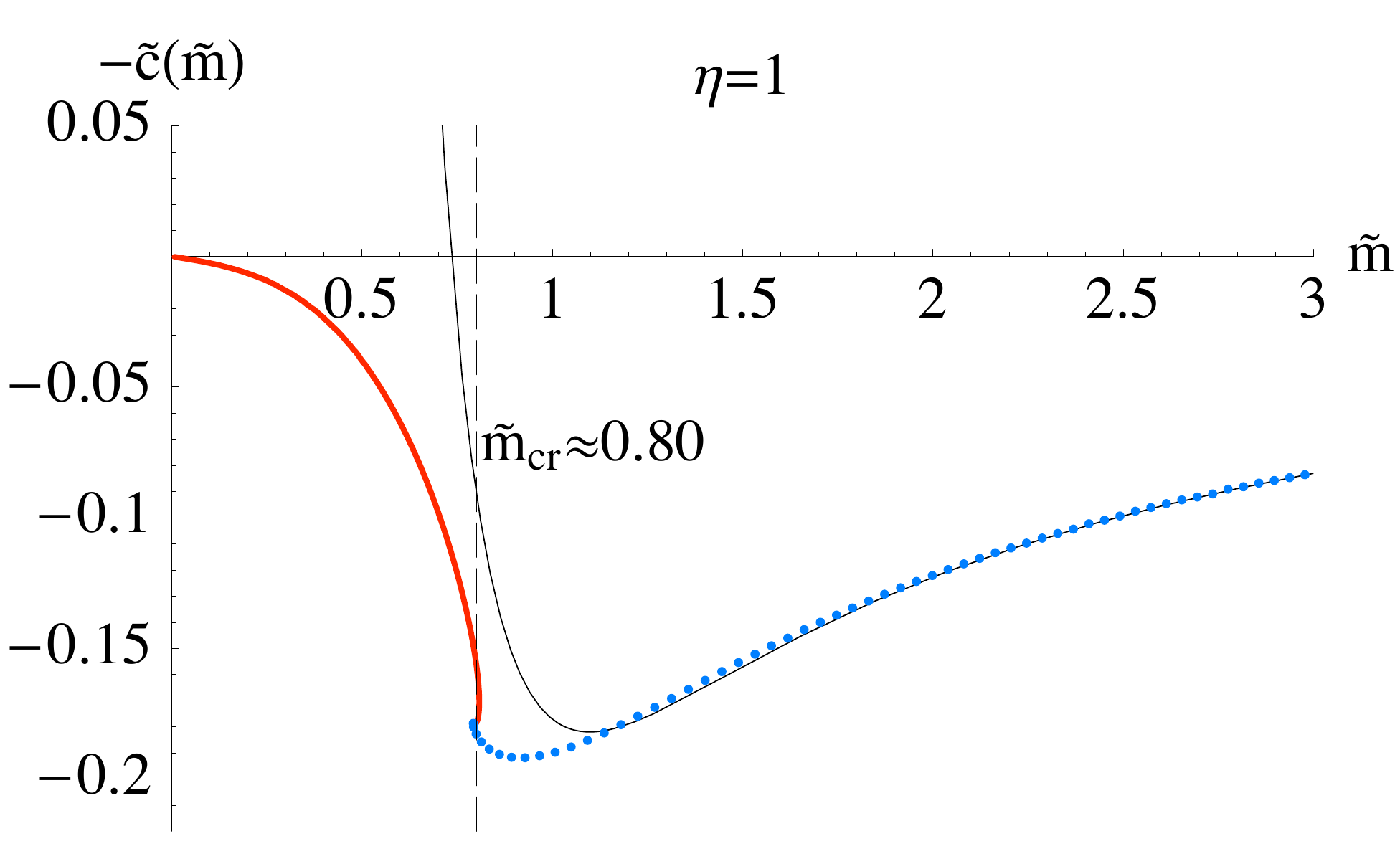}}
\subfigure[] {\includegraphics[angle=0,
width=0.45\textwidth]{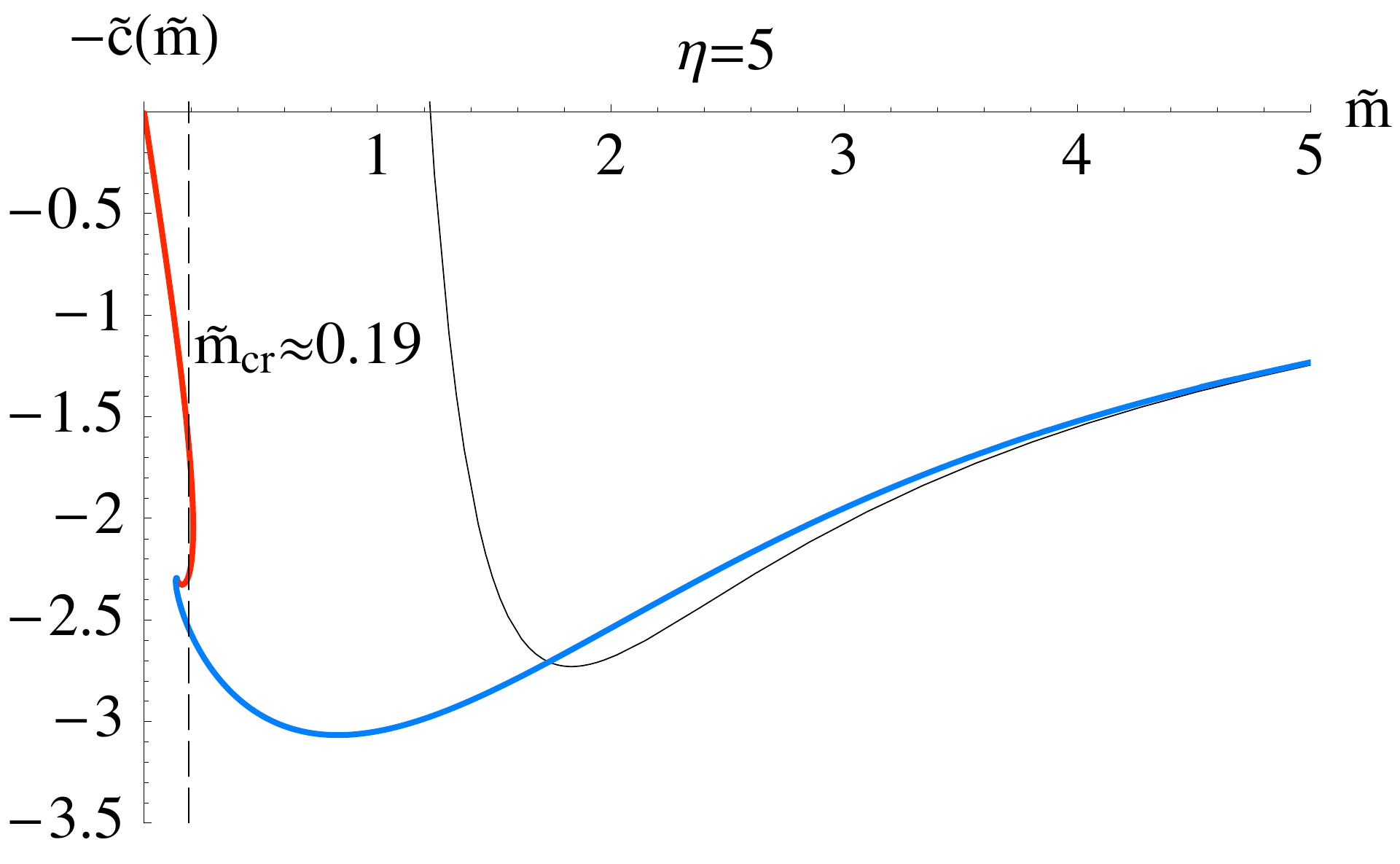}}
\caption{\small For strong magnetic  field the condensate is negative. The value of $\tilde m_{\rm cr}$ continues to drop as we increase $\eta$.}\label{fig:4}
\end{center}
\end{figure}

Let us turn on a magnetic field.  As one can see from Figure~\ref{fig:4}, the effect of the magnetic field is to decrease the
magnitude of $\tilde m_{\rm cr}$.  In addition, the condensate now becomes
negative for sufficiently large $\tilde m$ and approaches zero from
below as $\tilde m\to\infty$. It is also interesting that equation
(\ref{weakcond}) is still a good approximation for $\tilde m>\tilde
m_{\rm cr}$.

\begin{figure}[h]
   \centering
   \includegraphics[ width=11cm]{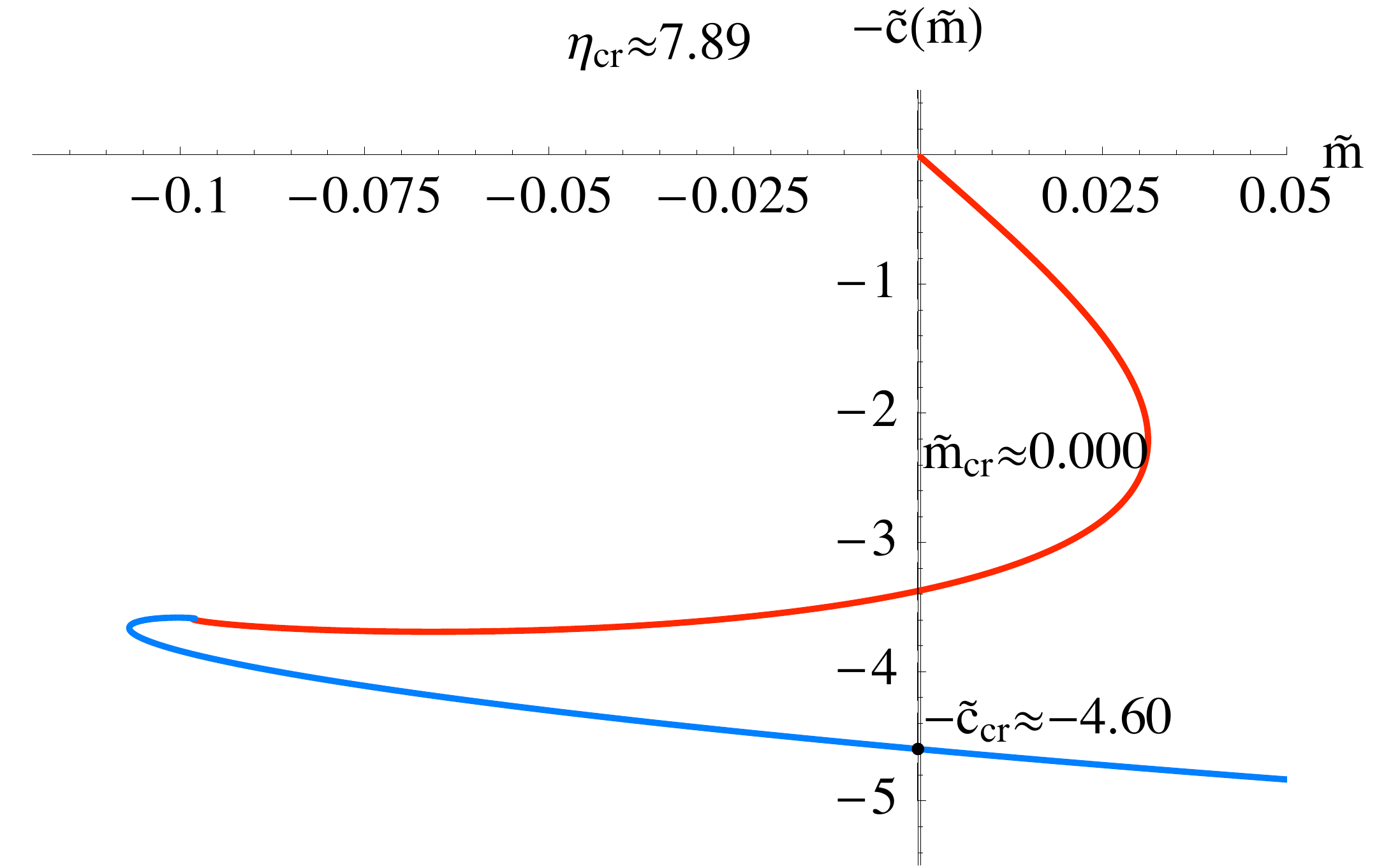}
   \caption{\small For $\eta=\eta_{\rm cr}$ the critical parameter $m_{\rm cr}$ vanishes. There are two $\tilde m=0$ states with equal energies, one of them has non--vanishing condensate $-\tilde c_{\rm cr}\approx -4.60$ and therefore spontaneously breaks the chiral symmetry. }
   \label{fig:6}
\end{figure}

For sufficiently strong magnetic field, the condensate has only
negative values and the critical value of $\tilde m$ continues to
decrease, as is presented in Figure~\ref{fig:4}.
   If we further increase the magnitude of the magnetic field, some
   states start having negative values of $\tilde m$, as shown in
   Figure~\ref{fig:6}. The negative values of $\tilde m$ do not mean
   that we have  negative bare quark masses; rather, it implies that
   the D7--brane embeddings have crossed $L = 0$ at least once. It was
   argued in ref.~\cite{Babington:2003vm} that such embeddings are not
   consistent with a holographic gauge theory interpretation and are
   therefore to be considered unphysical.  We will adopt this
   interpretation here, therefore taking as physical only the $\tilde
   m>0$ branch of the $-\tilde c$ vs $\tilde m$ plots.  However, the
   prescription for determining the value of $m_{\rm cr}$ continues to be
   valid, as long as the obtained value of $\tilde m_{\rm cr}$ is
   positive.  Therefore, we will continue to use it in order to
   determine the value of $\eta\equiv\eta_{\rm cr}$ for which $\tilde
   m_{\rm cr}=0$.

\begin{figure}[h]
   \centering
   \includegraphics[ width=11cm]{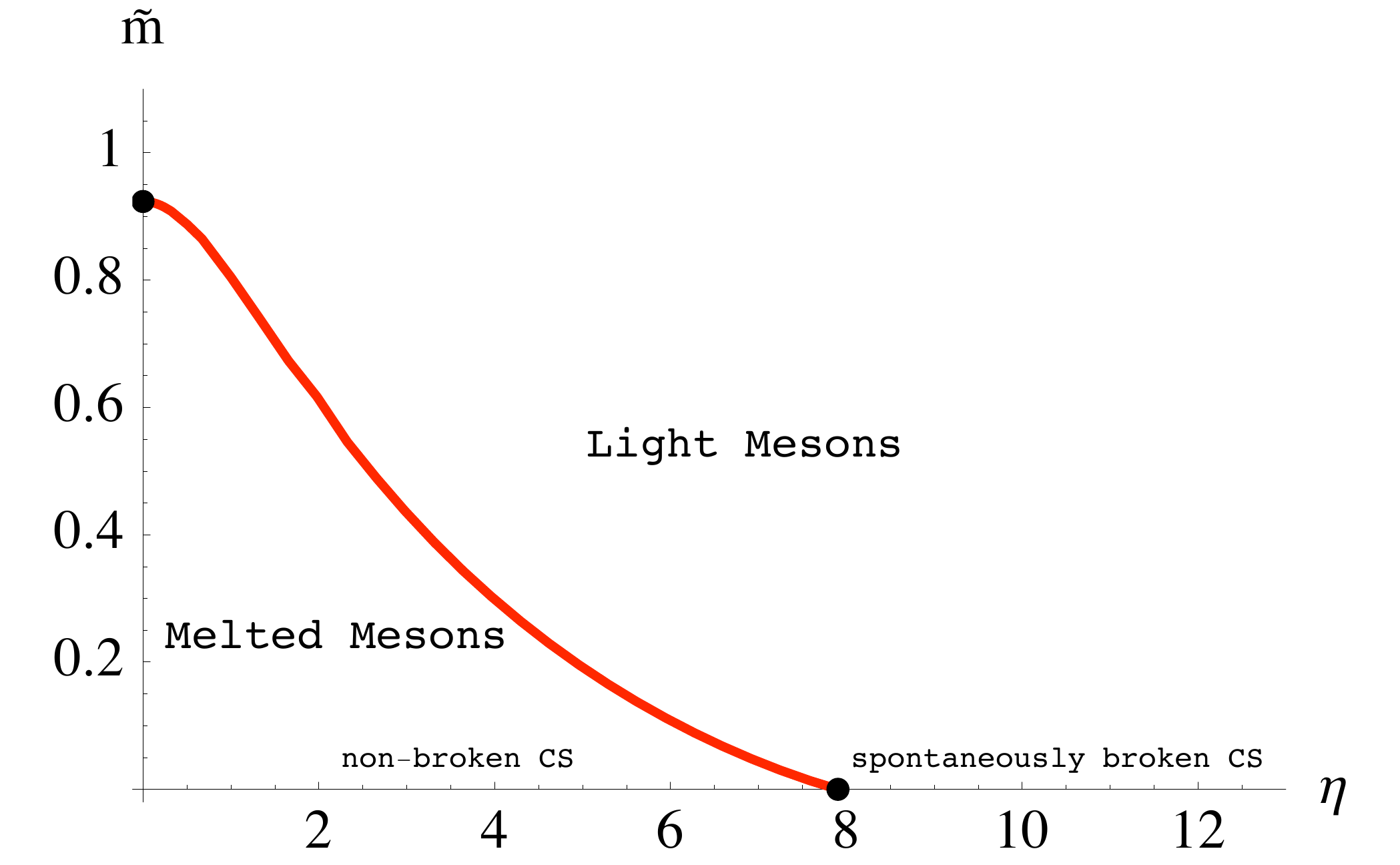}
   \caption{\small The curve separates the two phases corresponding to discrete meson spectrum (light mesons) and continuous meson spectrum (melted mesons). }
   \label{fig:phase diagram}
\end{figure}

As one can see in Figure~\ref{fig:6}, the value of $\eta_{\rm cr}$ that we
obtain is $\eta_{\rm cr}\approx 7.89$. Note also that, for this value of
$\eta$, the Minkowski $\tilde m=0$ embedding has a non--zero fermionic
condensate $\tilde c_{\rm cr}$, and hence the chiral symmetry is
spontaneously broken. For $\eta>\eta_{\rm cr}$, the stable solutions are
purely Minkowski embeddings, and the first order phase transition
disappears; therefore, we have only one class of solutions (the blue
curve) that exhibit spontaneous chiral symmetry breaking at zero bare
quark mass.  Some black hole embeddings remain meta--stable, but
eventually all black hole embeddings become unstable for large enough
$\eta$.  This is confirmed by our study of the meson spectrum which
we present in later sections of the chapter.
The above results can be summarized in a single two dimensional phase
diagram which we present in Figure~\ref{fig:phase diagram}.
The curve separates the two phases corresponding to a discrete meson
spectrum (light mesons) and a continuous meson spectrum (melted
mesons) respectively. The crossing of the  curve is associated with
the first order phase transition corresponding to the melting of the
mesons. If we cross the  curve along the vertical axis, we have the
phase transition described in refs.~\cite{Albash:2006ew,Babington:2003vm,
  Mateos:2006nu}. Crossing the curve along the
horizontal axis corresponds to a transition from unbroken to
spontaneously broken chiral symmetry, meaning the
parameter $\tilde c$ jumps from zero to $\tilde c_{\rm cr}\approx 4.60$,
resulting in non--zero quark condensate of the ground state.
It is interesting to explore the dependence of the fermionic
condensate at zero bare quark mass on the magnetic field. We can fix this from a simple
dimensional analysis:
\begin{equation}
c_{\rm cr}=b^3\tilde c_{\rm cr}(\eta)=\frac{\tilde c_{\rm cr}(\eta)}{\eta^{3/2}}R^3H^{3/2} \ .
\label{ccrit}
\end{equation}
In the $T\to 0$ limit, we should recover the result from ref.~\cite{Filev:2007gb}: $c_{\rm cr}\approx0.226R^3H^{3/2}$, which
implies that $\tilde c_{\rm cr}(\eta)\approx 0.226\eta^{3/2}$ for $\eta\gg
1$. This rather rich phase structure is accompanied by an equally rich thermodynamics. We refer the reader to the original paper.~\cite{Albash:2007bk} for an extensive account of the associated free energy, entropy and magnetization.

\section*{3.5 \hspace{2pt} Meson spectrum}
\addcontentsline{toc}{section}{3.5 \hspace{0.15cm} Meson spectrum}

In this section, we calculate the meson spectrum of the gauge theory.
The mesons we are considering are formed from quark--antiquark pairs,
so the relevant objects to consider are 7--7 strings.  In our
supergravity description, these strings are described by fluctuations
(to second order in $\alpha'$) of the probe branes' action about the
classical embeddings we found in the previous
sections \cite{Kruczenski:2003be}.  Studying the meson spectrum serves
two purposes.  First, tachyons in the meson spectrum from fluctuations
of the classical embeddings indicate the instability of the embedding.
Second, a massless meson satisfying a Gell-Mann-Oakes-Renner (GMOR)
relation \cite{GellMann:1968rz} will confirm that spontaneous chiral symmetry breaking has
occurred.

\subsection*{3.5.1 \hspace{2pt} The scalar fluctuation}
\addcontentsline{toc}{subsection}{3.5.1 \hspace{0.15cm} The scalar fluctuation}

In order to solve for the meson spectrum given by equation (\ref{eqt:eom_theta}), we consider an ansatz for the field $\chi$ of the form:
\begin{eqnarray}
\chi = h(\tilde{z}) \exp \left(- i \tilde{\omega} t \right) \ , \quad {\rm with} \quad z =  b^{-2} \tilde{z}  \ , \quad  \omega = R^{-2} b \ \tilde{\omega} \ .
\end{eqnarray}
In these coordinates, the event horizon is located at $\tilde{z} = 1$.

We have two different kinds of embeddings here, but we will consider only the massless black hole embedding. Our goal will be to demonstrate that the external magnetic field destabilizes this classical embedding and therefore gives rise to the chiral symmetry breaking phenomena observed in earlier sections. For the black hole embeddings, the appropriate fluctuation modes are the in--falling modes \cite{Hoyos:2006gb}, which yields the corresponding complex-valued quasinormal mode spectrum for $\tilde{\omega}$; the real part of $\tilde{\omega}$
corresponds to the mass of the meson before it melts, and the
imaginary part of $\tilde{\omega}$ is the inverse lifetime (to
a factor of 2) \cite{Hoyos:2006gb}.

\begin{figure}[!ht]
\begin{center}
\includegraphics[angle=0,
width=0.65\textwidth]{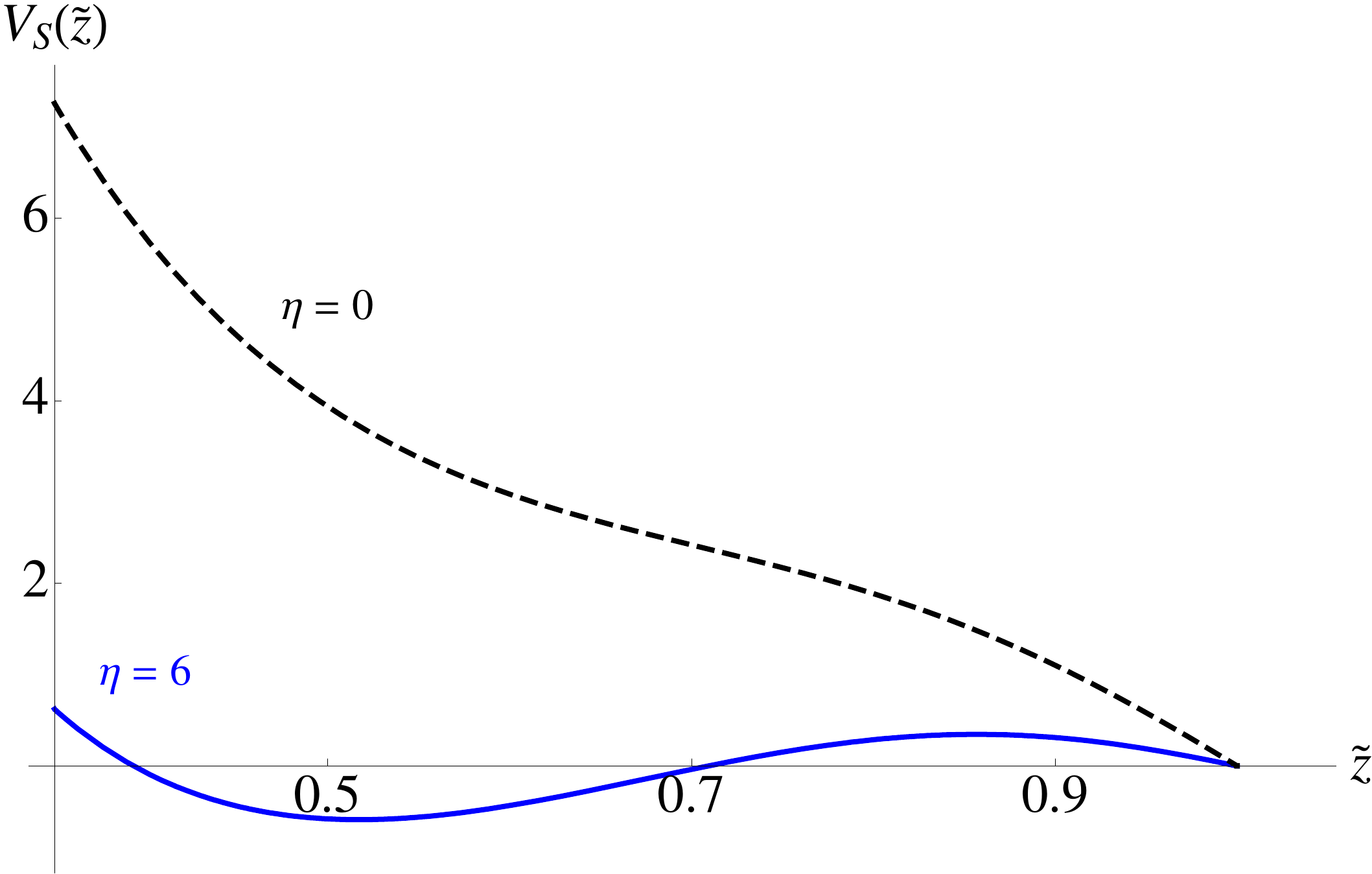}
\caption{\small The effective Schr\"{o}dinger potential for the scalar field fluctuation at two different values of the magnetic field.}
\label{fig:schrchi}
\end{center}
\end{figure}

The equation of motion (\ref{eqt:eom_theta}) for the fluctuation corresponding to the massless embedding, $\theta_0(\tilde{z}) = 0$, is given by :
\begin{eqnarray}
h''(\tilde{z}) + \left( \frac{2 \tilde{z}}{\tilde{z}^2 - 1} - \frac{1}{\tilde{z} \left(1+ \tilde{z}^2 \eta^2 \right)} \right) h'(\tilde{z}) + \frac{3 + \tilde{z} \left(-3 \tilde{z} + \tilde{\omega}^2 \right)}{ 4 \tilde{z}^2  \left(\tilde{z}^2 - 1\right)^2} h(\tilde{z}) &=& 0 \ .
\end{eqnarray}
Upon using the change of variables outlined in appendix C, we can obtain the effective Schr\"{o}dinger potential denoted by $V_S(\tilde{z})$. In this case, the exact form of this potential is not very illuminating, so we do not present it here. Alternatively, in figure \ref{fig:schrchi} we pictorially demonstrate the effect of the magnetic field on this potential. 

It is clear that a sufficiently large non-zero value of $\eta$ generates a locally negative potential well. This in turn suggests that the imaginary part of $\tilde{\omega}$ may become positive signaling the existence of tachyonic modes. Such tachyonic modes are purely sourced by the magnetic field which tends to disfavor the black hole embeddings. We will not explore the quasinormal modes in any more details, but refer to ref.~\cite{Albash:2007bk} where more extensive numerical studies have been performed.

\subsection*{3.5.2 \hspace{2pt} The vector fluctuation}
\addcontentsline{toc}{subsection}{3.5.2 \hspace{0.15cm} The vector fluctuation}

\begin{figure}[!ht]
\begin{center}
\includegraphics[angle=0,
width=0.65\textwidth]{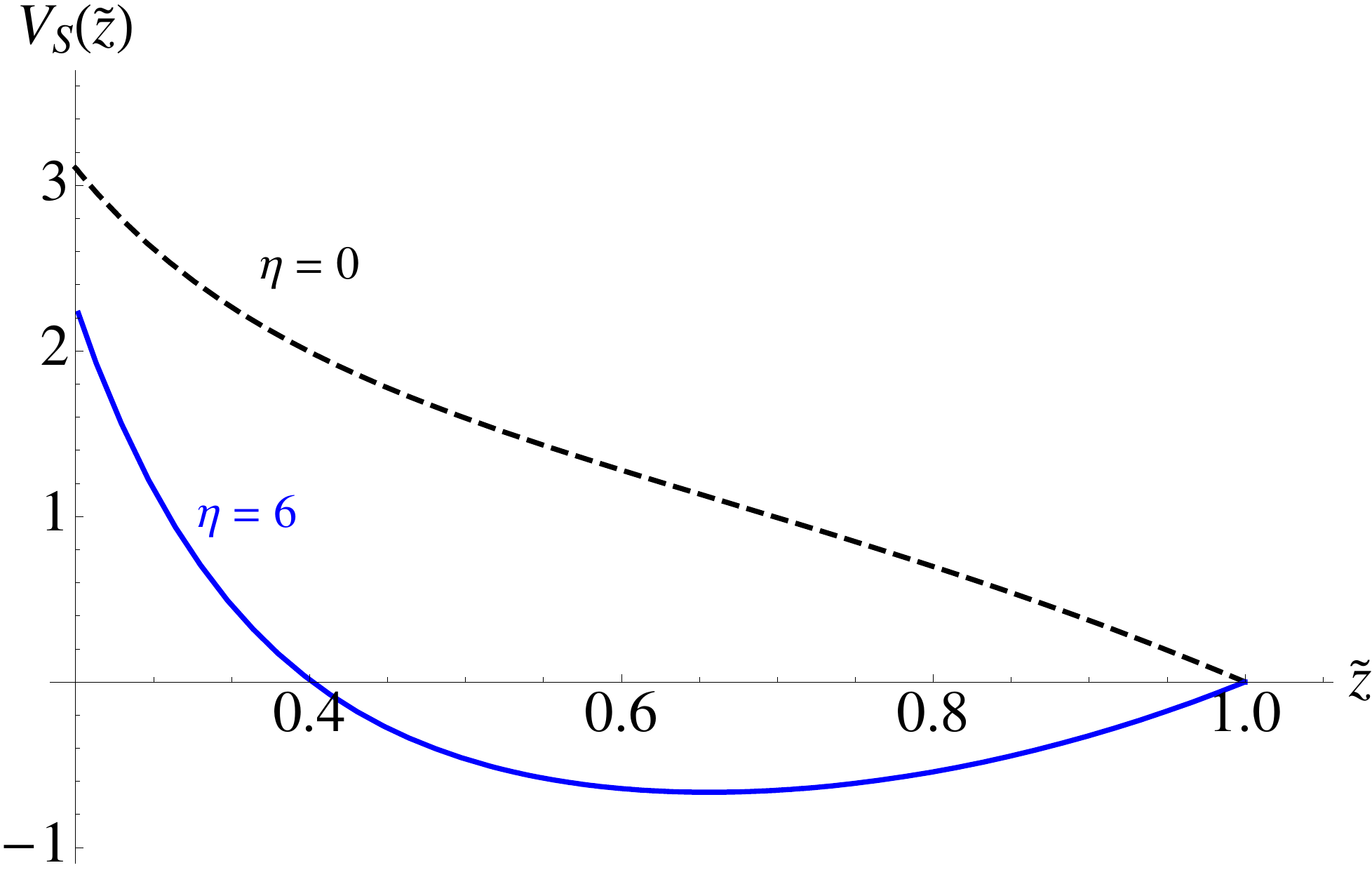}
\caption{\small The effective Schr\"{o}dinger potential for the gauge field fluctuation at two different values of the magnetic field.}
\label{fig:schrA1}
\end{center}
\end{figure}

In general the scalar fluctuation $\Phi$ and the vector fluctuation $A_1$ are coupled (as shown in
equations (\ref{eqt:eom_phi_A2}) and (\ref{eqt:eom_phi_A1})). However, for the massless embedding these coupled equations reduce to a single equation
\begin{eqnarray*}
A''(\tilde{z}) + \frac{\tilde{z} \left(2+\eta^2 \left( 3 \tilde{z}^2 -1 \right) \right)}{\left(\tilde{z}^2-1 \right) \left(1+ \tilde{z}^2 \eta^2 \right)} A'(\tilde{z}) + \frac{\tilde{\omega}^2}{4 \tilde{z} \left(\tilde{z}^2 - 1\right)^2 } A(\tilde{z})&=& 0 \ ,
\end{eqnarray*}
where we consider an ansatz (as before) of the form:
\begin{eqnarray*}
A_1 = A(\tilde{z}) \exp \left(- i \tilde{\omega} t \right) \ .
\end{eqnarray*}
Now we can again use the prescription in appendix C to obtain the effective Schr\"{o}dinger potential for the gauge field fluctuation. Again, instead of explicitly writing down the potential we pictorially demonstrate the effect of the magnetic field in figure~\ref{fig:schrA1}.

The potential has remarkable similarity to the potential of the scalar field fluctuation in figure~\ref{fig:schrchi} and we again see the local development of a negative energy potential well for sufficiently large magnetic fields. This further conforms to the magnetic field's role in destabilizing the black hole embeddings. More extensive numerical study has been carried out in ref.~\cite{Albash:2007bk}.

It is possible to numerically extract the meson spectrum for the Minkowski embeddings (when the scalar field $\Phi$ and the vector field $A$ are coupled) and obtain the analogue of Zeeman splitting and the Gell-Mann--Oakes--Renner relation. We will not elaborate on these aspects here, but point the interested reader to the relevant reference~\cite{Albash:2007bk}.

\section*{3.6 \hspace{2pt} Concluding remarks}
\addcontentsline{toc}{section}{3.6 \hspace{0.15cm} Concluding remarks}

We have extended the holographic study of large $N_c$ gauge theory in an
external magnetic field described in ref.~\cite{Filev:2007gb}, to the
case of finite temperature, allowing us to study the properties of the
quark dynamics when the theory is in the deconfined plasma phase.

The meson melting phase transition exists only below a critical value
of the applied field. This is the critical value above which
spontaneous chiral symmetry breaking is triggered (in the case of zero
mass). Above this value, regardless of the quark mass (or for fixed
quark mass, regardless of the temperature) the system remains in a
phase with a discrete spectrum of stable masses. Evidently, for these
values of the field, it is magnetically favorable for the quarks and
anti--quarks to bind together, reducing the degrees of freedom of the
system , as can be seen from our computation of the entropy.
Meanwhile, the magnetization is greater in this
un--melted phase.

There have been non--perturbative studies of fermionic models in
background magnetic field as briefly reviewed in chapter 2. Generally, those works use quite different methods to
examine aspects of the physics --- some primary non--perturbative
tools are the Dyson--Schwinger equations in various truncations). Our
results (and the zero temperature result obtained with these methods
in the zero temperature case \cite{Filev:2007gb}) are consistent with
the general expectations from those works, which is that strong
magnetic fields are generically expected to be a catalyst for
spontaneous chiral symmetry breaking in a wide class of models (see
{\it e.g.}, refs. \cite{Gusynin:1995nb,Semenoff:1999xv,Miransky:2002eb} for a discussion
of the conjectured universality of this result).


\chapter*{Chapter 4: \hspace{1pt} Phase structure of finite temperature large $N_c$ flavoured Yang--Mills theory in an external electric field}

\addcontentsline{toc}{chapter}{Chapter 4:\hspace{0.15cm}
Phase structure of finite temperature large $N_c$ flavoured Yang--Mills theory in an external electric field}

\section*{4.1 \hspace{2pt} Introductory remarks}
\addcontentsline{toc}{section}{4.1 \hspace{0.15cm} Introductory remarks}

The material presented in this chapter is based on the work in collaboration with Tameem Albash, Veselin Filev and Clifford Johnson\cite{Albash:2007bq}.

In this chapter, we study large $N_c$ ${\rm SU}(N_c)$ gauge theory with
non--backreacting hypermultiplet quark flavours in the
presence of a background electric field, at both zero and finite
temperature. The holographic description of this theory at finite baryon chemical potential was first studied in ref.~\cite{Karch:2007pd}, where the existence of a global electric current induced was demonstrated. Further study of this set up was considered in ref.~\cite{O'Bannon:2007in}. The phase structure of the theory at zero baryon chemical potential was studied in ref.~\cite{Albash:2007bq}.

We may expect the behavior of the theory to be quite different from the magnetic case considered in the previous chapter. In a sense the electric field is pulling the quarks apart, decreasing the bound energy and thus we may expect that the onset of the deconfinement phase transition will happen at lower temperature, than in the absence of an electric field. Furthermore, at zero temperature we may expect that for sufficiently strong electric field the binding energy of the quarks will be completely overcome and the mesons will dissociate into their constituent quarks.

It is worth noting that the phase transition will be driven by the quantum fluctuations, which will result in a quantum phase transition. On the other side, the mesons are electrically neutral, while the dissociated quarks aren't. Hence the dissociation can be seen as a {\it conductor/insulator} phase transition \cite{Albash:2007bq}.

\section*{4.2 \hspace{2pt} General set-up}
\addcontentsline{toc}{section}{4.2 \hspace{0.15cm} General set up}

We consider the form for the metric of the AdS$_5$--Schwarzschild$\times S^5$ background presented in equation (\ref{eqt: metricchange}):
\begin{eqnarray}
ds^2  &=& \frac{1}{4 r^2 R^2} \left(-\frac{f_1^2}{f_2} dt^2 + f_2 d \vec{x}^2\right) + \frac{R^2}{r^2}  dr^2 + R^2 \cos^2 \theta d \Omega_3^2 \nonumber \\
&& \hskip2cm + R^2 d \theta^2 + R^2 \sin^2 \theta d \phi^2 \ , \\  
f_1  &=& 4r^4 - b^4 \ ,\quad  
f_2 = 4r^4 + b^4 \nonumber \ . 
\end{eqnarray}

In order to introduce our background electric field (and current), we consider an ansatz for the world--volume gauge field of the form~\cite{Karch:2007pd}:
\begin{equation} 
A_1(r)  =- E t + F(r) \ .
\end{equation}
This ensures a constant electric field $F_{01}=E$ in the gauge theory; $F(r)$ sources a current $J^1$ on the world--volume \cite{Karch:2007pd}.  For the D7--brane embedding, we consider  an ansatz as before: $\theta \equiv  \theta(r)$.

The DBI action is then given by
\begin{eqnarray}\label{eqt:action}
S_\mathrm{D7} & = & - T_{\mathrm{D7}} N_f  \int d^8 \xi \ \frac{\cos^3 \theta(r)}{16 r^5} \left[ f_2^2 f_1^2 \left(1+ r^2 \theta'(r)^2\right) \right. \nonumber\\
&&  \left. - 4 \left(2 \pi \alpha' \right)^2 r^4 \left \{ - f_1 ^2 f_2 F'(r)^2 + 4 f_2^2 R^4 E^2 \left(1+ r^2 \theta'(r)^2\right)   \right\} \right]^{1/2} \ .  
\end{eqnarray}

The equations of motion derived from equation (\ref{eqt:action}) for $F(r)$ introduces a constant of motion~$K$.  We can invert the equation of motion to write a first order differential  equation for $F(r)$ in terms of  $K$: 
\begin{equation}
\frac{ \left(2 \pi \alpha'\right)^2 \cos^3 \theta f_1^2 f_2 F'}{ 16 r^5 \sqrt{f_1^2 f_2^2 \left(1 + r^2 {\theta'}^2 \right) - 4 \left(2 \pi \alpha'\right)^2 r^4 \left(- f_1^2 f_2 {F'}^2 + 4 f_2^2 R^4 E^2 \left( 1 + r^2 {\theta'}^2 \right)\right)}} = K \ ,
\label{eqt:Kpseudo}
\end{equation}
In the limit of $r \to \infty$, the asymptotic solution for $ F(r)$ is given by:
\begin{eqnarray}
\lim_{r \to \infty} F (r) & = & f_0 + \frac{K}{2 r^2} + \dots \nonumber
\end{eqnarray}
Normalizability of the solution requires us to take $f_0 = 0$.  The constant $K$ is related to the (vacuum expectation value) vev of the current by the following relation\cite{Karch:2007pd}:
\begin{equation}\label{eqt:current}
\langle J^1 \rangle = \langle \bar{\psi} \gamma^1 \psi \rangle = - 4 \pi^3 \alpha' V N_f T_{\mathrm{D7}}  K \ ,
\end{equation}
where $V$ is the spatial volume of the gauge theory.  It is convenient to exchange the field $F(r)$ in favour of the constant $K$ by performing a Legendre transformation:
\begin{eqnarray}
I_\mathrm{D7} &=& S_\mathrm{D7}  - \int d^8 \xi \  F'(r) \frac{\delta S_{D7}}{\delta F'(r)} \nonumber \\
&=& -2 \pi^2 V N_f T_{\mathrm{D7}} \int d t \int d r \left[ \frac{\sqrt{\left(1 + r^2 \theta'(r)^2\right)}}{16 r^5 f_1 \sqrt{f_2} } \right. \nonumber \\
&& \left. \times \sqrt{ \left(f_1^2 - 16 r^4 \left(2 \pi \alpha'\right)^2 E^2 \right) \left( -64 r^6  K^2 f_2^2  + f_1^2 f_2^3 \cos^6 \theta(r) \right)} \right] \ ,\label{eqt:legendre}
\end{eqnarray}
It is a simple check to show that ${\delta I_{\mathrm{D7}}}/{\delta K} = F'(r)$.  The expression under the square root in equation (\ref{eqt:legendre}) must be positive in order to keep the action real.  Either both terms should be negative or both should be positive, and they must change sign at the same radial distance~$r_\ast$.  This results in two conditions:
\begin{equation} \label{eqt:cond1}
\left(4 r_\ast^4 - b^4 \right)^2 - 16 r_\ast^4 \left(2 \pi \alpha' \right)^2 E^2  =  0 \quad \implies \quad r_\ast^2 = \frac{1}{2} \left[ \left( 2 \pi \alpha' E\right) + \sqrt{\left( 2 \pi \alpha' E\right)^2 + b^4}\right] \ ,
\end{equation}
\begin{equation} \label{eqt: ohm}
K^2 = \frac{\left( 4 r_\ast^4 - b^4 \right)^2 \left(4 r_\ast^4 + b^4 \right)^3 \cos^6 \theta(r_\ast)}{64 r_\ast^6 \left(4 r_\ast^4 + b^4 \right)^2} \ .
\end{equation}
Note that equation (\ref{eqt: ohm}) relates the current to the electric field in a form of Ohm's law, and this was used in ref.~\cite{Karch:2007pd} to determine the conductivity of the fundamental matter in the quark--gluon plasma.  We refer to $r_\ast$ as the ``vanishing locus''.  From the action $I_{D7}$ defined in equation (\ref{eqt:legendre}), we can derive the equation of motion for the embeddings $\theta(r)$. Its exact form is not illuminating, and so we do not display it here. In the limit of large $r$, the solution of the equation of motion asymptotes to the familiar form:
\begin{equation}
\theta(r) = \frac{m}{r} + \frac{c}{r^3} \ .
\end{equation}
The constants $m$ and $c$ are related to the bare quark mass and the condensate respectively, in a manner that was reviewed in Chapter 3.

\section*{4.3 \hspace{2pt} Properties of the solutions}
\addcontentsline{toc}{section}{4.3 \hspace{0.15cm} Properties of the Solutions}

\subsection*{4.3.1 \hspace{2pt} Exact results at large mass}
\addcontentsline{toc}{subsection}{4.3.1 \hspace{0.15cm} Exact results at large mass}

It is instructive to study the properties of the quark condensate as a function of the bare quark mass for large mass. This corresponds to $ m \gg \left( 2 \pi \alpha' E\right) R^2/b$ in terms of the dimensionfull quantities.  This limit constrains us to consider only the so--called ``Minkowski'' embeddings for the probe D7--brane for which the vev of the current (defined in equation (\ref{eqt:current})) vanishes by virtue of equation (\ref{eqt: ohm}).  To extract the behaviour of quark mass and condensate we linearize the equation of motion obtained from equation (\ref{eqt:legendre}) in the same way as described in ref.~\cite{Albash:2007bk}. Rewriting the result in terms of dimensionful parameters, we find the following analytic behaviour for the condensate:
\begin{eqnarray}
\langle \bar{\psi}\psi \rangle\propto -c=\frac{R^4\hat{E}^2}{4m}+\frac{b^8+4b^4R^4 \hat{E}^2+8R^8 \hat{E}^4}{96m^5}+O\left(\frac{1}{m^7}\right) \ ,
\end{eqnarray}
where $\hat{E} = 2 \pi \alpha' E$.  This may suggest that for high enough bare quark mass, the condensate vanishes.  However unlike the results for an external magnetic field~\cite{Albash:2007bk}, the condensate approaches zero on the positive side of the condensate axis.  We verify these results in the next section using numerical techniques.

\subsection*{4.3.2 \hspace{2pt} The case of vanishing temperature}
\addcontentsline{toc}{subsection}{4.3.2 \hspace{0.15cm} The case of vanishing temperature}

It is instructive to first study the zero temperature case, since a number of important features appear in this case.  For this simple case, the energy scale is set by $R\sqrt{\hat{E}}$, and therefore it is convenient to introduce the dimensionless quantities ${\hat r}$ and ${\hat m}$ {\it via}:
\begin{equation}
r=R\sqrt{E}\, {\hat r}\ ;\quad m=R\sqrt{E}\,{\hat m}\ .
\label{dimless}
\end{equation}
We solve the equation of motion for $\theta(\hat{r})$ using a shooting technique by imposing infrared boundary conditions described in refs.~\cite{Albash:2006ew,Albash:2006bs,Karch:2006bv}. For ``Minkowski'' embeddings, in order to avoid a conical singularity at $\theta = \pi/2$ \cite{Karch:2006bv}, we impose the following boundary condition:
\begin{equation}\label{eqt:minkow}
 \theta'(\hat r)|_{\theta = \pi/2}= -\infty \ .
\end{equation}
For embeddings reaching the pseudo-horizon we find it numerically convenient to shoot forward (towards infinity) and backward (towards the origin) from the vanishing locus ${\hat r}_\ast=1$.  The boundary condition ensuring the smoothness of solutions across the vanishing locus can be determined from the equation of motion itself, and it is found to be:
\begin{eqnarray}
&& \theta'(\hat{r}_\ast)= \frac{\partial \hat u}{\partial \hat r}\frac{\partial\theta}{\partial \hat u} \ , \quad \theta_0 \equiv \theta(\hat{r}_\ast)\ , \nonumber\\ 
&&\frac{\partial\theta}{\partial\hat u}=-\left.\frac{1}{\hat{u}}\tan\frac{{\theta}_0}{2}\right|_{\hat u(\hat r_\ast)}\ .
\end{eqnarray}
From the above boundary condition we can see that the embeddings reaching the vanishing locus at $\theta=\pi/2$ avoids the conical singularity by having a diverging derivative.  Next we proceed to discuss our numerical findings.

\begin{figure}[ht]
   \centering
   \includegraphics[width=11cm]{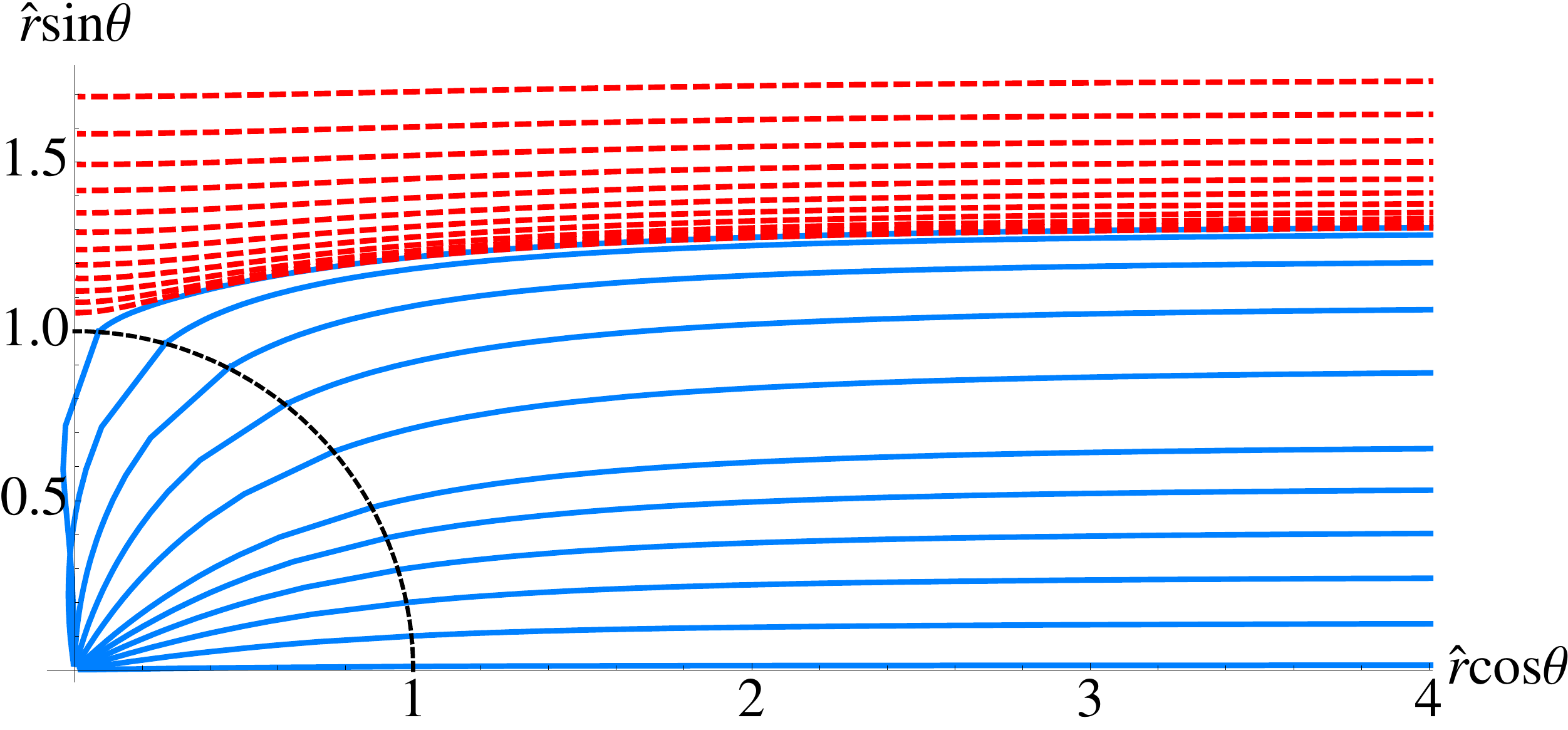}
   \caption{\small The solid curves  starting far left (red) represent solutions smoothly closing before reaching the vanishing locus.  The solid remaining curves (blue) correspond to embeddings passing the vanishing locus (denoted by a semi--circular dashed black curve). }
   \label{fig:embed0}
\end{figure}

\begin{figure}[h] 
  \centering \includegraphics[width=2.8in]{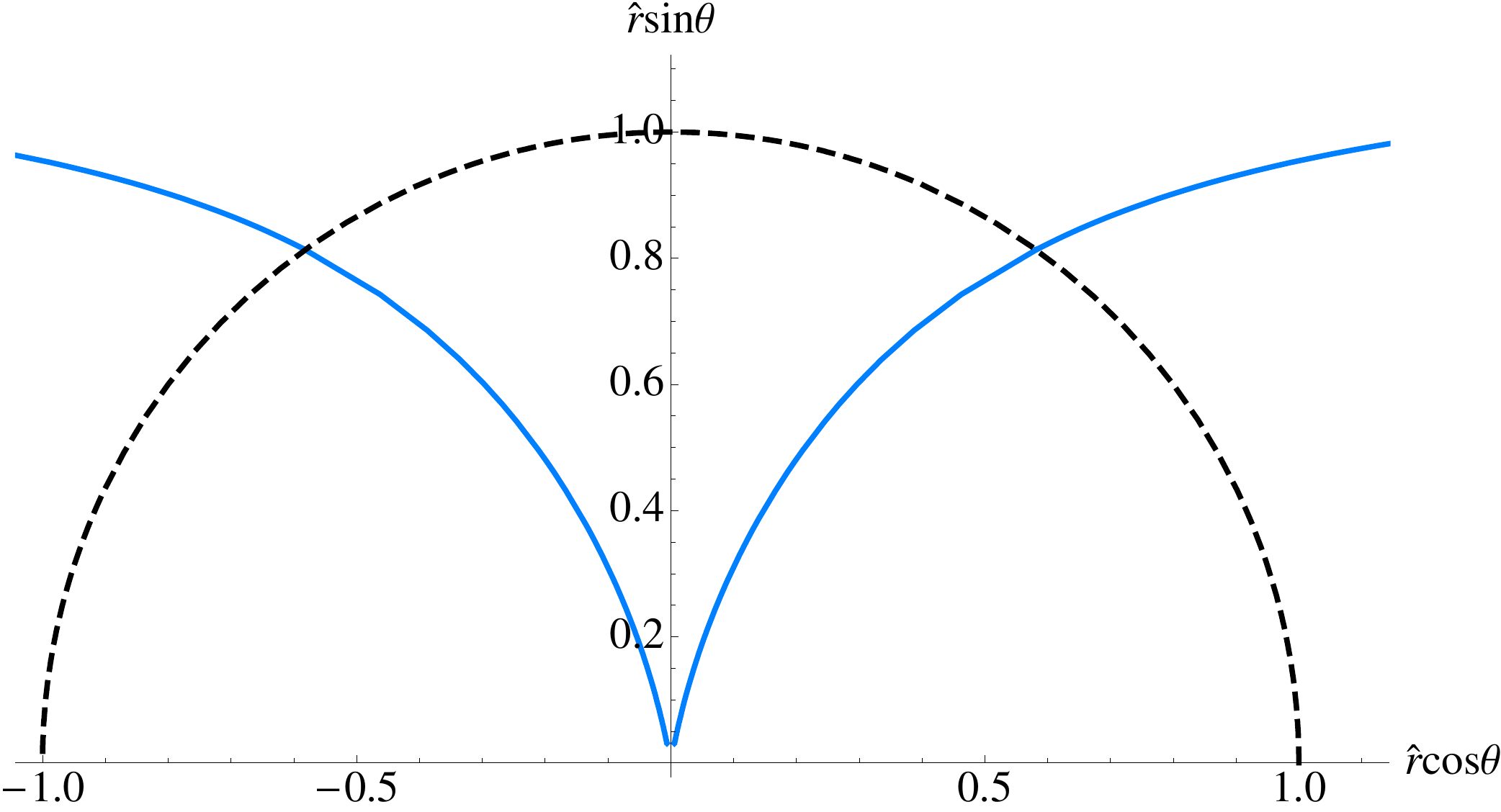}
  \includegraphics[width=2.8in]{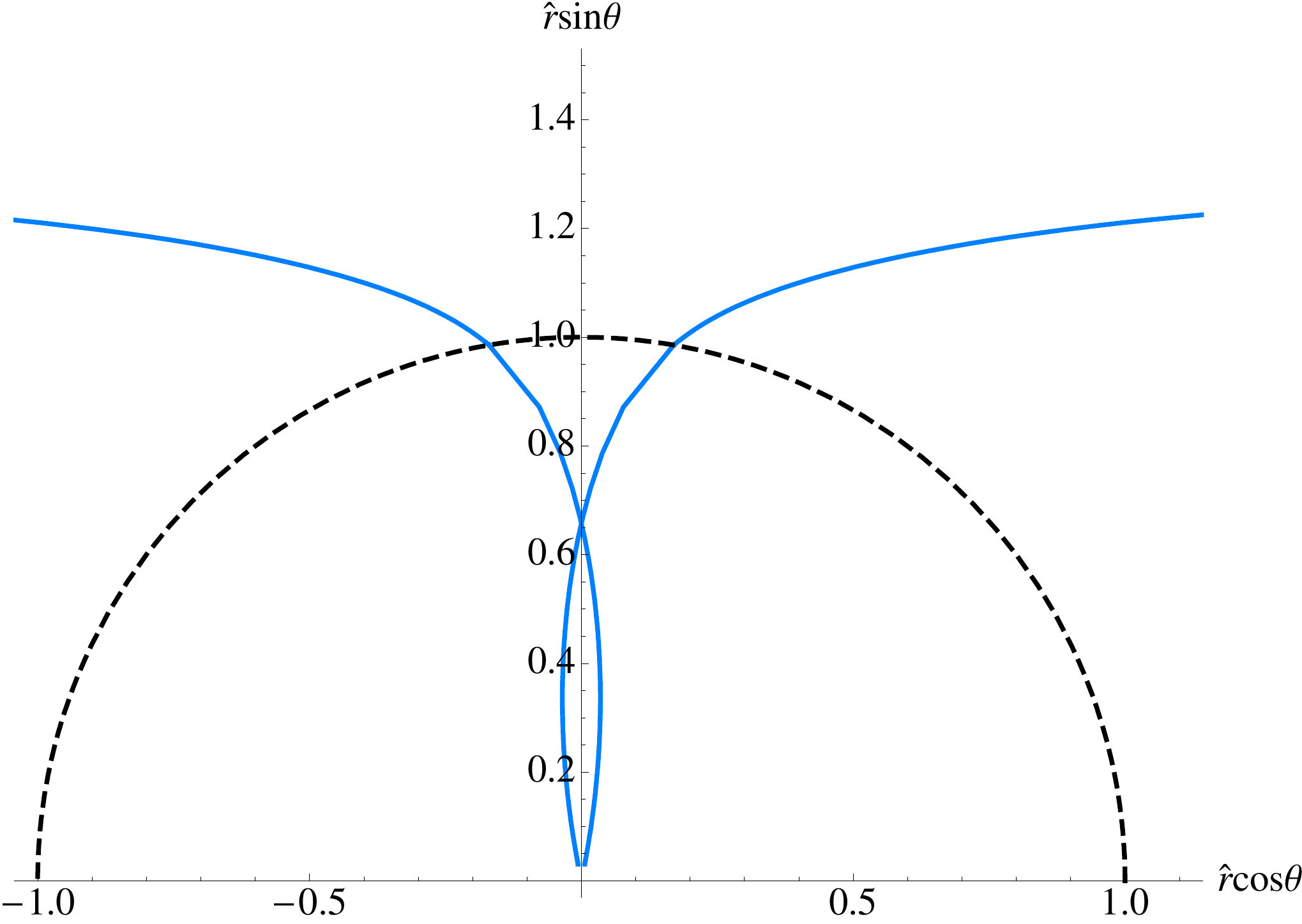}
     \caption{\small Two kinds of embeddings that pass the vanishing locus corresponding to D7--brane closing at the origin (left), and forming a conical singularity (right) respectively. The semi--circular (red) curve corresponds to the vanishing locus. }
     \label{fig:em02}
  \end{figure}

Numerical analysis shows that there are three different classes of embeddings, which can be classified by the topology of the probe D7--brane.  First, we have the smooth ``Minkowski'' embeddings that close above the vanishing locus, due to the shrinking of the $S^3$ that the probe brane wraps.  Second, we have the embeddings that reach the vanishing locus before the $S^3$ shrinks.  These embeddings can in turn be classified into two different categories. The first are singular solutions that cross the vanishing locus and close before reaching the origin. These have a conical singularity at the closing point because the derivative does not diverge.  We discuss these solutions further below. The second are smooth solutions that pass through the vanishing locus and reach the origin with no singular behaviour. These different embeddings are  summarized in figures \ref{fig:embed0} and \ref{fig:em02}.

  We can explicitly see the conical singularity appearing for the
  brane (second curve in Figure~\ref{fig:em02}) since it intersects
  the vertical axis at some non--zero value of $\theta$. Depending on
  the magnitude of the electric field this class of singular solutions
  starts appearing after a certain value of
  $\left(\theta_{0}\right)_{\rm min}$, close to $\pi/2$ (which
  corresponds to the critical embedding), is reached and persists
  until $\theta_0=\pi/2$.

  The existence of these embeddings suggests the possibility of a
  transition in topology of the probe brane as a function of the
  parameters, as happened for finite temperature. Here, it can happen
  as a result of the external electric field that we've applied. The
  Minkowski embeddings simply have a shrinking $S^3$, while the
  embeddings reaching the origin are distinguished by having an $S^3$
  shrinking as well as touching the AdS horizon. However, the presence
  of singular solutions makes the nature of the transition subtle, as
  we discuss later. We will later see that finite temperature case
  also reveals similar classification of embeddings.

\begin{figure}[ht]
  \centering \includegraphics[width=6.5cm]{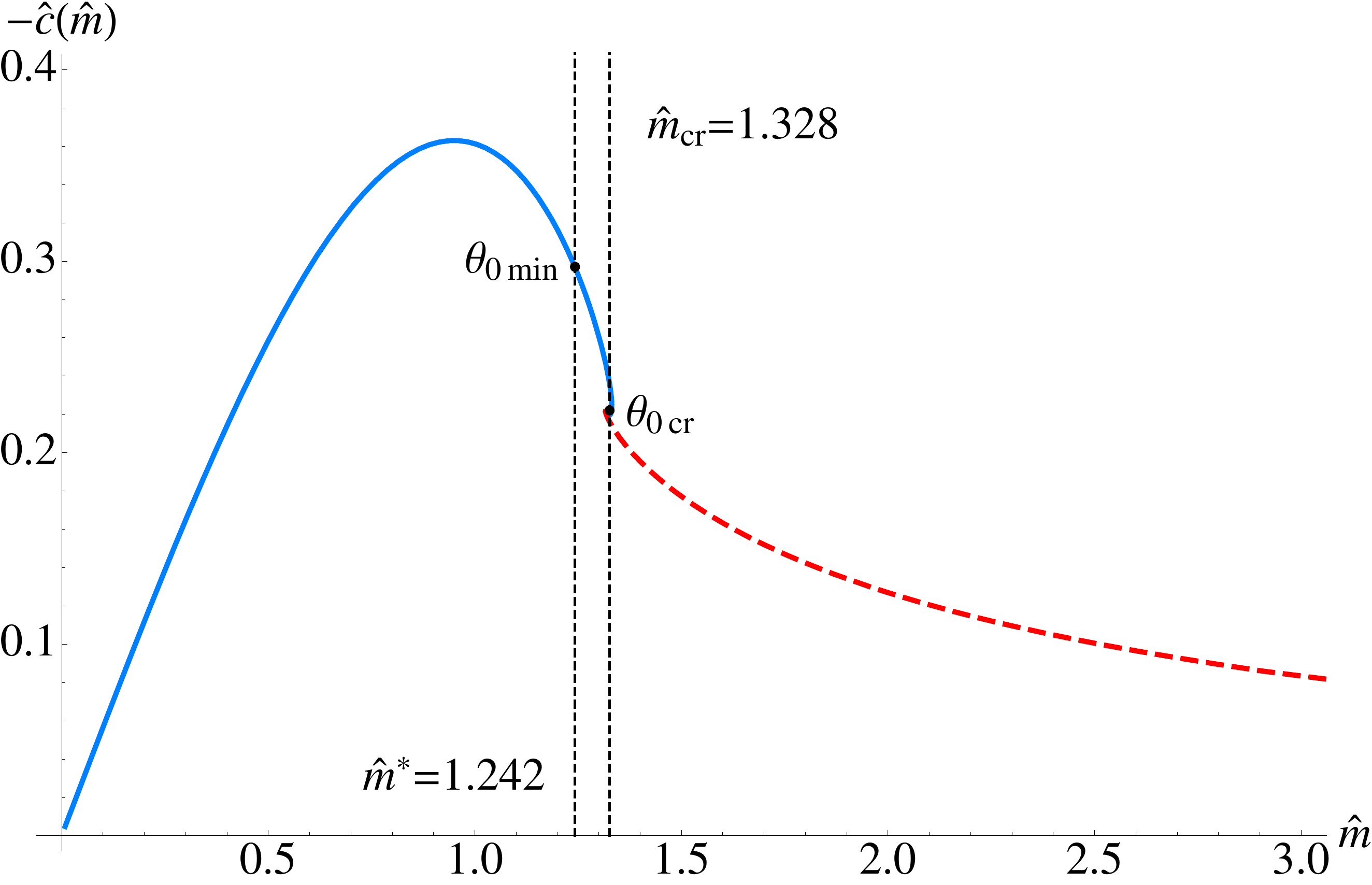}
  \includegraphics[width=6.5cm]{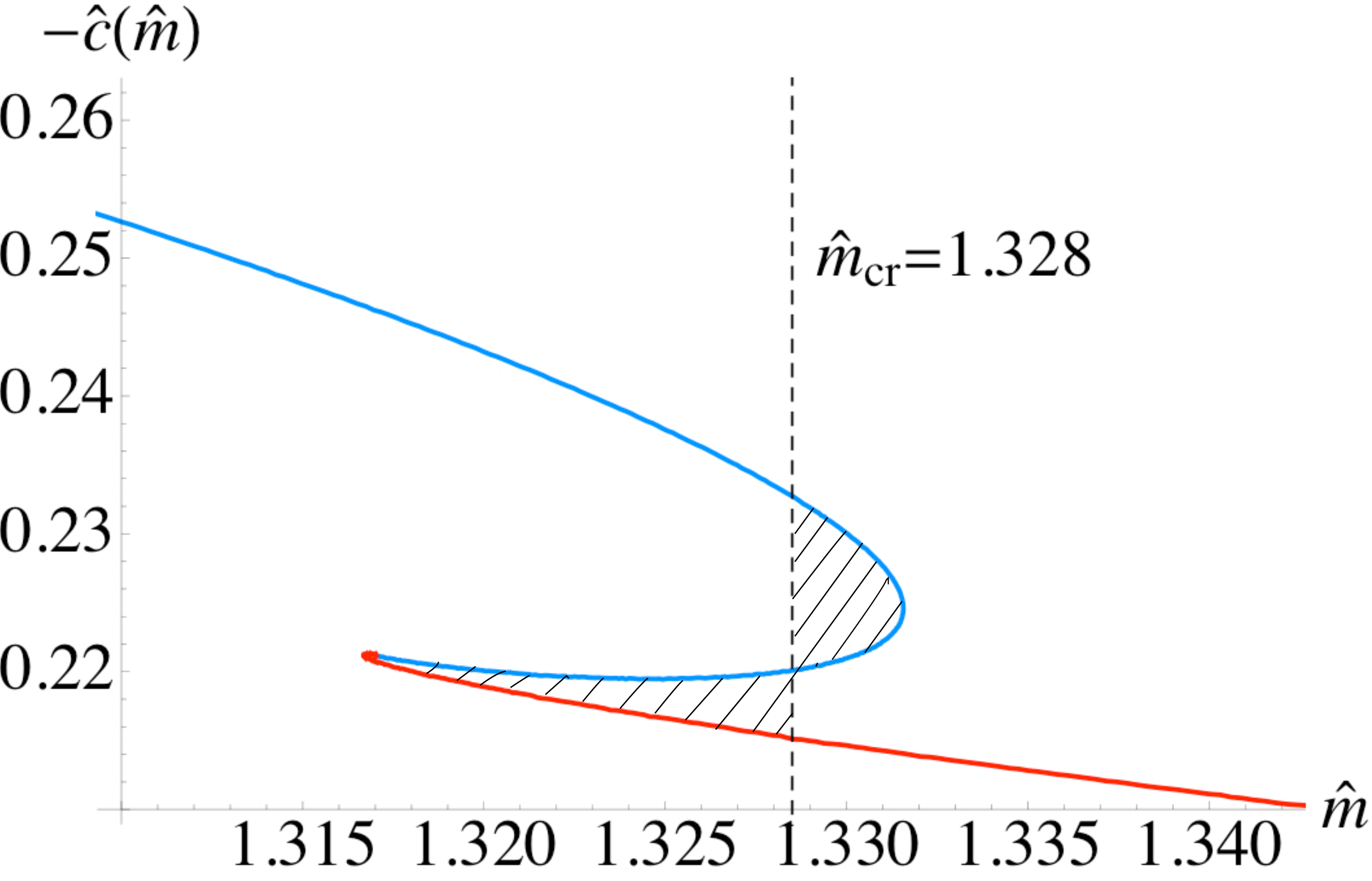}
   \caption{\small The condensate as a function of bare quark mass ${\hat m}$. The curve segment coming in from the left (blue)  corresponds to embeddings reaching the vanishing locus and the curve segments going out to the right (red) correspond to Minkowski embeddings. There is a family of  solutions that contain a conical singularity between the vanishing locus and the origin. They exist for $(\theta_0)_{\rm min}<{\theta}_0<(\theta_0)_{\rm cr}$. On the right is a magnification of the turn--around region where these segments join, showing multi--valuedness. An analysis of the free energy reveals a first order phase transition at ${\hat m}_{\rm cr}$ where there is a jump from one type of embedding to another. }
   \label{fig:cm0}
\end{figure}

  From the asymptotic behavior of these embeddings we can extract
  condensate as a function of the bare quark mass. In Figure~\ref{fig:cm0} we show this dependence. There are two important mass
  scales. One, ${\hat m}_{\rm cr}$, is where the phase transition
  between the two types of embedding occurs, and the other ${\hat
    m}^\ast$, is where the singular solutions first appear. These
  values are in turn determined by the parameters $(\theta_0)_{\rm
    cr}$ and $(\theta_0)_{\rm min}$, respectively. If ${\theta_0}_{\rm
    min}>(\theta_0)_{\rm cr}$, then the singular solutions are never
  thermodynamically favored.
We see the presence of a first order phase transition speculated based
on the general arguments. Furthermore, we observe that there is no
chiral symmetry breaking, since in all cases the condensate vanishes for vanishing
mass. This is to be contrasted with the case of external magnetic
field studied with these methods in the previous chapters.

This latter observation fits our intuition that the mesons in this
theory have a binding energy that grows with (it is proportional to)
the constituent quark mass~\cite{Kruczenski:2003be}. For a given quark
mass there should exist sufficiently high electric field that can
reduce the binding energy the quarks. This allows two things to happen:
First, this inhibits the formation of chiral condensate, and second,
this ultimately will dissociate the mesons into its constituent
quarks.  This dissociation is in fact a transition from an insulating
to a conducting phase, mediated by the external electric field. On the
dual gravity side this corresponds to a transition from Minkowski
embeddings that do not reach the origin  to embeddings that do.

We must note that the appearance of the singular solutions (those that
have a conical singularity) are not well understood at the moment.
There is therefore the possibility that there is an as yet to be
identified intermediate phase right after the meson dissociates. One simple possibility (but not the
only one) is that stringy corrections smooth out the singularity in
the interior, while preserving the asymptotic behavior.

The finite temperature version of the story rather similar to the vanishing temperature case. We will not elaborate on this case further, but refer the reader to the original paper.~\cite{Albash:2007bq}.

\section*{4.4 \hspace{2pt} Concluding  remarks}
\addcontentsline{toc}{section}{4.4 \hspace{0.15cm} Concluding remarks}

It is very encouraging that non--trivial non--perturbative phenomena
resulting from external fields such as those we have seen here (a
dissociation phase transition, metal--insulator transition and the
associated response current) can be so readily extracted in this kind
of holographic study. We found that since the electric field works
together with the presence of finite temperature, the resulting phase
diagram which accounts for the effects of both is rather simple
(ignoring the complication of the special conical solutions ---see
below).

Clearly, the story is not quite complete in this electric case, since
above a certain value of the electric field, some of our solutions
develop conical singularities in the interior. We are not entirely
sure about the nature of these solutions. As already stated, one
possibility is that the solutions are locally (near the singularity)
corrected by stringy physics, perhaps smoothing the conical points
into throats that connect to the horizon. We would expect in that case
that our phase diagram would largely remain intact, since the values
of the masses and condensates for each solution are read off at
infinity, far from where the conical singularity develops. The results
for the relative free energies of those solution (compared to the
other solutions at the same values of the mass) would be the same, and
so the complete story would be unaffected. A more drastic possibility
is simply that the physics is considerably modified by instabilities
associated with those conical solutions that our current study has not
revealed --- the appropriate part of the meson spectrum is
particularly difficult to study numerically here, and so the
tell--tale signs of tachyonic modes that represent an instability are
hard to check for in this case. Such a modification could remove quite
a significant part of the phase diagram for large enough values of the
electric field, allowing the possibility of a richer structure than we
have seen so far.


\chapter*{Chapter 5: \hspace{1pt} External Fields and Chiral Symmetry Breaking in the Witten--Sakai-Sugimoto Model}
\addcontentsline{toc}{chapter}{Chapter 5:\hspace{0.15cm}
External Fields and Chiral Symmetry Breaking in the Witten--Sakai-Sugimoto Model}

\section*{5.1 \hspace{2pt} Introductory remarks}
\addcontentsline{toc}{section}{5.1 \hspace{0.15cm} Introductory remarks}

We will change gears to discuss the dynamics of flavours in large $N_c$ gauge theories in the holographic model in Type IIA framework. We have reviewed this set-up briefly in the introduction, but we will recollect some key features and build up from there. We will be interested in very similar physics as we have discussed in the previous two chapters. The material presented in this chapter is based on the work in collaboration with Clifford Johnson\cite{Johnson:2008vna}.

The Witten--Sakai-Sugimoto model, as described in ref.~\cite{Sakai:2004cn} is
one construction which cleanly realizes chiral symmetry breaking
and confinement/deconfinement transition. The supergravity background of this model is
constructed of near-horizon geometry of $N_c$ D4-branes, following
ref.\cite{Witten:1998zw}. The study of $N_f$ flavour D8 branes in
this background when $N_f\ll N_c$ reveals a nice geometric realization
of chiral symmetry breaking. The flavour branes do not backreact on
the background geometry in this probe limit and therefore studying
their dynamics using the Dirac-Born-Infeld (DBI) action (including a
Wess-Zumino term, if necessary) suffices to capture the corresponding
gauge theory dynamics of fundamental flavours in an analogue of the
quenched approximation.

In this (DBI) regime it is possible to capture general gauge theory
features such as the phase diagram for temperature vs chemical
potential by considering probe brane in finite temperature
supergravity background and exciting specific gauge field on the
world--volume of the probe brane. Previous such studies including the
non--zero chemical potential in this model have been carried out in
{\it e.g.},
refs.~\cite{Horigome:2006xu,Rozali:2007rx,Bergman:2007wp,Davis:2007ka}.
Here, we will introduce an external magnetic and electric field.

A clear method for introducing a background magnetic field has been
previously discussed in the chapters~3 and 4.
We considered pure gauge $B$--field in the supergravity
background, which is equivalent to exciting a gauge field on the
world-volume of the flavour branes corresponding to a magnetic field.

We find that the presence of magnetic field promotes the spontaneous
breaking of chiral symmetry as found previously in chapter~3 in a different model. This is expected from the field theory
perspective and is widely recognized as a sort of ``magnetic
catalysis'' for chiral symmetry breaking (see {\it e.g.},
ref.~\cite{Miransky:2002eb}). Further study of the phase structure of
this model reveals the existence of a finite critical temperature (for
restoration of chiral symmetry) for large magnetic field. We
 analyze a number of other physical quantities  such as
 the latent heat and relative magnetisation associated to the phase
transition.

We also find that our phase structure is rather generic for the
Witten--Sakai-Sugimoto type holographic models where we consider the dynamics
of probe D$p$-brane in D4-brane background. We address some of the
physics of an external electric field in the model. We find that in
the symmetry--restored phase an external electric field drives a
current in the gauge theory due to pair creation, and the symmetry--broken
phase does not conduct. However we have not considered the presence of baryons in our set-up, which could give rise to a non-zero current in the phase where chiral symmetry is broken.

\section*{5.2 \hspace{2pt} The Witten--Sakai-Sugimoto Construction}
\addcontentsline{toc}{section}{5.2 \hspace{0.15cm} The Witten--Sakai-Sugimoto Construction}

The Witten--Sakai-Sugimoto model\cite{Sakai:2004cn} consists of near-horizon limit
of $N_c$ D4-branes wrapped on a circle of radius $R$ in the $x^4$
direction with anti-periodic boundary condition for fermions. The
D4-branes are intersected in the compact $x^4$ direction by $N_f$
$\overline{\rm D8}$-branes at $x^4=-\frac{L}{2}$ and $N_f$ D8-branes at
$x^4=\frac{L}{2}$ (with the constraint that $L\le \pi R$). This is
dual to a $(4+1)$-dimensional ${\rm SU}(N_c)$ Yang-Mills theory with gauge
coupling constant $g_5$; left and right handed quarks are introduced
by the flavour $\overline{\rm D8}$ and D8-branes in the probe limit that share three spatial directions with the D4-branes. The
flavour branes introduce a global flavour symmetry ${\rm U}(N_f)_L\times
{\rm U}(N_f)_R$ as seen from the $(4+1)$-dimensional D4-brane worldvolume
gauge theory. This global symmetry is identified with the chiral symmetry (non-abelian) of the effective $(3+1)$-dimensional gauge theory with chiral fermions. In the probe limit, the dynamics of the flavour branes
is described by the DBI action in the background of $N_c$ D4-brane
geometry. The background metric of D4-brane is obtained from Type
IIA supergravity and is given by
\begin{eqnarray}\label{eqt: metsakai}
&& ds^2 = \left(\frac{u}{R}\right)^{3/2}\left(-dt^2+dx_idx^i+f(u)(dx^4)^2\right)+\left(\frac{u}{R}\right)^{-3/2}\left(\frac{du^2}{f(u)}+u^2d\Omega_4^2\right)\ ,   \nonumber\\
&&   e^{\Phi}=g_s\left(\frac{u}{R}\right)^{3/4}\ , \quad F_{(4)}=\frac{2\pi N_c}{V_4}\epsilon_4\ ,\quad f(u)=1-\left(\frac{U_{\rm KK}}{u}\right)^3\ , \quad R^3=\pi g_sN_c\ell_s^3\ . \nonumber\\
\end{eqnarray}
Here $x^i$ are the flat $3$-directions, $t$ is the time coordinate,
$x^4$ is the spatial compact circle, $\Omega_4$ are the $S^4$
directions and $u$ is the radial direction. $\ell_s$ is the string
length, $g_s$ is the string coupling; $V_4$ and $\epsilon_4$ are the
volume and volume form of $S^4$ respectively. Also, $\Phi$ is the
dilaton and $F_{(4)}$ is the RR four-form field strength. To avoid a
conical singularity in the $\{x^4,u\}$ plane one should make periodic
identification:
\begin{equation}
\delta x^4=\frac{4\pi}{3}\left(\frac{R^3}{U_{\rm KK}}\right)^{1/2}=2\pi R_4\ .
\end{equation}
This endows the background with a smooth cigar geometry in the $\{x^4,u\}$ plane.

On the other hand, the finite temperature Euclidean background is given by
\begin{eqnarray}\label{eqt: highmet1}
&& ds^2=\left(\frac{u}{R}\right)^{3/2}\left(dx_idx^i+f(u)dt_{\rm E}^2+(dx^4)^2\right)+\left(\frac{u}{R}\right)^{-3/2}\left(\frac{du^2}{f(u)}+u^2d\Omega_4^2\right)\ ,\nonumber\\
&& t_{\rm E}=t_{\rm E}+\frac{4\pi R^{3/2}}{3U_T^{1/2}}\ ,\quad T=\frac{1}{\beta}=\left(\frac{4\pi R^{3/2}}{3U_T^{1/2}}\right)^{-1}\ ,\quad f(u)=1-\left(\frac{U_T}{u}\right)^3\ .
\end{eqnarray}
All the parameters are given by the same formula as equation~(\ref{eqt: metsakai}). The dilaton, RR $4$-form, the radius $R$ are also given by the same formula as equation~(\ref{eqt: metsakai}).

It is clear that at finite temperature we have two candidate supergravity solutions: the Euclidean version of (\ref{eqt: metsakai}) and the background in (\ref{eqt: highmet1}). Depending on the temperature of the background one of these two possible backgrounds is energetically favoured. At low temperatures, $T_d< 1/(2\pi R_4)$, the Euclidean version of (\ref{eqt: metsakai}) is the favoured background whereas for high temperatures, $T_d> 1/(2\pi R_4)$ the background in (\ref{eqt: highmet1}) is energetically favourable. At $T_d = 1/(2\pi R_4)$ there is a first order phase transition, which corresponds to the confinement/deconfinement transition in the dual gauge theory. Henceforth we will sometimes refer (\ref{eqt: metsakai}) as the confining background and (\ref{eqt: highmet1}) as the deconfining background.

Now one can introduce the flavour brane--anti-brane system in the probe limit, namely $N_f\ll N_c$. In this limit the probe branes do not backreact on the geometry and the classical profile of the probe is solely determined by the Dirac-Born-Infeld action. We will consider the following ansatz for the flavour $\overline{\rm D8}$-D8 branes:
\begin{equation}\label{eqt: ansatz}
\{t,x_i,x^4=\tau,\Omega_4,u=u(\tau)\}\ ,
\end{equation}
where we note that the coordinates in the parenthesis should be understood as the worldvolume coordinates of the $\overline{\rm D8}$/D8-brane.
\begin{figure}[!ht]
\begin{center}
\includegraphics[angle=0,
width=0.85\textwidth]{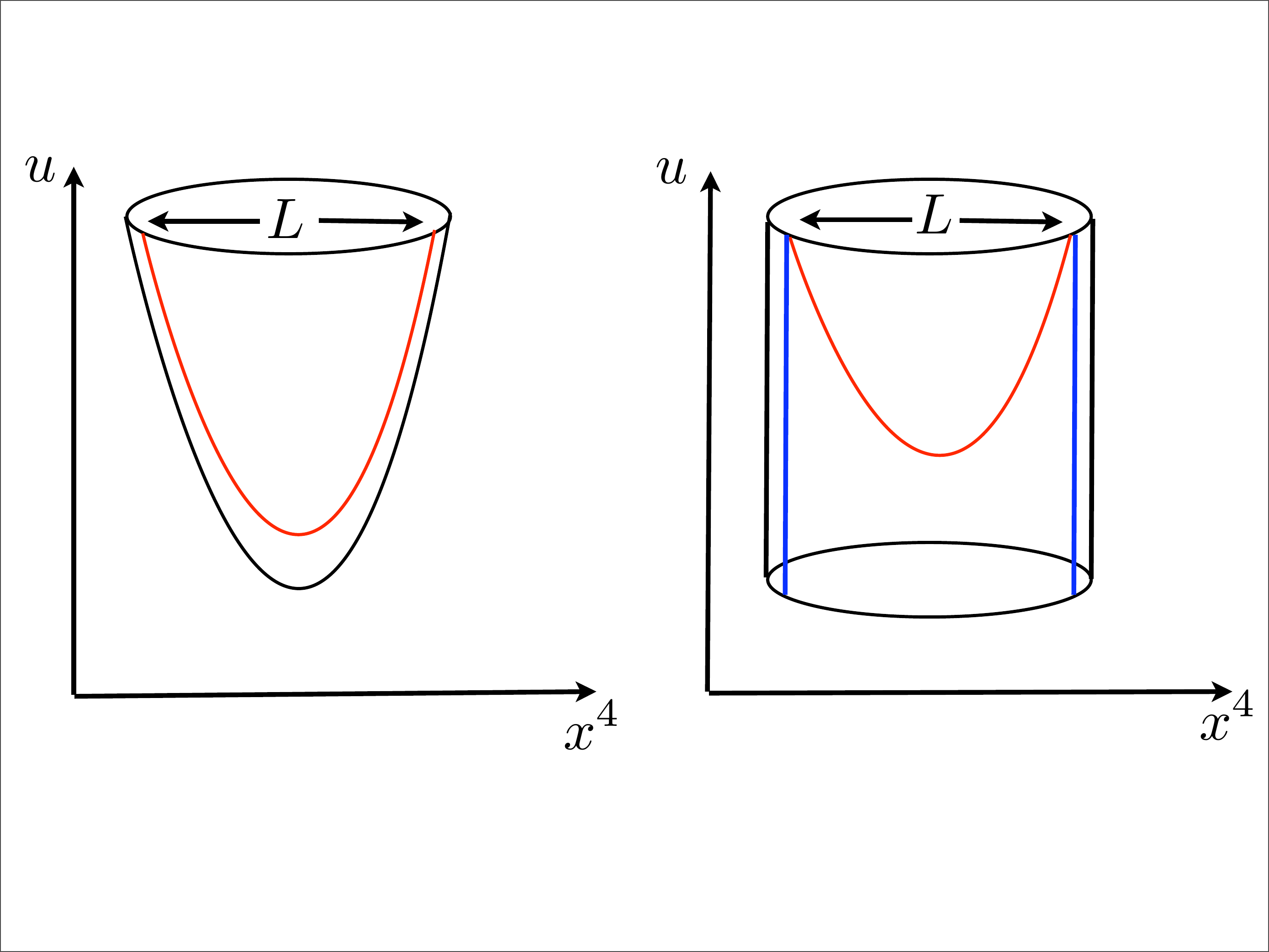}
\caption{\small The possible shape of embeddings of the probe brane. The left diagram shows the only possible (U)-shape of the probe brane in the confining background. On the right, we have shown two candidate embeddings of the probe brane in the deconfined phase: the U-shaped (in red) and the straight (in blue).}
\label{fig: shaped4d8}
\end{center}
\end{figure}
Before discussing the effect of the external magnetic field, let us review the physics when temperature is the only external parameter. We can understand the main qualitative feature without resorting to actual calculation.

The profile of the probe brane is clearly a function denoted by $x^4(u)$ or $u(x^4)$. In the low temperature regime, the relevant background has a smooth cigar-shape in the $\{x^4, u\}$-plane. Therefore the $\overline{\rm D8}$/D8-pair separated by a distance $L$ at the boundary is bound to join together in the bulk. This is illustrated in the left of figure~\ref{fig: shaped4d8}. This implies that the chiral symmetry group ${\rm U}(N_f)_L\times {\rm U}(N_f)_R$ is broken down to a ${\rm U}(N_f)_{\rm diag}$, which is a rather simple geometric realization of chiral symmetry breaking. Thus in this model, in the low temperature regime chiral symmetry is dynamically broken.

The physics of chiral symmetry breaking is different in the high temperature regime. In this case, the relevant background does not have a cigar shape in $\{x^4, u\}$-plane (this is illustrated in the right of figure \ref{fig: shaped4d8}). Therefore for a given asymptotic brane--anti-brane separation $L$, it is possible to have two candidate solutions. The brane--anti-brane pair can join smoothly as before which realizes the spontaneous breaking of chiral symmetry. We will refer such embeddings as the ``curved" or the ``U-shaped" embeddings. The brane--anti-brane pair can also end separately on the bulk horizon which corresponds to restoration of the full asymptotic chiral symmetry group. We will refer such embeddings as the ``straight" embeddings. For a fixed background temperature, depending on the asymptotic separation $L$ one of these two possible embeddings are energetically favourable. It can be shown that for ``low" temperatures the ``curved" embeddings are the favoured one, which means that the chiral symmetry is dynamically broken. At ``high" temperature, the ``straight" embeddings are favoured and therefore the broken symmetry is restored. This is a first order phase transition. We will make precise sense of what we mean by ``low" and ``high" temperatures next.

We have two first order phase transitions taking place in this set-up. The background undergoes a confinement/deconfinement transition and the probe flavours undergo a chiral symmetry restoring transition. The first transition is controlled by the radius $R_4$ of the compact $x^4$-direction and the second transition is controlled by varying $L$, the asymptotic separation of the brane--anti-brane pair. Therefore the only relevant dimensionless quantity is the ratio $L/ R_4$.

If the separation $L$ obeys the bound that $L/R_4<0.97$, then there exists a phase which is deconfined with broken chiral symmetry. We will sometimes refer this as the ``intermediate" phase. In this case, the chiral symmetry restoring temperature is given by $T_{\chi \rm SB}\simeq 0.154/ L$. Thus in this phase there is a separation of scales between the gluon deconfinement temperature and the chiral symmetry restoring temperature. If however $L/R_4>0.97$, then deconfinement and chiral symmetry restoration happens together. This has been discussed in refs.~\cite{Aharony:2006da, Parnachev:2006dn}.

We can also set $R_4\to\infty$ so that the spatial circle direction now becomes a flat extended direction; this particular limit has been studied in detail in refs.~\cite{Antonyan:2006vw, Antonyan:2006qy, Antonyan:2006pg}. It turns out that in this particular limit the dual gauge theory is described by the Nambu-Jona-Lasinio model (when probing with D8-branes) or the Gross-Neveu model (when probing with D6-branes) with a non-local four-fermi interaction. The strength of the four-fermi interaction is set by the asymptotic separation $L$. In this case, the confining phase does not exist and therefore the chiral symmetry restoring transition is the only phase transition.

\section*{5.3 \hspace{2pt} The probe brane analysis}
\addcontentsline{toc}{section}{5.3 \hspace{0.15cm} The probe brane analysis}

Many aspects of the finite temperature physics have been studied
before, {\it e.g.} in refs.~\cite{Aharony:2006da},\cite{Parnachev:2006dn}
and references therein. To introduce external magnetic field we excite a gauge field $A_3=H x^2$ on the
worldvolume of the probe brane.

\subsection*{5.3.1 \hspace{2pt} The low temperature background}
\addcontentsline{toc}{subsection}{5.3.1 \hspace{0.15cm} The low temperature background}

For notational convenience we rename $x^4$ coordinate to be $\tau$. In this case the relevant background is given by equation~(\ref{eqt:
  metsakai}). It is straightforward to check that in this case the Wess-Zumino term does not contribute to the probe brane action, which is therefore completely determined by the DBI action:
\begin{eqnarray}\label{eqt: dbi}
S_{D8} &=& \mu_8\int d^9\xi e^{-\Phi}\sqrt{{\rm det}(P[G_{\mu\nu}+B_{\mu\nu}])}
     = C\int d\tau \mathcal{L}(u,u')\ ,
\end{eqnarray}
where $C=\mu_8V_{S^4}V_{R^3}/g_sT$ and $u' = du/ d\tau$. Recall that we put the flavour branes ($\overline{D8}/D8$) with the asymptotic condition that as $u\to \infty$, $\tau\to\pm L/2$; where the $\pm$ corresponds to the $\overline{\rm D8}$ and D8-brane respectively. Also note that the coordinate $\tau$ is restricted from $-\pi R$ to $+\pi R$.

One can immediately note from equation~(\ref{eqt: dbi}) that $\mathcal{L}(u,u')$ is independent of $\tau$, therefore the Hamiltonian corresponding to $\tau$ will be a constant of motion. Carrying out the following Legendre transformation we should have
\begin{equation}\label{eqt: hamil}
\mathcal{H}_{\tau}=u'\frac{\partial\mathcal{L}(u,u')}{\partial u'}-\mathcal{L}(u,u')={\rm const.}\\
\end{equation}
So the first integral of motion that follows from equation~(\ref{eqt: hamil}) is given by
\begin{equation}\label{eqt: firstint}
u^4\frac{\left(1+H^2\left(\frac{R}{u}\right)^3\right)^{\frac{1}{2}}f(u)}{\left(f(u)+\left(\frac{R}{u}\right)^3\frac{u'^2}{f(u)}\right)^{\frac{1}{2}}}=U_0^4\left(1+H^2\left(\frac{R}{U_0}\right)^3\right)^{\frac{1}{2}}\sqrt{f(U_0)}\ .\\
\end{equation}

We have rewritten the constant in the right hand side in a convenient
way (in terms of a new parameter $U_0$). Note that $U_0$ is the minimum value of $u$ that the probe brane
can reach satisfying $u'|_{u=U_0}=0$. We assume that the brane--anti-brane pair join smoothly, which implies that there is no resultant force present at the point where they meet. Typically this would mean that there is no other source ({\it e.g.}, a baryon vertex or a bunch of F-strings) present at this point. For zero background magnetic
field this set up reduces to the low temperature case analyzed in
ref.~\cite{Aharony:2006da}. Let us now focus on the solution for the
probe brane profile.

We will compare the behaviour of the brane profile in the presence of
magnetic field to the case when it is turned off. For notational
convenience we define the following:
\begin{eqnarray}\label{eqt: rescale}
y=\frac{u}{U_0}\ , \quad y_{\rm KK}=\frac{U_{\rm KK}}{U_0}\ , \quad R=U_0d\ , \quad L=U_0l\ . \nonumber
\end{eqnarray}
With the above redefinitions we can obtain the difference in slope of the profile in presence and in absence of magnetic field as
\begin{eqnarray}\label{eqt: diffslope}
u_H'^2-u_{H=0}'^2=f(y)^2\frac{f(y)}{f(1)}\left(\frac{y}{d}\right)^3y^8\frac{H^2d^3}{1+H^2d^3}\left(\frac{1}{y^3}-1\right)\ , \nonumber
\end{eqnarray}
with $y\in [1,\infty]$. So we get that $|u_H'|\le |u_{H=0}'|$ for each value of $y$. This in turn means that the magnetic field bends the profile of the D8/$\overline{\rm D8}$ brane and therefore forces the brane--anti-brane pair to join closer to the boundary (and hence break chiral symmetry) for fixed asymptotic separation.

We can study this explicitly as follows. The brane--anti-brane separation at the boundary ($u\to~\infty$) is given by
\begin{eqnarray}\label{eqt: sepmag}
\frac{L}{2}=\int d\tau =\int_{U_{0(H)}}^\infty\frac{du_H}{u_H'}
           &=&\frac{R^{3/2}}{\sqrt{U_{0(H)}}}\int_1^\infty \frac{y^{-3/2}dy}{f(y)\left[\frac{1+H^2\left(\frac{d}{y}\right)^3}{1+H^2d^3}\frac{f(y)}{f(1)}y^8-1\right]^{1/2}}\nonumber\\
           &=&\int_1^\infty \mathcal{I}_{(H)}(y)dy\ .
\end{eqnarray}
Clearly putting $H=0$ we get the corresponding separation when the background magnetic field is switched off.
\begin{eqnarray}\label{eqt: sepnomag}
\frac{L}{2}=\frac{R^{3/2}}{\sqrt{U_0}}\int_1^\infty\frac{y^{-3/2}dy}{f(y)\left(\frac{f(y)}{f(1)}y^8-1\right)^{1/2}}
           =\int_1^\infty dy \mathcal{I}_{H=0}\ .
\end{eqnarray}

\begin{figure}[!ht]
\begin{center}
\includegraphics[angle=0,
width=0.65\textwidth]{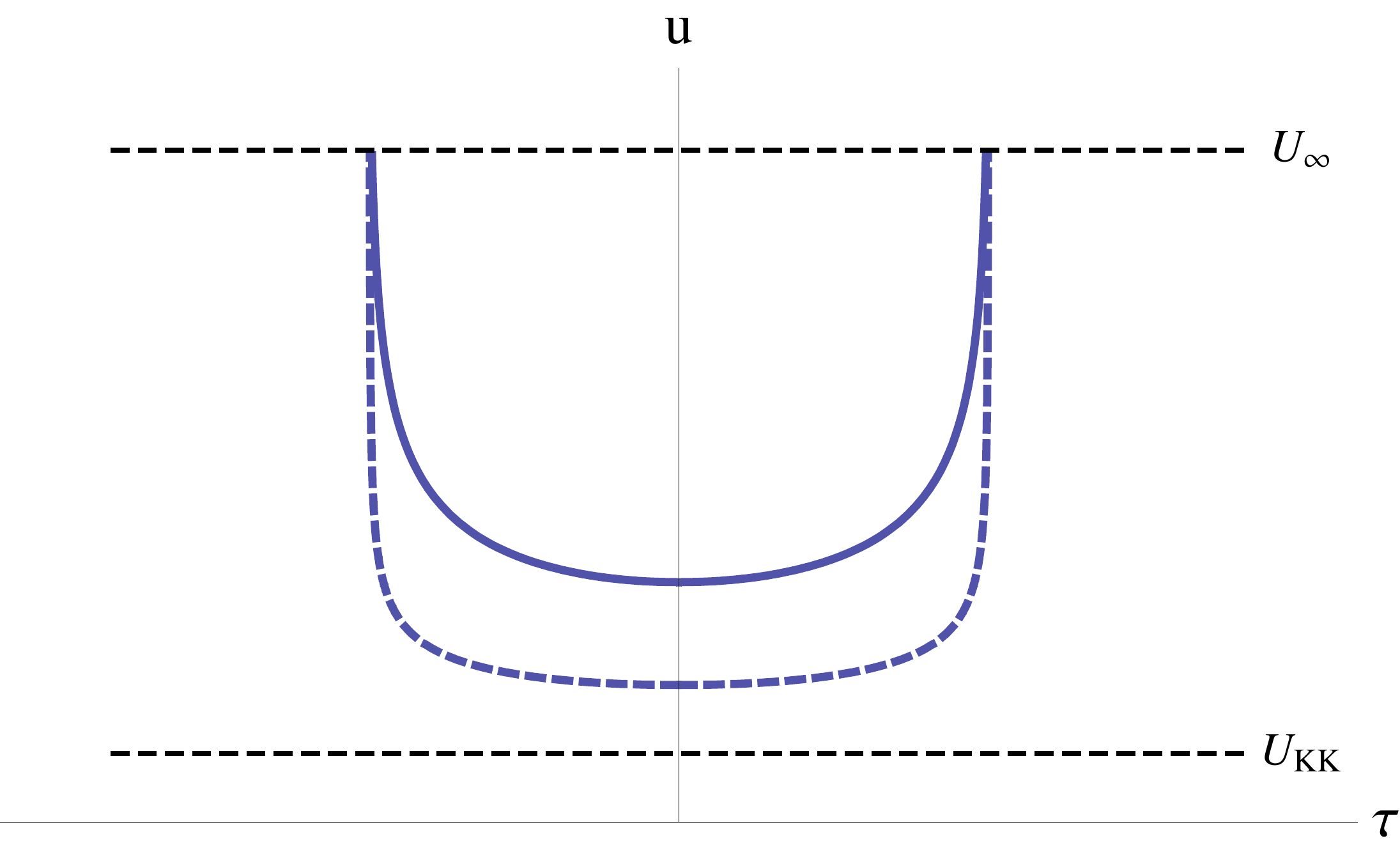}
\caption{\small The dashed U-shaped curve represents a profile in vanishing background field and the solid U-shaped curve represents a profile when a non-zero magnetic field is present. These profiles are obtained by numerically solving the equation of motion for the probe brane.}
\label{fig: plow}
\end{center}
\end{figure}

For the same asymptotic separation magnetic field changes the brane profile's point of closest approach $U_{0(H)}$. We can compare $U_{0(H)}$ and $U_{0}$. Equating equation~(\ref{eqt: sepmag}) and equation~(\ref{eqt: sepnomag}) one gets
\begin{equation}\label{eqt: compa}
\sqrt{\frac{U_{0(H)}}{U_{0}}}=\frac{\int_1^\infty \mathcal{I}_{H=0}dy}{\int_1^\infty \mathcal{I}_{(H)}dy}\ .\\
\end{equation}\\
Some algebra shows that $\mathcal{I}_{H=0}\ge \mathcal{I}_{(H)}$ for all $y$, so the ratio on the right hand side is greater than or equal to one, which also means that $U_{0(H)}\ge U_{0}$. Therefore for the same asymptotic separation the magnetic field can only help to join the brane--anti-brane pair favouring chiral symmetry breaking. This is pictorially represented in figure \ref{fig: plow}.
We can extract more qualitative features in appropriate limits. To do so, let us rewrite equation~(\ref{eqt: sepmag}) with the change to variable $z=y^{-3}$. With this equation~(\ref{eqt: sepmag}) becomes
\begin{equation}\label{eqt: sepz}
\frac{L}{2}=\frac{R^{3/2}}{3\sqrt{U_{0}}}\sqrt{\left(1-y_{\rm KK}^3\right)\left(1+H^2d^3\right)}\int_0^1 \frac{\left(1-y_{\rm KK}^3z\right)^{-1} z^{-5/6}dz}{\sqrt{\frac{\left(1-y_{\rm KK}^3z\right)\left(1+H^2d^3z\right)}{z^{8/3}}-\left(1-y_{\rm KK}^3\right)\left(1+H^2d^3\right)}}\ . 
\end{equation}

Now small asymptotic separation corresponds to large values of $U_{0}$ which means $y_{\rm KK} \ll 1$. So for small $L$ and weak magnetic field ($1/d^{3/2}\gg H$), the leading behaviour of the separation is given by (using equation~(\ref{eqt: sepz})), $L\sim R^{3/2}/\sqrt{U_{0}}$. This is same as the leading behaviour in zero magnetic field case in ref.~\cite{Aharony:2006da}. However, for strong magnetic field ($1/d^{3/2} \ll H$), the leading behaviour obtained from equation~(\ref{eqt: sepz}) is given by, $L\sim R^3 H/U_{0}^2$. So for fixed value of $U_{0}$ the asymptotic separation scales with the applied magnetic field strength $H$. This is however true only in the $y_{\rm KK}\ll 1$ limit.

Note that in this model the bare quark mass is always zero as there is no separation between the flavour and the colour branes at the boundary; however since the branes join at some length scale $U_0 \ge U_{\rm KK}$ in the core one can consider a string stretching from $u=U_{\rm KK}$ to $u=U_0$. The mass associated with the string can be identified to be the effective constituent quark mass as argued in ref.~\cite{Aharony:2006da}. If we denote this mass by $M_{q}$, then $M_{q}=\frac{1}{2\pi\alpha'}\int_{U_{\rm KK}}^{U_0}\sqrt{g_{tt}g_{uu}}$\ , where $U_0$ has to be determined from equation~(\ref{eqt: sepz}) for given~$L$. This turns out to be a self-consistency equation for $U_0$. This is analogous to the Gap equation in the field theory context ({\it e.g.}, in ref.~\cite{Antonyan:2006vw}). The self-consistency equation turns out to be
\begin{eqnarray}\label{eqt: lowgap}
&& U_0=\frac{4}{9}\frac{R^3}{L^2}\left(1-y_{\rm KK}^3\right)\left(U_0^3+H^2R^3\right) I(U_0,H)^2\ ,\quad {\rm where} \quad\nonumber\\
&& I(U_0,H)=\int_0^1\frac{\left(1-y_{\rm KK}^3z\right)^{-1}z^{-5/6}dz}{\left(z^{-8/3}\left(1-y_{\rm KK}^3z\right)\left(U_0^3+H^2R^3\right)-\left(1-y_{\rm KK}^3\right)\left(U_0^3+H^2R^3\right)\right)^{1/2}}\ .\nonumber\\
\end{eqnarray}

Now equation~(\ref{eqt: lowgap}) can be solved perturbatively, {\it i.e.},
starting with an initial value for the parameter $U_0$ we can
determine the next order approximation to $U_0$ using
equation~(\ref{eqt: lowgap}); and we continue until the desired
accuracy has been achieved. It is straightforward to guess the initial
value of $U_0$. Plugging in $H=0$ in equation~(\ref{eqt: lowgap}) we
should get the constituent mass for the low temperature case. This can
serve as the initial guess for small magnetic fields. Once $U_0$ is
known for small magnetic fields, it can be used as the initial guess
for successively higher values of magnetic fields. Thus we obtain the
dependence which is shown in figure below.
\begin{figure}[!ht]
\begin{center}
\includegraphics[angle=0,
width=0.55\textwidth]{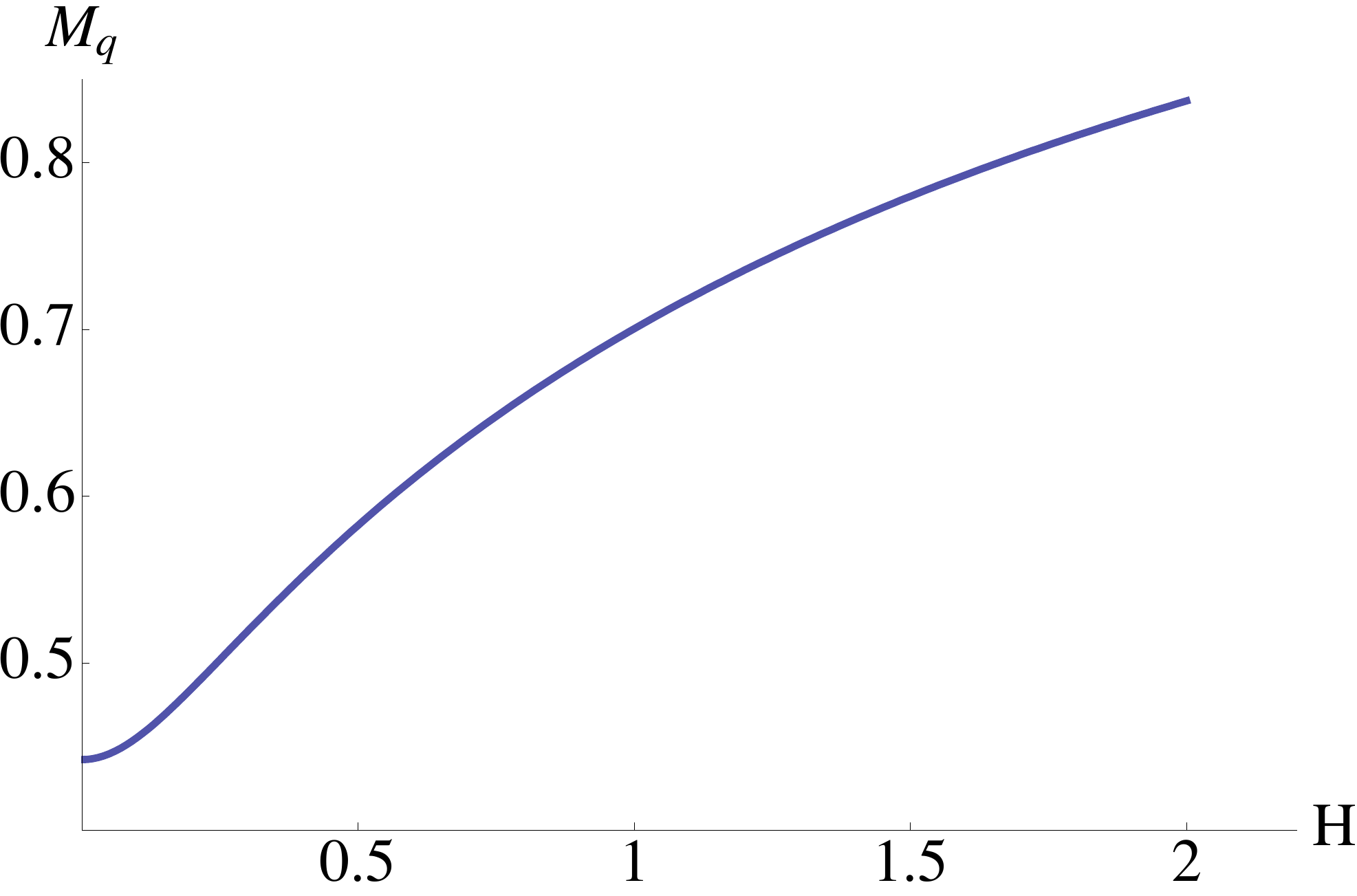}
\caption{\small The dependence of constituent quark mass (measured in units of $(2\pi\alpha')^{-1}$) on the external magnetic field in the low temperature phase. We have fixed $U_{\rm KK}=0.4$ and $R=1$.}
\label{fig: mdyn0}
\end{center}
\end{figure}
It is expected that the mass of the vector and pseudoscalar meson
would monotonically increase as the constituent quark mass increases.
Figure \ref{fig: mdyn0} therefore should capture the behaviour of
massive vector and pseudoscalar meson spectra in presence of magnetic
field.

Note that for this background disjoint brane pair do not exist: The
constant $\tau$-solutions namely, $\tau=-\pi R_4/2$ and $\tau=+\pi R_4/2$
join at $u=U_{\rm KK}$. This is because of the cigar geometry of the
background in the $\{\tau,u\}$ submanifold and the fact that $\tau$-circle
is wrapped by the probe branes. The brane pair must join together
since there is no place in the geometry for them to end separately. So
the only configuration possible in the low temperature phase breaks
chiral symmetry by reducing the global ${\rm U}(N_f)_L\times {\rm U}(N_f)_R$ to the
diagonal ${\rm U}(N_f)$. This is geometrically understood as the joining of
the flavour $8$-branes.

\subsection*{5.3.2 \hspace{2pt} The high temperature background}
\addcontentsline{toc}{subsection}{5.3.2 \hspace{0.15cm} The high temperature background}

\begin{figure}[!ht]
\begin{center}
\includegraphics[angle=0,
width=0.65\textwidth]{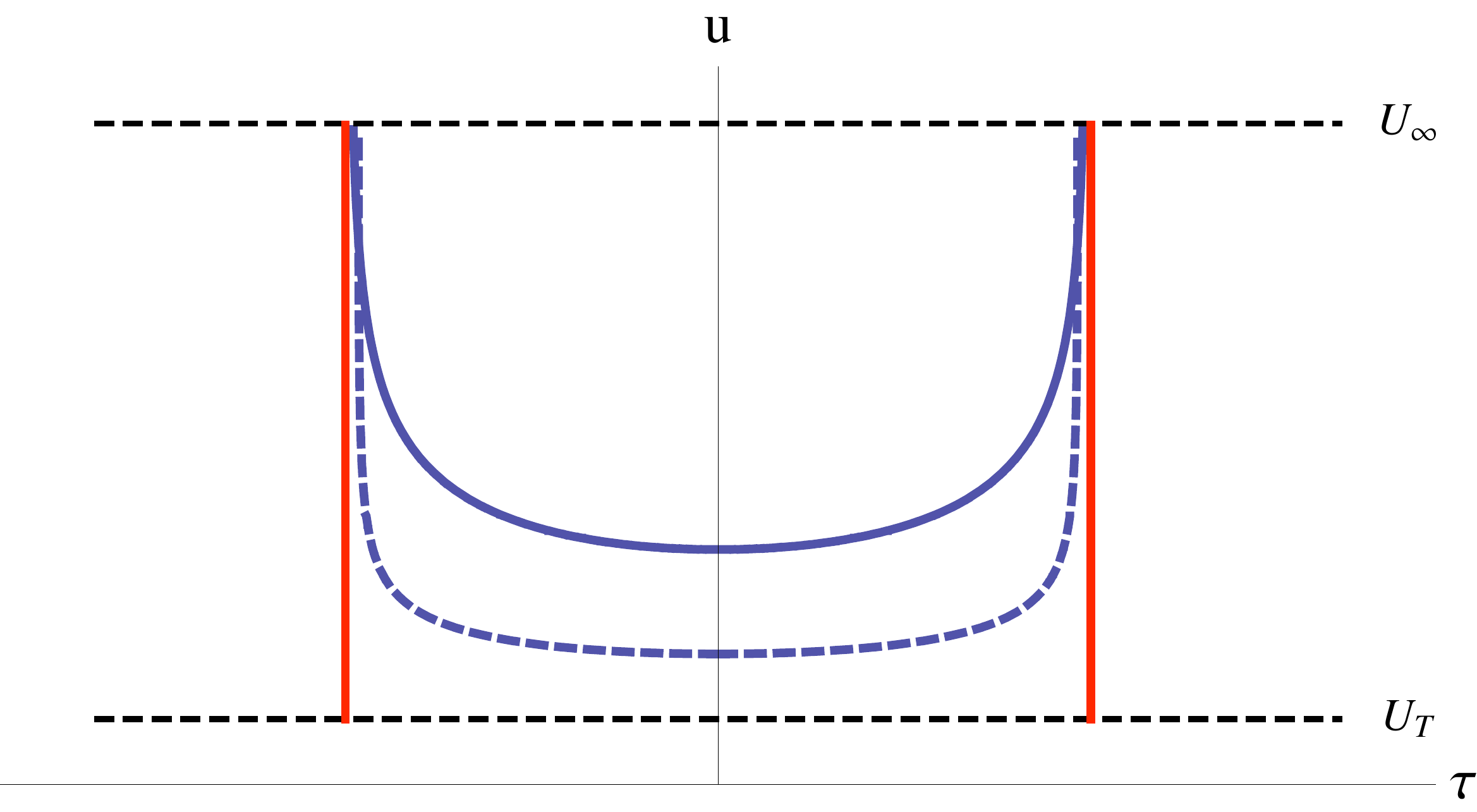}
\caption{\small The dashed U-shaped curve represents a profile in vanishing background field and the solid U-shaped curve represents a profile when a non-zero magnetic field is present. The straight (red) solution does not have any qualitative change in presence or absence of the external field. These profiles are obtained by numerically solving the equation of motion for the probe brane.}
\label{fig: phigh}
\end{center}
\end{figure}

Recall that the high temperature background is given by
equation~(\ref{eqt: highmet1}); and we again use the same ansatz for the probe given
by equation~(\ref{eqt: ansatz}). As before the profile of the probe brane is completely determined by the DBI action, from which a first integral of motion can be readily obtained to be
\begin{eqnarray}\label{eqt: highfirstint}
u^4\frac{\left(1+H^2\left(\frac{R}{u}\right)^3\right)^{\frac{1}{2}}f(u)}{\left(f(u)+\left(\frac{R}{u}\right)^3u'^2\right)^{\frac{1}{2}}}=U_0^4\left(1+H^2\left(\frac{R}{U_0}\right)^3\right)^{\frac{1}{2}}\sqrt{f(U_0)}\ .
\end{eqnarray}
For convenience we use the dimensionless variables defined in equation~(\ref{eqt: rescale}) along with the new variable $y_T=U_T/U_0$. Starting from the first integral of motion in equation~(\ref{eqt: highfirstint}) it is easy to verify that the finite temperature analogue to equation~(\ref{eqt: diffslope}) takes the following form
\begin{eqnarray}\label{eqt: newcom}
u_H'^2-u_{H=0}'^2=\left(\frac{y}{d}\right)^3\frac{f(y)^2}{f(1)}y^8\frac{H^2d^3}{1+H^2d^3}\left(\frac{1}{y^3}-1\right)\ .
\end{eqnarray}
This also suggests that $|u_H'|\le |u_{H=0}'|$, leading us to the same conclusion that the magnetic field helps bending the branes. The analogue to equation~(\ref{eqt: sepmag}) and~(\ref{eqt: sepnomag}) now take the following forms
\begin{eqnarray}\label{eqt: newlen}
&& \frac{L}{2}=\frac{R^{3/2}}{\sqrt{U_{0(H)}}}\int_1^\infty \frac{y^{-3/2}dy}{\sqrt{f(y)}\left[\frac{f(y)}{f(1)}\frac{1+H^2\left(\frac{d}{y}\right)^3}{1+H^2d^3}y^8-1\right]^{1/2}}=\int_1^\infty \mathcal{I}_H dy\ ,\nonumber\\
&& \frac{L}{2}=\frac{R^{3/2}}{\sqrt{U_{0(H)}}}\int_1^\infty \frac{y^{-3/2}dy}{\sqrt{f(y)}\left[\frac{f(y)}{f(1)}y^8-1\right]^{1/2}}=\int_1^\infty \mathcal{I}_0 dy\ ,
\end{eqnarray}
leading us to a similar conclusion as the low temperature case. This is pictorially represented in figure \ref{fig: phigh}.
As before one can extract the dependence of the asymptotic separation in the small $y_T$ limit. Again with the change to variable $z=y^{-3}$ we get
\begin{equation}\label{eqt: sepzmag}
\frac{L}{2}=\frac{R^{3/2}}{3\sqrt{U_{0(H)}}}\sqrt{\left(1-y_T^3\right)\left(1+H^2d^3\right)}\int_0^1 \frac{\left(1-y_T^3z\right)^{-1/2} z^{-5/6}dz}{\sqrt{\frac{\left(1-y_T^3z\right)\left(1+H^2d^3z\right)}{z^{8/3}}-\left(1-y_T^3\right)\left(1+H^2d^3\right)}}\ .
\end{equation}
So from equation~(\ref{eqt: sepzmag}) one can see that for small $y_T$ and weak magnetic field $L\sim R^{3/2}/\sqrt{U_{0(H)}}$; whereas for small $y_T$ and strong magnetic field we get $L\sim R^3H/U_{0(H)}^2$.

\begin{figure}[!ht]
\begin{center}
\includegraphics[angle=0,
width=0.55\textwidth]{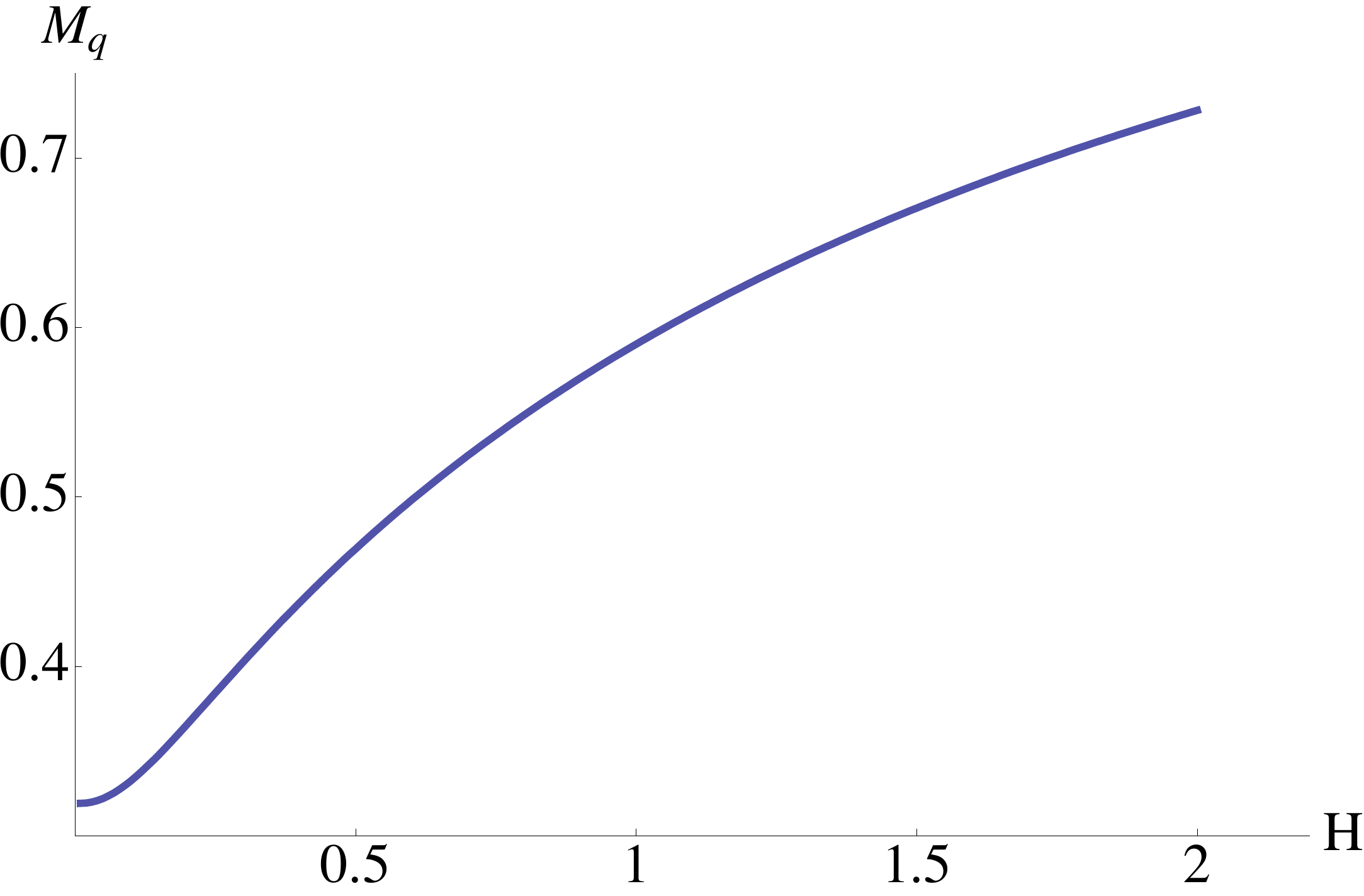}
\caption{\small The dependence of constituent quark mass (measured in units of $(2\pi\alpha')^{-1}$) on the external magnetic field in the high temperature phase. We have fixed $U_{T} = 0.3$ and $R = 1$.}
\label{fig: mdynTT}
\end{center}
\end{figure}

The joining of the flavour branes inside the core can be associated
with the effective constituent quark mass. This corresponds to a
self-consistency equation for $U_0$ as in the low temperature case.
The equation can again be solved using the same perturbative approach
and the results are summarised in the figure \ref{fig: mdynTT}. This
behaviour of constituent mass is valid when the curved solutions are
the lowest energy solutions (the chiral symmetry broken phase). In the
presence of finite temperature there will be a first order transition
to chiral symmetry restored phase. We will study this transition later
in the next section.

Now the trivial solution of equation~(\ref{eqt: highfirstint}), which
is given by $\tau'=0$ has a different physical meaning from the point
of chiral symmetry breaking. Since the $\{t,u\}$ submanifold has the
cigar geometry, the solutions $\tau=\pm L/2$ can remain disjoint and
end at $u=U_T$ separately. Therefore the trivial solutions in the high
temperature case preserve the full ${\rm U}(N_f)_L\times {\rm U}(N_f)_R$ symmetry
by remaining disjoint. In order to determine the true minimum energy
configuration we need to compare the energies of the curved and the
straight branes. We pursue this in the next section.

\subsection*{5.3.3 \hspace{2pt} Chiral symmetry breaking}
\addcontentsline{toc}{subsection}{5.3.3 \hspace{0.15cm} Chiral symmetry breaking}

To determine the true vacuum
we consider the difference between the energies of the curved and straight branes, which is given by
\begin{eqnarray}\label{eqt: dissmag}
\Delta S =\frac{S_{\rm curved}-S_{\rm straight}}{C U_0^5d^{\frac{3}{2}}}
         &=& \int_1^\infty dy y\left(y^3+H^2d^3\right)^{1/2}\left[\frac{1}{\left(1-\frac{f(1)(1+H^2d^3)}{f(y)(y^3+H^2d^3)}y^{-5}\right)^{1/2}}-1\right]\nonumber\\
       & - & \int_{y_T}^1 dy y\left(y^3+H^2d^3\right)^{1/2}\ .
\end{eqnarray}
Here $\Delta S<0$ would mean chiral symmetry breaking, $\Delta S>0$ would mean chiral symmetry restoration and $\Delta S=0$ would characterize a transition from symmetry broken phase to a symmetry restored one.
We employ numerical analysis to study this.

It is known from, {\it e.g.},
ref.~\cite{Aharony:2006da} that for high temperature and zero magnetic
field there exists a critical temperature beyond which the straight
branes are energetically favoured, implying that in the dual gauge
theory chiral symmetry is restored. Below the temperature, however,
the symmetry is broken by energetically favoured curved brane pair
that join together.

\begin{figure}[!ht]
\begin{center}
\includegraphics[angle=0,
width=0.65\textwidth]{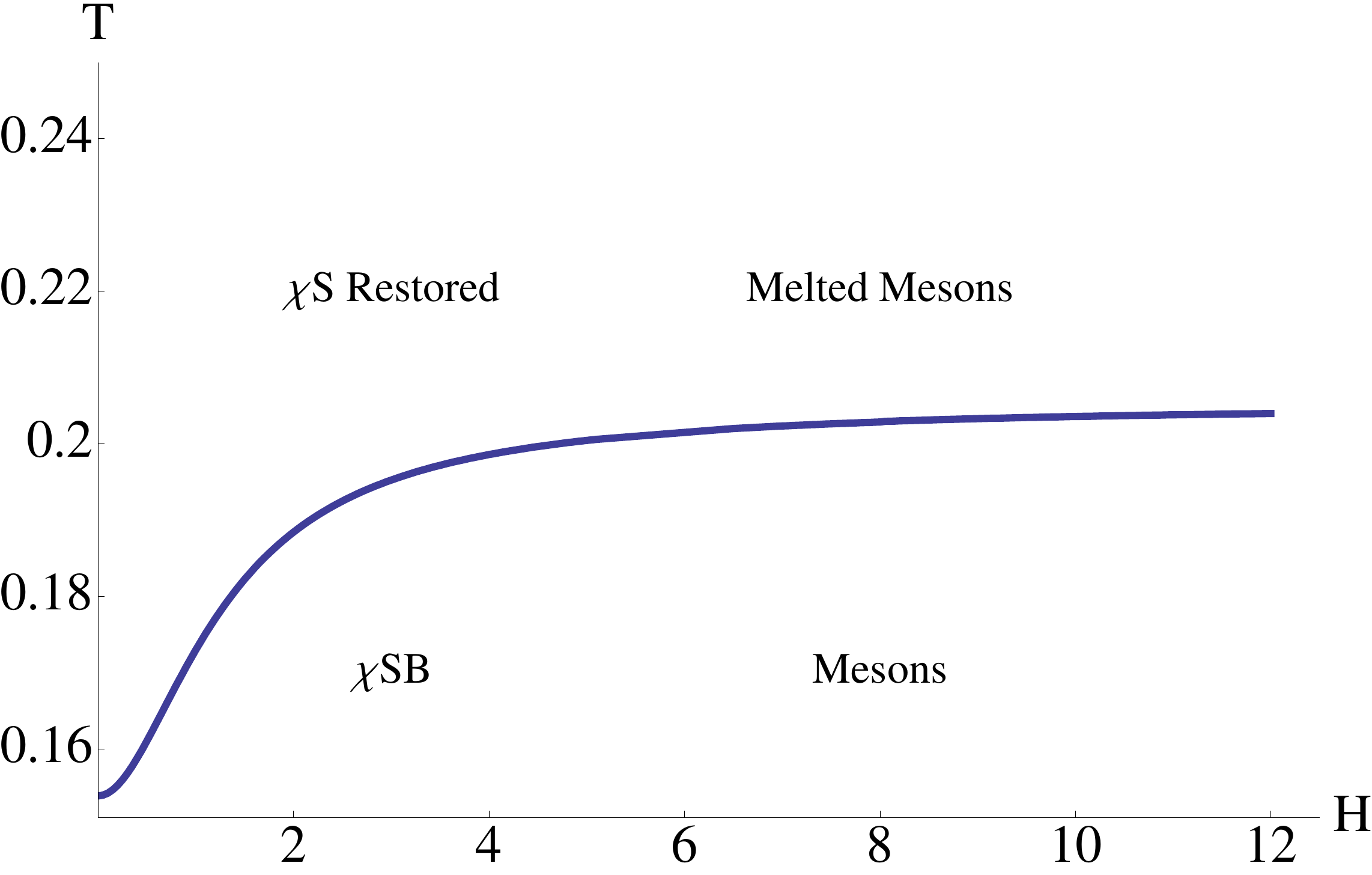}
\caption{\small The phase diagram in $T-H$ plane. The vertical axis is expressed in units of $1/L$ and the horizontal axis is expressed in units of $R$.}
\label{fig: phase}
\end{center}
\end{figure}

From the zero of $\Delta S$ we can find out
the critical value of $y_T$ for which the symmetry restoration occurs.
Now to represent the phase diagram in terms of physical quantities, we
recall that there is a length scale $L$ corresponding to the
separation of the brane--anti-brane pair at the boundary. So we
express the chiral symmetry restoring temperature in units of $1/L$
using the critical value of $y_T$ in equation~(\ref{eqt: sepzmag}). The
resulting phase diagram is shown in figure~\ref{fig: phase}.
It is interesting to note that the presence of magnetic field
increases the symmetry restoring temperature. In other words it
promoted the spontaneous breaking of chiral symmetry. This fits with
the general expectations from field theory (see {\it e.g.} refs.\cite{Miransky:2002eb}) and
the supergravity/probe intuition that introducing a magnetic field
places more energy into the system; therefore in order to minimize the
energy, condensates are formed (the branes bend more) resulting in
more readily broken chiral symmetry. (It should be noted, however,
that in this specific holographic model the identification of a quark
condensate is a rather subtle issue (see {\it e.g.},
ref.~\cite{Bergman:2007pm})). An interesting and compelling identification of this condensate was made in ref.~\cite{Aharony:2008an} in terms of an open Wilson line (or in short, the OWL) operator. It was further demonstrated in ref.~\cite{Argyres:2008sw} that this OWL operator does have the right dependence with respect to the magnetic field which is consistent with the ``magnetic catalysis" phenomenon.

We can extract some more information about the transition by studying
certain thermodynamic quantities at the phase transition. To that end,
let us note that the first order phase transition is accompanied by
entropy density that jumps at $T=T_c$ yielding a non--zero latent heat
as reported in ref.~\cite{Parnachev:2006dn}, also a change in
magnetization
\begin{eqnarray}
&& \Delta s=-\frac{1}{V_{R^3}}\frac{\partial\left(S_{\rm curved}-S_{\rm straight}\right)}{\partial T}\ , \quad C_{\rm latent}=T_c\Delta s\ ,\nonumber\\
&& \Delta\mu=-\frac{1}{V_{R^3}}\frac{\partial\left(S_{\rm curved}-S_{\rm straight}\right)}{\partial H}\ .
\end{eqnarray}

\begin{figure}[h!]
\begin{center}
\subfigure[] {\includegraphics[angle=0,
width=0.45\textwidth]{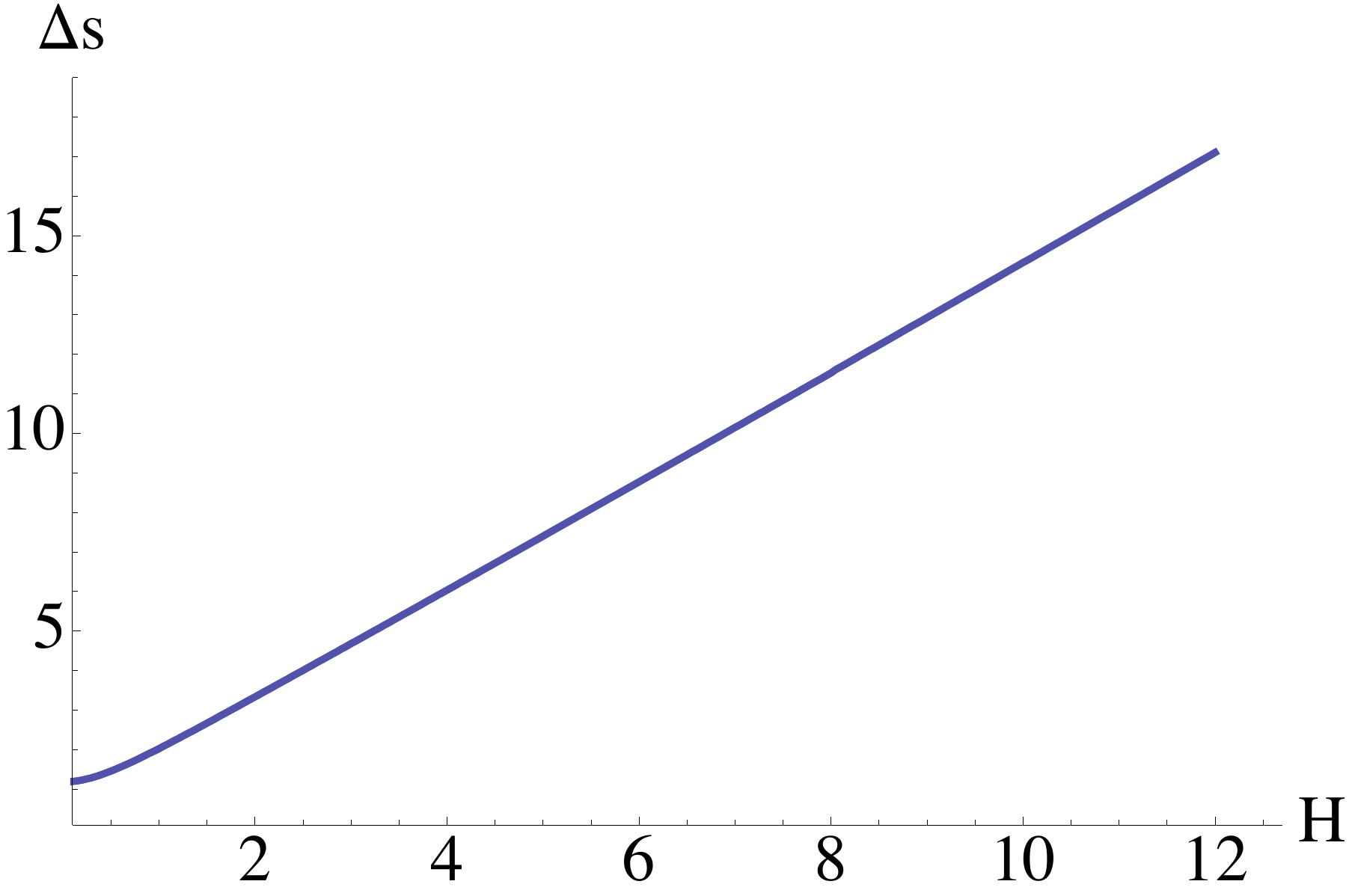} \label{fig: scrit}}
\subfigure[] {\includegraphics[angle=0,
width=0.45\textwidth]{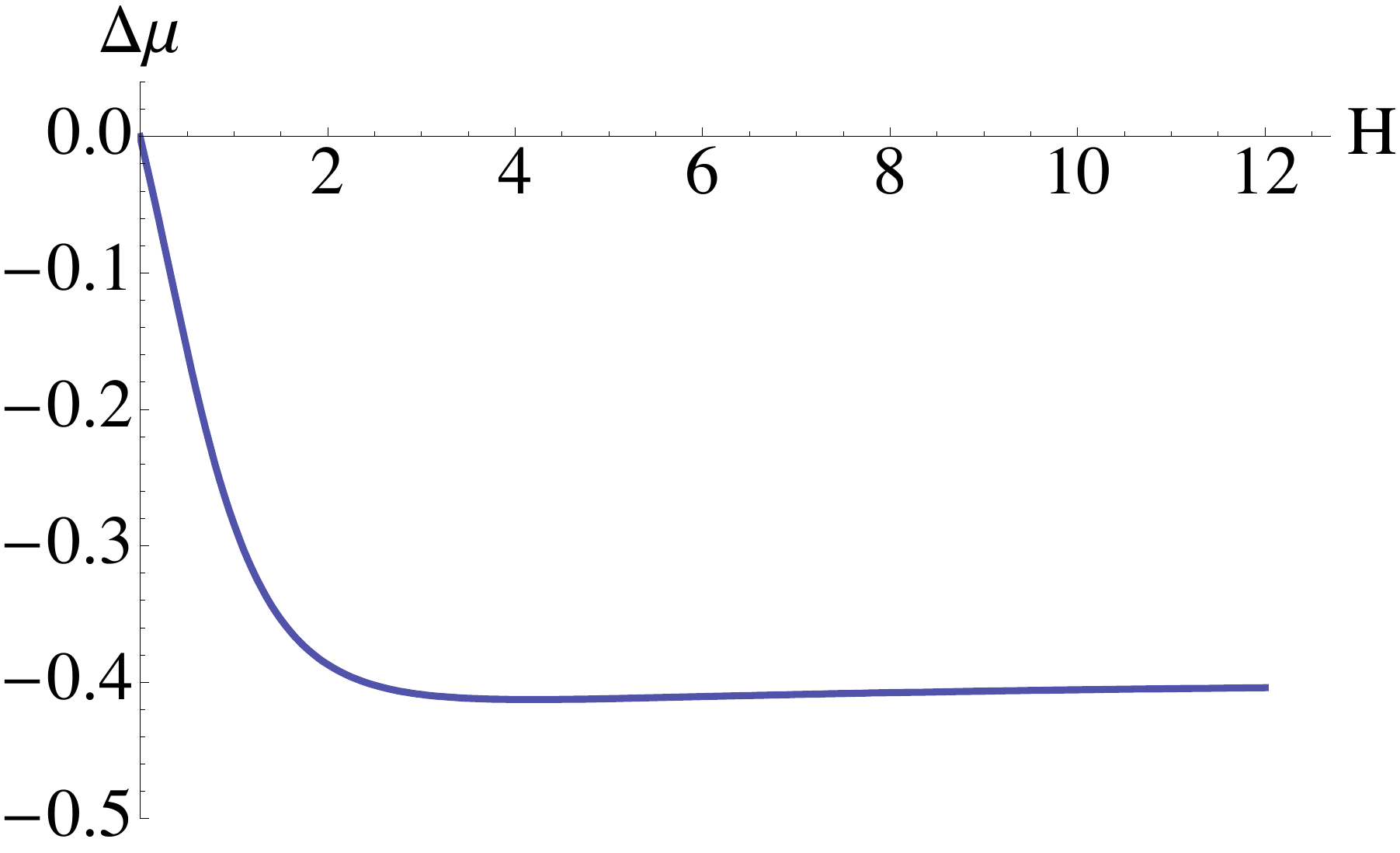} \label{fig: mucrit}}
\caption{\small The behaviour of jump in entropy and change in magnetisation at the critical temperature where chiral symmetry is being restored. The vertical axis is evaluated in units of $CU_0^{7/2}R^{3/2}$ and we have set $d=1$.}
\end{center}
\end{figure}

The absolute free energy and any thermodynamic quantity obtained from it (such as the absolute magnetization) for the two classes of embeddings (the straight and the curved branes respectively) are formally divergent quantities. Hence we compute the relative quantities which are finite. We studied the dependence of the change in entropy density and the relative magnetization numerically, and the
results are shown in figures \ref{fig: scrit} and \ref{fig: mucrit}
respectively. The relative magnetization also shows a similar
saturation behaviour for high enough magnetic field. The straight branes correspond to the melted phase where quarks are free whereas the curved branes correspond to the mesonic phase where quarks exists in the form of bound states or chiral condensates. Therefore it is expected that the chiral symmetry restored phase (corresponding to the straight branes) is more ionized than the chiral symmetry broken phase (corresponding to the curved branes). This is in accord with our observation that the relative magnetization is negative in figure \ref{fig: mucrit}.
%

\section*{5.4 \hspace{2pt} A note on background electric field}
\addcontentsline{toc}{section}{5.4 \hspace{0.15cm} A note on background electric field}

For this section we again consider the $8$-branes as flavours. We can
introduce a background electric field by considering the following
form of world--volume gauge field\cite{Karch:2007pd,Filev:2007gb}:
\begin{equation}
A_1(t,u)=(-Et+h(u))\ .\\
\end{equation}
This means we have a non-zero constant electric field along $x^1$. The
function $h(u)$ encodes the response of the fundamental flavours to
the external field, namely it encodes the information of the non-zero
current when flavours are free to move and therefore conduct. Note
that here we do not introduce any chemical potential, therefore in the
gauge theory there is no a priori candidate for carrying the charge.
However, there could still be current caused by pair creation in
presence of the electric field. We comment on some observations and
expectations henceforth.

For simplicity we assume $2\pi\alpha'=1$. We choose the same ansatz
for the probe brane profile as equation~(\ref{eqt: ansatz}). The DBI
action can be computed to be
\begin{eqnarray}\label{eqt: se}
S_{D8}&=& C\int du dt e^{-\phi} G_{xx}\left({\rm det}S_4\right)^{\frac{1}{2}}\left[G_{tt}G_{xx}g_{uu}+G_{tt}h'^2-g_{uu}E^2\right]^{1/2}\ , \nonumber\\
g_{uu}&=&G_{uu}+\tau'^2G_{\tau\tau}\ , 
\end{eqnarray}
where $C=(N_f \mu_8)/(V_{\mathbb{R}^3}V_{S_4})$, $G_{\mu\nu}$ is the
background metric and $g_{uu}$ is the induced metric component on the
world-volume of the D8-brane along $u$-direction.

Now we will have two constants of motion corresponding to the function
$\tau(u)$ and $h(u)$ as follows
\begin{eqnarray}\label{eqt: Eeom}
\frac{Ce^{-\phi}\sqrt{ {\rm det} S_4}G_{xx}(G_{tt}G_{xx}-E^2)G_{\tau\tau}\tau'}{\sqrt{G_{tt}h'^2-(E^2-G_{tt}G_{xx})(G_{uu}+G_{\tau\tau}\tau'^2)}} & = & B \ ,\\
\frac{Ce^{-\phi}\sqrt{ {\rm det} S_4}G_{xx} G_{tt}h'}{\sqrt{G_{tt}h'^2-(E^2-G_{tt}G_{xx})(G_{uu}+G_{\tau\tau}\tau'^2)}} & = & J \ .
\end{eqnarray}
The constants $B$ and $J$ are related to the minimum radial distance
along the world-volume of the probe brane ($U_0$ as before) and the
gauge theory current (a similar identification made in
ref.~\cite{Karch:2007pd} holds here also).

Solving equation~(\ref{eqt: Eeom}) numerically to look for all
possible solutions is a difficult problem. Nevertheless, it can be
shown from equation~(\ref{eqt: Eeom}) that for confined phase there
can be only the joined solution (curved as before), for which the
current vanishes if we impose the condition that the brane--anti-brane
pair join smoothly at $U_0$. This can be shown by formally solving
equation~(\ref{eqt: Eeom}) to obtain $\tau(u)$ and $h(u)$ as a
function of $u$ and the constants $B$ and $J$. If we expand the
solution around the joining point $u=U_0$ and demand that
$\tau'\to\infty$ as $u\to U_0$, we obtain that the current identically
vanishes. However, if the branes do join but not smoothly, this is no
longer necessarily true. One could, in this case, have cusp-like solutions for which the brane--anti-brane pair join at an angle and in order to stabilize the system it is necessary to consider the inclusion of a bunch of fundamental strings extending from the joining point $u=U_0$ to $u=U_{\rm KK}$. This leads us to a construction much like in refs.~\cite{Rozali:2007rx, Bergman:2007wp, Davis:2007ka}, where the effect of baryons was considered. Therefore the symmetry broken phase may have a non-zero current carried by the baryons. The chiral symmetry is always broken in this
phase forced by the topology of the background.

For the deconfined phase, it can also be shown in a similar way that
there exists curved solutions joining smoothly at some $U_0$, which
has zero current modulo the caveat mentioned in the last paragraph. This is intuitive from the gauge theory point of
view, since we are in a chiral symmetry broken phase therefore there is
no charge carrier present to conduct (ignoring the possibility of a baryon current). The possible effect of pair
creation is diminished by the existence of quark bound states in the
symmetry broken phase. However, for straight branes we expect non-zero
current to flow.

A familiar fact from studying flavours in electric field tells us that
the presence of electric field induces a so called ``vanishing locus"
for the probe DBI action. The ``healing" procedure ({\it e.g.}, in
ref.~\cite{Filev:2007gb}) is to give a non-zero vev to the current. In
practice this is obtained by substituting the functions $\tau(u)$ and
$h(u)$ from equation~(\ref{eqt: Eeom}) in favour of the constants $B$
and $J$ in the action in equation~(\ref{eqt: se}) and demanding the
reality condition for the action for $u\in [U_T,\infty)$. The
condition leads to the following two equations
\begin{eqnarray}\label{eqt: cond}
&& \left(G_{tt}G_{xx}-E^2\right)^2=0\ ,\nonumber\\ 
 && B^2 e^{2\phi}G_{tt}-\left(G_{tt}G_{xx}-E^2\right)G_{\tau\tau}\left(C^2({\rm det} S_4)G_{tt}G_{xx}^2-e^{2\phi}J^2\right)=0\ . 
\end{eqnarray}
The two expressions are the terms in the on--shell action that go to
zero in the numerator and the denominator respectively. It can be
shown from equation~(\ref{eqt: cond}) that in order to have $J\not=0$
one has to have $B=0$, which corresponds to the straight brane
solutions.

For a given electric field, we can determine the position of the
vanishing locus $u_{eh}$ from the first condition in
equation~(\ref{eqt: cond}). Knowing $u_{eh}$ we can then extract the
current $J$, and therefore the conductivity using $J=\sigma E$. To
express the conductivity in terms of physical quantities let us recall
that
\begin{eqnarray}
\mu_8=(2\pi)^{-8}\alpha'^{-\frac{9}{2}}\ , \quad T=\frac{3\sqrt{U_T}}{4\pi R^{3/2}}\ .\nonumber
\end{eqnarray}
Now we will restore the factors of $(2\pi\alpha')$ and also set
$R^3=\lambda_5\alpha'/4=1$. Combining everything we get the
following expression
\begin{eqnarray}
\sigma=\frac{1}{27\pi}\lambda_5 N_fN_c T^2\left(1+\frac{27}{8}\frac{E}{\lambda_5\pi^2T^3}\right)^{1/3}.
\end{eqnarray}
This conductivity is due to the melting of mesons at high temperature
and pair creation mediated by the electric field. Since we get this
formula using DBI action, it captures non-linear behaviour of the
conductivity with respect to the electric field.

The effect of the electric field on chiral symmetry breaking should be
to reduce the symmetry restoring temperature by polarizing the bound
states into constituent quarks. The analysis above points to the fact
that electric field works as expected from field theory perspective.
However, the energy consideration does not lead to the expected
result, because as electric field increases $u_{\rm eh}$ also increases,
therefore the straight branes which extend all the way down to $U_T$
has more DBI action energy as compared to their curved counterparts,
which can only extend down to $u_{\rm eh}$. We hope to address this issue in future.

\section*{5.6 \hspace{2pt} Concluding remarks}
\addcontentsline{toc}{section}{5.6 \hspace{0.15cm} Concluding remarks}

We have extended the study of Witten--Sakai-Sugimoto model to include the
presence of external electric and magnetic fields, examining the
dynamics of the flavour sector in (an analogue of) the ``quenched''
approximation. We have seen that external magnetic field helps in
chiral symmetry breaking. This particular effect of an external
magnetic field has been referred to as magnetic catalysis in field
theory literature (see {\it e.g.}, refs.~\cite{Miransky:2002eb}). Our
results and observations are consistent with results from those
approaches. We found that the chiral symmetry restoring temperature
increases with increasing magnetic field. We have further observed
that such holographic models have an upper bound for the symmetry
restoring temperature depending on the dimension of the gauge theory
and the probe brane.

We briefly studied the effect of external electric field, and though
we have not explored all of the details, we expect the dynamics to be
also consistent with the field theory intuition, although (as we did
for magnetic field here) the precise details should be interesting to
uncover.

Another important avenue would be to study the meson spectra in the
presence of these external fields, which we pursue in the next chapter focussing on the magnetic field case.


\chapter*{Chapter 6:  \hspace{1pt} The Mesons in the Magnetic Field}
\addcontentsline{toc}{chapter}{Chapter 6:\hspace{0.15cm} The Mesons in the Magnetic Field}

\section*{6.1 \hspace{2pt} Introductory remarks}
\addcontentsline{toc}{section}{6.1 \hspace{0.15cm} Introductory remarks}

In this chapter we will investigate the effect of an external magnetic field on finite temperature the meson spectrum in the Witten--Sakai-Sugimoto model. The material presented in this chapter is based on the work done in collaboration with Clifford Johnson\cite{Johnson:2009ev}.

Following the field theory literature, {\it e.g.}
ref.~\cite{Miransky:2002eb}, we have briefly reviewed in Chapter 2 that for $(3+1)$ and $(2+1)$-dimensional
field theories there is a universal phenomenon at work that is
sometimes referred to as the magnetic catalysis of chiral symmetry
breaking.

As we have seen in the previous chapters, this physics (that of a magnetic field catalyzing chiral symmetry
breaking) can now be studied using the alternative technique of
gauge/gravity duals (the magnetic field maps to a background B--field
in the string theory and causes the flavour branes in the geometry to
bend, generically), and it has been observed\cite{Johnson:2008vna} to
persist in all the Witten--Sakai-Sugimoto type models, where the flavour
degrees of freedom are inserted as defects in the dual geometry for
several $(2+1)$-- and $(3+1)$--dimensional gauge theories. We encountered a similar
effect in chapter~3 in a different model, where some of the
structure of the meson spectrum was also studied, revealing phenomena
such as Zeeman splitting,  level mixing, {\it etc} in ref.~\cite{Albash:2007bk}. Here we would like
to study the meson spectrum in the presence of an external magnetic field
in the Witten--Sakai-Sugimoto model supplementing the study of the phase
diagram discussed in the previous chapter.

Previous studies of the meson spectrum in this particular model have
been carried out in {\it e.g.}, ref. \cite{Peeters:2006iu} for both
low spin and high spin mesons. A comprehensive review on holographic studies of mesons can be found in ref.~\cite{Erdmenger:2007cm}. We uncover the role of the magnetic
field in these spectra in this particular Type IIA model. For the low spin mesons we restrict ourselves
to the high temperature phase only and therefore study the quasinormal
modes of the scalar and a subset of the vector mesons. This comes with
a caveat. The precise identification between the quasinormal frequency
associated with the supergravity fluctuations in this model and the
meson spectrum in the dual field theory is rather subtle and poorly
understood. For a detailed analysis regarding the issue of
  precise identification, see ref.~\cite{Paredes:2008nf}. In the
Witten--Sakai-Sugimoto type models the mesons transform under an ${\rm U}(N_f)_{\rm
  diag}$, whereas the quasinormal modes transform under either an
${\rm U}(N_f)_L$ or an ${\rm U}(N_f)_R$. This fact alone readily makes it
difficult to precisely define what gauge theory quantities we would be
studying by extracting the quasinormal modes. Nevertheless the
quasinormal modes are naturally associated to the embeddings falling
into the black hole of the gravity dual and therefore we study those
in their own rights.

Moreover, since the embedding function of the probe D8--brane is
trivial ($\tau_0(u)={\rm const}$, where $\tau_0(u)$ denotes the
profile function) in the high temperature phase, the scalar and the
vector fluctuations remain decoupled even in the presence of a
background anti--symmetric field such as a constant magnetic
field. This is to be contrasted with the D3--D7 model in a background
magnetic field\cite{Albash:2007bq} where the profile function of the
probe D7--brane is non-trivial and therefore the scalar fluctuations
couple with the vector fluctuations. Thus studying quasinormal modes
in presence of a background anti--symmetric field in general is more
difficult in the later model.

The study of high spin mesons is another avenue that we pursue. We
find that the presence of the magnetic field enhances the stability of
such mesons by increasing their angular momentum. This is turn increases
the dissociation temperature at which they fall apart into their
constituents. We also uncover a simple realization of the Zeeman
effect in the presence of a background field. It can be readily seen
that such effects exist in all the Witten--Sakai-Sugimoto type models with
probe branes of diverse dimensions.

\section*{6.2 \hspace{2pt} Mesons with small spin}
\addcontentsline{toc}{section}{6.2 \hspace{0.15cm} Mesons with small spin}

Mesons with low spin in the dual gauge theory correspond to the small fluctuations of the classical profile of the probe brane. The fluctuation in the geometric shape (which is denoted by the function $\tau(u)$) of the probe branes corresponds to the scalar (or pseudo-scalar) meson mode and the fluctuation of the Maxwell field on the worldvolume of the D8/$\overline{\rm D8}$-brane corresponds to the vector (or axial vector) mesons.

\subsection*{6.2.1 \hspace{2pt} Scalar fluctuation}
\addcontentsline{toc}{subsection}{6.2.1 \hspace{0.15cm} Scalar fluctuation}

Let us consider a fluctuation of the probe brane embedding function $\tau(u)=\tau_0(u)+\chi$, where $\tau_0(u)$ is the classical profile. Before proceeding further let us recall that the function $\tau_0(u)$ is non-trivial only for the U-shaped configuration (for the straight branes, this function is just a constant). Here we will consider the high temperature phase only for which $\tau_0(u)={\rm const.}$ and the scalar fluctuation is always decoupled from the vector fluctuations.

To recast the equation of motion for the fluctuation mode into an equivalent Schr\"{o}dinger equation we define, $f_{\chi}(u)=\sigma(u)g(u)$, where $\sigma(u)$ is defined in eqn. (\ref{eqt: schg}) (the details of the coordinate changes are given in appendix C). The equivalent Schr\"{o}dinger equation is given by (to avoid notational clutter we choose $R=1$ and $U_T=1$)
\begin{eqnarray}\label{eqt: scalar}
&& \partial_{\tilde u}^2g(\tilde u)+\omega^2g(\tilde u)-V_S(\tilde u)g(\tilde u)=0\ , {\rm where} \quad d\tilde{u}=\frac{u^{3/2}}{u^3-1}du\ , \nonumber\\ 
&& {\rm and}\quad V_S(u)=\frac{\left(u^3-1\right) \left(5 H^4 \left(7u^3+5 \right) +2 H^2 u^3 \left(71 u^3+7 \right) +16 u^6
   \left(5 u^3+1\right)\right)}{16 u^5 \left(H^2 +u^3\right)^2}\ . \nonumber\\
\end{eqnarray}
Here $\tilde{u}$ is the ``tortoise" coordinate defined in appendix C. For simplicity, we continue to express the Schr\"{o}dinger potential in the original $u$-variable.
\begin{figure}[!ht]
\begin{center}
\includegraphics[angle=0,
width=0.55\textwidth]{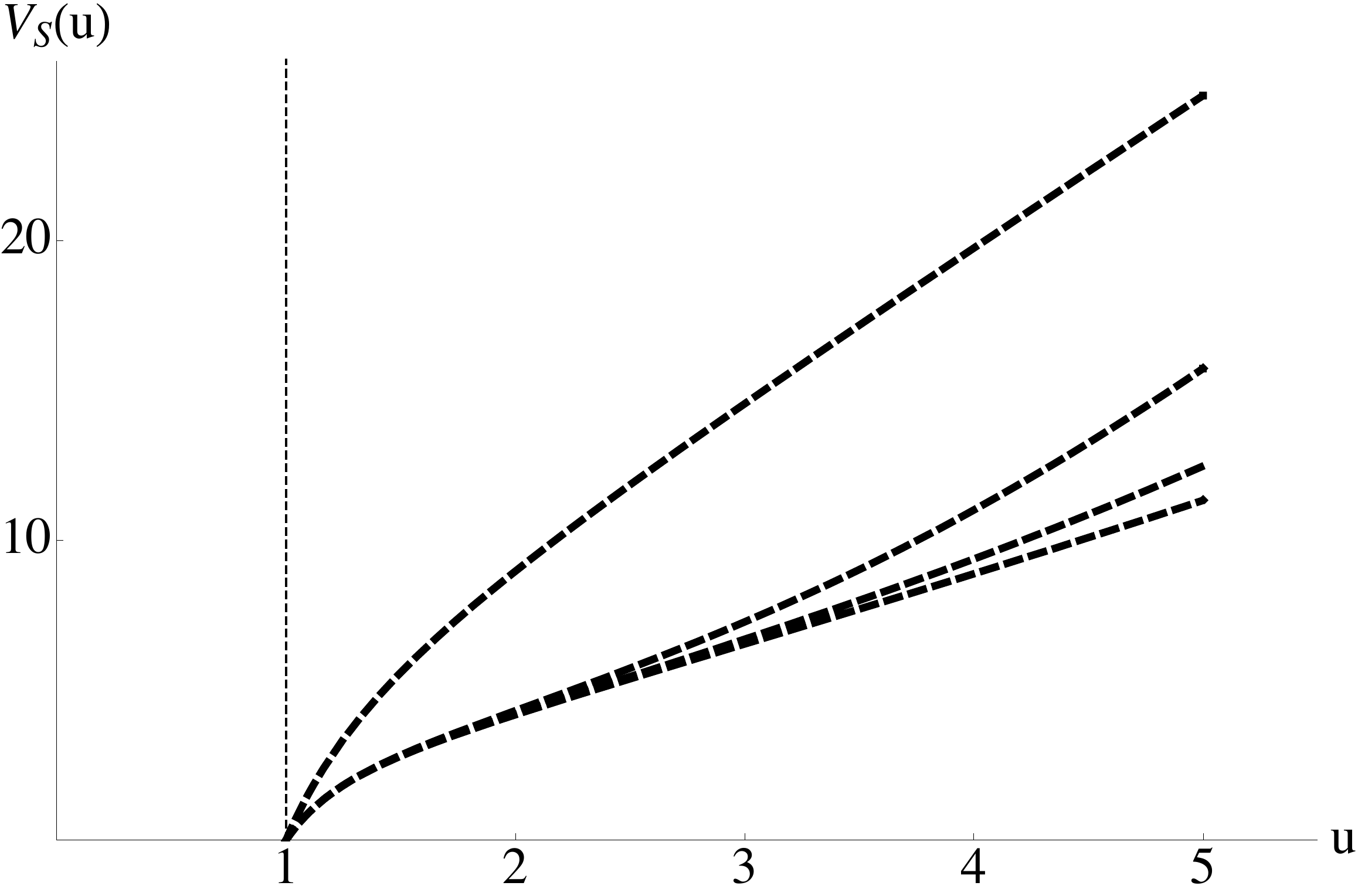}
\caption{\small The Schr\"{o}dinger potential for different values of external magnetic field. The dashed vertical black line represents the position of the horizon, the top-most curve represents the potential at zero external field. As the field increases we find the potential grows more slowly with increasing~$u$.}
\label{fig: vscalarT}
\end{center}
\end{figure}
Now we can understand qualitative features of imaginary part of quasinormal frequency by studying this effective potential. We resort to numerics to study the potential $V_s(u)$. This potential vanishes at $u=1$ because of the presence of the horizon and diverges as $u\to\infty$ reflecting infinite gravitational potential of the background geometry.

The important feature that we find from figure \ref{fig: vscalarT} is that the Schr\"{o}dinger potential is always positive, therefore from eqn. (\ref{eqt: vsch}) we can conclude that the imaginary part of the quasinormal frequency is strictly negative. Moreover, the potential does not develop a negative energy well, therefore there exists no negative energy bound states in this phase. This is consistent with the fact that the probe brane configuration is stable and tachyon free and chiral symmetry restoration is accompanied by meson melting transition. Tachyonic modes in the scalar field fluctuation are likely to appear in the overheated low temperature phase which is bypassed by the transition. Also we see that the potential vanishes at the horizon and therefore the spectrum is continuous, which is consistent with the melting transition.

\begin{figure}[!ht]
\begin{center}
\subfigure[] {\includegraphics[angle=0,
width=0.45\textwidth]{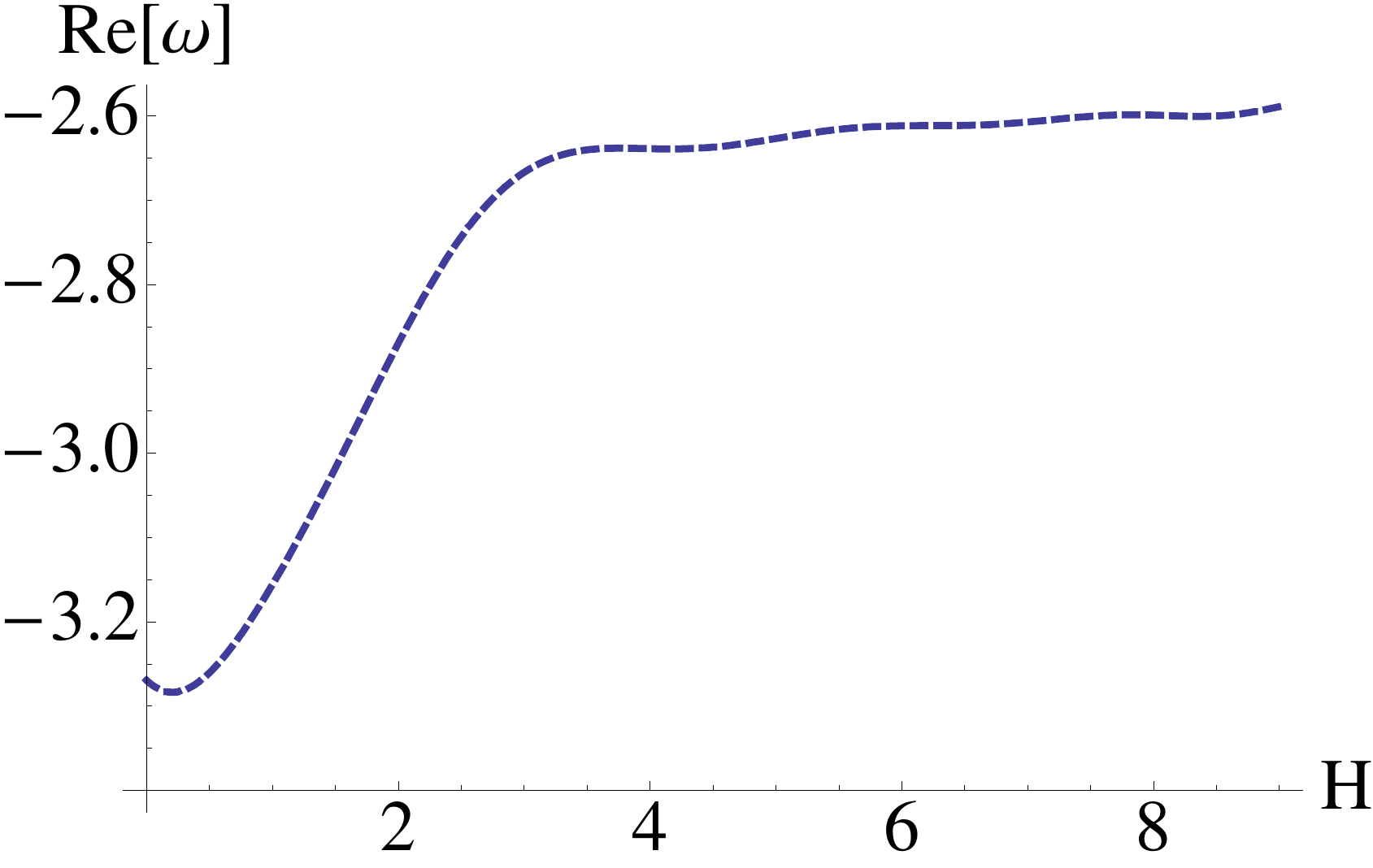} \label{fig: sR}}
\subfigure[] {\includegraphics[angle=0,
width=0.45\textwidth]{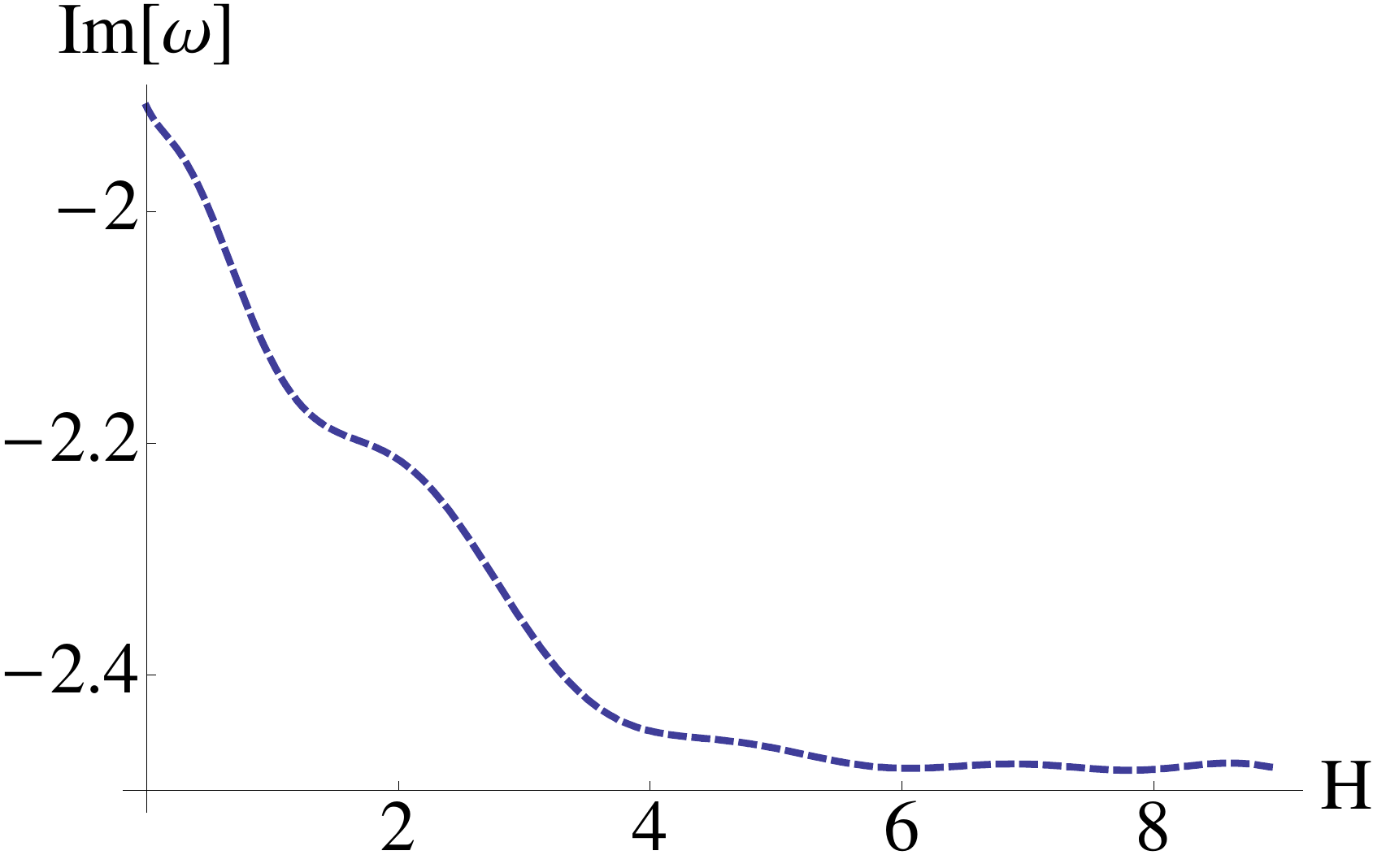} \label{fig: sI}}
\caption{\small The real ($\omega_{\rm R}={\rm Re}[\omega]$) and imaginary ($\omega_{\rm I}={\rm Im}[\omega]$) part of the quasinormal modes for the scalar fluctuation as a function of the external field. The quasinormal frequency is measured in units of the background temperature $T$. The small wiggles are due to numerical errors; here we have plotted the polynomial which fits the data. As expected before, we confirm that ${\rm Abs}[\omega_{\rm R}]> {\rm Abs}[\omega_{\rm I}]$, however they are roughly of the same order.}
\end{center}
\end{figure}

Now we can observe that the Schr\"{o}dinger potential increases monotonically as we move away from the horizon. This implies that ${\rm Re}[\omega^2]>0$ and hence ${\rm Abs}[\omega_{\rm R}]> {\rm Abs}[\omega_{\rm I}]$. By explicit numerical calculation we will find that this is indeed the case (we point to figure \ref{fig: sR} and \ref{fig: sI}).

To extract more of the physics, we now compute the quasinormal frequency corresponding to this scalar fluctuation. We find it extremely convenient to perform a change of coordinates as implemented in ref.~\cite{Evans:2008tv}. Let us introduce a new variable $\rho=u/U_T$ and $v=t+\alpha(\rho)$, where $\alpha(\rho)$ is determined from the condition that 
\begin{eqnarray}\label{eqt: vt}
dv=dt+\frac{1}{\rho^3f(\rho)}d\rho\ .
\end{eqnarray}
This coordinate system $\{v,\rho\}$ makes the numerics conveniently stable to find the quasinormal modes. We refer the reader to ref.~\cite{Evans:2008tv} for more details. In presence of an external field this equation does not take the general form described in ref.~\cite{Evans:2008tv}. This is due to the breaking of the ${\rm SO}(3,1)\to {\rm SO}(1,1)\times {\rm SO}(2)$ by the presence of the magnetic field as a result of which the Laplacian in the $\{x^2, x^3\}$-direction gets squashed. The equation of motion for the fluctuation now takes the form
\begin{eqnarray}
&& \partial_\rho\left(\rho^{11/2}(1+H^2\rho^{-3})^{1/2}ff_{\chi}'\right) + \rho^{5/2}(1+H^2\rho^{-3})^{1/2}f_{\chi} \nonumber\\
& - & i\omega\left[\partial_\rho\rho^{5/2}(1+H^2\rho^{-3})^{1/2}f_{\chi}+2\rho^{5/2}(1+H^2\rho^{-3})^{1/2}f_{\chi}'\right] =0 \ . 
\end{eqnarray}
It is now straightforward to check that this equation admits the existence of quasinormal modes. To find how it depends on the background magnetic field we use a shooting technique to solve the fluctuation equation and pick the appropriate value of $\omega$ for which the equation admits a normalizable solution (this is achieved by imposing the condition that $f_{\chi}\to 0$ as $\rho\to\infty$). As boundary conditions we impose $f_{\chi}(1+\epsilon)=1$ and $f_{\chi}'(1+\epsilon)$ equal to a value obtained by requiring the equation of motion to be regular near $\rho=1$, the event horizon. Here the parameter $\epsilon$ is a vanishingly small number in our numerical scheme. 
\label{fig: can}
The result of this pursuit is summarised in figure \ref{fig: sR} and \ref{fig: sI}. Since we are in the high temperature phase the quasinormal mode is comprised of both real and imaginary parts. It would be expected that the absolute value of the real part corresponds to the mass of the meson before it melts and the absolute value of the imaginary part corresponds to its inverse lifetime analogous to ref.~\cite{Hoyos:2006gb} for the D3--D7 system. Figure \ref{fig: sR} then tells us that the magnetic field increases the scalar meson mass and figure \ref{fig: sI} hints that the inverse lifetime gets bigger as the field is dialed up. However the precise identification of the meson mass and inverse lifetime with the quasinormal modes of the scalar fluctuation is a subtle issue in this particular model (see ref.~\cite{Paredes:2008nf}).

\subsection*{6.3.2 \hspace{2pt} Vector fluctuation}
\addcontentsline{toc}{subsection}{6.3.2 \hspace{0.15cm} Vector fluctuation}

The vector meson spectrum in general is determined by a set of coupled equations of motion. It is however possible to completely decouple the transverse and the longitudinal modes if we restrict ourselves to the oscillations parallel to the magnetic field. If we restrict ourselves to the oscillation in one of the perpendicular directions ({\it e.g.} along the $x^3$-direction) to the applied magnetic field, then it is also possible to decouple one of the transverse modes (in this case the $A_2$-vector mode). However, the remaining transverse mode and the longitudinal mode remain coupled. The details are given in appendix B. Here we study the decoupled sectors only.

\subsubsection*{6.3.2a \hspace{2pt} The transverse mode}
\addcontentsline{toc}{subsubsection}{6.3.2a \hspace{0.15cm} The transverse mode}

We begin analyzing the transverse vector meson spctra. To extract qualitative features of the spectra we study the Schr\"{o}dinger potential. The equation of motion for the $A_2$ vector modes is given by
\begin{equation}
\partial_u\left[e^{-\phi}\sqrt {-\det \left(E^{(0)}\right)} \mathcal{S}^{22}\mathcal{S}^{uu} A_2'(u)\right]+\left(e^{-\phi}\sqrt {-\det \left(E^{(0)}\right)}\right)\omega^2\mathcal{S}^{tt}\mathcal{S}^{22}A_2(u)=0\ , 
\end{equation}
where we chose an ansatz of the form $A_2=A_2(u){\rm exp}(-i\omega t)$. It is straightforward to turn this equation in the form of a Schr\"{o}dinger equation with an effective potential given by
\begin{eqnarray}
&& V_S(u)=\frac{\left(u^3-1\right) \left(5 \left(7 u^3+5\right) H^4-2 \left(u^6-43 u^3\right) H^2+8 \left(u^9+2 u^6\right)\right)}{16 u^5\left(u^3+H^2\right)^2}\ , \nonumber\\
&& {\rm with} \quad d\tilde{u}=\frac{u^{3/2}}{u^3-1}du\ ,
\end{eqnarray}
where $\tilde{u}$ is the ``tortoise" coordinate. Clearly the horizon is located at $\tilde{u}\to\infty$ and the boundary is located at $\tilde{u}\to 0$.
\begin{figure}[!ht]
\begin{center}
\includegraphics[angle=0,
width=0.55\textwidth]{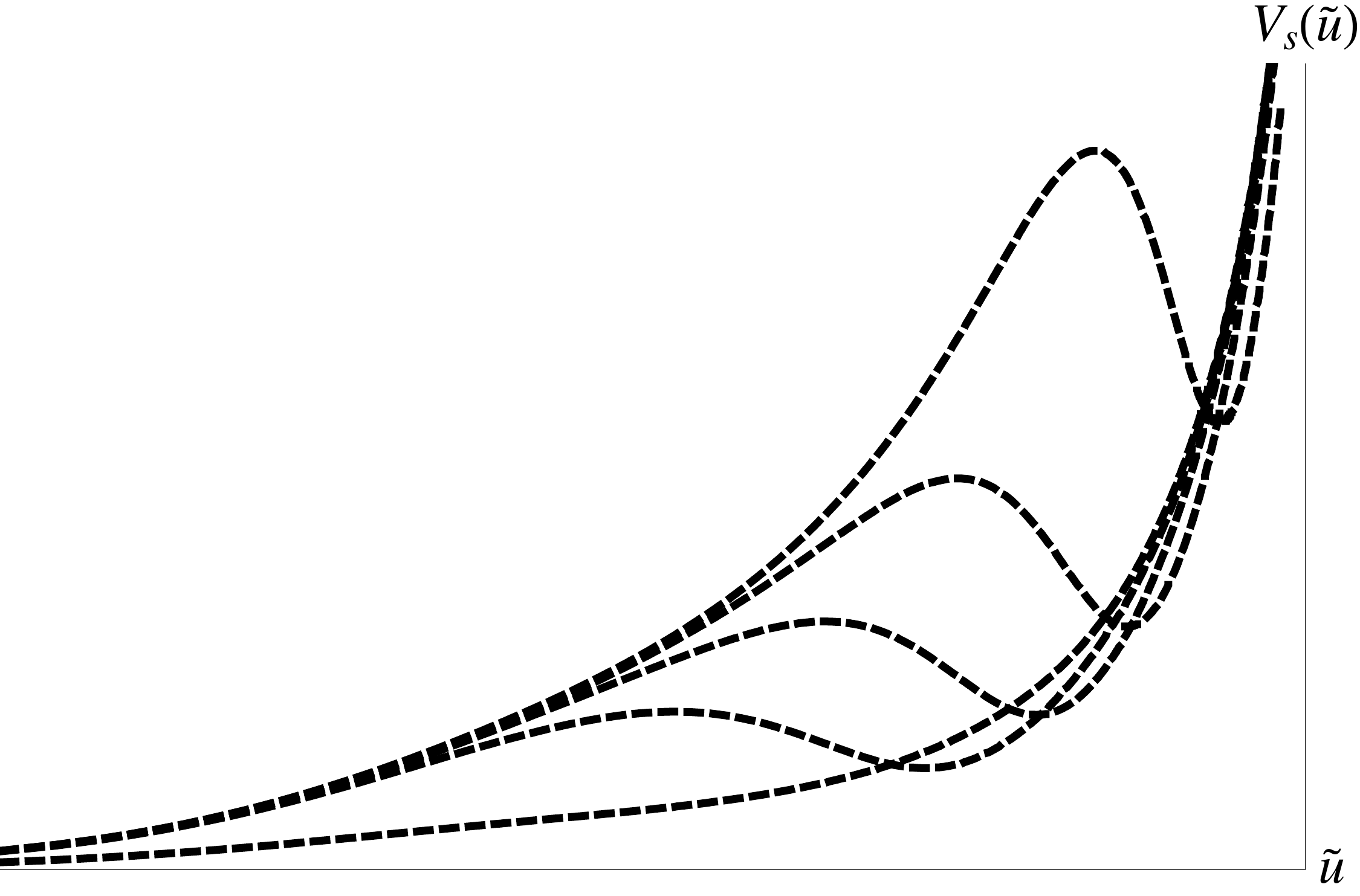}
\caption{\small The Schr\"{o}dinger potential for the vector meson for different values of external magnetic field. Here $\tilde{u}$ denotes the ``tortoise" coordinate mentioned in appendix C. As the magnetic field increases we observe that a small well develops in the effective Schr\"{o}dinger potential. The depth of this positive well is set by the background field.}
\label{fig: A2sch}
\end{center}
\end{figure}
The corresponding potential is shown in figure \ref{fig: A2sch}. The main qualitative feature that emerges from this is the fact that there is no instability due to existence of bound states of positive imaginary frequency modes since the potential remains positive definite. We observe, however, that a non-vanishing magnetic field can create a dimple in the otherwise smooth and rather featureless potential and therefore create local (positive) potential well. Therefore it is possible to have long-lived bound states with ${\rm Abs}[\omega_{\rm R}]\gg {\rm Abs}[\omega_{\rm I}]$. In this case, ${\rm Abs}[\omega_{\rm R}]$ is set by the depth of the well, which in turn is set by the background field. If we increase the magnetic field even higher then the dimple gets bigger and bigger and ultimately swallows the well. At even higher magnetic field the potential again takes a shape similar to the vanishing external field case. Therefore the imaginary part of the quasinormal frequency may have an interesting behaviour with increasing magnetic field. We may expect that  ${\rm Abs}[\omega_{\rm I}]$ starts decreasing with increasing magnetic field, but after a critical value it starts increasing again. Next we turn to the numerics.

The equation of motion (for the vector mode $A_2$ with non-zero spatial momentum) in this case is given by
\begin{eqnarray}
\partial_a\left[e^{-\phi}\sqrt {-\det \left(E^{(0)}\right)} S^{22}\left(S^{a\rho}F_{\rho2}+S^{av}F_{v2}+S^{ai}F_{i2}\right)\right]=0\ , \quad i\in \{1,3\}\ ,
\end{eqnarray}
where $S^{ab}$ are the analogue of $\mathcal{S}^{ab}$ in eqn. (\ref{eqt: comp}) when the coordinate change described in eqn.~(\ref{eqt: vt}) has been performed.

The background field breaks the full ${\rm SO}(3,1)$ Lorentz symmetry and there is an unbroken ${\rm SO}(2)$ symmetry corresponding to the rotation in the plane perpendicular to the magnetic field. Therefore the transverse vector meson spectra with a non-zero momentum will have two distinct branches. Mesons can have momentum along the direction of the background field (corresponding to $i=1$) or in the perpendicular plane (corresponding to $i=3$).

Without any loss of generality we can consider studying the spectrum of the $A_2$ vector meson. We can give this meson a momentum along the $x^1$-direction or along the $x^3$-direction (but no momentum along $x^2$-direction since this is the transverse mode). The spectrum would be entirely equivalent to the spectrum of the $A_3$ vector modes having momentum along the $x^1$-direction or along the $x^2$-direction.

Now we follow the same numerical approach. We impose $A_2(1+\epsilon)=1$ and fix $A_2'(1+\epsilon)$ from the equation of motion near the horizon. With these boundary conditions we look for normalizable solutions for $A_2$, which gives the quasinormal modes.
\begin{figure}[!ht]
\begin{center}
\subfigure[] {\includegraphics[angle=0,
width=0.45\textwidth]{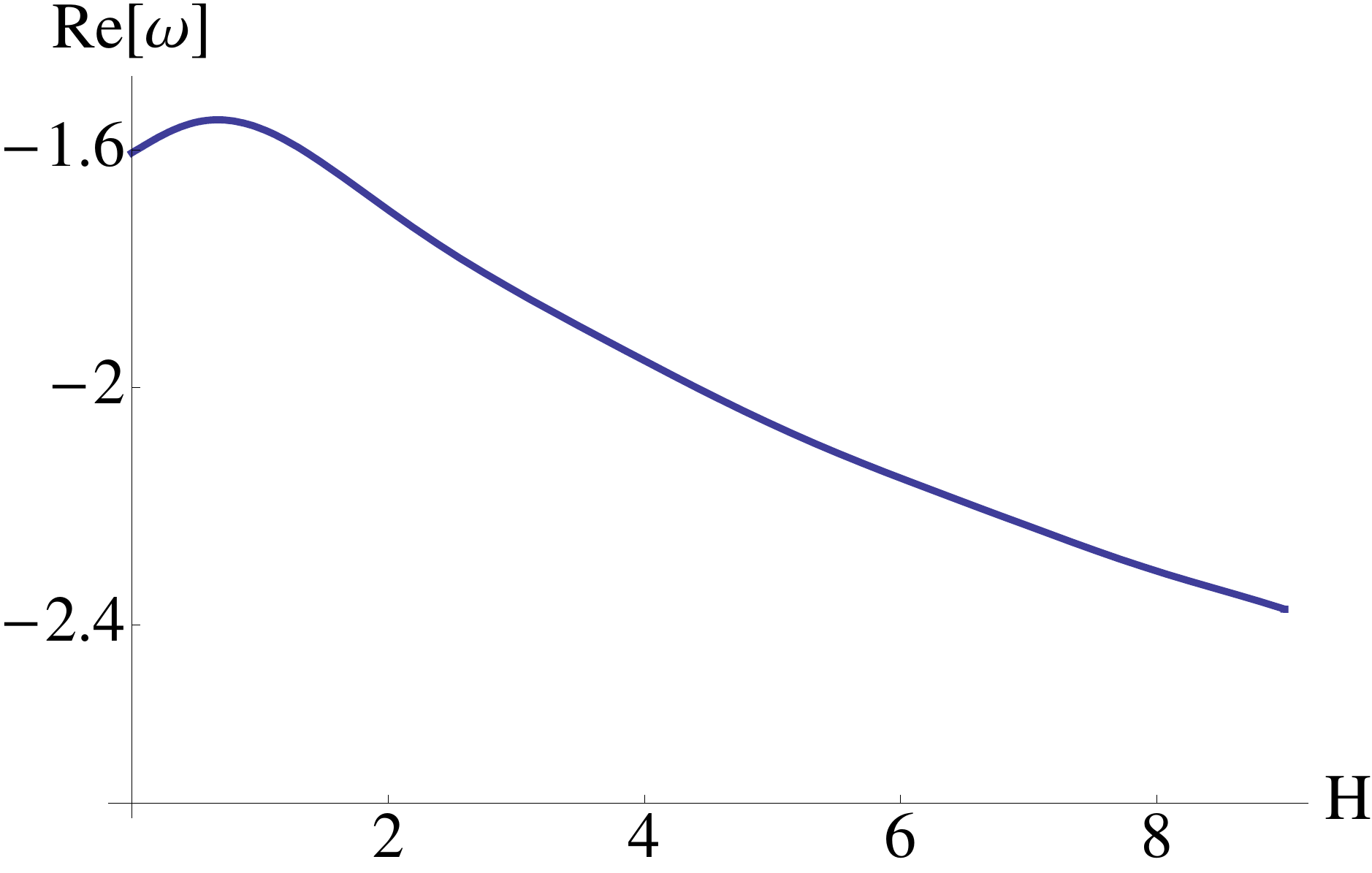} \label{fig: reb}}
\subfigure[] {\includegraphics[angle=0,
width=0.45\textwidth]{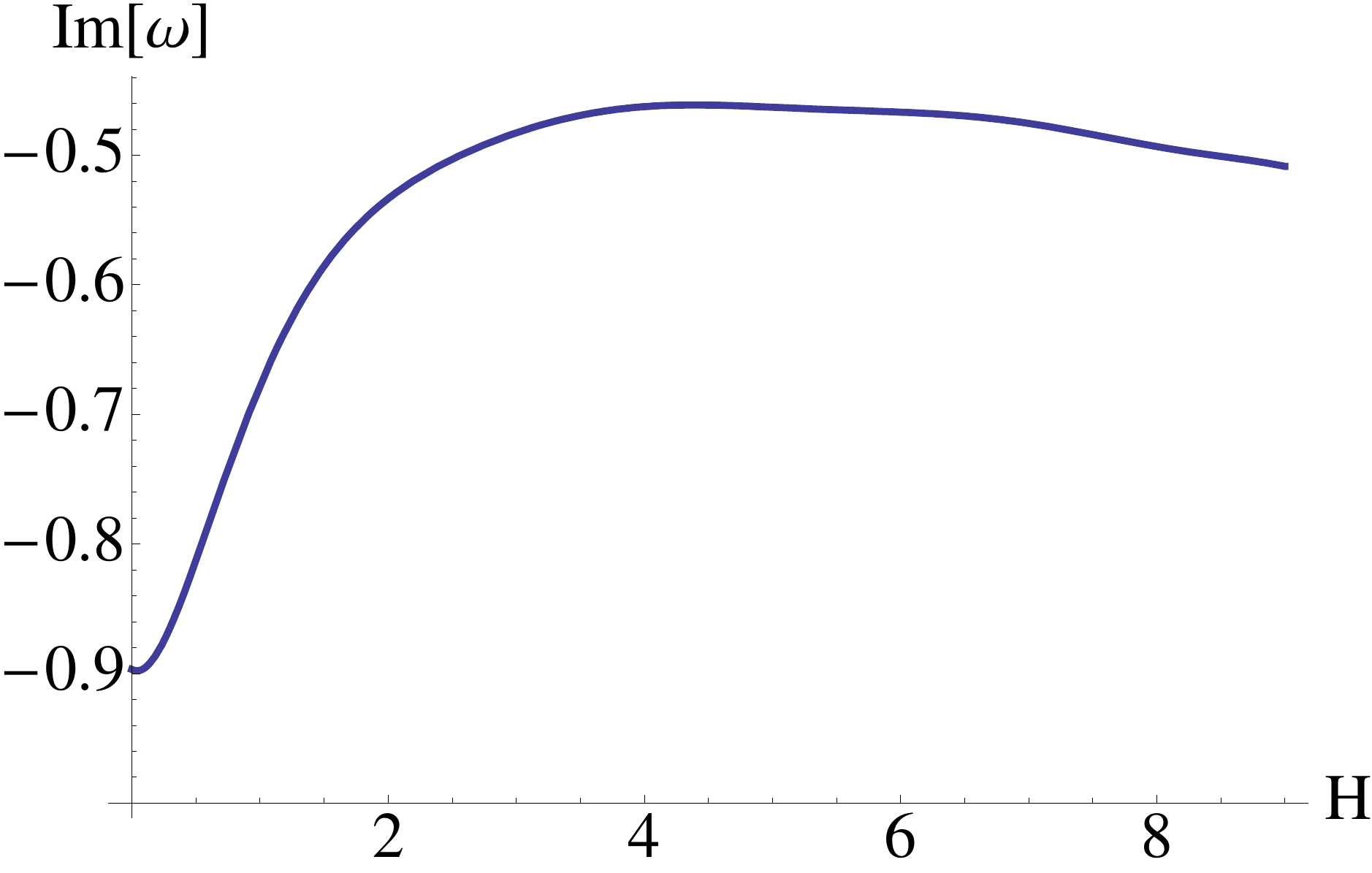} \label{fig: imb}}
\caption{\small The real and imaginary part of the quasinormal modes for the vector fluctuation as a function of the external field. The frequency is expressed in units of background temperature $T$ and the magnetic field is expressed in units of $R$. We find a local minima in ${\rm Abs}[\omega_{\rm I}]$ in the parameter range we have explored. This is qualitatively similar to the behaviour anticipated from analyzing the effective Schr\"{o}dinger potential. We also observe that in the most of the parameter space ${\rm Abs}[\omega_{\rm R}]\gg {\rm Abs}[\omega_{\rm I}]$.}
\end{center}
\end{figure}
\begin{figure}[!ht]
\begin{center}
\subfigure[] {\includegraphics[angle=0,
width=0.45\textwidth]{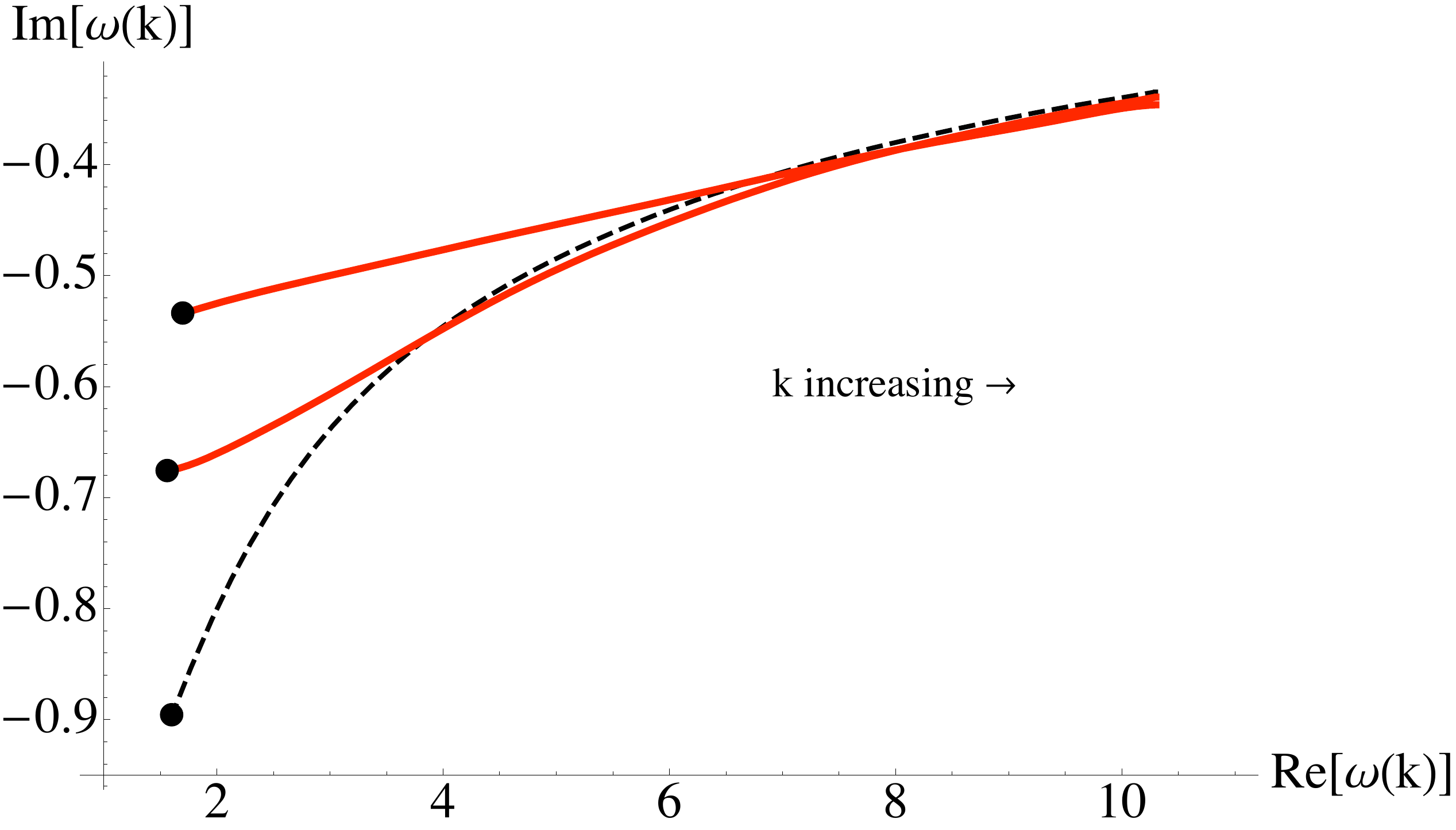} \label{fig: dis}}
\subfigure[] {\includegraphics[angle=0,
width=0.45\textwidth]{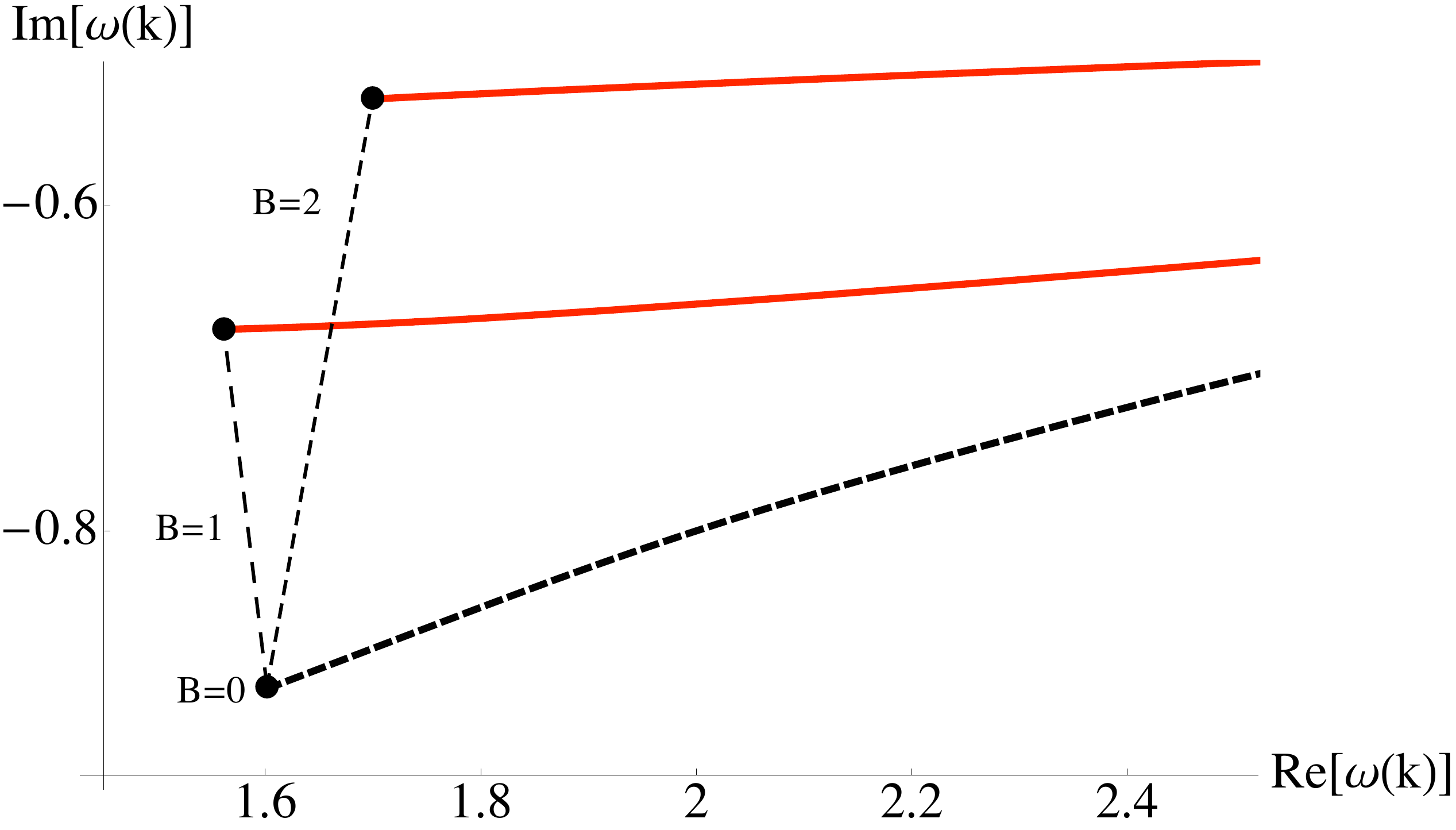} \label{fig: dissplit}}
\subfigure[]{\includegraphics[angle=0,
width=0.45\textwidth]{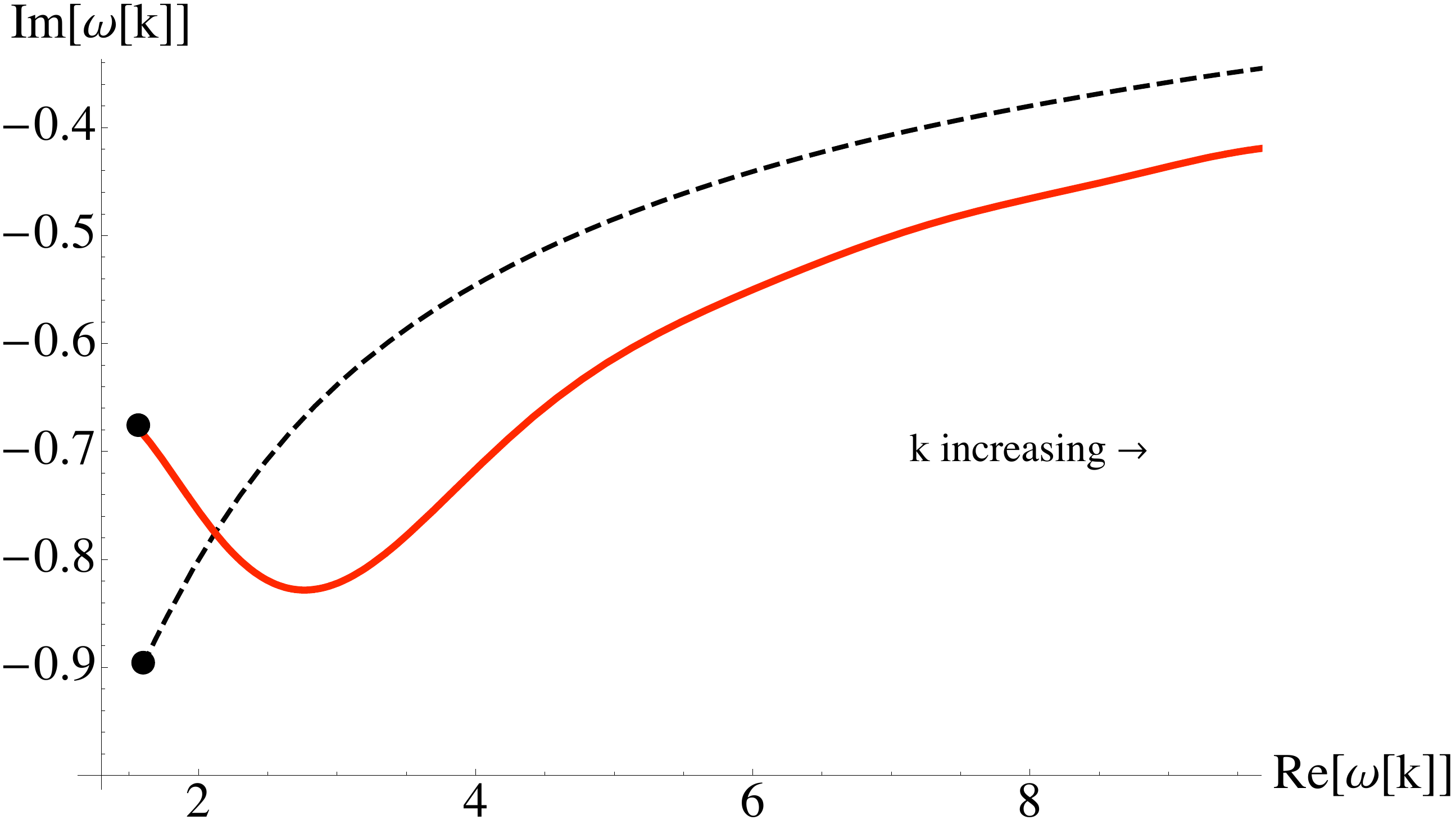} \label{fig: dis3}}
\caption{\small The dispersion relation between the real and imaginary part of the quasinormal frequency for two different values of magnetic field. The red (solid) curves correspond to the dispersion relation for non-zero magnetic field. The black (dashed) curve corresponds to the dispersion relation when there is no magnetic field. The black dots represent the corresponding values of quasinormal frequency for vanishing spatial momentum. Figure \ref{fig: dis} and \ref{fig: dissplit} correspond to vector mesons having momenta along $x^1$-direction and figure \ref{fig: dis3} corresponds to vector meson with momentum along $x^3$-direction.}
\end{center}
\end{figure}
In figure \ref{fig: reb} and \ref{fig: imb} we have shown the dependence of the real and imaginary part of quasinormal frequency for the particular choice of $k=0$. We observe that the inverse lifetime of the vector meson gets shorter as the external field is dialed up and therefore enhances the stability of the meson. This behaviour is opposite to what we observed for the scalar meson fluctuation.

Next we obtain the dispersion relation when $k\not=0$. In figure \ref{fig: dis} we demonstrate this dispersion relation for a wide range of values for $k$ and in figure \ref{fig: dissplit} we have shown a magnified version of figure \ref{fig: dis} to more clearly see the role an external magnetic field plays in this case. These two figures correspond to the dispersion relation when the momentum is parallel ({\it i.e.} along $x^1$-direction) to the magnetic field. We observe that for high enough spatial momentum the dispersion relation curve approximates the zero field dispersion curve quite well. We have shown in figure \ref{fig: dis3} the dispersion relation when the momentum is perpendicular ({\it i.e.} along $x^3$-direction) to the background field. The qualitative behaviour between these two dispersion curves are clearly different; however for large enough spatial momenta these curves tend to become insensitive to the background field. This is because the heavier the meson becomes the less sensitive  it is to the background field.

\subsubsection*{6.3.2b \hspace{2pt} The longitudinal mode}
\addcontentsline{toc}{subsubsection}{6.3.2b \hspace{0.15cm} The longitudinal mode}

Now we study the longitudinal mode. Bearing in mind the change of variables in equation (\ref{eqt: vt}), we define the gauge invariant longitudinal mode to be given by
\begin{equation}
\omega A_1+k A_v= \mathbb{E}\ .
\end{equation}
Now the equation of motion for this longitudinal mode is given by
\begin{eqnarray}\label{eqt: longEv}
\mathbb{E}''(\rho)+\mathcal{Z}_1(\rho)\mathbb{E}'(\rho)+\mathcal{Z}_2(\rho)\mathbb{E}(\rho)=0\ ,
\end{eqnarray}
where $\mathcal{Z}_1$ and $\mathcal{Z}_2$ are known functions of $\rho$. The explicit expressions of $\mathcal{Z}_1$ and $\mathcal{Z}_2$ are not very illuminating, thus we do not provide them here.

Equation (\ref{eqt: longEv}) now can be numerically solved by using the boundary conditions that $\mathbb{E}(1+\epsilon)=1$ and $\mathbb{E}'(1+\epsilon)$ equal to a value obtained from the equation of motion itself. The condition of normalizability then fixes the quasinormal modes.

\begin{figure}[!ht]
\begin{center}
\subfigure[] {\includegraphics[angle=0,
width=0.45\textwidth]{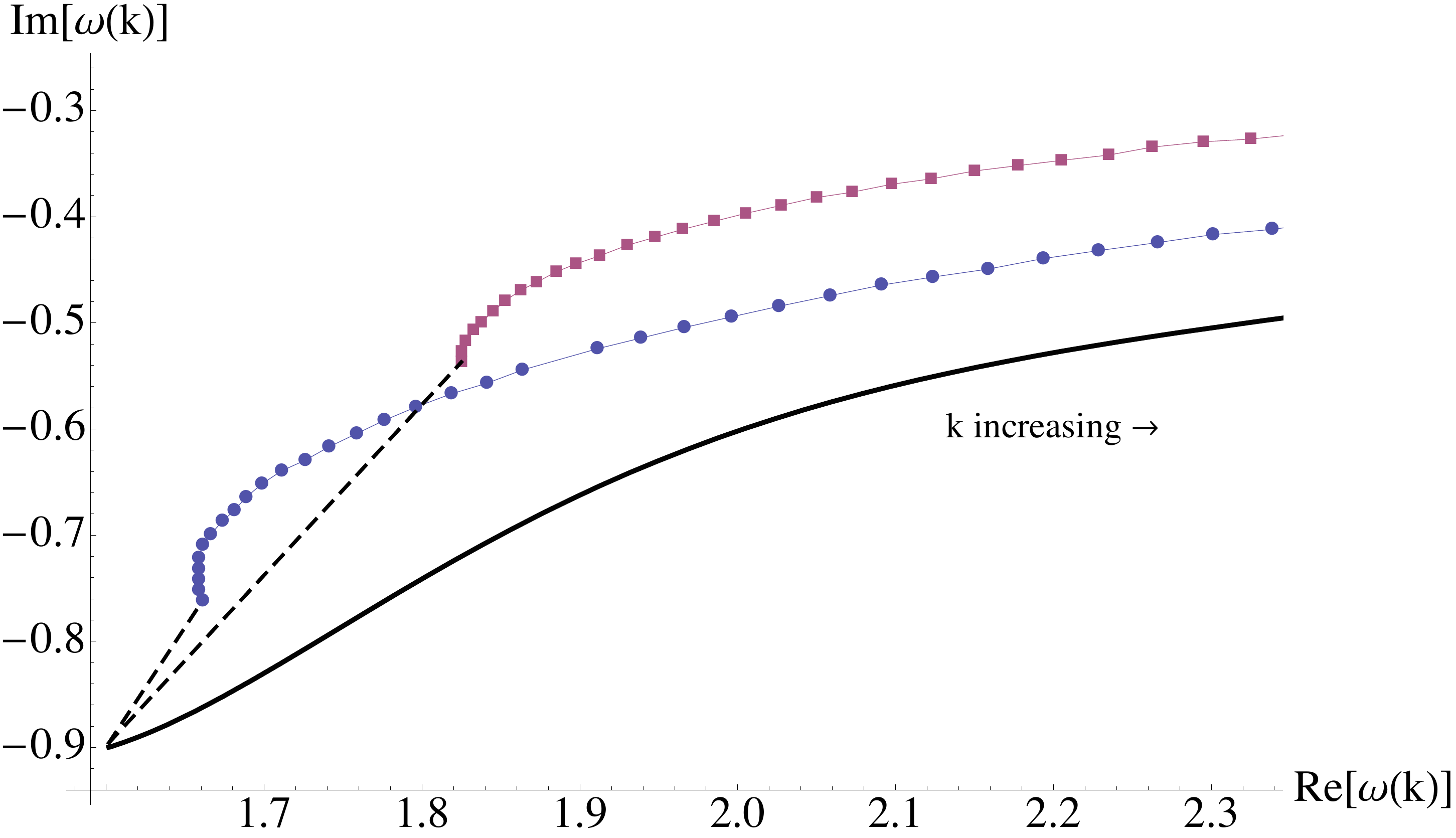} \label{fig: quasilong}}
\subfigure[] {\includegraphics[angle=0,
width=0.45\textwidth]{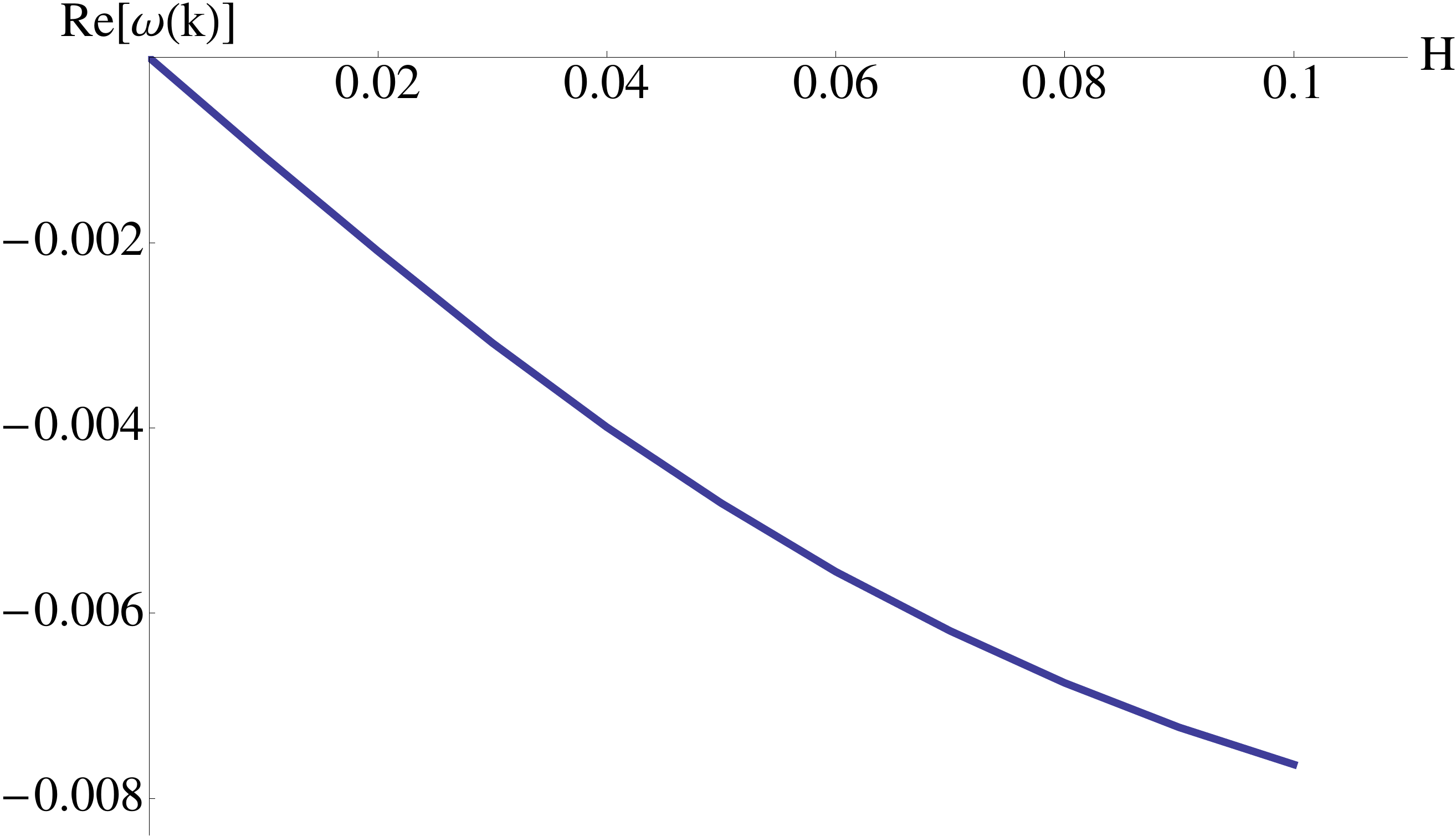} \label{fig: hydrolong}}
\caption{\small The dispersion relation of longitudinal oscillation in presence of a background magnetic field. The quasinormal frequency is measured in units of the background temperature. In figure \ref{fig: quasilong} the black solid curve represents the dispersion relation when the magnetic field is absent, the blue (circular) dots correspond to $H = 5\times 10^{-2}$ and the purple (square) dots correspond to $H = 10^{-1}$. $H$ has been expressed in units of $R$. In figure \ref{fig: hydrolong} we have shown the dependence of the lowest quasinormal frequency corresponding to the hydrodynamic mode with the magnetic field for a fixed momentum $k=0.01$. The imaginary part of the quasinormal frequency does not depend on the magnetic field in the hydrodynamic limit.}
\end{center}
\end{figure}

The numerical result is shown in figure \ref{fig: quasilong}. Note that here we numerically explore only for positive values of $H$. The equation of motion for the longitudinal modes have a linear term in $H$, and therefore the quasinormal modes may depend on the sign of $H$. However, we found numerically that the qualitative features of the quasinormal modes are similar also for negative values of $H$. The qualitative behaviour of the dispersion curve is similar to the transverse modes. It is evident that increasing magnetic field brings quasinormal frequency closer to the real axis in the complex plane. However, for a fixed magnetic field, the dispersion curves tend to approach the zero magnetic field dispersion curve as the momentum increases.

\section*{6.4 \hspace{2pt} Dispersion relation in the hydrodynamic limit}
\addcontentsline{toc}{section}{6.4 \hspace{0.15cm} Dispersion relation in the hydrodynamic limit}

In this section we wish to explore the dispersion relation of the longitudinal mode in the hydrodynamic limit. Recall that the longitudinal gauge field fluctuations correspond to the diffusive mode in the dual gauge theory.

Going back to the original $\{t, u\}$-coordinate and defining the gauge invariant longitudinal mode $\cE ( u ) = \omega A_1 + k A_t$, the equation of motion is given by ({\it a la} equation (\ref{eqt: eomA1}))
\begin{eqnarray} \label{eqt: eomf}
\cE''(u) + \cZ_1(u) \cE'(u) + \cZ_2 (u) \cE(u) = 0 \ ,
\end{eqnarray}
where
\begin{eqnarray}
\cZ_1 ( u ) & = & \frac{\omega ^2 \left( \cS^{tt} \right)^2 \left( \cL \cS^{uu} \cS^{xx} \right) ' + k^2 \left( \cS^{xx} \right)^2 \left( \cL \cS^{uu} \cS^{tt} \right) ' }{ \cL \cS^{tt} \cS^{uu} \cS^{xx} \left( \cS^{xx} k^2 + \omega ^2 \cS^{tt} \right)} \ , \nonumber\\ \nonumber\\
\cZ_2 (u) & = & -\frac{ \cS^{xx} k^2 + \omega ^2 \cS^{tt}} { \cS^{uu}} \nonumber\\
& + & \frac{8 H k \sigma  \left( \cS^{xx} \left( 8 H k \sigma - \omega  \cL \cS^{uu} \left( \cS^{tt}\right) ' \right) + \omega  \cS^{tt} \left( 8 H \sigma  \omega / k + \cL \cS^{uu} \left(\cS^{xx}\right) ' \right)\right)}{ \cL^2 \cS^{tt} \cS^{xx} \left(\cS^{uu}\right)^2 
   \left( \cS^{xx} k^2 + \omega ^2 \cS^{tt} \right)} \ , \nonumber\\
\end{eqnarray}
with 
\begin{eqnarray}
\cL(u) = g_s e^{-\Phi} \sqrt{ - { \rm det } \left ( E^{(0)} \right ) } \ , \quad \sigma = \frac{1}{2} \ .
\end{eqnarray}
We also have the following constraint
\begin{eqnarray}\label{eqt: constr}
\cL(u) \cS^{uu} \left( \omega A_t' \cS^{tt} - k A_1' \cS^{xx} \right) + 8 \sigma H \cE = 0 \ .
\end{eqnarray}

Now we venture for the dispersion relation in the hydrodynamic limit. In this particular set-up, for simplicity, we will set the radius of the near-horizon $D4$-brane background to unity ($U_T = 1$) and also set the horizon radius to unity ($R = 1$). We will restore these factors at the very end. To facilitate the hydrodynamic limit, we will assume $\omega \ll 1$ and $k \ll 1$ and further assume that $H \ll 1$. The last condition is imposed assuming that the length scale associated with the magnetic field is small compared to the plasma temperature. To further quantify this limit, we will parametrize $H = k h$ holding $h$ fixed in the hydrodynamic limit.

In this limit we will attempt to solve the equation (\ref{eqt: eomf}) perturbatively in $k$. To this end, we write the following expansion\footnote{To directly compare to the notation of Appendix B of ref.~\cite{Johnson:2009ev}, we need to identify $\tilde{\omega} = \omega / 3$.}
\begin{eqnarray} \label{eqt: solpert}
\cE(u) = ( u - 1)^{- i \omega/3} \left[ \cF_{(0)} (u) + k \cF_{(1)} (u) + k^2 \cF_{(2)} (u) \right] + \cO \left( k^3 \right) \ .
\end{eqnarray}
Motivated from the hydrodynamic result in (\ref{eqt: hsound}) we parametrize the dispersion relation in the following manner
\begin{eqnarray} \label{eqt: dish}
\omega = - i \gamma_1 k^2 - i k^2 h^2 \gamma_2 + k^2 h \gamma_3 + \cO \left( k^3 \right) \ ,
\end{eqnarray}
where $\gamma_1$, $\gamma_2$ and $\gamma_3$ are three constants to be determined.

We will attempt to determine these constants by perturbatively solving the differential equation (\ref{eqt: eomf}) subject to the following Dirichlet boundary conditions
\begin{eqnarray}
\lim_{u\to 1} \cE(u) = 1 \ ,  \quad \lim _{u\to\infty} \cE(u) = 0 \ .
\end{eqnarray}
These boundary conditions determine the so-called quasinormal modes of the corresponding fluctuation spectrum. Recall that when $h=0$, we have the known dispersion relation $\omega = - i (2/3) k^2$ \cite{Johnson:2009ev} and hence we expect to obtain $\gamma_1 = 2/3$.

Before proceeding further we have to fix the boundary conditions on the functions $\cF_{(i)}$. Since the differential equation in (\ref{eqt: eomf}) is homogeneous, we can impose the following boundary conditions without any loss of generality
\begin{eqnarray} \label{eqt: bc}
&& \lim_{u\to 1} \cF_{(0)} = 1 \ , \quad \lim_{u\to 1} \cF_{(1),(2)} = 0 \ , \quad \nonumber\\
&& \lim_{u\to\infty} \cF_{(0),(1),(2)} = 0 \ .
\end{eqnarray}

Now we substitute the expressions in (\ref{eqt: solpert}) and (\ref{eqt: dish}) in the fluctuation equation (\ref{eqt: eomf}) and analyze the corresponding equations at various orders of the expansion parameter $k$.

At the leading order in $k$, we get the following equation
\begin{eqnarray}
\cF_{(0)}''(u) + \partial_u \left[ \log \left( \cL \cS^{tt} \cS^{uu} \right)\right] \cF_{(0)}' (u) = 0 \ .
\end{eqnarray}
Note that there is no $\cF_{(0)}$ term present in the above equation. The solution of this equation subject to the boundary condition in (\ref{eqt: bc}) is given by
\begin{eqnarray}
\cF_{(0)} = \frac{1}{u^{3/2}} \ ,
\end{eqnarray}
where we have used the specific functional forms of $\cL$, $\cS^{tt}$ and $\cS^{uu}$. Clearly at this order we do not get any information about the dispersion relation. 

At the next order, we get the following equation
\begin{eqnarray}
\cF_{(1)}''(u) + \partial_u \left[ \log \left( \cL \cS^{tt} \cS^{uu} \right)\right] \cF_{(1)}' (u) = 0 \ ,
\end{eqnarray}
which is exactly similar to the equation satisfied by the zeroth order solution $\cF_{(0)}$. However, as we have arranged, $\cF_{(1)}$  satisfies a different boundary condition near the horizon as compared to the solution $\cF_{(0)}$. It is straightforward to check that in this case we get $\cF_{(1)} = 0$. At this order, we do not get any non-trivial information. 

The differential equation at the next order is the most interesting one. The general structure of the equation is obtained to be
\begin{eqnarray} \label{eqt: eomf2}
\cF_{(2)}''(u) + \partial_u \left[ \log \left( \cL \cS^{tt} \cS^{uu} \right)\right] \cF_{(2)}' (u) + \cJ \left( \cF_{(0)}, \cF_{(0)}' , \cF_{(1)}, \cF_{(1)}' \right)= 0 \ ,
\end{eqnarray}
where $\cJ$ sources an inhomogeneous piece constructed from the solutions obtained at lower orders \footnote{The full expression of $\cJ$ is rather complicated, therefore we do not present it here.}. Unlike the lower orders, equation (\ref{eqt: eomf2}) is not exactly solvable, however we can solve it analytically near the horizon and near the boundary. 

Near the boundary, as $u\to \infty$, we can always obtain a solution for $\cF_{(2)}$ which is consistent with the boundary condition in (\ref{eqt: bc}). The non-trivial information comes from solving the above equation near the horizon. In this region the function $\cF_{(2)}$ takes the following form
\begin{equation}
\cF_{(2)} = {\rm finite} + ( \gamma_1 + h (h \gamma_2 + i \gamma_3 )) \left( 3 \gamma_2 h^2 + i (3 \gamma_3 + 16 \sigma ) h + 3 \gamma_1 - 2 \right) \log (u-1) \ ,
\end{equation}
where the ``finite" piece consists of one undetermined constant of integration and can be arranged to be zero. Thus we need to separately impose the coefficient of the log-divergent term to be zero. This gives the following algebraic condition
\begin{eqnarray}
&& {\rm either} \quad \gamma_1 + h \left (h \gamma_2 + i \gamma_3 \right) = 0 \ , \nonumber\\
&& {\rm or} \quad 3 \gamma_2 h^2 + i \left (3 \gamma_3 + 16 \sigma \right) h + 3 \gamma_1 - 2 = 0 \ .
\end{eqnarray}
Interestingly, if we treat $h$ to be an independent parameter, then the algebraic constraints impose numerical values for $\gamma_i$'s at different orders of $h$. Clearly the first algebraic condition implies $\gamma_1 = 0 = \gamma_2 = 0 = \gamma_3$ and hence gives $\omega = 0$ by virtue of (\ref{eqt: dish}). However, the second algebraic condition imposes
\begin{eqnarray}
\gamma_1 = \frac{2}{3} \ , \quad \gamma_2 = 0 \ , \quad \gamma_3 = - \frac{16\sigma}{3} \ ,
\end{eqnarray}
which ultimately leads to the following dispersion relation
\begin{eqnarray} \label{eqt: hdiffgrav}
\omega = - i \frac{2}{3} k^2 \pm \frac{16 \sigma}{3} k H \ .
\end{eqnarray}
Comparing (\ref{eqt: hdiffgrav}) with (\ref{eqt: hsound}) we get 
\begin{eqnarray}
\frac{\eta}{\epsilon + P} = \frac{2}{3} \ , \quad \frac{\zeta_w}{2 \left(\epsilon + P \right)} = \frac{16 \sigma}{3} \ .
\end{eqnarray}
Thus we find that the magnetic field sources a propagating (sound) mode in the otherwise diffusive channel. In the hydrodynamic paradigm, such a mode is clearly sourced by the anomaly associated with a global U(1) symmetry as demonstrated in appendix E.  It is also obvious from the presence of the Chern-Simons coefficient $\sigma$ in the above formulae. Finally restoring the factors of temperature, we get the following dispersion relation (after setting $R = 1$)
\begin{eqnarray}
\omega = - i \frac{1}{2 \pi T} k^2 \pm \frac{72}{\left( 4 \pi T \right)^3} k H \ .
\end{eqnarray}
%

\section*{6.5 \hspace{2pt} Mesons with large spin}
\addcontentsline{toc}{section}{6.5 \hspace{0.15cm} Mesons with large spin}

Mesons with large spin can be described by strings having both its end points on the flavour brane and rotating on a two plane along the directions where the gauge theory lives. For large enough angular momentum, the length of the string is much larger compared to $\ell_s$ and we can use classical Nambu-Goto action to describe the dynamics of the rotating string. Mesons with large spin have previously been analyzed in, {\it e.g.}, refs. \cite{Peeters:2006iu, Kruczenski:2003be}. Here we will follow the same framework, now including the effect of an external magnetic field.

In absence of magnetic field due to rotational symmetry, the spectrum does not depend on the direction of the angular momentum $J$. An external magnetic field breaks this symmetry and introduces a splitting of the energy levels (Zeeman effect). Here we will focus on the two cases, namely when the magnetic field is parallel or anti-parallel to the angular momentum. In this case there should be a Zeeman splitting in energy for mesons with non-zero magnetic moment.

This idea has been implemented in ref.~\cite{Jensen:2008yp} in a different model. We adopt similar system and consider two flavour brane--anti-brane pair very close to each other so that in the bulk they join leaving two diagonal ${\rm U}(1)$s (call this ${\rm U}(1)_A$ and ${\rm U}(1)_B$ corresponding to the two pair of flavour branes). We consider now turning on fields having equal opposite charge under these two ${\rm U}(1)$s. Now $AA$ and $BB$ strings will have equal and opposite charges but $AB$ and $BA$ strings will have equal charges. In the first case, the total orbital magnetic moment of the meson will vanish whereas in the second case the meson will have a non-zero magnetic moment. The magnetic field will induce a quadratic correction to the meson energy for the former and a linear Zeeman splitting for the later. Here we will analyze the later case only. It is useful to note that this configuration is symmetric under reflection around the midpoint of the string (at $\rho=0$). We will refer to this as the symmetric configuration.

To proceed we write the relevant part of the background metric in the following form (rewriting the $\{x^2,x^3\}$-two plane in polar coordinate $\{\rho,\phi\}$)
\begin{eqnarray}
ds^2&=&\left(\frac{u}{R}\right)^{3/2}\left(f(u)dt_E^2+d\rho^2+\rho^2d\phi^2\right)+\left(\frac{R}{u}\right)^{3/2}\frac{du^2}{f(u)}\ ,\nonumber\\
f(u)&=&1-\left(\frac{U_T}{u}\right)^3\ .
\end{eqnarray}
Here we have explicitly written down the metric corresponding to the high temperature phase only. The background metric corresponding to the low temperature phase can likewise be written in an analogous form. We recover the zero temperature background by setting $U_T=0$. We will work with the Nambu-Goto action for the string with the following ansatz for the string profile
\begin{eqnarray}
t=\tau\ , \quad \rho=\rho(\sigma)\ , \quad u=u(\sigma)\ , \quad \phi=\omega\tau\ .
\end{eqnarray}
From now on the parameters $\{\tau,\sigma\}$ will always refer to the worldsheet coordinates for the string. We also assume that $\omega$ is positive ({\it i.e.} only clockwise rotation).

In the presence of an external constant magnetic field the Nambu-Goto action is accompanied by a boundary term coupled to the magnetic potential $A_{\phi}= H \rho^2/2$. This potential $A_{\phi}$ gives rise to a magnetic field $B_{(2)}= H \rho d\rho\wedge d\phi= H dx^2\wedge dx^3$. Taking this boundary term into account the action for the string is given by
\begin{eqnarray}\label{eqt: action}
&& S = \frac{1}{2\pi\alpha'}\int d\tau d\sigma \left[\left(\frac{u}{R}\right)^3\left(f(u)-\rho^2\omega^2\right)\left(\rho'^2+\frac{u'^2}{f(u)}\left(\frac{R}{u}\right)^3\right)\right]^{1/2}+\Delta S_{B}\ ,\nonumber\\
&& \Delta S_B = \frac{1}{2\pi\alpha'}\left[\int \left. A\right |_{\sigma^+} + \int \left. A\right |_{\sigma^-}\right]\ , \quad A=\frac{H \rho^2}{2}d\phi\ ,
\end{eqnarray}
where $\sigma^{\pm}$ represents the right and the left boundaries and the relative positive sign between the two boundary contributions coming from the magnetic potential is due to the  symmetric configuration. For the non-symmetric configuration the there will be a relative negative sign between these two terms. We refer to ref. \cite{Jensen:2008yp} for further details on this construction. For convenience we introduce the rescaled variables
\begin{equation}
\tilde{\rho}=\rho\omega\ , \quad \tilde{u}=u\omega\ , \quad \tilde{R}=R\omega\ , \quad \tilde{\sigma}=\sigma\omega\ .
\end{equation}
The expression for the energy and the angular momentum for the spinning string in the rescaled variables can simply be obtained to give
\begin{eqnarray}\label{eqt: ej}
&& E = \omega\frac{\partial L}{\partial\omega}-L=\frac{1}{2\pi\alpha'}\left(\frac{1}{\omega^{5/2}R^{3/2}}\right)\int d\tilde{\sigma}\tilde{u}^{3/2}\left(\tilde{\rho}'^2+\tilde{u}'^2\left(\frac{\tilde{R}}{\tilde{u}}\right)^3\right)^{1/2}\frac{1}{\sqrt{1-\tilde{\rho}^2}}\ , \nonumber\\
&& J = \frac{\partial L}{\partial\omega}=\frac{1}{2\pi\alpha'}\left(\frac{1}{\omega^{7/2}R^{3/2}}\right)\int d\tilde{\sigma}\frac{\tilde{\rho}^2}{\sqrt{1-\tilde{\rho}^2}}\left[\tilde{u}^3\left(\tilde{\rho}'^2+\tilde{u}'^2\left(\frac{\tilde{R}}{\tilde{u}}\right)^3\right)\right]^{1/2}+\Delta J_B\ ,\nonumber\\
&& \Delta J_B=\frac{1}{2\pi\alpha'\omega^2}\left(\left. H\frac{\tilde{\rho}^2}{2}\right | _{\sigma^+} + \left. H\frac{\tilde{\rho}^2}{2}\right | _{\sigma^-}\right)\ .
\end{eqnarray}
We observe that the external field can enhance angular momentum of the meson if it is aligned in parallel (positive values of $H$); and can reduce the angular momentum if it is anti-parallel (negative values of $H$).

To proceed we need to solve for the string embeddings. The equation of motion obtained from the Nambu-Goto action has to be supplemented by the boundary condition
\begin{eqnarray}
\left. \pi_X^1\right |_{\partial\Sigma}=\left. \frac{\partial L}{\partial (X')^M}\delta X^M \right |_{\partial\Sigma}=0\ .
\end{eqnarray}
Since $\left. \delta u\right |_{\partial\Sigma}=0$ and $\left. \delta\rho\right |_{\partial\Sigma}$ is arbitrary, we have to impose the Neumann boundary condition (in absence of any external field) $\left. (\partial L/\partial \rho')\right |_{\partial\Sigma}=0$. This condition gives us the following constraint
\begin{eqnarray}\label{eqt: bc}
\left. \pi_{\tilde{\rho}}^1\right |_{\partial\Sigma}=\left. \tilde{\rho}'\left[\frac{\left(\tilde{u}/R\omega\right)^3(1-\tilde{\rho}^2)}{\tilde{\rho}'^2+\tilde{u}'^2\left(\omega R/\tilde{u}\right)}\right]^{1/2}\right |_{\partial\Sigma}=0\ .
\end{eqnarray}
So we need to impose the boundary condition $\left. \tilde{\rho}'\right |_{\partial\Sigma}=0$, which means that the string is ending perpendicularly on the flavour brane--anti-brane pair. We will come back to the details of our numerical scheme in a later section.

The presence of an external magnetic field changes this boundary condition. It can be easily seen that the boundary condition to be satisfied is now given by
\begin{eqnarray}\label{eqt: bcb}
\left. \pi_{\tilde{\rho}}^1\right |_{\partial\Sigma} - \left(- H \tilde{\rho}\right)=0\ ,
\end{eqnarray}
where $\left. \pi_{\tilde{\rho}}^1\right |_{\partial\Sigma}$ is given in equation (\ref{eqt: bc}). Therefore we need to solve the equation of motion obtained from the action in eqn. (\ref{eqt: action}) subject to the boundary condition given in eqn. (\ref{eqt: bc}) or (\ref{eqt: bcb}) in absence or presence of a magnetic field respectively.

We will begin by obtaining some analytical results in the large angular frequency limit in the zero temperature phase. We will choose $\sigma=\rho$ gauge and send $\omega\to\infty$; from the reality condition of the Nambu-Goto action in eqn. (\ref{eqt: action}) this limit sets a bound $|\rho|<1/\omega\to 0$; and therefore equivalently we study the short string limit.

\subsection*{6.5.1 \hspace{2pt} Analytical results}
\addcontentsline{toc}{subsection}{6.5.1 \hspace{0.15cm} Analytical results}

In this section we consider the zero temperature background and spinning strings with large angular frequency only.

\subsubsection*{6.5.1a \hspace{2pt} Vanishing magnetic field}
\addcontentsline{toc}{subsubsection}{6.5.1a \hspace{0.15cm} Vanishing magnetic field}

In this section we closely follow the approach adopted in ref.~\cite{Kruczenski:2003be}. In the large angular frequency limit the relevant string profile is well approximated by the local geometry very close to the point from where the string hangs. In this limit the string hangs from the point where the brane--anti-brane pair join (the radial position denoted as $U_0$). We therefore pick the following ansatz to approximate the string profile
\begin{eqnarray}\label{eqt: oansat}
\tilde{u}(\tilde{\rho})=\omega U_0+\frac{f(\tilde{\rho})}{\omega}\ .
\end{eqnarray}
It is easy to check that this ansatz is consistent with the equation of motion for the string in $\omega\to\infty$ limit. To obtain the leading order behaviour we have to solve for the function $f(\tilde{\rho})$ with the boundary conditions that $f(0)=0=f'(0)$. At $\tilde{u}=\omega U_0$ we require $\tilde{u}'(\tilde{\rho})=0$, which sets the latter boundary condition for the function $f(\tilde{\rho})$. Substituting the ansatz given in eqn.~(\ref{eqt: oansat}) in the equation of motion for the string and keeping only the leading order terms we obtain a differential equation for the function $f(\tilde{\rho})$ which can be integrated to obtain the following analytic form
\begin{eqnarray}\label{eqt: anprofile}
f(\tilde{\rho})=\frac{3U_0^2}{8 R^3}\left(\tilde{\rho}^2+\arcsin^2(\tilde{\rho})\right)\ .
\end{eqnarray}
Plugging back $f(\tilde{\rho})$ from eqn. (\ref{eqt: anprofile}) to eqn. (\ref{eqt: ej}) and simplifying we get the following relations
\begin{equation}\label{eqt: approx}
E=\frac{\pi U_0^{3/2}\lambda}{2 R^{9/2}}\frac{1}{\omega}+\mathcal{O}(\omega^{-5/2})\ ,\quad
J = \frac{\pi U_0^{3/2}\lambda}{4 R^{9/2}}\frac{1}{\omega^2}+\mathcal{O}(\omega^{-3})\ , \quad
E=\frac{\sqrt{\pi}U_0^{3/4}}{R^{9/4}}\sqrt{\lambda J} \ .
\end{equation}
From this we can see that $\omega\to\infty$ limit is equivalent to $J\ll \lambda$ limit. Also we find that the mesons with large angular frequency follow a Regge trajectory with an effective tension 
\begin{equation}\label{eqt: regge}
\tau_{\rm eff}=E^2/(2\pi J)=\lambda U_0^{3/2}/(2 R^{9/2})\ . 
\end{equation}
The radial position $U_0$ is related to the asymptotic separation $L$ between the flavours via $U_0^{1/2}=R^{3/2}/(4L) B(9/16,1/2)$, where $B(a,b)$ is the Beta function. This length $L$ sets the coupling strength of the non-local four Fermi interaction for the dual NJL model.

\subsubsection*{6.5.1b \hspace{2pt} Non-vanishing Magnetic Field}
\addcontentsline{toc}{subsubsection}{6.5.1b \hspace{0.15cm} Non-vanishing Magnetic Field}

\begin{figure}[!ht]
\begin{center}
\subfigure[] {\includegraphics[angle=0,
width=0.45\textwidth]{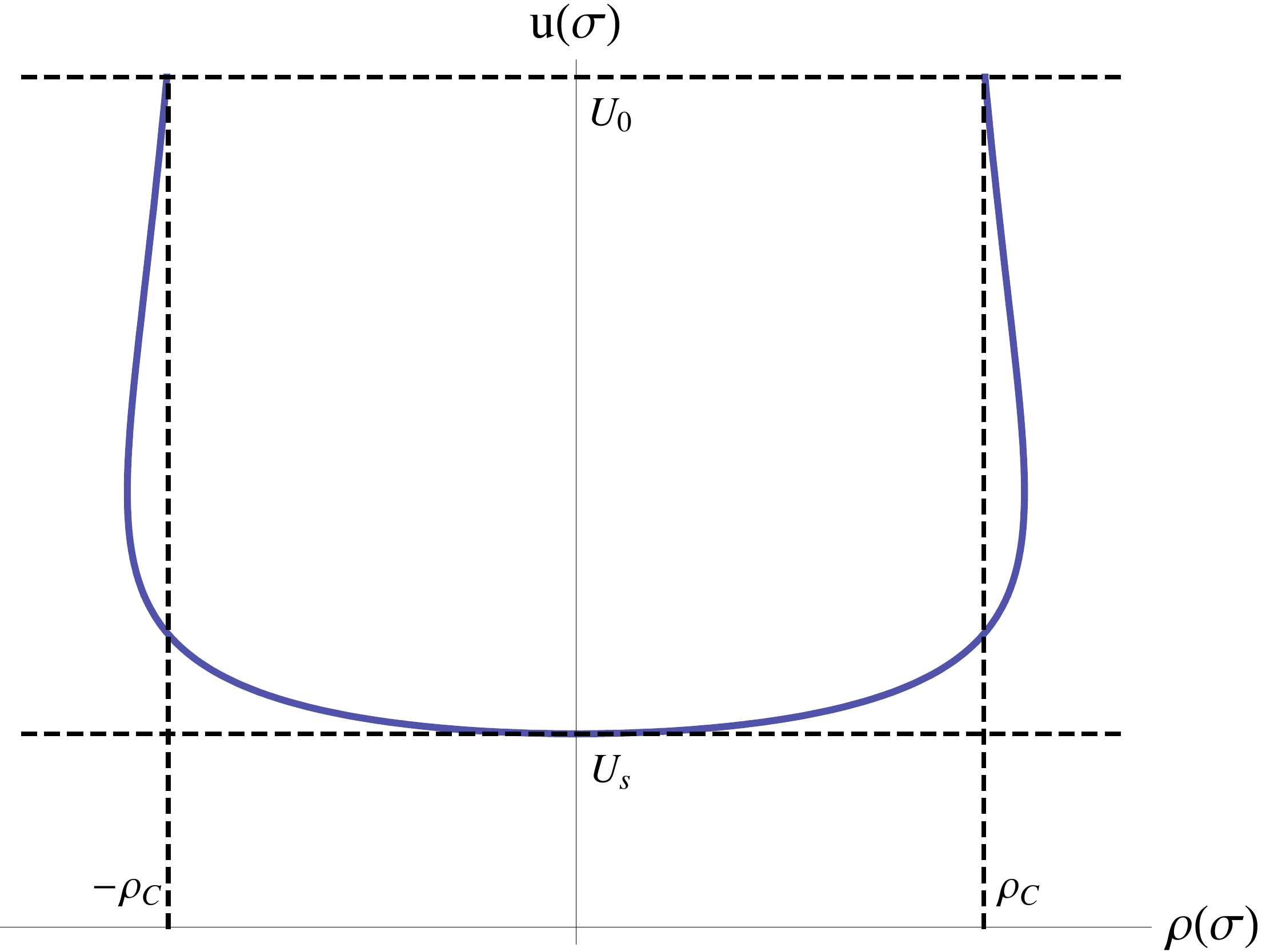} \label{fig: prohi}}
\subfigure[] {\includegraphics[angle=0,
width=0.45\textwidth]{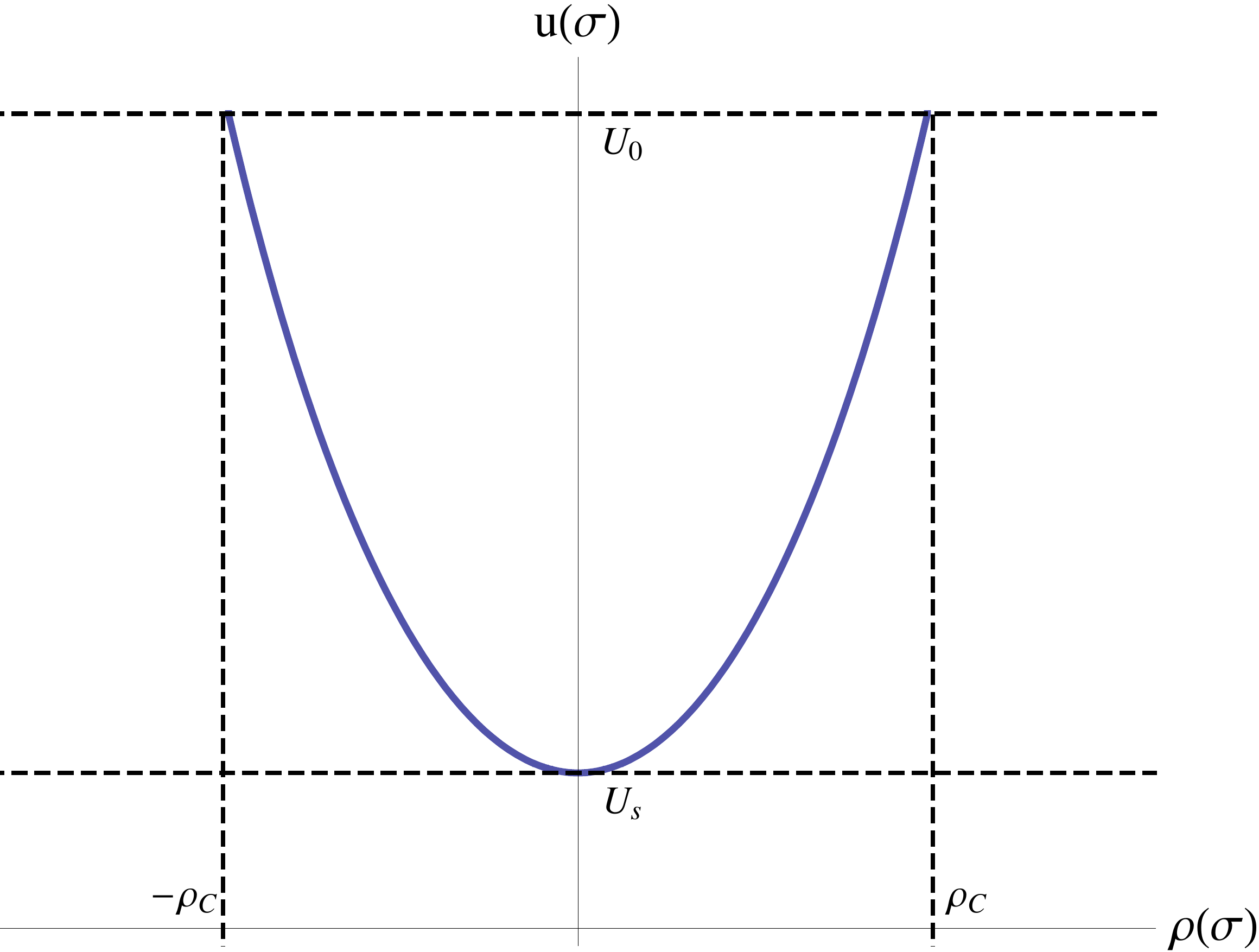} \label{fig: prolo}}
\caption{\small The two possible profiles are presented; $U_s$ is the radial turnaround position for the string. These are obtained by solving the equations of motion numerically, although we come back to the numerical results in a later section. The profile in the left hand side has more energy than the one in the right hand side.}
\end{center}
\end{figure}

In presence of a magnetic field the leading order solution for $\tilde{u}(\tilde{\rho})$ remains the same as in eqn. (\ref{eqt: anprofile}); however the boundary condition needs to be modified as given by eqn. (\ref{eqt: bcb}). Using the expression for $\left. \pi_{\tilde{\rho}}^1\right |_{\partial\Sigma}$ from eqn. (\ref{eqt: bc}) (in the gauge $\sigma=\tilde{\rho}$) in the modified boundary condition given in eqn. (\ref{eqt: bcb}) we can determine the value of $\tilde{\rho}$ at the boundary (meaning at $u=U_0$ where the string ends on the flavour brane). In the limit of small magnetic field this critical value of $\tilde{\rho}$ (denoted as $\tilde{\rho}_C$) can be obtained to be
\begin{eqnarray}
\tilde{\rho}_C=1-\frac{H ^2}{2}\left(\frac{R}{U_0}\right)^3\ .
\end{eqnarray}
With this $\tilde{\rho}_C$, there will be two different string profiles which we have shown in figure \ref{fig: prohi} and \ref{fig: prolo} respectively.

These two different profiles result in two different energies (let us denote these energies by $E_{\pm}$). The one with the lower energy ({\it i.e.}, with $E_{-}$) is extended from $\tilde{\rho}=0$ to $\tilde{\rho}_C$ and the one with the higher energy ({\it i.e.}, with $E_{+}$) is extended from $\tilde{\rho}=0$ to $\tilde{\rho}=1$ and then folds back to $\tilde{\rho}=\tilde{\rho}_C$.
Using the formulae for meson energy $E$ and angular momentum $J$ given in eqn. (\ref{eqt: ej}) we get the following linear order corrections (Zeeman splitting) introduced by the magnetic field
\begin{eqnarray}\label{eqt: splite}
E_{-} &=& \frac{\sqrt{\pi}U_0^{3/4}}{R^{9/4}}\sqrt{\lambda J}-\frac{2}{\sqrt{\pi}}\sqrt{\frac{J}{\lambda}}\frac{H R^{9/4}}{U_0^{3/4}}\ ,\nonumber\\
E_{+} &=& \frac{\sqrt{\pi}U_0^{3/4}}{R^{9/4}}\sqrt{\lambda J}+\frac{2}{\sqrt{\pi}}\sqrt{\frac{J}{\lambda}}\frac{H R^{9/4}}{U_0^{3/4}}\ .
\end{eqnarray}
Recall that the parameter $U_0$ is also related to the constituent quark mass. For a study of how the constituent quark mass behaves (for a fixed value of $L$) with the magnetic field see ref. \cite{Johnson:2008vna}.

It is easy to check that this splitting is consistent with the general formula for energy (obtained in ref.~\cite{Abouelsaood:1986gd}) of a Nambu-Goto string with an effective tension $\tau_{\rm eff}$ moving in a background magnetic field. The energy spectrum (for large $J$) is given by the following formula
\begin{eqnarray}\label{eqt: exact}
E^2=(2\pi\tau_{\rm eff})(1-\kappa) J\ , \quad \kappa=\frac{2}{\pi}\arctan\left(\frac{H}{\tau_{\rm eff}}\right)\ .
\end{eqnarray}
Recalling the expression for $\tau_{\rm eff}$ from eqn. (\ref{eqt: regge}) and using the general result above the linear order correction to the energy of the string can be obtained which matches exactly with what is obtained in eqn. (\ref{eqt: splite}).

\subsection*{6.5.2 \hspace{2pt} Numerical results}
\addcontentsline{toc}{subsection}{6.5.2 \hspace{0.15cm} Numerical results}

In order to extract the relevant physics we need to solve for the string embedding. To that end we choose the parametrization $\sigma=\rho+u$ since it provides better numerical stability as discussed in {\it e.g.}, ref.~\cite{Jensen:2008yp}.

To solve the equation of motion first we fix the maximum radial position the string can attain (this is the position where the brane--anti-brane pair join, which is denoted by $U_0$). This in turn fixes the asymptotic separation between the brane--anti-brane pair. Fixing $U_0$~is equivalent to fixing the constituent quark mass of the dual gauge theory and fixing $L$, the asymptotic separation of the brane--anti-brane pair, is equivalent to keeping the four fermi coupling of the dual NJL model. Now for a given $U_0$ we look for a radial turnaround position for the string (call it $U_s$) such that shooting from $U_s$ with the IR boundary conditions $\left. u(\sigma)\right |_{U_s}=U_s$ and $\left. u'(\sigma)\right |_{U_s}=0$ would satisfy the UV boundary condition $\left. u'(\sigma)\right |_{U_0}=1$ (up to numerical accuracy). The existence of the turnaround position is guaranteed by the symmetric configuration that we consider here. Below we show two such representative profiles for $\omega =0.5$ and $\omega=3$ respectively (corresponding to the blue and the red curve).
\begin{figure}[!ht]
\begin{center}
\includegraphics[angle=0,
width=0.65\textwidth]{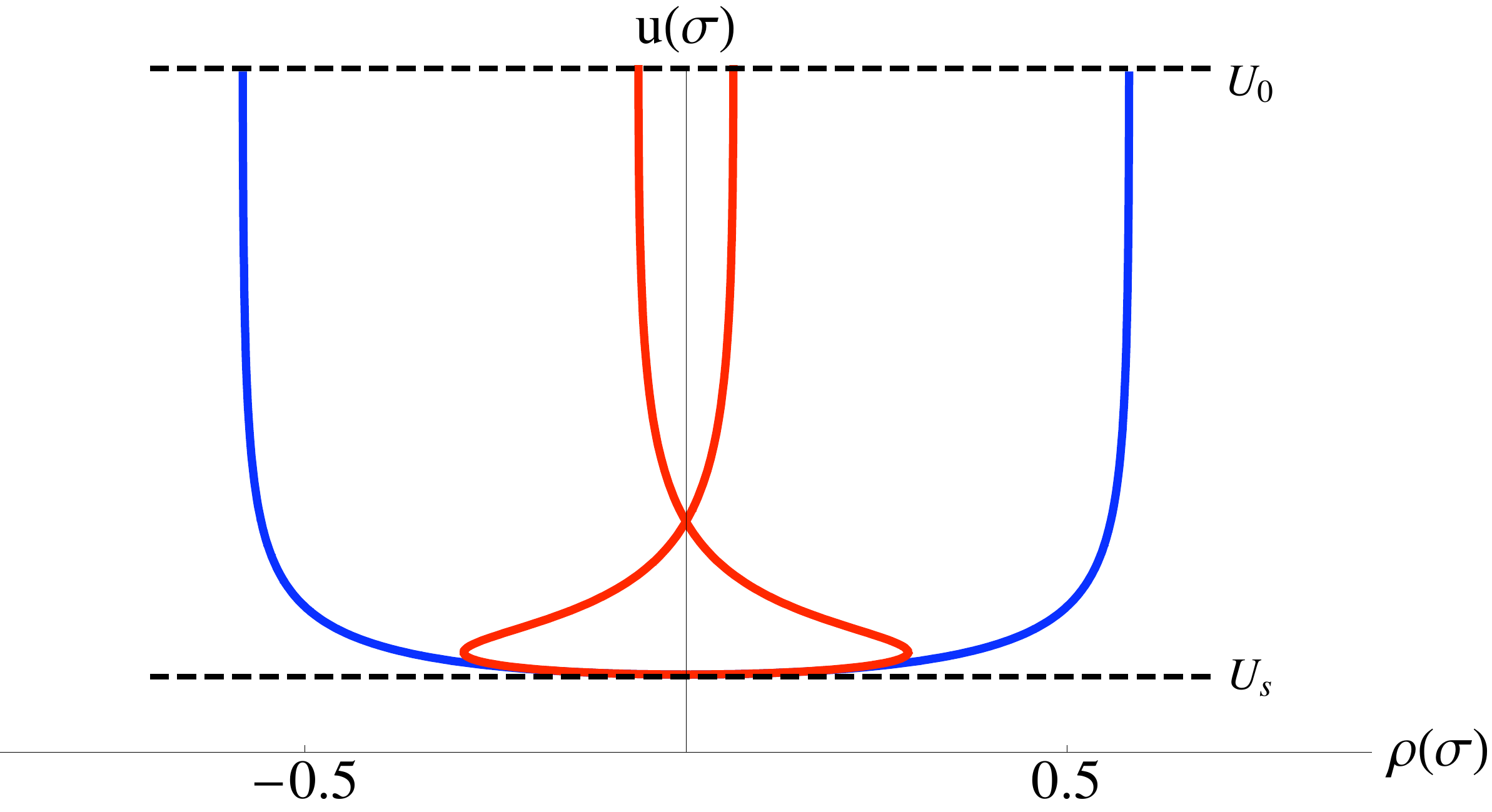}
\caption{\small Two representative spinning string profiles. We have set $R=1$.}
\label{fig: mdynT}
\end{center}
\end{figure}

Such profiles were discussed in ref.~\cite{Kruczenski:2003be} in a different but closely related model. In brief, we find that for a given value of $\omega$ there are classes of solutions distinguished by the number of times they cross zero (referred to as ``nodes" in ref.~\cite{Kruczenski:2003be}) along the horizontal axis. Here we will constrain ourselves to the $n=1$ case only.

We will now explore the behaviour of high spin mesons in presence of a background magnetic field. The key feature that we will observe for both zero and finite temperature phases is the existence of a shift in physical quantities such as energy and angular momentum in presence of a non-zero magnetic field. We begin with our numerical results for the zero temperature phase.

\subsubsection*{6.5.2a \hspace{2pt} Low temperature phase}
\addcontentsline{toc}{subsubsection}{6.5.2a \hspace{0.15cm} Low temperature phase}

%
\begin{figure}[!ht]
\begin{center}
\includegraphics[angle=0,
width=0.65\textwidth]{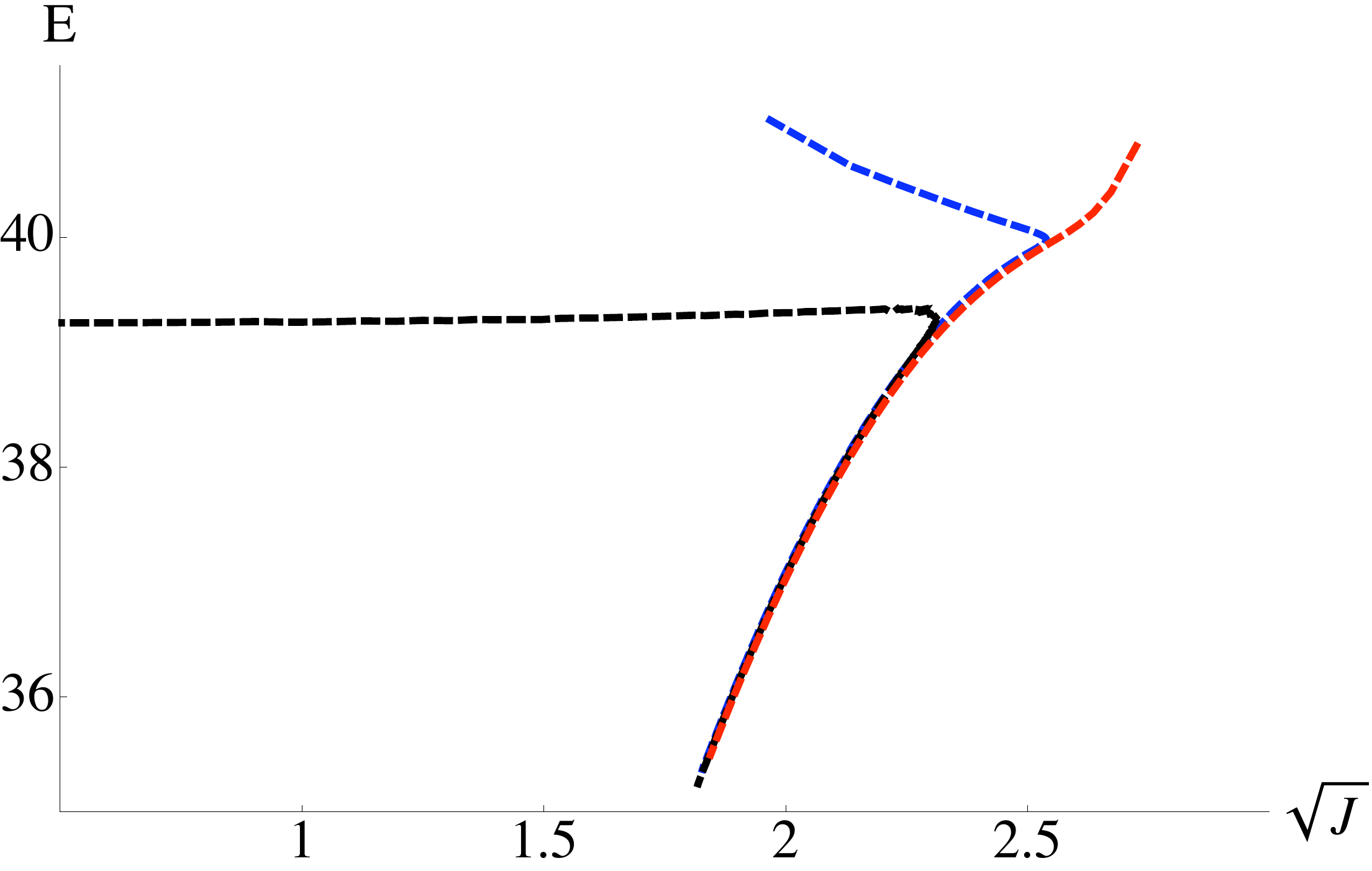}
\caption{\small The effect of an external magnetic field in the zero temperature phase. The black dashed lines represent $H=0$, blue dashed lines represent $H=-0.1$ and the red dashed lines represent $H =0.1$; $E$ and $J$ have been evaluated in units of $1/(2\pi\alpha')$. $H$ have been evaluated in units of $R$.}
\label{fig: ej}
\end{center}
\end{figure}

In figure \ref{fig: ej} we have shown the dependence of the meson energy $E$ as a function of the angular momentum of the meson. For zero external magnetic field we recover the behaviour discussed in ref.~\cite{Peeters:2006iu}. We find that the external magnetic field shifts the energy and the angular momentum of the meson for generic values of the parameters of the system; however with increasing angular frequency these observables tend to become insensitive to the external field. In view of the approximate analytic results obtained in eqn. (\ref{eqt: splite}) it is straightforward to observe that an external field contributes only at order $H/\omega$ and hence has vanishingly small effects for large enough angular frequency.

The existence of two energy branches for the same given value of $J$ seems to persist (as in vanishing external field) even in the presence of a magnetic field, although as the magnetic field is increased we observe a tendency that this multi-valuedness in energy starts disappearing. Therefore any possible unstable upper branch in figure \ref{fig: ej} can get promoted to a stable one by having sufficiently high angular momentum.

Moreover there exists a maximum angular momentum beyond which the spinning meson dissociates. This dissociation is mediated by the acceleration of the spinning string. However the presence of magnetic field can again stabilize the mesons by raising the maxima. Physically this simply corresponds to the fact that the magnetic field enhances the angular momentum of the meson and thus makes it stable.
\begin{figure}[!ht]
\begin{center}
\includegraphics[angle=0,
width=0.55\textwidth]{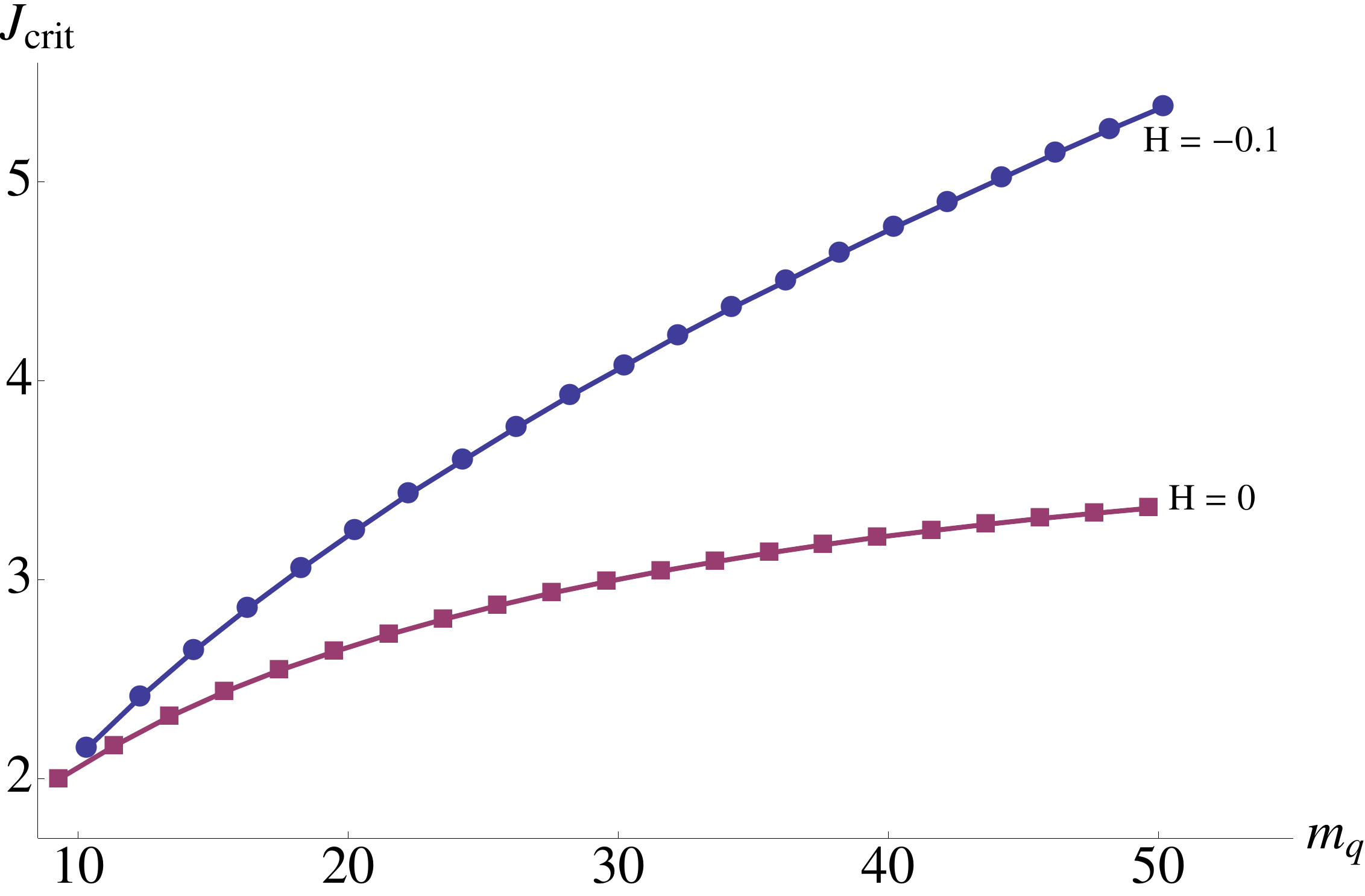}
\caption{\small Dependence of maximum spin with magnetic field for varying constituent quark mass. $J_{\rm crit}$ and $m_q$ have been calculated in units of $1/(2\pi\alpha')$. We observe that even for a certain negative value of $H$, {\it i.e.} when the magnetic field is aligned anti-parallel to its angular momentum, the meson can be more stable than its zero field counterpart.}
\label{fig: jcrit}
\end{center}
\end{figure}

To demonstrate this effect we observe the plot in figure \ref{fig: jcrit}. Clearly the $H=-0.1$ branch is higher than the $H = 0$ branch; an even higher value of $H$ would inhibit any dissociation at all. Also the heavier the constituent quark mass is the more $J_{\rm crit}$ becomes making it less likely for the heavy mesons to dissociate.
\begin{figure}[!ht]
\begin{center}
\includegraphics[angle=0,
width=0.55\textwidth]{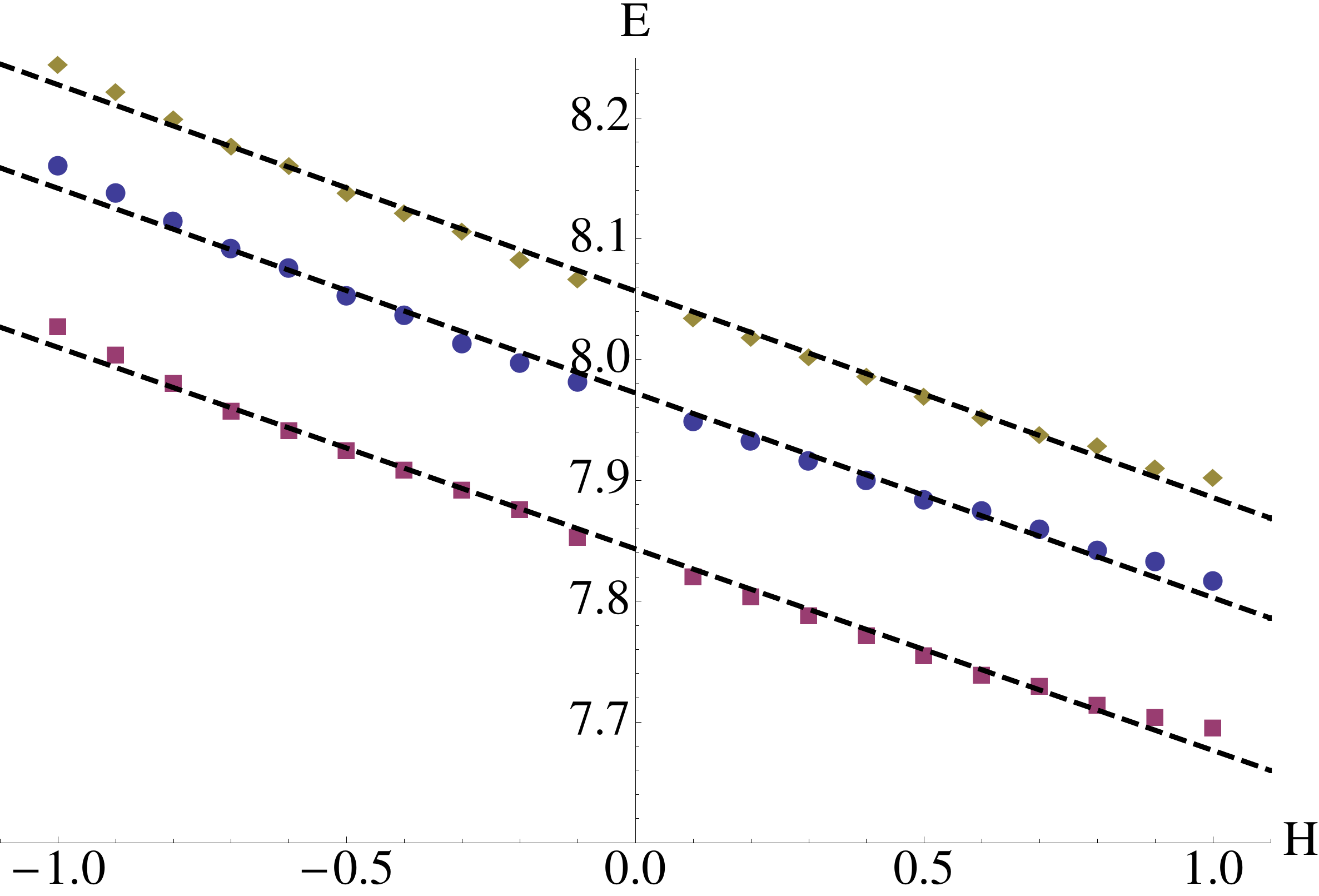}
\caption{\small Linear correction to energy for the equal charge case. From top to bottom the values of angular momenta are $J=0.937,$ $0.907,$ $0.862$ respectively. The dashed straight lines are the best fit lines. $E$ and $J$ are both presented in units of $1/(2\pi\alpha')$. We observe that all the three best fit lines have same (up to numerical accuracy) slope, which we expect from the very nature of Zeeman splitting.}
\label{fig: zee}
\end{center}
\end{figure}

A key feature of the high spin meson spectrum in presence of an external magnetic field is the linear Zeeman splitting. In figure \ref{fig: zee} we have shown such linear correction in energy for fixed values of the meson angular momentum.

\subsubsection*{6.5.2b \hspace{2pt} High temperature phase}
\addcontentsline{toc}{subsubsection}{6.5.2b \hspace{0.15cm} High temperature phase}

We now perform a similar numerical study of high spin mesons in this intermediate temperature phase.
\begin{figure}[!ht]
\begin{center}
\includegraphics[angle=0,
width=0.65\textwidth]{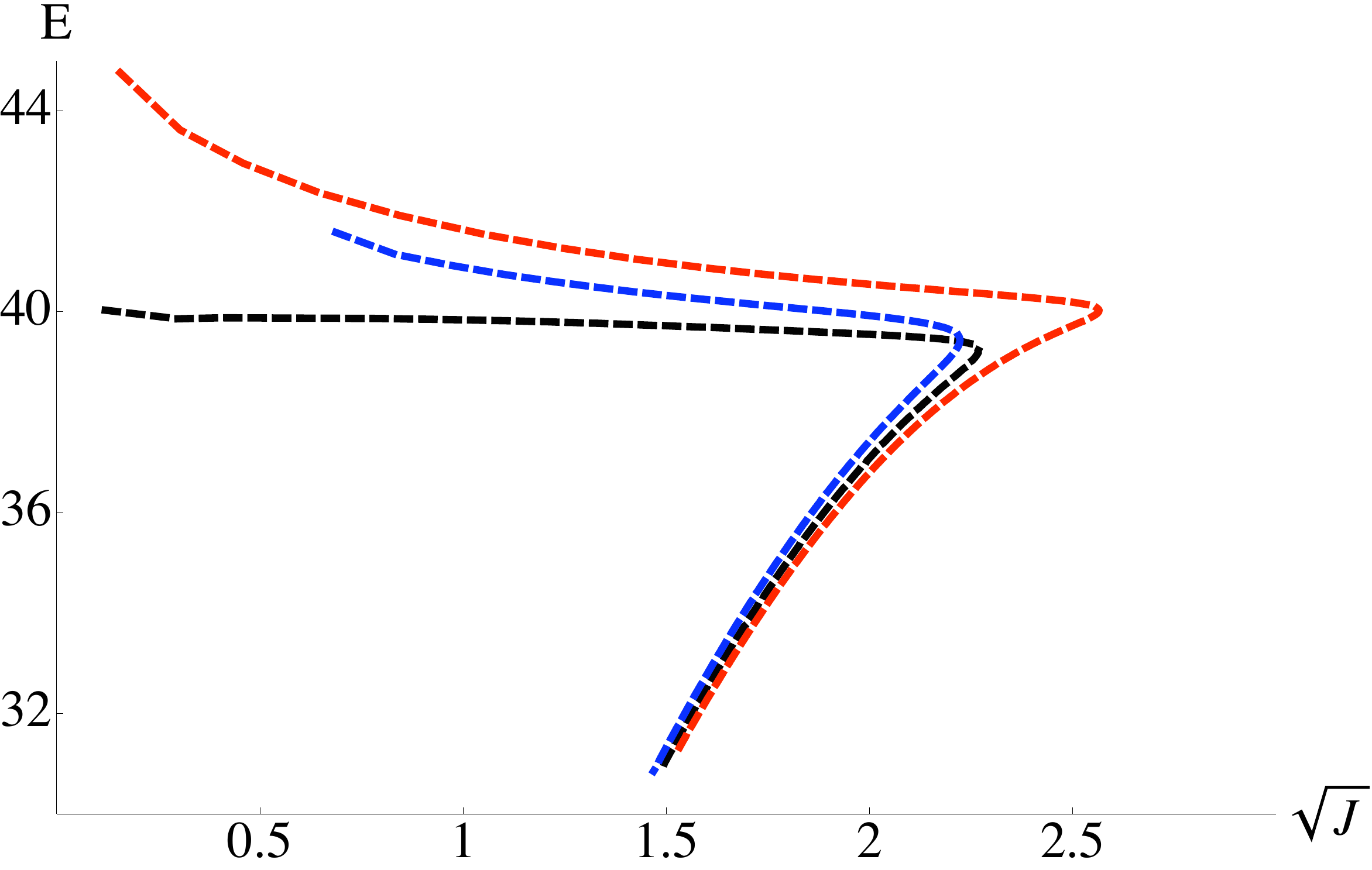}
\caption{\small The shifts in presence of an external magnetic field. $E$ and $J$ have been computed in units of $1/(2\pi\alpha')$. $H$ has been evaluated in units of $R$. The black curve corresponds to vanishing magnetic field, the blue curve corresponds to setting $H =-1$ and the red curve corresponds to setting $H =1$, where $H$ has been expressed in units of $R$.}
\label{fig: ejt}
\end{center}
\end{figure}

In figure \ref{fig: ejt} we have shown the dependence of meson energy as a function of its angular momentum. The qualitative features of finite temperature physics are similar to that of the zero temperature physics. There is however one distinction as compared to the energy spectrum at zero temperature phase. The presence of magnetic field at finite temperature does not promote the possible unstable upper branch in figure \ref{fig: ejt} to a stable one within the range of values for the magnetic field that we have explored using our numerical approach (we expect however that for sufficiently high values of $H$ this unstable mode will be promoted to be a stable one). A possible thermally enhanced decay channel therefore remains open at finite temperature for a rather high value of the external field. The other familiar role of the magnetic field, we again find, is to introduce shifts in the physical quantities such as the energy and angular momentum of the meson. 
\begin{figure}[!ht]
\begin{center}
\subfigure[] {\includegraphics[angle=0,
width=0.45\textwidth]{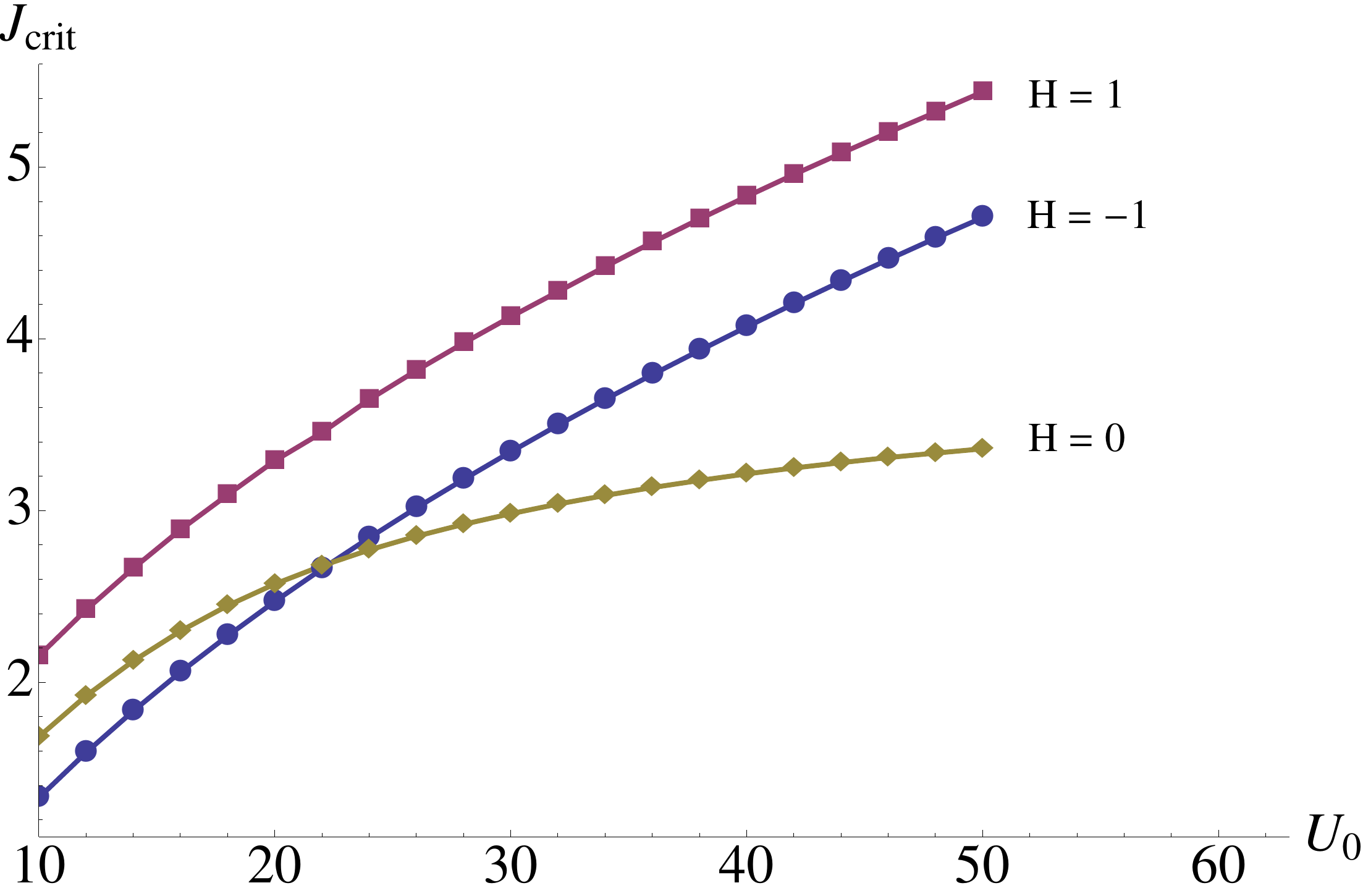} \label{fig: jcritt}}
\subfigure[] {\includegraphics[angle=0,
width=0.45\textwidth]{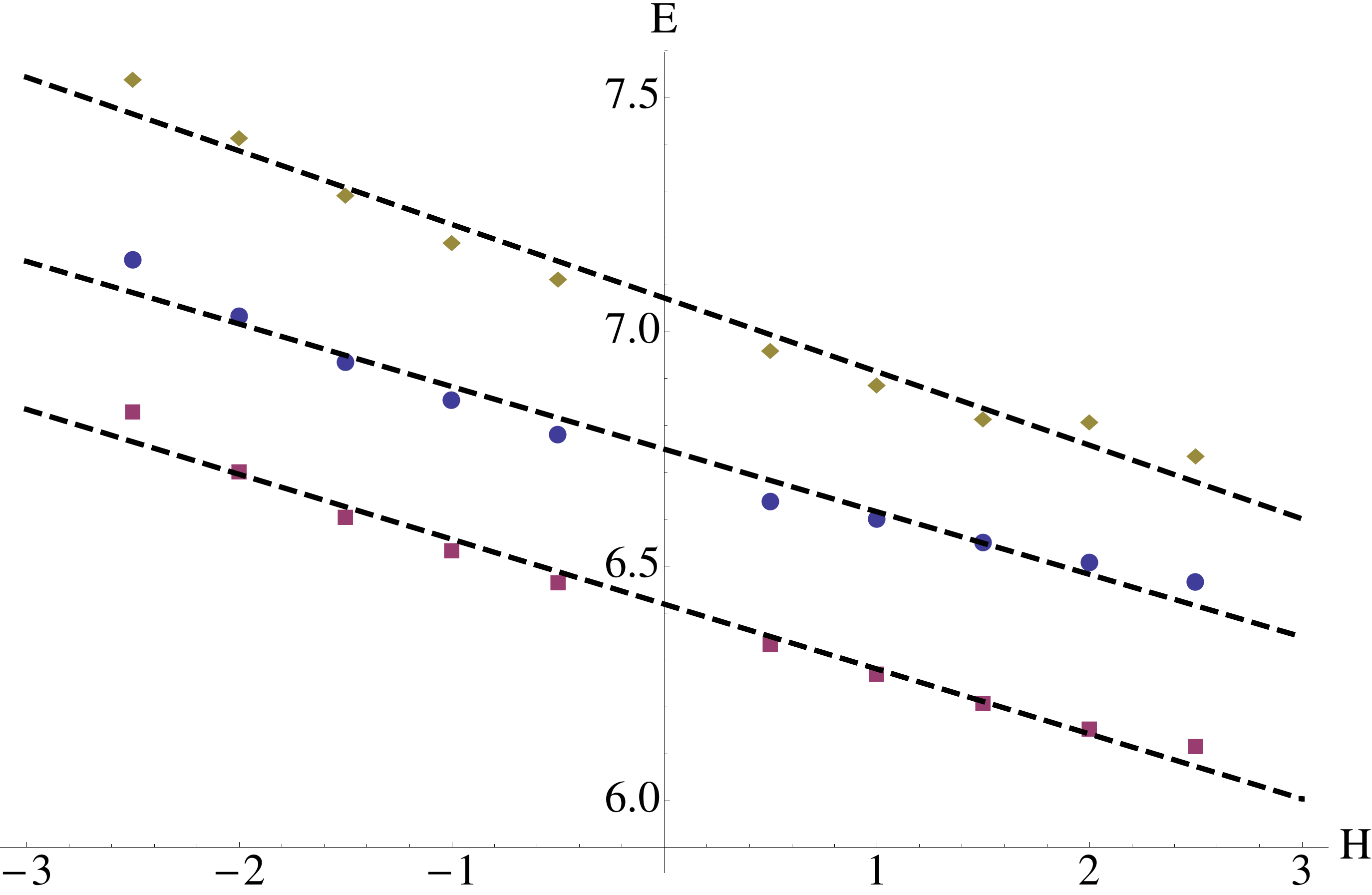} \label{fig: zeeT}}
\caption{\small Figure \ref{fig: jcritt} shows the dependence of maximum spin calculated in units of $1/(2\pi\alpha')$ with the applied magnetic field. We can observe that there exists a certain range of values for $U_0$, which in turn fixes the value of the constituent quark mass, where the angular momentum is lowered by the magnetic field. Figure \ref{fig: zeeT} shows the linear correction to meson energy for the equal charge case at finite temperature. From top to bottom the values of angular momenta are $J=0.630,$ $0.560$ and $0.490$ evaluated in units of $1/(2\pi\alpha')$ respectively. The dashed lines are the best fit lines.}
\end{center}
\end{figure}

The meson dissociation due to spin should now be enhanced due to thermal fluctuation. We have previously observed that the magnetic field adds extra angular momentum to the system and stabilizes the mesons. We study in figure \ref{fig: jcritt} the effect of two competing parameters on the critical angular momentum of the meson. Finally we observe the familiar Zeeman splitting in figure \ref{fig: zeeT}.

\section*{6.6 \hspace{2pt} Concluding remarks}
\addcontentsline{toc}{section}{6.6 \hspace{0.15cm} Concluding remarks}

We have uncovered a sub--sector of the meson spectrum in the presence
of an external magnetic field, complementing the phase structure that
we obtained in ref.~\cite{Johnson:2008vna}. Mesons with small spin that
can be obtained by studying the quadratic fluctuations of the probe
brane classical configuration are harder to study in absolute
generality. In general the task is to solve a set of coupled
differential equations, which we have here analysed in the situations
where it is possible to decouple a subset of the modes. We have also
analysed the large spin meson spectra in presence of a magnetic field
to realize the well--known Zeeman effect. We also found that an external magnetic field enhances the stability of mesons and inhibits the dissociation. In appendix D we have
presented a model calculation analyzing spinning strings in Rindler
space to capture key features of the large spin meson dissociation.

There are several directions to pursue in future work. It is of
interest to obtain the low spin meson spectra in the low temperature
phase ({\it i.e.,} when chiral symmetry is broken) and their
dispersion relations. In the present work, we have focussed on the
effect of a magnetic field. The analysis of the meson spectra and
quasinormal modes in the presence of an electric field could be
pursued in a similar spirit. The effect on the spectra of the presence
of a chemical potential would also be of great interest. These
external fields can all be realized as the non--normalizable modes of
anti--symmetric fields on the world volume of the probe brane, and so
we would expect that in all these cases in the high temperature
chirally symmetric phase, the scalar and vector fluctuations always
remain decoupled, as we have seen here.


\chapter*{Chapter 7:  \hspace{1pt} Conclusion}
\addcontentsline{toc}{chapter}{Chapter 7:\hspace{0.15cm} Conclusion}

We have extensively used the gauge-gravity duality to learn about the strong coupling dynamics of fundamental flavours. In particular we have used the framework of two different holographic models which share some rather generic features of QCD and are therefore expected to teach us about physics of the real world. In this thesis we have focussed entirely on the fundamental flavour sector. Our studies mainly included the effect of an external electric-magnetic field on the phase structure of the flavours at finite temperature, associated thermodynamics, chiral symmetry breaking and the associated meson spectrum.

The two holographic models we have discussed are structurally quite different. In retrospect, let us make a comparative note. In the Type IIB model we studied the dynamics of a probe D$7$-brane in the background of the near-horizon limit of a stack of $N_c$ D$3$-branes. At zero temperature this construction preserves $\mathcal{N} = 2$ supersymmetry\cite{Karch:2002sh}, however finite temperature breaks supersymmetry completely. In this model, we have identified the ${\rm U}(1)_{\rm R}$ symmetry corresponding to the rotation group in the transverse 2-plane of the probe D7-brane as the chiral symmetry. At zero temperature the presence of supersymmetry is enough to guarantee that the quark condensate must vanish at zero bare quark mass as mentioned in ref.~\cite{Babington:2003vm}. Thus there is no spontaneous breaking of chiral symmetry in this model at zero temperature. Moreover, at zero temperature and zero bare quark mass this theory is still conformal in the limit $N_f\ll N_c$. Therefore there is no scale in this system and thus no phase transition can occur. The zero temperature physics is somewhat trivial, although we can analyze many aspects such as the meson spectrum completely analytically\cite{Kruczenski:2003be, Kruczenski:2003uq}. At zero temperature, the chiral symmetry can be broken explicitly by adding a mass term for the quark. In the gravity dual picture this mass term corresponds to a separation of the probe D7-branes from the D3-branes in the transverse 2-plane, which can be introduced naturally in this model. This mass term corresponds to a non-normalizable mode in the profile of the probe brane and using the AdS/CFT dictionary we can read off the corresponding normalizable mode as the condensate.

Let us contrast this with the Type IIA model. Unlike the Type IIB case, this is not a conformal theory. In this case, we study the dynamics of a probe D8/$\overline{{\rm D} 8}$-brane in the background of the near-horizon limit of a stack of $N_c$ D4-branes compactified on a spatial circle along the direction denoted by $x^4$. By construction we impose anti-periodic boundary condition for the adjoint fermions which breaks supersymmetry. The global symmetry group is ${\rm U}(N_f)_L\times {\rm U}(N_f)_R$, which we continue to refer as the chiral symmetry\footnote{Note that our definition of chiral symmetry in the Type IIB and the Type IIA set-up are quite different, however in an abuse of language we often do not make this distinction.}. The probe brane--anti-brane pair intersect the D4-branes at $x^4 = \pm L/2$, therefore the quarks are massless by construction. Introducing mass term for the fundamental flavours and identifying the chiral condensate is rather subtle in this model. Work along this direction was initially carried out in refs.~\cite{Bergman:2007pm, Dhar:2008um} and later in ref.~\cite{Aharony:2008an}.

Let us recall that for all practical purposes the so called instructive limit of flattening out the $x^4$-direction (by sending $R_4\to\infty$) has almost no consequences in our gravity calculations. In this limit the zero temperature background does not break supersymmetry. However, introducing flavours by adding the brane--anti-brane probe immediately breaks all the supersymmetry. Therefore in this model the flavour physics is not influenced by supersymmetry at all.

In the low temperature phase ({\it i.e.} the dual gauge theory is in the confining phase), when $x^4$ is a compact direction the brane--anti-brane pair join together forced by the cigar topology of the background in the $\{x^4,u\}$-plane. This implies that the asymptotic global symmetry group ${\rm U}(N_f)_L\times {\rm U}(N_f)_R$ breaks down to a ${\rm U}(N_f)_{\rm diag}$, which is very similar to the spontaneous chiral symmetry breaking in QCD. In the limit $R_4\to\infty$, the confining phase disappears. Furthermore, the background topology in the $\{x^4,u\}$-submanifold is not cigar-shaped anymore. Therefore the flavour brane--anti-brane pair can either join together or end separately in the bulk. However it was shown in ref.~\cite{Antonyan:2006vw} that it is energetically favourable for the probe brane--anti-brane pair to join together. Thus in this model, in the low temperature regime chiral symmetry is spontaneously broken. This is in sharp contrast with the Type IIB model.

It was observed in ref.~\cite{Filev:2007gb} that in the Type IIB set-up an external magnetic field breaks all supersymmetry and therefore can induce spontaneous chiral symmetry breaking. Furthermore this external magnetic field provides a natural scale for this system at zero temperature given by $R\sqrt{H}$, where $R$ is the radius of the AdS-space and $H$ is the strength of the magnetic field. Thus conformal invariance is completely lost even at zero temperature. A naive dimensional analysis can be used to fix the behaviour of the chiral condensate $c\sim R^3 H^{3/2}$, which we have discussed in the $T\to 0$ limit in chapter 3.

The background temperature is another important physical scale. In the absence of any external magnetic field we discussed the temperature driven meson-melting transition and associated meson spectrum in chapter 3. This transition can simply be described as follows. Below a certain critical temperature quarks exist in bound states such as mesons. Above this critical temperature, these bound states melt and form a plasma of the constituents. In chapter~3 we also investigated the case when both temperature and magnetic field are present. The phase structure is now controlled by two physical scales: the temperature and the chiral symmetry breaking scale induced by the external magnetic field. We have found that these two physical scales have competing interplay and discussed the relevant phase structure in detail in chapter~3. The main feature that emerged from this study is the so called effect of ``magnetic catalysis" of chiral symmetry breaking. Namely, for a given temperature sufficiently high magnetic field induces chiral symmetry breaking by promoting the stability of quark bound states such as mesons or the chiral condensate. This means that for a given temperature we can increase the strength of the magnetic field so that the meson melting transition disappears altogether. In this model, this effect exists for arbitrarily large temperature.

In the Type IIA model, the low temperature phase breaks chiral symmetry spontaneously. Therefore introducing a magnetic field is expected to further enhance this effect. This is what we observed in chapter 4. However, the finite temperature physics is much richer in content. Sufficiently high temperature can trigger the restoration of chiral symmetry, which is similar to the meson-melting transition in the Type IIB set-up. Interestingly we found that an external magnetic field and temperature again appear as two competing parameters resulting in a non-trivial phase diagram. An external magnetic field promotes the breaking of chiral symmetry and avoids the melting transition. However, in this model there seems to exist an upper bound of the chiral symmetry restoring temperature for arbitrarily large magnetic field. The existence of this upper bound in temperature is also in sharp contrast with what we observed in the Type IIB set-up.

On a related note, in chapter 6 we have discussed the effect of the magnetic field on associated spectrum of mesons with both small and large spins. We have observed that the magnetic field increases the lifetime of the melting meson and enhances the angular momentum of mesons with large spin. Both these effects favour the bound states over the free constituent fundamental matter.

There is an interesting point to note. As we have discussed, both these gravity duals are formally quite different. Even the breaking of the symmetry group that we have studied under the name of chiral symmetry is rather different. However, the qualitative feature of the ``magnetic catalysis" seems to be a robust one in both these models. It is tempting to conjecture that this ``magnetic catalysis" in chiral symmetry breaking is indeed an universal phenomenon. Similar conclusion has been arrived previously by studying several field theory models in various non-perturbative frameworks ({\it e.g.} using the Dyson--Schwinger equation) in {\it e.g.} refs.~\cite{Gusynin:1995nb, Semenoff:1999xv, Miransky:2002eb}. It is perhaps not impossible to imagine probing some of these properties of the real quark-gluon plasma within an experimental set-up with an external magnetic field.

Emboldened by the interesting physics observed in the presence of an external magnetic field, it is natural to consider the effect of an external electric field. We have analyzed the effect of a constant electric field (in the Type IIB model) in chapter 4 and briefly towards the end of chapter 5 (in the Type IIA model). Intuitively, an electric field tries to dissociate any quark bound state into its constituents and drives a current. This is what we observed in chapter 4. The probe brane analysis in both these models seems to have an interesting feature. It is the existence of a vanishing locus or the so called pseudo-horizon (above the actual horizon) at which the probe brane action vanishes. In order for the probe brane to extend beyond this point, an additional gauge field is excited on the worldvolume. This additional gauge field gives the expectation value of the current that is caused by the electric field. We have computed the conductivity in both the Type IIB and the Type IIA models.

However, much work is still needed to understand the curious features in presence of an electric field. In chapter 4, we found that depending on the value of the polar angle of entering the vanishing locus the probe brane develops a conical singularity. Similar conical singularity was observed in the Type IIA model in ref.~\cite{Bergman:2008sg}. We have conjectured that stringy corrections may possibly cure this singularity. Another relevant avenue that we have not pursued in this thesis is the analysis of the meson spectrum in presence of the electric field.

Following our general approach, we can further analyze the effect of a chemical potential in both these holographic models. The phase structure of thermal $\mathcal{N} = 4$ Yang-Mills was analyzed in presence of a finite baryon density in ref.~\cite{Kobayashi:2006sb}; similar study was carried out in the Type IIA model first in ref.~\cite{Horigome:2006xu} and later in refs.~\cite{Parnachev:2006ev, Davis:2007ka, Bergman:2007wp, Rozali:2007rx, Parnachev:2007bc, Aharony:2007uu}. It is natural to further introduce an external electric-magnetic field and investigate their effect. It turns out to have a rich phenomenology. Turning on both magnetic field and chemical potential has many interesting features, {\it e.g.} it induces an axial current in the Type IIA model. Work along this direction has been discussed in refs.~\cite{Bergman:2008qv, Thompson:2008qw, Lifschytz:2009si, Lifschytz:2009sz}. This phenomenological richness is mainly sourced by the non-vanishing Chern-Simons term supported by the worldvolume of the probe branes. This non-vanishing Chern-Simons term is equivalent to the inclusion of the axial anomaly in the flavour sector of the dual field theory. It is further noteworthy that the effect of such axial anomaly terms within QCD has been recently studied in the context of the RHIC physics and new phenomena, {\it e.g.} the Chiral Magnetic effect\cite{Fukushima:2008xe}, have been predicted. It is rather encouraging that similar effects are also observed within the Type IIA model.

The role of finite chemical potential was studied within the Type IIB model in {\it e.g.} refs.~\cite{Kobayashi:2006sb, Mateos:2007vc} and later generalized to isospin chemical potential in ref.~\cite{Erdmenger:2008yj}. Interestingly this model also shows remarkable phenomenological richness when a magnetic field is further introduced. This has recently been studied in detail in refs.~\cite{Evans:2010iy, Jensen:2010vd}, where upon a further analysis a holographic Berezinskii-Kosterlitz-Thouless transition has also been observed\cite{Jensen:2010ga, Evans:2010hi}.

Undoubtedly both these models have remarkable rich features. Let us now go back and comment on the main theme of this thesis: the physics of chiral symmetry breaking. As we have noted earlier what we have called chiral symmetry in the models we discussed are quite different. We also noted that the symmetry breaking pattern in the Type IIA model, {\it i.e.} the U$(N_f)_L \times$U$(N_f)_R \to$U$(N_f)_{\rm diag}$, is much closer to the chiral symmetry breaking in QCD which we ultimately aim to understand better. This therefore poses a question: whether one can realize similar symmetry breaking pattern within some other brane-construction or supergravity background. To slightly rephrase this question, we can focus on the fact that despite the beautiful realization of chiral symmetry breaking in the Witten-Sakai-Sugimoto model it has some serious drawbacks: {\it e.g.} the string coupling is not constant and diverges in the UV. Furthermore, the dual field theory is really a $(4+1)$-dimensional gauge theory compactified on a circle and not a true $(3+1)$-dimensional gauge theory. One is thus led to consider other possible supergravity models which might realize the chiral symmetry breaking pattern as the Witten-Sakai-Sugimoto model but do not suffer from its drawbacks.

Recently in ref.~\cite{Kuperstein:2008cq} a similar chiral symmetry breaking mechanism has been proposed by considering probe D7--anti-D7 branes in the Klebanov-Witten background\cite{Klebanov:1998hh}\footnote{This background was originally found by Romans\cite{Romans:1984an}.}. In ref.~\cite{Dymarsky:2009cm}, this construction has been further generalized for the Klebanov-Strassler background\cite{Klebanov:2000hb}. Note that both these models are within the Type IIB supergravity framework. Evidently, these constructions provide us with the opportunity to explore more. It would be very interesting to consider the effect of an external electric-magnetic field excited on the worldvolume of the probe brane--anti-brane system. Although the finite temperature Klebanov-Strassler background is not analytically known (for more details see {\it e.g.} ref.~\cite{Aharony:2007vg}) and therefore hard to analyze, the zero temperature physics may still contain a rich physics.

A crucial ingredient of our analysis is the large $N_c$ and the $N_c \gg N_f$ limit. In the real world we have $N_c \sim N_f$, thus a probe computation is not sufficient. Instead one needs to consider the effect of the backreaction of the probe flavour branes on the background. Early attempts to include backreaction in the Type IIB model has been discussed in {\it e.g.} ref.~\cite{Erdmenger:2004dk}. Some recent work has also been done to include the backreaction of the probe branes in the Witten-Sakai-Sugimoto model in ref.~\cite{Burrington:2007qd}. However, this remains a much less controllable scenario. Similar attempts have also been made to include the backreaction in {\it e.g.} refs.~\cite{Benini:2006hh, Benini:2007gx} in the flavoured Klebanov-Witten and the Klebanov-Strassler model that we have not discussed here.

Nonetheless, we hope we made a convincing case in favour of using the techniques of gauge-gravity duality to shed light on strong coupling dynamics. Let us now offer some comments from a broader perspective. String theory eventually emerged as a potential candidate theory for quantum gravity, however many aspects of this theory still remain poorly-understood. On the other hand, quantum field theories are much better understood. Thus, as proposed in ref.~\cite{Maldacena:1997re}, the gauge-gravity duality can itself serve as a definition of a theory of quantum gravity. The concrete examples of this gauge-gravity duality within string theory is therefore an exciting prospect. It is also natural to construct more such interesting examples.

Continuing on this thought, it is tempting to ask whether we can realize the gravity duals of everyday systems. There are many condensed matter systems which are described by strongly coupled, scale invariant theories. These include high $T_c$ superconductors near their quantum critical point, fermions at unitarity {\it etc.} At the time of writing, constructing and analyzing gravity duals of various strongly coupled condensed matter phenomena is an extremely active and developing field of research. We refer to the refs.~\cite{Hartnoll:2009sz, Herzog:2009xv} for a comprehensive review of the current progress.

In conclusion we hope our approach of learning strong coupling dynamics by using the gauge-gravity duality, in one way or the other,  will teach us interesting and useful lessons.

%
%
%
%
%
%
%
%
%


\chapter*{Appendices}

\addcontentsline{toc}{chapter}{Appendices}

\section*{\hspace{2pt} Appendix A: The fluctuation action of the probe in Type IIB model}
\appendix
\renewcommand{\theequation}{A.\arabic{equation}}
\setcounter{equation}{0}
\addcontentsline{toc}{section}{\hspace{0 cm} Appendix A: The fluctuation action of the probe in Type IIB model}

The relevant pieces of the action to second order in $\alpha'$ are:
\begin{eqnarray} \label{eqt:fluctuation}
S & = & -  T_{D7} \int d^8 \xi \sqrt{g_{ab} + B_{a b}+ 2 \pi \alpha' F_{a b}} + \left(2\pi \alpha'\right) \mu_7 \int_{\mathcal{M}_8} F_{(2)} \wedge B_{(2)} \wedge P\left[\tilde{C}_{(4)} \right]  \nonumber \\
&&  +\left(2\pi \alpha' \right)^2 \mu_7 \frac{1}{2} \int_{\mathcal{M}_8} F_{(2)} \wedge F_{(2)} \wedge P\left[C_{(4)}\right] \ , \\
C_{(4)} &=& \frac{1}{g_s} \frac{u^4}{R^4} dt \wedge dx^1 \wedge dx^2 \wedge dx^3  \ , \nonumber\\
\tilde{C}_{(4)} &=& -\frac{R^4}{g_s}  \left(1- \cos^4\theta \right) \sin\psi \cos\psi\ d \psi \wedge d \phi_2 \wedge d \phi_3 \wedge d\phi _1\ ,
\end{eqnarray}
where $P\left[C_{(4)}\right]$ is the pull--back of the 4--form
potential sourced by the stack of $N_c$ D3--branes,
$P\left[\tilde{C}_{(4)}\right]$ is the pull--back of the 4--form
magnetic dual to $C_{(4)}$, and $F_{(2)}$ is the Maxwell 2--form on
the D7--brane world--volume.  At this point, we resort to a different
set of coordinates than we have been using.  Instead of using the
coordinates $(\rho, L)$ introduced in
equation~(\ref{eqt:changeofcoordinates}), we return to the coordinates
$(z=1/u^2, \theta)$ because the analysis is simpler.
We consider fluctuations of the form:
\begin{eqnarray}
\theta = \theta_0(z) + 2 \pi \alpha' \chi(\xi^a) \ , \quad \phi_1 = 2 \pi \alpha' \Phi(\xi^a) \ ,\label{eqt:ansatz1} 
\end{eqnarray}
where the indices $a, b = 0 \dots 7$ run along the world--volume of the
D7--brane.  $\theta_0(z)$ corresponds to the classical embedding from
the classical equations of motion.  Plugging the ansatz in
equations~(\ref{eqt:ansatz1}) into the action
and expanding to second order in $\left(2 \pi \alpha' \right)$, we get
as second order terms in the lagrangian:
\begin{eqnarray}
\mathcal{L}_{\chi} &=& - \frac{1}{2} \sqrt{-E} S^{a b} R^2 \partial_a \chi \partial_b \chi + \frac{1}{2} \sqrt{-E} R^4 \left(\theta_0' \right)^2 E^{z z} S^{a b} \partial_a \chi \partial_b \chi \\
&& - \frac{1}{2} \chi^2 \left[ \partial_\theta^2 \sqrt{-E} - \partial_z \left(E^{zz} R^2 \theta_0' \partial_\theta \sqrt{-E} \right) \right] \nonumber \ , \\
\mathcal{L}_{\Phi} &=& - \frac{1}{2} \sqrt{-E} S^{a b} R^2 \sin^2 \theta_0 \partial_a \Phi \partial_b \Phi \ , \nonumber \\
\mathcal{L}_{F} &=& - \frac{1}{4} \sqrt{-E} S^{a b} S^{c d} F_{b c} F_{a d}  \nonumber \ , \\
\mathcal{L}_{\chi F} &=& - \chi F_{23} \left[ \partial_z \left(\sqrt{-E} R^2 \theta_0' E^{zz} J^{23} \right) - J^{23} \partial_\theta \sqrt{-E} \right] = - \chi F_{2 3} f \nonumber \ , \\
\mathcal{L}_{F}^{\mathrm{WZ}} &=&  \frac{1}{8} \frac{1}{z^2 R^4} F_{m n} F_{o p} \epsilon^{m n o p} \nonumber \ , \\
\mathcal{L}_{\Phi F}^{\mathrm{WZ}} &=& -\Phi F_{01} H  R^4 \sin\psi \cos\psi \partial_z \left( 1- \cos^4 \theta_0 \right) = - \Phi F_{0 1} H R^4 \sin \psi \cos \psi \partial_z K \ .\nonumber
\end{eqnarray}
We have taken $E_{a b} = g^{(0)}_{a b} + B_{a b}$ to be the zeroth
order contribution from the DBI action.  In addition, we use that
$E^{a b} = S^{a b} + J^{a b}$, where $ S^{a b} = S^{ b a}$ and $J^{a
  b} = - J^{b a}$.  We use this notation for brevity.  The indices $m,
n, o, p = 4 \dots 9$ run in the transverse directions to the
D3--branes.

To proceed, we impose $A_m = 0$ and $\partial_m A_a = 0$. We further assume that $\partial_i \chi = 0$, in which case it is possible to consistently set $A_2 = 0 = A_3$.  Finally we get the following equations to consider
\begin{eqnarray}
0 &=& \partial_a \left[\sqrt{-E} S^{a b} R^2 \left(\frac{1+4b^4 z^4 \left(\theta_0'\right)^2}{1+4 z^2 \left(\theta'_0\right)^2} \right) \partial_b \chi \right] -\chi  \left[ \partial_\theta^2 \sqrt{-E} \right. \nonumber \\
&& \left. - \partial_z \left(E^{zz} R^2 \theta_0' \partial_\theta \sqrt{-E} \right) \right]  \ , \label{eqt:eom_theta} \\
0 &=&- \partial_0 \left(\sqrt{-E} S^{00} S^{11} \partial_0 A_1 \right) + \partial_z K H \partial_0 \Phi - \partial_z \left(\sqrt{-E} S^{zz} S^{11} \partial_z A_1 \right)  \ , \label{eqt:eom_phi_A2} \\
0 &=& \partial_a \left( \sqrt{-E} S^{a b} R^2 \sin^2 \theta_0 \partial_b \Phi \right) - F_{01} H  R^4 \sin\psi \cos\psi \partial_z K  \ . \label{eqt:eom_phi_A1}
\end{eqnarray}
Clearly we end up with one decoupled equation for $\chi$ and a set of coupled equations for $\Phi$ and $A$- fluctuations.

\newpage

\section*{\hspace{2pt} Appendix B: The fluctuation action of the probe in Type IIA model}
\appendix
\renewcommand{\theequation}{B.\arabic{equation}}
\setcounter{equation}{0}
\addcontentsline{toc}{section}{\hspace{0 cm} Appendix B: The fluctuation action of the probe in Type IIA model}

Here we provide the quadratic action for fluctuations of the probe brane. We consider the following fluctuation of the embedding function
\begin{eqnarray}
\tau(u)=\tau_0(u)+(2\pi\alpha')\chi(\xi^a)\ ,
\end{eqnarray}
where $\tau_0(u)$ is the classical embedding of the probe brane and $\{\xi^a\}$ refer to its worldvolume coordinates where $a = 0, \ldots, 8$.

Now we can determine the effective Lagrangian corresponding to small fluctuations by expanding the DBI action up to quadratic orders. We use $(2 \pi \alpha')$ as a bookkeeping parameter for this expansion. The contribution at order $(2\pi\alpha')^2$ is summarised below
\begin{eqnarray}
&& S_{\rm DBI} = \int d^9\xi \mathcal{L}_{\rm total}\ , \quad \mathcal{L}_{\rm total}=-\mu_8\left[\mathcal{L}_{\chi}+\mathcal{L}_{F}+\mathcal{L}_{\chi F}\right]\ ,\nonumber\\
&& \mathcal{L}_{\chi}=\frac{1}{2}G_{\tau\tau}\left(\mathcal{S}^{uu}G_{\tau\tau}\tau_0'(u)^2-1\right)e^{-\Phi}\sqrt{-{\rm det}\left(E^{(0)}\right)}\mathcal{S}^{ab}\left(\partial_a\chi\right)\left(\partial_b\chi\right)\ ,\nonumber\\
&& \mathcal{L}_{F}=\frac{1}{4}e^{-\Phi}\sqrt{-{\rm det}\left(E^{(0)}\right)}\mathcal{S}^{aa'}\mathcal{S}^{bb'}F_{ab}F_{a'b'}\ ,\nonumber\\
&& \mathcal{L}_{\chi F}=\frac{1}{2}\partial_u\left[e^{-\Phi}\tau_0'(u)G_{\tau\tau}\mathcal{S}^{uu}\sqrt{-{\rm det}\left(E^{(0)}\right)} \mathcal{A}^{ab}\right]\chi F_{ab}\ ,
\end{eqnarray}
where
\begin{eqnarray}\label{eqt: comp}
&& \left(E^{(0)}\right)^{-1}=\mathcal{S}^{ab}+\mathcal{A}^{ab} \ ,\nonumber\\
&& \mathcal{S}^{ab}={\rm diag}\left\{-G^{tt},G^{xx},\frac{G_{xx}}{G_{xx}^2+ H ^2},\frac{G_{xx}}{G_{xx}^2+ H^2},g_{uu}^{-1}\right\}\times \left |\left |\Omega_4\right |\right |\ ,\nonumber\\
&& \mathcal{A}^{ab}=\frac{H}{G_{xx}^2+ H^2}\left(\delta_3^a\delta_2^b-\delta_2^a\delta_3^b\right)\ , \quad g_{uu}=G_{uu}+\tau_0'(u)^2G_{\tau\tau}\ , \nonumber\\
&& \left |\left |\Omega_4\right |\right |={\rm diag}\left\{\left(u/R\right)^{-3/2}u^2, \left(u/R\right)^{-3/2}u^2,\left(u/R\right)^{-3/2}u^2,\left(u/R\right)^{-3/2}u^2\right\}\ ,\nonumber\\
&& \sqrt{-{\rm det}\left(E^{(0)}\right)}=\sqrt{G_{tt}G_{xx}g_{uu}\left(G_{xx}^2+ H^2\right)\left({\rm det} \left |\left |\Omega_4\right |\right |\right)}\ ,
\end{eqnarray}
where $\mathcal{S}$ and $\mathcal{A}$ are the symmetric and anti-symmetric part respectively and $\left |\left |\Omega_4\right |\right |$ denotes the diagonal metric for the $S_4$.

The term $\mathcal{L}_{\chi F}$ is the interaction term that couples the scalar and the vector modes. Clearly if $\tau_0'(u)=0$ then the coupling vanishes. Therefore for the high temperature phase ({\it i.e.}, when $\tau_0(u)={\rm const.}$) the scalar and vector meson modes are always decoupled.

Now we determine the contribution coming from the Wess-Zumino term. The total contribution coming from the Wess-Zumino term at order $(2\pi\alpha')^2$ is given by
\begin{eqnarray}\label{eqt: lwz}
S_{\rm WZ}&=&\frac{1}{2}\mu_8\int_{\mathcal{M}_9}\left({\rm P}[C_5]\wedge F_2\wedge F_2-\frac{1}{3} F_4\wedge A\wedge F_2\wedge B_2\right)\ , \quad {\rm with}\nonumber\\
F_4&=&dC_3\, \quad F_2=dA\ ,
\end{eqnarray}
In what follows we will focus on the special case of $\tau_0'(u)=0$ in which case the first term in eqn. (\ref{eqt: lwz}) does not contribute at the leading order.

The equations of motion for the gauge fields then decouple from the scalar fluctuation and is given by (using $R=1$)
\begin{eqnarray}\label{eqt: aeom}
\partial_a\left[e^{-\Phi}\sqrt{-{\rm det}\left(E^{(0)}\right)}\mathcal{S}^{aa'}\mathcal{S}^{bb'}F_{a'b'}\right]+ 4 g_s^{-1} H \epsilon^{ba'b'23}\partial_{b'} A_{a'}=0\ .
\end{eqnarray}
where now $a, a', b, b'\in \{\mathbb{R}^{1,3}\}\bigcup \{u\}$, on the probe brane worldvolume. It is now possible to consistently set all the $A_\alpha=0$ where $\alpha\in S^4$. This choice imposes a constraint of the following form
\begin{eqnarray}\label{eqt: constr2}
\mathcal{S}^{\mu\nu}\partial_\alpha\partial_{\mu}A_{\nu}=0\ .
\end{eqnarray}
This constraint can be trivially satisfied by looking at the gauge field fluctuations independent of the spherical directions (in other words focussing only on the ${\rm SO}(5)$ singlet states).

It is straightforward to show that with the following ansatz for the gauge field fluctuations decouple into two sectors
\begin{eqnarray}
A_t  =  A_t(u) e^{-i\omega t+ik x^1}\ , \quad A_i  =  A_i(u) e^{-i\omega t+ik x^1}\ ,  \quad A_u = 0 \ .
\end{eqnarray}

The choice $A_u=0$ imposes the following constraint
\begin{equation}\label{eqt: constrpara}
e^{-\Phi}\sqrt{-{\rm det}\left(E^{(0)}\right)}\mathcal{S}^{uu}\left[\mathcal{S}^{tt}(\omega A_t')-\mathcal{S}^{11}(kA_1')\right]+4 g_s^{-1} H (\omega A_1+kA_t)=0\ .
\end{equation}
The equations of motion are given by
\begin{eqnarray}\label{eqt: eomA1}
&&  \partial_u\left[e^{-\Phi}\sqrt{-{\rm det}\left(E^{(0)}\right)}\mathcal{S}^{tt}\mathcal{S}^{uu}A_t'\right] + 4 g_s^{-1} H A_1' \nonumber\\
 & - & e^{-\Phi}\sqrt{-{\rm det}\left(E^{(0)}\right)} \mathcal{S}^{tt}k\mathcal{S}^{11}(\omega A_1+kA_t)  =  0 \ , \\
&& \partial_u\left[e^{-\Phi}\sqrt{-{\rm det}\left(E^{(0)}\right)}\mathcal{S}^{11}\mathcal{S}^{uu}A_1'\right] - 4 g_s^{-1} H A_t'  \nonumber\\
& - & e^{-\Phi}\sqrt{-{\rm det}\left(E^{(0)}\right)} \mathcal{S}^{11}\omega\mathcal{S}^{tt}(\omega A_1+kA_t) = 0 \ , 
\end{eqnarray}
and
\begin{eqnarray}
&& \partial_u\left[e^{-\Phi}\sqrt{-{\rm det}\left(E^{(0)}\right)}\mathcal{S}^{22}\mathcal{S}^{uu}A_{2, 3}'\right] \nonumber\\
& - & e^{-\Phi}\sqrt{-{\rm det}\left(E^{(0)}\right)} \mathcal{S}^{22}\left(k^2\mathcal{S}^{11}+\omega^2\mathcal{S}^{tt}\right)A_{2, 3}  =  0 \ .
\end{eqnarray}
In this case, the longitudinal modes ($A_t, A_1$ components oscillating along $\{t,x^1\}$ plane) and the transverse modes ($A_2, A_3$ components oscillating along $\{t,x^1\}$ plane) clearly decouple. It is useful to note that $A_2$, $A_3$ transform as vectors and $\{A_t, A_1\}$ transforms as a scalar under the ${\rm SO}(2)$ rotation group in the plane perpendicular to the magnetic field.

\newpage

\section*{\hspace{2pt} Appendix C: To obtain the effective Schr\"{o}dinger equation}\label{schrodinger}
\appendix
\renewcommand{\theequation}{C.\arabic{equation}}
\setcounter{equation}{0}
\addcontentsline{toc}{section}{\hspace{0 cm} Appendix C: To obtain the effective Schr\"{o}dinger equation}

Here we outline the general variable changes to obtain the corresponding Schr\"{o}dinger equation from the original equation for fluctuation modes. Without any loss of generality we can write the equation for any fluctuation mode as follows
\begin{eqnarray}\label{eqt: seom}
a_1(u)f_\chi''(u)+a_2(u)f_\chi'(u)+\omega^2a_3(u)f_\chi(u)=0\ ,
\end{eqnarray}
where, $a_1(u)$, $a_2(u)$ and $a_3(u)$ are known functions and prime denotes derivative with respect to $u$, $f_{\chi}(u)$ denotes the radial profile of the fluctuation mode and $\omega$ is the corresponding oscillatory frequency (in the finite temperature case this is the quasinormal mode). Let us rewrite $f_\chi(u)=\sigma(u)g(u)$ with 
\begin{eqnarray}\label{eqt: schg}
\frac{\sigma'(u)}{\sigma(u)}=-\frac{1}{2}\left[\frac{a_2(u)}{a_1(u)}+\frac{1}{2}\left(\frac{a_1(u)}{a_3(u)}\right)\partial_u\left(\frac{a_3(u)}{a_1(u)}\right)\right]\ .
\end{eqnarray}
The Schr\"{o}dinger equation is then obtained to be
\begin{eqnarray}\label{eqt: scsch}
 \sqrt{\frac{a_1(u)}{a_3(u)}}\partial_u\left(\sqrt{\frac{a_1(u)}{a_3(u)}}\left(\partial_u g(u)\right)\right)+\omega^2 g(u)-V_s(u) g(u)=0\ , \label{tortoise} 
\end{eqnarray}
where
\begin{eqnarray}
 V_s(u) & = & - \frac{1}{B(u)^2}\left[\frac{1}{4}\left(A(u)+\frac{B'(u)}{B(u)}\right)^2 - \frac{1}{2}\partial_u\left(A+\frac{B'(u)}{B(u)}\right) \right. \nonumber\\
&& \left. - \frac{A(u)}{2}\left(A(u)+\frac{B'(u)}{B(u)}\right)\right]\ , \\
 A(u) & = & \frac{a_2(u)}{a_1(u)}\ , \quad B(u)=\left(\frac{a_3(u)}{a_1(u)}\right)^{1/2}\ .
\end{eqnarray}
A more conventional form of the Schr\"{o}dinger equation can be obtained after changing variables to ``tortoise" coordinate $d\tilde{u}=B(u) du$; the horizon is then located at $\tilde{u}\to\infty$. The main goal of this exercise is to obtain the potential $V_S(u)$ and extract qualitative features of the meson spectrum.

To determine the quasinormal modes we consider only in-falling modes at the horizon; such a solution can be written as $g(\tilde{u})={\rm exp}(-i\omega t)\psi(\tilde{u})$. Now multiplying equation in (\ref{tortoise}) by $\psi^*(\tilde{u})$, integrating by parts from the horizon to the boundary and using the equation of motion we get
\begin{eqnarray}\label{eqt: vsch}
\int_{\tilde{u_b}}^\infty d\tilde{u}\left(|\psi'(\tilde{u})|^2+V_s(\tilde{u})|\psi(\tilde{u})|^2\right)=-\frac{|\omega|^2|\psi(\infty)|^2}{{\rm Im}[ \omega]}\ .
\end{eqnarray}
Note that the left hand side is positive definite as long as $V_S$ is positive. This implies that ${\rm Im} [\omega] < 0$ and therefore the corresponding mode is stable. On the other hand, a sufficiently strong negative support of $V_S$ would indicate the presence of tachyonic modes in the fluctuation spectrum and hence an instability.

Let us now represent the quasinormal frequency as $\omega=\omega_{\rm R}+i\omega_{\rm I}$. It can be noted from the effective Schr\"{o}dinger equation that there is a $\mathbb{Z}_2$ symmetry in $\omega_{\rm R}$ (namely, if $g(u)$ is a solution with a frequency $\omega$ then $g^*(u)$ is also a solution with frequency $-\omega^*$\cite{Hoyos:2006gb}).

Here we have explicitly assumed that the fluctuation does not have any momentum mode. The fluctuation equation for the meson having a momentum can also be recast in the form of a Schr\"{o}dinger equation. The changes of variables are exactly similar as already mentioned in eqn. (\ref{eqt: scsch}), but the effective potential receives a positive contribution coming from the momentum 
\begin{eqnarray}
V_S(u,k)=V_S(u,0)+k^2\left(\frac{|a_4|}{a_3}\right)\ ,
\end{eqnarray}
where the left hand side of the equation (\ref{eqt: seom}) is now accompanied by another term of the form $-|a_4(u)|k^2 f_{\chi}$, the relative negative sign is due to the fact that $k$-corresponds to a spatial oscillation.

\newpage

\section*{\hspace{2pt} Appendix D: Lowest hydrodynamic mode in vanishing magnetic field}
\appendix
\renewcommand{\theequation}{D.\arabic{equation}}
\setcounter{equation}{0}
\addcontentsline{toc}{section}{\hspace{0 cm} Appendix D: Lowest hydrodynamic mode in vanishing magnetic field}

In holographic contexts the study of the lowest-lying (hydrodynamic) quasinormal modes of various fluctuation fields in a given background has become a vast literature by now. Initial work along this direction was done in {\it e.g.} ref.~\cite{Starinets:2002br, Policastro:2002se, Policastro:2002tn, Son:2006em} and later reviewed in {\it e.g.} ref.~\cite{Starinets:2008fb}. These studies were done for the $\mathcal{N} = 4$ SYM theory and hence within the Type IIB supergravity. Similar analysis was also done for the Witten-Sakai-Sugimoto model in {\it e.g.} ref.~\cite{Benincasa:2005iv, Benincasa:2006ei}.

Here we analyze the gauge field fluctuations when there is no magnetic field. We provide an analytical derivation for the dispersion relation of the lowest hydrodynamic mode of the longitudinal oscillation, which has been obtained by other methods in refs. \cite{Myers:2007we, Evans:2008tv}. Similar analytical study of a ${\rm U}(1)$ fluctuation field in AdS$_{p+2}$ background was done in refs. \cite{Karch:2008fa, Starinets:2008fb}.

The longitudinal mode is defined to be $\mathcal{E}(u)=\omega A_1+ kA_t$. Using this definition, the constraint equation in (\ref{eqt: constrpara}) and the equations of motion in (\ref{eqt: eomA1}) (after setting $B=0$) we obtain the equation of motion for the longitudinal mode to be
\begin{eqnarray}\label{eqt: h0eom}
\frac{d^2\mathcal{E}}{dx^2} & + & \frac{5\left(1-x^{-3}\right)\left(\omega^2-k^2\left(1-x^{-3}\right)\right)+6\omega^2x^{-3}}{2x\left(1-x^{-3}\right)\left(\omega^2-k^2\left(1-x^{-3}\right)\right)}\frac{d\mathcal{E}}{dx} \nonumber\\
& + &\frac{R^3\left(\omega^2-k^2\left(1-x^{-3}\right)\right)}{U_T x^3\left(1-x^{-3}\right)^2}\mathcal{E}(x)=0\ , 
\end{eqnarray}
where $x= u/U_T$.

The near-horizon limit $u\to U_T$ is now achieved by taking the $x\to 1$ limit. In this limit, the equation of motion takes the form
\begin{eqnarray}
\frac{d^2\mathcal{E}}{dx^2}+\frac{1}{x-1}\frac{d\mathcal{E}}{dx}+\frac{R^3\omega^2}{9U_T(x-1)^2}\mathcal{E}(x)=0\ .
\end{eqnarray}
The general solution of this equation is given by a linear combination of the incoming and the outgoing modes
\begin{eqnarray}
\mathcal{E}(x)=C_1 \cos (\tilde{\omega} \log (x-1))+C_2 \sin (\tilde{\omega}\log (x-1))\ , \quad \tilde{\omega}=\sqrt{\frac{R^3\omega^2}{9 U_T}}=\frac{\omega}{4\pi T}\ .
\end{eqnarray}
The incoming boundary condition singles out the solution with $C_1=-C_2$. Near the horizon $\mathcal{E}(x)$ is therefore obtained to be
\begin{eqnarray}\label{eqt: Ehori}
\mathcal{E}(x)=C_L {\rm Exp}\left(-i\tilde{\omega}\log(x-1)\right)\ ,
\end{eqnarray}
where $C_L$ is an yet undetermined constant.

Let us also define $\tilde{k}=k/(4\pi T)$. Now, for small enough $\omega$ and $k$ (such that $\tilde{\omega}\ll 1$ and $\tilde{k}\ll 1$), we can ignore the last term in equation (\ref{eqt: h0eom}). It turns out that in this limit the equation of motion for the longitudinal mode is also exactly solvable.
\begin{eqnarray}\label{eqt: eomE}
\mathcal{E}(x)=M_2+\frac{1}{3} M_1 \left(-\frac{2 \tilde{k}^2}{x^{3/2}}+\tilde{\omega} ^2 \log \left[\frac{x^{3/2}+1}{x^{3/2}-1}\right]\right)\ ,
\end{eqnarray}
where $M_1$ and $M_2$ are constants of integration. Near the boundary, {\it i.e.}, $x\to\infty$, this reduces to
\begin{eqnarray}
\mathcal{E}(x)=M_2-\frac{2}{3}M_1\left(\tilde{k}^2-\tilde{\omega}^2\right)\left(\frac{1}{x^{3/2}}\right)\ .
\end{eqnarray}
Therefore normalizability of $\mathcal{E}(x)$ forces us to impose $M_2=0$. The quasi-normal modes are therefore obtained to be the solution of this constraint.

On the other hand, in the vicinity of the horizon the solution in (\ref{eqt: eomE}) takes the following form
\begin{eqnarray}\label{eqt: expan1}
\mathcal{E}(x)&=&\frac{1}{3} \left(-2 M_1 \tilde{k}^2+3 M_2+M_1 \tilde{\omega} ^2 \log \left(\frac{4}{3}\right)\right)-\frac{1}{3}M_1 \tilde{\omega} ^2 \log (x-1)+\nonumber\\
&& \frac{1}{6} M_1 \left(6 \tilde{k}^2+\tilde{\omega} ^2\right) (x-1)-\frac{1}{144} M_1 \left(180 \tilde{k}^2+\tilde{\omega} ^2\right) (x-1)^2+ \ldots \nonumber\\
\end{eqnarray}

Now, for sufficiently small values of $\tilde{\omega}$, we can expand the solution obtained in (\ref{eqt: Ehori}) to get
\begin{eqnarray}\label{eqt: expan2}
\mathcal{E}(x)=C_L-i C_L\tilde{\omega}\log(x-1)-\frac{1}{2}C_L\tilde{\omega}^2\left(\log(x-1)\right)^2\ldots
\end{eqnarray}
Comparing equation (\ref{eqt: expan1}) and (\ref{eqt: expan2}) we get
\begin{eqnarray}
&& \frac{1}{3} \left(-2 M_1 \tilde{k}^2+3 M_2+M_1 \tilde{\omega} ^2 \log \left(\frac{4}{3}\right)\right)=C_L\ , \nonumber\\
&& \frac{1}{3}M_1 \tilde{\omega} ^2=i C_L\tilde{\omega}\ .
\end{eqnarray}
Setting $M_2=0$, the solution of this equation is given by
\begin{eqnarray}\label{eqt: dis0}
M_1=\frac{3iC_L}{\tilde{\omega}}\ , \quad \tilde{\omega}=-2i \tilde{k}^2+\dots 
\end{eqnarray}
Restating this result in dimensionful parameters, we get the lowest hydrodynamic quasinormal frequency
\begin{eqnarray}
\omega=-i D_R k^2\ , \quad D_R=\frac{1}{2\pi T}\ ,
\end{eqnarray}
where $D_R$ is the R-charge diffusion constant. This value for the diffusion constant was obtained numerically by studying the spectral functions of holographic flavours in ref.~\cite{Myers:2007we} and was also confirmed in ref.~\cite{Evans:2008tv} by numerically obtaining the lowest quasinormal mode of the longitudinal oscillation. Not surprisingly we reproduce the $(3+1)$-dimensional result of ref. \cite{Starinets:2008fb} since the flavour branes describe an effective $(3+1)$-dimensional field theory.

On the other hand, it is easy to show that the transverse mode does not have any solution compatible with the hydrodynamic limit. See, {\it e.g.} ref.~\cite{Kovtun:2005ev}.

\newpage

\section*{\hspace{2pt} Appendix E: Global anomalies and dispersions from first order hydrodynamics}
\appendix
\renewcommand{\theequation}{E.\arabic{equation}}
\setcounter{equation}{0}
\addcontentsline{toc}{section}{\hspace{0 cm} Appendix E: Global anomalies and dispersions from first order hydrodynamics}

Here we will briefly review the dispersion relations resulting from ideal hydrodynamics in the presence of a quantum anomaly\cite{Son:2009tf}. Our goal here will be to demonstrate the existence of the ``anomalous sound" mode (observed in section 6.4) within the hydrodynamic equations. Let us consider a $(3+1)$-dimensional relativistic fluid with one (classically) conserved U(1) which results in an U(1)$^3$ anomaly (after quantization). One example of such will be the QCD triangle anomaly coming from the chiral ${\rm U}(1)$ which is a true symmetry of the system at the classical level.

We also turn on a slowly varying non-dynamical background gauge field $A_\mu$ which couples to the current $J^\mu$. We assume that $A_\mu \sim \cO(k^0)$, where $k$ is the momentum, and therefore $F_{\mu\nu}\sim \cO(k)$. We will work within the framework of first order hydrodynamics and therefore keep terms of order $\cO(k)$ in the constitutive equations for $T^{\mu\nu}$ and $J^\mu$. We will closely follow the set-up and conventions of ref.~\cite{Son:2009tf}.

In the presence of external fields (such as electric-magnetic fields) the hydrodynamic equations take the following form
\begin{eqnarray}\label{eqt: heom}
\partial_\mu T^{\mu\nu} = F^{\nu\lambda} J_{\lambda}\ , \quad \partial_\mu J^\mu = C E^\mu B_\mu\ , 
\end{eqnarray}
where $C$ is the anomaly coefficient and we have also defined 
\begin{eqnarray}
E^\mu = F^{\mu\nu} u_\nu\ , \quad B^\mu = \frac{1}{2} \epsilon^{\mu\nu\rho\sigma} u_\nu F_{\rho\sigma}\ ,
\end{eqnarray}
where $u^\mu$ is the fluid velocity vector normalized as $u^\mu u_\mu = -1$.

Note that $B^\mu$ is identically zero for lower dimensional theories. The non-zero right hand sides of the hydrodynamic equations in (\ref{eqt: heom}) reflect the work done by the external field on the system and on the anomaly\cite{Son:2009tf}.

The constitutive equations for $T^{\mu\nu}$ and $J^\mu$ are given by
\begin{eqnarray}
T^{\mu\nu} &=& \epsilon  u^\mu u^\nu + P \Delta^{\mu\nu} + \Pi^{\mu\nu}\ , \nonumber\\
J^\mu &=& \rho u^\mu + \nu^\mu\ ,
\end{eqnarray}
where 
\begin{eqnarray}
\Delta^{\mu\nu} = \eta^{\mu\nu} + u^\mu u^\nu\ , \quad \eta^{\mu\nu} = {\rm diag}(-1, 1, 1, 1)\ ,
\end{eqnarray}
and $\epsilon$, $P$ and $\rho$ denote the energy density, the pressure and the charge density respectively. The terms $\Pi^{\mu\nu}$ and $\nu^{\mu}$ are of $\cO(k)$ and capture the dissipative effects. These dissipative terms are obtained from the requirement of the existence of an entropy current with non-negative derivative. The result of such an exercise is given by\cite{Son:2009tf}
\begin{eqnarray}
\Pi^{\mu \nu} &=& -\eta \sigma^{\mu\nu} - \zeta \Delta^{\mu\nu} \left(\partial_\alpha u^\alpha \right)\ , \nonumber\\
\sigma^{\mu\nu} &=& \Delta^{\mu\alpha} \Delta^{\nu\beta} \left(\partial_\alpha u_\beta + \partial_\beta u_\alpha \right) - \frac{2}{d-1} \Delta^{\mu\nu} \left(\partial_\alpha u^\alpha\right)
\end{eqnarray}
and 
\begin{eqnarray}
\nu^\mu &=& \sigma_Q \Delta^{\mu\nu} \left(-\partial_\nu \mu + F_{\nu \alpha} u^\alpha + \frac{\mu}{T} \partial_\nu T \right) + \zeta_w w^\mu + \zeta_B B^\mu\ , \nonumber\\
w^\mu &=& \frac{1}{2} \epsilon^{\mu\nu\rho\sigma} u_\nu \partial_\rho u_\sigma\ ,
\end{eqnarray}
where $d$ is the spacetime dimension (which in this case is $4$), $\eta$ is the shear viscosity, $\zeta$ is the bulk viscosity and $\zeta_w$, $\zeta_B$ are the new transport coefficients sourced by the anomaly.

We will consider linear fluctuations around the equilibrium configuration. The fluctuation variables will be the velocity vector $\delta u ^\mu$ as well as the temperature $\delta T$ and the chemical potential $\delta \mu$. The normalization condition of the fluid velocity vector requires $\delta u^\mu u_\mu = 0$, {\it i.e.}, the fluctuations be orthogonal to the equilibrium configuration.

We choose the following equilibrium configuration
\begin{eqnarray}
&& F_{0i} = 0 \ , \quad i = 1, 2, 3\ , \quad F_{23} = H = - F_{32}\ , \quad F_{12}= 0 = F_{31} \ , \nonumber\\
&& u^\mu = \left(1, 0, 0, 0 \right)\ , \quad T= {\rm const.} \ , \quad \mu = {\rm const.} 
\end{eqnarray}
Therefore our fluctuation variables are: $\delta u_x$, $\delta u_y$, $\delta u_z$, $\delta T$ and $\delta \mu$ and all fluctuations are of the plane-wave form ${\rm exp}(- i \omega t + i k x)$. The equation for the fluctuations modes are given by
\begin{eqnarray}
\partial_\mu \delta T^{\mu\nu} = F^{\nu\lambda} \delta J_\lambda\ , \quad \partial_\mu \delta J^\mu = C \delta \left(E^\mu B_\mu \right) \ ,
\end{eqnarray}
where 
\begin{eqnarray}
\delta T^{\mu\nu} &=& \left(\delta \epsilon\right) u^\mu u^\nu + \epsilon \delta\left(u^\mu u^\nu\right) + \left(\delta P\right) \Delta^{\mu\nu} + P \delta \Delta^{\mu \nu} + \delta \Pi^{\mu\nu}\ , \nonumber\\
\delta J^\mu &=& \left(\delta \rho \right) u^\mu + \rho \delta u^\mu + \delta \nu^\mu\ .
\end{eqnarray}
From these we can obtain the equations for the fluctuations to be given by
\begin{eqnarray}
\omega \delta \epsilon - k \left( \epsilon + P \right) \delta u _x & = & 0 \ , \\
\omega \left( \epsilon + P \right) \delta u_x - k \delta P + i k^2 \left( \zeta + \frac{4}{3} \eta \right) \delta u_x & = & 0 \ , \\
\omega \delta \rho + \left( \omega \zeta_B H + k \rho \right) \delta u_x - i k^2 \sigma_Q \delta \mu + i k^2 \sigma_Q \frac{\mu}{T} \delta T & = & 0 \ ,
\end{eqnarray}
and
\begin{eqnarray}
\left[ \omega \left(\epsilon + P \right) + i k^2 \eta + i \sigma_Q H^2 + \frac{1}{2} \zeta_w k H \right]\delta u _y - i \rho H \delta u_z & = & 0 \ , \\
\left[ \omega \left( \epsilon + P \right) + i k^2 \eta + i \sigma_Q H^2 + \frac{1}{2} \zeta_w k H \right] \delta u_z + i \rho H \delta u_y & = & 0 \ .
\end{eqnarray}
Clearly the fluctuation sectors $\{\delta \mu, \delta T, \delta u_x\}$ and $\{\delta u_y, \delta u_z\}$ are decoupled.

Let us first we consider the case when $H=0$, $\rho=0$ and $\mu = 0$. The equations of motion for the fluctuation modes take the following form
\begin{eqnarray}\label{eqt: sound}
 \omega \delta \epsilon - k \left( \epsilon + P \right) \delta u _x & = & 0 \ ,  \nonumber\\
 \omega \left( \epsilon + P \right) \delta u_x  - k \delta P + i k^2 \left( \zeta + \frac{4}{3} \eta \right) \delta u_x & = & 0 \ , 
\end{eqnarray}
and
\begin{eqnarray}\label{eqt: shear}
 \omega \delta \rho - i k^2 \sigma_Q \delta \mu & = & 0 \ , \nonumber\\ 
 \left[ \omega \left(\epsilon + P \right) + i k^2 \eta \right]\delta u _{y, z} & = & 0 \ .
\end{eqnarray}
The dispersion relation resulting from the equations in (\ref{eqt: sound}) is given by
\begin{eqnarray}
\omega = \pm c_s k - i k^2 \frac{\zeta + \frac{4}{3}\eta}{2 \left( \epsilon + P \right)} + \ldots\ , \quad c_s^2 = \frac{\partial P}{\partial \epsilon}\ ,
\end{eqnarray}
where $c_s$ is the speed of sound (more generally the speed of the propagating degree of freedom). On the other hand, the dispersion relation resulting from the last equation in (\ref{eqt: shear}) is given by
\begin{eqnarray}
\omega = - i k^2 \frac{\eta}{\epsilon + P} + \ldots\ ,
\end{eqnarray}
whereas  the first equation in (\ref{eqt: shear}) describes a diffusive mode for $\delta \mu$ given by
\begin{eqnarray}
\omega =  i k^2 \sigma_Q \left( \frac{\partial \mu}{\partial \rho} \right) + \ldots\ 
\end{eqnarray}
Therefore equation (\ref{eqt: sound}) is known as the sound mode equation and (\ref{eqt: shear}) is known as the shear mode equation.

Let us now focus on a bit more general case when $H\not = 0$, but $\rho = 0 = \mu$. In this case, the diffusive mode corresponding to $\delta \mu$ couples to the sound mode equation and thus it does not make sense to address them separately. The other diffusive mode, corresponding to $\delta u_y$ or $\delta u_z$, however does remain decoupled. Let us write down the equations below:\\

1. The coupled mode:
\begin{eqnarray}
\omega \delta \epsilon - q \left( \epsilon + P \right) \delta u _x & = & 0 \ , \nonumber\\ 
\omega \left( \epsilon + P \right) \delta u_x - q \delta P + i q^2 \left( \zeta + \frac{4}{3} \eta \right) \delta u_x & = & 0 \ , \nonumber\\ 
\omega \delta \rho +  \omega \zeta_B H  \delta u_x - i q^2 \sigma_Q \delta \mu  & = & 0 \ .
\end{eqnarray}

2. The shear mode:
\begin{eqnarray}
\left[ \omega \left(\epsilon + P \right) + i q^2 \eta + i \sigma_Q H^2 + \frac{1}{2} \zeta_w q H \right]\delta u _y & = & 0 \ , \nonumber\\ 
\left[ \omega \left(\epsilon + P \right) + i q^2 \eta + i \sigma_Q H^2 + \frac{1}{2} \zeta_w q H \right]\delta u _z & = & 0 \ .
\end{eqnarray}
From the shear channel, we readily obtain the following dispersion relation
\begin{eqnarray}\label{eqt: hsound}
\omega = - i k^2 \frac{\eta}{\epsilon + P} - i H^2 \frac{\sigma_Q}{\epsilon + P} \pm k H \frac{\zeta_w}{2 \left(\epsilon + P \right)}\ .
\end{eqnarray}
Assuming that $H$ is small in the hydrodynamic limit (specifically if we focus on the regime when $H\sim k$), we get a new propagating mode whose velocity is now given by
\begin{eqnarray}
c_s^{\rm anomalous} = \frac{\zeta_w}{2 \left(\epsilon + P \right)} H\ .
\end{eqnarray}
Therefore the presence of an external magnetic field can source a propagating mode in the otherwise diffusive channel. Note that this effect is completely universal, {\it i.e.}, present for both conformal and non-conformal systems obeying arbitrary equation of state (since we did not use the equation of state anywhere). The magnetic field also modifies the diffusive mode.

On the other hand, the coupled modes have a simple behaviour. The corresponding characteristic equation factorizes in the following diffusive and sound channels
\begin{eqnarray}
\omega = i k^2 \frac{\sigma_Q}{\epsilon + P}\ , \quad \omega = \pm c_s k - i k^2 \frac{\zeta + \frac{4}{3} \eta}{2 \left(\epsilon + P \right)}\ .
\end{eqnarray}
In other words, the magnetic field does not affect the already existing shear and sound channels.

Let us now consider the case when all the external parameters are present {\it i.e.} $H \not = 0$ and $\rho \not = 0$. As alluded to earlier we will consider only $\{\delta T, \delta \mu, \delta u_x\}$ and $\{\delta u_y, \delta u_z\}$ as the independent fluctuation modes, which incidentally do remain decoupled.

The shear channel dispersion relation now takes the form
\begin{eqnarray}
\omega = - i k^2 \frac{\eta}{\epsilon + P } - i H^2 \frac{\sigma_Q}{\epsilon +  P } \pm \frac{\zeta_w}{2 \left( \epsilon + P \right)} k H \pm \frac{\rho H }{\epsilon + P}\ .
\end{eqnarray}
This has the same qualitative features as before, namely the propagating mode still shows up. The only difference now is the new term proportional to the charge density and linear in $H$. Note that this dispersion relation should also be rather general, since we have not assumed any equation of state specific to any system.

The other set of coupled mode can also be solved to obtain the corresponding dispersion relation, but this is cumbersome for the most general case. We need to know the equation of state to obtain the specific dispersion relation of a particular system.

\newpage

\section*{\hspace{2pt}  Appendix F: Spinning strings and background field in Rindler space, a model calculation}
\appendix
\renewcommand{\theequation}{F.\arabic{equation}}
\setcounter{equation}{0}
\addcontentsline{toc}{section}{\hspace{0 cm} Appendix F: Spinning strings and background field in Rindler space, a model calculation}

Here we would like to solve a toy problem to study the effect of the magnetic field on meson dissociation. We implicitly assume the framework where we can have two equal charges at the two ends of the string and therefore analyze the symmetric configuration only. Rindler spacetime provides an useful arena where many qualitative (if not quantitative) features of an event horizon can be realized in a simple set up. Such study has been previously carried out in ref.~\cite{Peeters:2007ti} without any background field. The main result of this exercise is that the high spin meson dissociates once a critical value of the acceleration is reached. This acceleration is in turn determined by the angular momentum of the meson.

There are two ways to physically interpret the results of such pursuits ({\it e.g.}, see ref.~\cite{Peeters:2007ti} and ref. \cite{Berenstein:2007tj}). We consider accelerating the rotating string in one of the space-like directions and therefore a Rindler horizon forms in the four dimensional spacetime. Alternatively, we can consider accelerating the string in the holographic direction and therefore a bulk spacetime horizon forms which sets the temperature of the dual gauge theory. To begin with we take the former point of view.

The metric for the Rindler space can be written as
\begin{eqnarray}\label{eqt: rmet}
ds^2=-\xi^2\kappa^2d\eta^2+d\xi^2+d\rho^2+\rho^2d\phi^2\ ,
\end{eqnarray}
where $\{\rho,\phi\}$-plane represents the plane where we would rotate the string. To add a background magnetic field we consider similar potential $A_\phi$ given before. Now the ansatz for the string in terms of its worldvolume coordinates $\{\tau,\sigma\}$ is given by
\begin{eqnarray}\label{eqt: ransatz}
\eta=\tau\ , \quad \xi=\xi(\sigma)\ ,\quad \rho=\rho(\sigma)\ , \quad \phi=\omega \tau\ .
\end{eqnarray}
This ansatz implies that the direction of acceleration of the string is orthogonal to the direction of its rotation. We can readily see that there are two conserved charges associated with the string, namely the angular momentum $J$, which is associated with the rotation along $\phi$-direction and the boost charge, which is associated with the translation in $\eta$-direction.

The Nambu-Goto action for the string now will be accompanied by a boundary term exactly similar to the term $\Delta S_B$ given in eqn. (\ref{eqt: action}). Therefore the full action is given by
\begin{eqnarray}
S=\frac{1}{2\pi\alpha'}\int d\tau d\sigma \sqrt{(\xi^2\kappa^2-\omega^2\rho^2)(\xi'^2+\rho'^2)}+\Delta S_B\ .
\end{eqnarray}
Varying this action we can obtain the equation of motion for the string profile and the boundary conditions we should impose. In absence of any external field this exercise has been carried out in details in ref.~\cite{Peeters:2007ti}, which we do not repeat here. The presence of the magnetic field enforces us to impose the following boundary condition
\begin{eqnarray}\label{eqt: bcr}
\left. \pi_\rho\right|_{\pm \ell/2}=\left.\frac{\partial L}{\partial \rho'}\right|_{\pm \ell/2}=- H \rho\omega\ ,
\end{eqnarray}
where $\ell$ is the distance between the two end points of the string. The other boundary term corresponding to the variation of $\xi$ is satisfied imposing the Dirichlet boundary condition $\delta\xi=0$. This corresponds to a situation where the end points move with a constant acceleration $a=\xi(\pm\ell/2)^{-1}$.

Now choosing a gauge $\rho=\sigma$ we can solve\footnote{The analytical solution of the equations of motion obtained from the Nambu-Goto action remains the same as in ref.~\cite{Peeters:2007ti}. We refer to ref.~\cite{Peeters:2007ti} for further details about solving the equations of motion. We use the same analytical solution and impose the boundary condition given in (\ref{eqt: bcr}) to obtain the results in eqn. (\ref{eqt: rsol}).} the equation of motion to obtain the profile for the string given by
\begin{eqnarray}\label{eqt: rsol}
\xi(\rho)&=&\frac{C}{\kappa}\cosh\left[\frac{\kappa}{\omega}\arcsin\left(\frac{\omega\rho}{C}\right)\right]\ \quad {\rm with} \quad C^2=\frac{\omega^2\ell^2}{4}(1+ H^2)\ .\nonumber\\
 J&=&\frac{1}{2\pi\alpha'}\frac{\ell^2}{4}\left(1+H^2\right)\arctan\left(\frac{1}{H}\right)\ , \quad T_s=\frac{1}{2\pi\alpha'}\ .
\end{eqnarray}
As the notation suggests the parameter $J$ represents the angular momentum of the meson.

As we have seen before, the presence of the background magnetic field does not change the equations of motion for the Nambu-Goto string. Therefore the functional form of the string profile given in equation (\ref{eqt: rsol}) is entirely determined by the Rindler geometry\cite{Peeters:2007ti}. We notice that the presence of the magnetic field does induce an effective length $\ell_{\rm eff}=\ell^2(1+H^2)$ by changing the boundary condition of the string end-points.

A non-zero value of $H$ brings about two key changes in the evaluation of the angular momentum. The profile of the string modifies in order to satisfy the boundary condition in eqn. (\ref{eqt: bcr}) and therefore the effective length $\ell_{\rm eff}$ arises in the formula in (\ref{eqt: rsol}). There is also an additional explicit contribution $\Delta J_B$ of the form shown in eqn. (\ref{eqt: ej}), which finally gives a functional factor of $\tan^{-1}(1/H)$ in the expression for the angular momentum.

\begin{figure}[!ht]
\begin{center}
\includegraphics[angle=0,
width=0.65\textwidth]{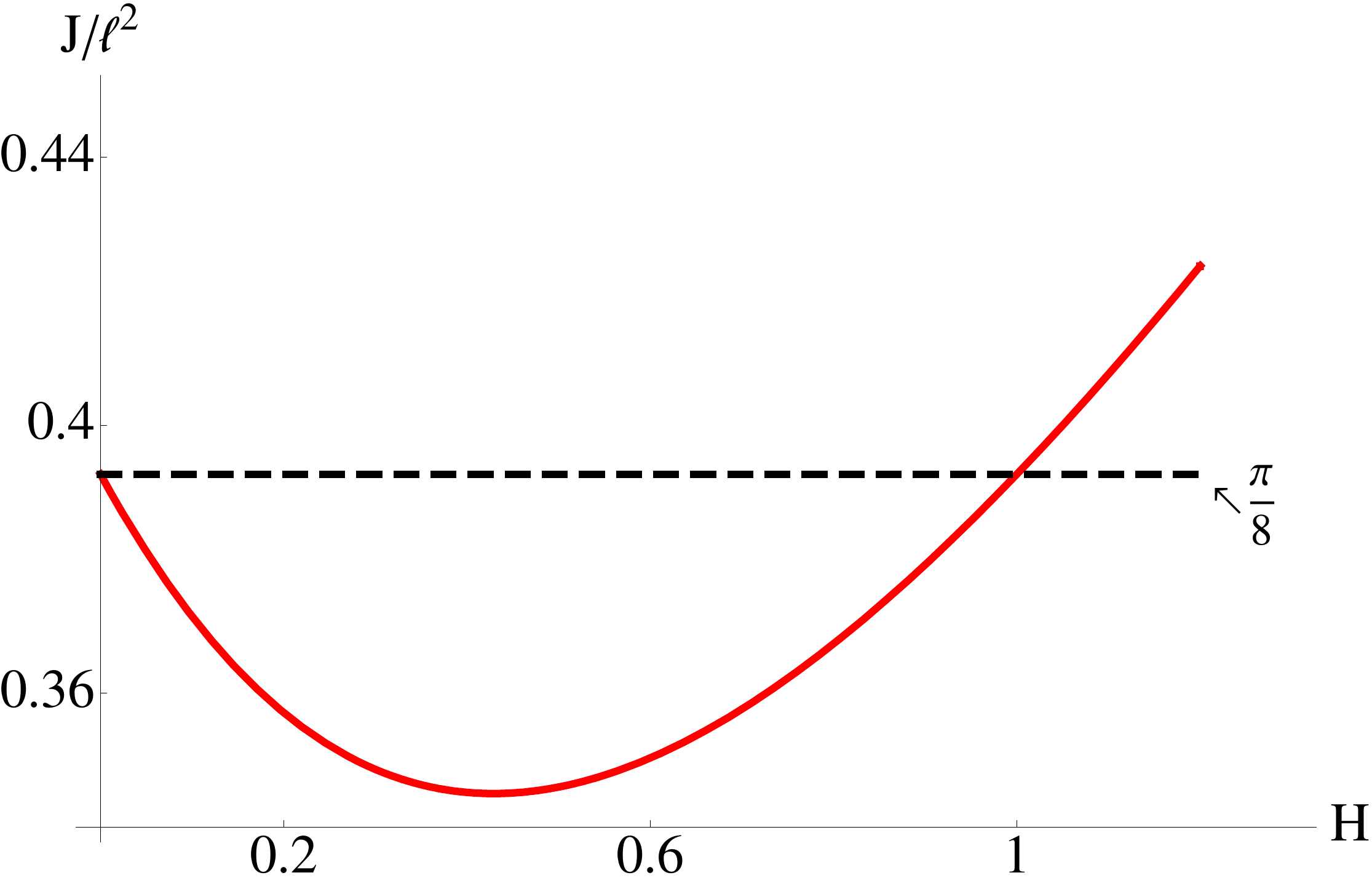}
\caption{\small The total angular momentum of the string in units of $(2\pi\alpha')$ in the background field. The horizontal dashed line represents the value of angular momentum of the string when there is no background magnetic field. We observe that the $B$-field initially reduces and then increases the angular momentum creating a local minima. However, we will observe that the stability of the meson is always enhanced by the $B$-field.}
\label{fig: jomegab}
\end{center}
\end{figure}

It can be shown that the explicit contribution of the of the magnetic field to the angular momentum, denoted by $\Delta J_B$, is actually cancelled by a term coming from integrating over the string profile. The net resulting angular momentum therefore can be lower than the angular momentum of the string in absence of the field. This is pictorially summarised in figure \ref{fig: jomegab}.

To find the maximum acceleration we identify $a^{-1}=\xi(\ell/2)$ and look for minima of the right hand side of the equation as a function of the parameter $C$. For now we content ourselves with only positive values of $H$, {\it i.e.} when the string angular momentum and the background field are parallel. It is a straightforward exercise to show that such minima corresponds to values of $C$ which satisfy the following equation (obtained from taking the first derivative of $\xi$ and setting it equal to zero at $\rho=\pm \ell/2$)
\begin{eqnarray}\label{eqt: solve}
&&x=m\tanh\left(\frac{m}{x}\right)\ , \quad {\rm where}\nonumber\\
&&x=\frac{C}{\kappa}\frac{1}{\sqrt{2\pi\alpha' J}}\ , \quad m=\arcsin\left(\frac{1}{\sqrt{1+H ^2}}\right)\left[\arctan\left(\frac{1}{H}\right)\right]^{-1/2}\ .
\end{eqnarray}
Clearly the parameter $H$ generates a set of such values of $C$ which satisfy the above equation and therefore promotes the maximum value of acceleration to a function of the magnetic field. Differentiating the eqn. in the first line of (\ref{eqt: solve}) and using identities for the trigonometric hyperbolic functions we obtain that the family of roots of the equation can be simply given by the relation $x=\alpha m$ where $\alpha=0.834$ is a constant.

Using these roots the maximum acceleration can be obtained as a function of the background magnetic field which is given by
\begin{eqnarray}
a_{\rm max}\propto \sqrt{\arctan\left(\frac{1}{H}\right)}\frac{1}{\arcsin\left(\frac{1}{\sqrt{1+H^2}}\right)}\ .
\end{eqnarray}

Alternatively we could imagine the Rindler space to be the near-horizon approximation of a blackhole background and the acceleration is along the holographic coordinate (we make the identification that $\xi=u$, where $u$ is the radial coordinate in the holographic set up). The parameter $\kappa$ (in eqn. (\ref{eqt: rmet})) in this case represents the surface gravity and sets the temperature of the dual gauge theory to be $T=\kappa/(2\pi)$.

Now we imagine the string end points to sit on the flavour brane at some value of $\xi_0=U_0$, which fixes the constituent quark mass; and we consider a similar ansatz for the string profile as presented in eqn. (\ref{eqt: ransatz}). The interpretation of the set up is different from the earlier one; however this brings about no change as far as the mathematics is concerned. We therefore again obtain the string profile as given by eqn. (\ref{eqt: rsol}).

However in this case the physical parameter that captures the dissociation of the meson is the size $\ell$ (which sets the angular momentum $J$ via the relation in eqn. (\ref{eqt: rsol})) of it. The length is obtained to be
\begin{eqnarray}\label{eqt: rlen}
\ell=\frac{2C}{\kappa}\frac{{\rm arccosh}\left(\kappa U_0/C\right)}{\sqrt{1+ H^2}}\left[\arcsin\left(1/\sqrt{1+ H^2}\right)\right]^{-1}\ .
\end{eqnarray}
We would hope to see that for a given value of $U_0$, there exists a maximum admissible size of the spinning meson beyond which it dissociates.

We can treat as before the parameter $C$ to be independent with respect to which we will consider maximizing the length function $\ell$. From eqn. (\ref{eqt: rlen}) it is clear that the external magnetic field does not play any role to determine the value of $C$ corresponding to the maximum of $\ell$; however it determines the the function $\ell_{\rm max}(H)$ and therefore also $J_{\rm max}(H)$ to be given by
\begin{eqnarray}\label{eqt: ljmax}
&& \ell_{\rm max}\propto \frac{1}{\sqrt{1+H^2}}\left[\arcsin\left(1/\sqrt{1+H^2}\right)\right]^{-1}\ , \\
&& J_{\rm max}\propto \arctan\left(\frac{1}{H}\right) \left[\arcsin\left(1/\sqrt{1+H^2}\right)\right]^{-2} \ ,
\end{eqnarray}
where the constants of proportionality depends on the background temperature $T$ and the parameter $U_0$.

For the background considered in the main text the Rindler spacetime does emerge when we zoom in the region close to the horizon\cite{Frolov:2006tc, Mateos:2006nu}. The toy model here can therefore be identified with a corresponding study of high spin mesons in the overheated phase. So if we want to make a direct connection between the Rindler background dynamics to the specific system studied in the text, we should remember that such a background appears precisely in the over-heated phase of the system, where any small fluctuation is likely to destroy the meson and drive the over-heated phase to the melted phase. Our emphasis here however is not to make this connection, we view this exercise as a more tractable analysis of some of the physics we studied in the more thorough holographic calculations.


\begin{thebibliography}{100}


\bibitem{Abouelsaood:1986gd}
A.~Abouelsaood, C.~G.~.~Callan, C.~R.~Nappi and S.~A.~Yost.
\newblock {Open Strings In Background Gauge Fields}.
\newblock {\em  Nucl.\ Phys.\  B}, 280:599 (1987). 


\bibitem{Aharony:2007vg}
O.~Aharony, A.~Buchel and P.~Kerner.
\newblock {The black hole in the throat - thermodynamics of strongly coupled cascading gauge theories}.
\newblock {\em  Phys.\ Rev.\  D}, 76:086005 (2007). 


\bibitem{Aharony:1999ti}
Ofer Aharony, Steven~S. Gubser, Juan~Martin Maldacena, Hirosi Ooguri, and Yaron Oz.
\newblock {Large N field theories, string theory and gravity}.
\newblock {\em Phys. Rept.}, 323:183--386, 2000.


\bibitem{Aharony:2008an}
O.~Aharony and D.~Kutasov.
\newblock {Holographic Duals of Long Open Strings}.
\newblock {\em  Phys.\ Rev.\  D}, 78:026005 (2008).


\bibitem{Aharony:2007uu}
O.~Aharony, K.~Peeters, J.~Sonnenschein and M.~Zamaklar.
\newblock {Rho meson condensation at finite isospin chemical potential in a holographic model for QCD}.
\newblock {\em  JHEP}, 0802:071 (2008).


\bibitem{Aharony:2006da}
Ofer Aharony, Jacob Sonnenschein, and Shimon Yankielowicz.
\newblock {A holographic model of deconfinement and chiral symmetry
  restoration}.
\newblock {\em Annals Phys.}, 322:1420--1443, 2007.


\bibitem{Albash:2006ew}
Tameem Albash, Veselin~G. Filev, Clifford~V. Johnson, and Arnab Kundu.
\newblock {A topology-changing phase transition and the dynamics of flavour}.
\newblock {\em Phys. Rev.}, D77:066004, 2008.


\bibitem{Albash:2006bs}
Tameem Albash, Veselin~G. Filev, Clifford~V. Johnson, and Arnab Kundu.
\newblock {Global Currents, Phase Transitions, and Chiral Symmetry Breaking in
  Large $N_c$ Gauge Theory}.
arXiv:hep-th/0605175.


\bibitem{Albash:2007bk}
Tameem Albash, Veselin~G. Filev, Clifford~V. Johnson, and Arnab Kundu.
\newblock {Finite Temperature Large N Gauge Theory with Quarks in an External
  Magnetic Field}.
JHEP {\bf 0807}, 080 (2008)
  [arXiv:0709.1547 [hep-th]].


\bibitem{Albash:2007bq}
Tameem Albash, Veselin~G. Filev, Clifford~V. Johnson, and Arnab Kundu.
\newblock {Quarks in an External Electric Field in Finite Temperature Large N
  Gauge Theory}.
JHEP {\bf 0808}, 092 (2008)
  [arXiv:0709.1554 [hep-th]].  


\bibitem{Antonyan:2006vw}
E.~Antonyan, J.~A. Harvey, S.~Jensen, and D.~Kutasov.
\newblock {NJL and QCD from string theory}.
\newblock 2006.


\bibitem{Antonyan:2006qy}
E.~Antonyan, J.~A.~Harvey and D.~Kutasov.
\newblock {The Gross-Neveu model from string theory}.
\newblock {\em  Nucl.\ Phys.\  B}, 776:93 (2007).


\bibitem{Antonyan:2006pg}
E.~Antonyan, J.~A.~Harvey and D.~Kutasov.
\newblock {Chiral symmetry breaking from intersecting D-branes}.
\newblock {\em  Nucl.\ Phys.\  B}, 784:1 (2007).






\bibitem{Argyres:2008sw}
P.~C.~Argyres, M.~Edalati, R.~G.~Leigh and J.~F.~Vazquez-Poritz.
\newblock {Open Wilson Lines and Chiral Condensates in Thermal Holographic QCD}.
\newblock {\em  Phys.\ Rev.\  D}, 79:045022 (2009). 


\bibitem{Babington:2003vm}
J.~Babington, J.~Erdmenger, Nick~J. Evans, Z.~Guralnik, and I.~Kirsch.
\newblock {Chiral symmetry breaking and pions in non-supersymmetric gauge /
  gravity duals}.
\newblock {\em Phys. Rev.}, D69:066007, 2004.


\bibitem{Balasubramanian:1999re}
Vijay Balasubramanian and Per Kraus.
\newblock {A stress tensor for anti-de Sitter gravity}.
\newblock {\em Commun. Math. Phys.}, 208:413--428, 1999.


\bibitem{Benincasa:2006ei}
P.~Benincasa and A.~Buchel.
\newblock {Hydrodynamics of Sakai-Sugimoto model in the quenched approximation}.
\newblock {\em  Phys.\ Lett.\  B}, 640:108 (2006). 


\bibitem{Benincasa:2005iv}
P.~Benincasa, A.~Buchel and A.~O.~Starinets.
\newblock {Sound waves in strongly coupled non-conformal gauge theory plasma}.
\newblock {\em  Nucl.\ Phys.\  B}, 733:160 (2006). 


\bibitem{Benini:2006hh}
F.~Benini, F.~Canoura, S.~Cremonesi, C.~Nunez and A.~V.~Ramallo.
\newblock {Unquenched flavours in the Klebanov-Witten model}.
\newblock {\em  JHEP}, 0702:090 (2007).


\bibitem{Benini:2007gx}
F.~Benini, F.~Canoura, S.~Cremonesi, C.~Nunez and A.~V.~Ramallo.
\newblock {Backreacting Flavors in the Klebanov-Strassler Background}.
\newblock {\em  JHEP}, 0709:109 (2007).


\bibitem{Berenstein:2007tj}
D.~Berenstein and H.~J.~Chung.
\newblock {Aspects of open strings in Rindler Space}.




\bibitem{Bergman:2007wp}
O.~Bergman, G.~Lifschytz and M.~Lippert.
\newblock {Holographic Nuclear Physics}.
\newblock {\em  JHEP}, 0711:056 (2007).


\bibitem{Bergman:2008sg}
O.~Bergman, G.~Lifschytz and M.~Lippert.
\newblock {Response of Holographic QCD to Electric and Magnetic Fields}.
\newblock {\em  JHEP}, 0805:007 (2008).


\bibitem{Bergman:2008qv}
O.~Bergman, G.~Lifschytz and M.~Lippert.
\newblock {Magnetic properties of dense holographic QCD}.


\bibitem{Bergman:2007pm}
O.~Bergman, S.~Seki and J.~Sonnenschein.
\newblock {Quark mass and condensate in HQCD}.
\newblock {\em  JHEP}, 0712:037 (2007).


\bibitem{Bhattacharyya:2008jc}
 S.~Bhattacharyya, V.~E.~Hubeny, S.~Minwalla and M.~Rangamani.
\newblock {Nonlinear Fluid Dynamics from Gravity}.
\newblock {\em  JHEP}, 0802:045 (2008).




\bibitem{Born:1934ab}
{M. Born and L. Infeld}.
\newblock {Foundations of the new field theory}.
\newblock {\em Proc. Roy. Soc. Lond.}, A150:465, 1934.


\bibitem{Bousso:2002ju}
R.~Bousso.
\newblock {The holographic principle}.
\newblock {\em Rev.\ Mod.\ Phys.}, 74:825, 2002.


\bibitem{Buchel:2003tz}
 A.~Buchel and J.~T.~Liu.
\newblock {Universality of the shear viscosity in supergravity}.
\newblock {\em Phys.\ Rev.\ Lett.}, 93:090602 (2004).


\bibitem{Buchel:2004qq}
 A.~Buchel.
\newblock {On universality of stress-energy tensor correlation functions in supergravity}.
\newblock {\em Phys.\ Lett.\ B}, 609:392 (2005).


\bibitem{Burrington:2007qd}
B.~A.~Burrington, V.~S.~Kaplunovsky and J.~Sonnenschein.
\newblock {Localized Backreacted Flavor Branes in Holographic QCD}.
\newblock {\em JHEP}, 0802:001 (2008).




\bibitem{Chamblin:1999tk}
Andrew Chamblin, Roberto Emparan, Clifford~V. Johnson, and Robert~C. Myers.
\newblock {Charged AdS black holes and catastrophic holography}.
\newblock {\em Phys. Rev.}, D60:064018, 1999.


\bibitem{Chamblin:1999hg}
Andrew Chamblin, Roberto Emparan, Clifford~V. Johnson, and Robert~C. Myers.
\newblock {Holography, thermodynamics and fluctuations of charged AdS black
  holes}.
\newblock {\em Phys. Rev.}, D60:104026, 1999.




\bibitem{Coleman:1973jx}
S.~R.~Coleman and E.~J.~Weinberg.
\newblock {Radiative Corrections As The Origin Of Spontaneous Symmetry Breaking}.
\newblock {\em Phys.\ Rev.\  D}, 7:1888, 1973.






\bibitem{creutz}
{M. Creutz}.
\newblock {Quarks, Gluons and Lattices}.
\newblock {\em Cambridge, UK: Univ. Press}, 1983.


\bibitem{Cvetic:1999ne}
Mirjam Cvetic and Steven~S. Gubser.
\newblock {Phases of R-charged black holes, spinning branes and strongly coupled gauge theories}.
\newblock {\em JHEP}, 9904:024, 1999.


\bibitem{Cvetic:1999rb}
Mirjam Cvetic and Steven~S. Gubser.
\newblock {Thermodynamic stability and phases of general spinning branes}.
\newblock {\em JHEP}, 07:010, 1999.




\bibitem{Davis:2007ka}
J.~L.~Davis, M.~Gutperle, P.~Kraus and I.~Sachs.
\newblock {Stringy NJL and Gross-Neveu models at finite density and temperature}.
\newblock {\em  JHEP}, 0710:049 (2007).


\bibitem{DD}
{Degrand and De Tar}.
\newblock {Lattice Methods for Quantum Chromodynamics}.
\newblock {\em World Scientific}, 2006.


\bibitem{deForcrand:2010ys}
P.~de Forcrand.
\newblock {Simulating QCD at finite density}.
\newblock {\em  PoS LAT2009}, 010 (2009).


\bibitem{DeTar:2009ef}
C.~DeTar and U.~M.~Heller.
\newblock {QCD Thermodynamics from the Lattice}.
\newblock {\em  Eur.\ Phys.\ J.\  A}, 41 (2009).


\bibitem{Dhar:2008um}
A.~Dhar and P.~Nag.
\newblock {Tachyon condensation and quark mass in modified Witten--Sakai-Sugimoto model}.
\newblock {\em  Phys.\ Rev.\  D}, 78:066021 (2008).


\bibitem{Dirac:1960ab}
{P. A. M. Dirac}.
\newblock {A Reformulation of the Born-Infeld Electrodynamics}.
\newblock {\em Proc. Roy. Soc. Lond.}, A257, 1960.


\bibitem{DiRenzo:2006nh}
{F.~Di Renzo, M.~Laine, V.~Miccio, Y.~Schroder and C.~Torrero}.
\newblock {The leading non-perturbative coefficient in the weak-coupling expansion  of hot QCD pressure}.
\newblock {\em JHEP}, 0607:026 (2006).


\bibitem{Dymarsky:2009cm}
{A.~Dymarsky, S.~Kuperstein and J.~Sonnenschein}.
\newblock {Chiral Symmetry Breaking with non-SUSY D7-branes in ISD backgrounds}.
\newblock {\em JHEP}, 0908:005 (2009).


\bibitem{Erdmenger:2004dk}
{J.~Erdmenger and I.~Kirsch}.
\newblock {Mesons in gauge / gravity dual with large number of fundamental fields}.
\newblock {\em JHEP}, 0412:025 (2004).


\bibitem{Erdmenger:2007cm}
Johannaa Erdmenger, Nick Evans, Ingo Kirsch, and Ed~Threlfall.
\newblock {Mesons in Gauge/Gravity Duals - A Review}.
\newblock {\em Eur. Phys. J.}, A35:81--133, 2008.


\bibitem{Erdmenger:2008yj}
J.~Erdmenger, M.~Kaminski, P.~Kerner and F.~Rust.
\newblock {Finite baryon and isospin chemical potential in AdS/CFT with flavor}.
\newblock {\em  JHEP}, 0811:031 (2008).




\bibitem{Erdmenger:2007bn}
Johanna Erdmenger, Rene Meyer, and Jonathan~P. Shock.
\newblock {AdS/CFT with Flavour in Electric and Magnetic Kalb-Ramond Fields}.
\newblock {\em JHEP}, 12:091, 2007.




\bibitem{Evans:2010iy}
N.~Evans, A.~Gebauer, K.~Y.~Kim and M.~Magou.
\newblock {Holographic Description of the Phase Diagram of a Chiral Symmetry Breaking Gauge Theory}.
\newblock {\em  JHEP}, 1003:132 (2010). 


\bibitem{Evans:2010hi}
N.~Evans, A.~Gebauer, K.~Y.~Kim and M.~Magou.
\newblock {Phase diagram of the D3/D5 system in a magnetic field and a BKT transition}.


\bibitem{Evans:2008tv}
N.~Evans and E.~Threlfall.
\newblock {Mesonic quasinormal modes of the Witten--Sakai-Sugimoto model at high temperature}.
\newblock {\em  Phys.\ Rev.\  D}, 77:126008 (2008). 


  
  


\bibitem{Filev:2007gb}
Veselin~G. Filev, Clifford~V. Johnson, R.~C. Rashkov, and K.~S. Viswanathan.
\newblock {Flavoured large N gauge theory in an external magnetic field}.
JHEP {\bf 0710}, 019 (2007)
  [arXiv:hep-th/0701001].
  
  
\bibitem{Fradkin:1985qd}
E.~S.~Fradkin and A.~A.~Tseytlin.
\newblock {Nonlinear Electrodynamics From Quantized Strings}.
\newblock {\em  Phys.\ Lett.\  B}, 163:123 (1985). 




\bibitem{Frolov:2006tc}
Valeri~P. Frolov.
\newblock {Merger transitions in brane-black-hole systems: Criticality,
  scaling, and self-similarity}.
\newblock {\em Phys. Rev.}, D74:044006, 2006.


\bibitem{Fukushima:2008xe}
K.~Fukushima, D.~E.~Kharzeev and H.~J.~Warringa.
\newblock {The Chiral Magnetic Effect}.
\newblock {\em Phys.\ Rev.\  D}, 78:074033, 2008.


\bibitem{Gao:2006up}
Yi-hong Gao, Wei-shui Xu, and Ding-fang Zeng.
\newblock {NGN, QCD(2) and chiral phase transition from string theory}.
\newblock {\em JHEP}, 08:018, 2006.


\bibitem{GellMann:1968rz}
Murray Gell-Mann, R.~J. Oakes, and B.~Renner.
\newblock {Behavior of current divergences under SU(3) x SU(3)}.
\newblock {\em Phys. Rev.}, 175:2195--2199, 1968.




\bibitem{Gibbons:1979xm}
G.~W. Gibbons and S.~W. Hawking.
\newblock {Classification of Gravitational Instanton Symmetries}.
\newblock {\em Commun. Math. Phys.}, 66:291--310, 1979.


\bibitem{Green:1987sp}
Michael~B. Green, J.~H. Schwarz, and Edward Witten.
\newblock {SUPERSTRING THEORY. VOL. 1: INTRODUCTION}.
\newblock Cambridge, Uk: Univ. Pr. ( 1987) 469 P. ( Cambridge Monographs On
  Mathematical Physics).
  

\bibitem{Green:1987mn}
Michael~B. Green, J.~H. Schwarz, and Edward Witten.
\newblock {SUPERSTRING THEORY. VOL. 2: LOOP AMPLITUDES, ANOMALIES AND
  PHENOMENOLOGY}.
\newblock Cambridge, Uk: Univ. Pr. ( 1987) 596 P. (Cambridge Monographs On
  Mathematical Physics).


\bibitem{Gross:1973id}
{D. J. Gross, F. Wilczek}.
\newblock {Ultraviolet behavior of non-abelian gauge theories}.
\newblock {\em Phys. Rev. Lett.}, 30:1343, 1973.


\bibitem{Gubser:1998jb}
{S.~S.~Gubser}.
\newblock {Thermodynamics of spinning D3-branes}.
\newblock {\em Nucl.\ Phys.\  B}, 551:667, 1999.


\bibitem{Gubser:2009md}
{S.~S.~Gubser and A.~Karch}.
\newblock {From gauge-string duality to strong interactions: a Pedestrian's Guide}.
\newblock {\em Ann.\ Rev.\ Nucl.\ Part.\ Sci.}, 59:145, 2009.





\bibitem{Gubser:1996de}
S.~S.~Gubser, I.~R.~Klebanov and A.~W.~Peet.
\newblock {Entropy and Temperature of Black 3-Branes}.
\newblock {\em Phys.\ Rev.\  D}, 54:3915 (1996).


\bibitem{Gubser:1998bc}
S.~S. Gubser, Igor~R. Klebanov, and Alexander~M. Polyakov.
\newblock {Gauge theory correlators from non-critical string theory}.
\newblock {\em Phys. Lett.}, B428:105--114, 1998.




\bibitem{Gubser:1997yh}
Steven~S. Gubser, Igor~R. Klebanov, and Arkady~A. Tseytlin.
\newblock {String theory and classical absorption by three-branes}.
\newblock {\em Nucl. Phys.}, B499:217--240, 1997.


\bibitem{Gusynin:1995nb}
V.~P. Gusynin, V.~A. Miransky, and I.~A. Shovkovy.
\newblock {Dimensional reduction and catalysis of dynamical symmetry breaking
  by a magnetic field}.
\newblock {\em Nucl. Phys.}, B462:249--290, 1996.




\bibitem{Hartnoll:2009sz}
S.~A.~Hartnoll.
\newblock {Lectures on holographic methods for condensed matter physics}. 2009.
\newblock {Class. Quant. Grav.}, 26:224002, 2009.


\bibitem{Heinz:2004qz}
{U.~W.~Heinz}.
\newblock {Concepts of heavy-ion physics}.


\bibitem{Herzog:2009xv}
C.~P.~Herzog.
\newblock {Lectures on Holographic Superfluidity and Superconductivity}. 2009.
\newblock{J. Phys. A}, 42:343001, 2009. 






\bibitem{Horigome:2006xu}
Norio Horigome and Yoshiaki Tanii.
\newblock {Holographic chiral phase transition with chemical potential}.
\newblock {\em JHEP}, 01:072, 2007.


\bibitem{Hoyos:2006gb}
Carlos Hoyos-Badajoz, Karl Landsteiner, and Sergio Montero.
\newblock {Holographic Meson Melting}.
\newblock {\em JHEP}, 04:031, 2007.


\bibitem{Itzhaki:1998dd}
{N.~Itzhaki, J.~M.~Maldacena, J.~Sonnenschein and S.~Yankielowicz}.
\newblock {Supergravity and the large N limit of theories with sixteen supercharges}.
\newblock {\em  Phys.\ Rev.\  D}, 58: 046004 (1998).


\bibitem{Jensen:2008yp}
K.~D.~Jensen, A.~Karch and J.~Price.
\newblock {Strongly bound mesons at finite temperature and in magnetic fields from AdS/CFT}.
\newblock {\em  JHEP}, 0804:058 (2008). 


\bibitem{Jensen:2010ga}
K.~Jensen, A.~Karch, D.~T.~Son and E.~G.~Thompson.
\newblock {Holographic Berezinskii-Kosterlitz-Thouless Transitions}.


\bibitem{Jensen:2010vd}
K.~Jensen, A.~Karch and E.~G.~Thompson.
\newblock {A Holographic Quantum Critical Point at Finite Magnetic Field and Finite Density}.
\newblock {\em  JHEP}, 1005:015 (2010). 


\bibitem{Johnson:2003gi}
C.~V. Johnson.
\newblock {D-branes}.
\newblock Cambridge, USA: Univ. Pr. (2003) 548 p.


\bibitem{Johnson:2008vna}
C.~V.~Johnson and A.~Kundu.
\newblock {External Fields and Chiral Symmetry Breaking in the Witten--Sakai-Sugimoto Model}.
\newblock {\em  JHEP}, 0812:053 (2008).


\bibitem{Johnson:2009ev}
C.~V.~Johnson and A.~Kundu.
\newblock {Meson Spectra and Magnetic Fields in the Witten--Sakai-Sugimoto Model}.
\newblock {\em  JHEP}, 0907:103 (2009).


\bibitem{Kajantie:2002wa}
{K.~Kajantie, M.~Laine, K.~Rummukainen and Y.~Schroder}.
\newblock {The pressure of hot QCD up to g**6 ln(1/g)}.
\newblock {\em Phys.\ Rev.\  D}, 67:105008 (2003).


\bibitem{Kanitscheider:2008kd}
I.~Kanitscheider, K.~Skenderis and M.~Taylor.
\newblock {Precision holography for non-conformal branes}.
\newblock {\em JHEP}, 0809:094 (2008).


\bibitem{Karch:2002sh}
Andreas Karch and Emanuel Katz.
\newblock {Adding flavour to AdS/CFT}.
\newblock {\em JHEP}, 06:043, 2002.




\bibitem{Karch:2005ms}
Andreas Karch, Andy O'Bannon, and Kostas Skenderis.
\newblock {Holographic renormalization of probe D-branes in AdS/CFT}.
\newblock {\em JHEP}, 04:015, 2006.


\bibitem{Karch:2006bv}
Andreas Karch and Andy O'Bannon.
\newblock {Chiral transition of N = 4 super Yang--Mills with flavour on a
  3-sphere}.
\newblock {\em Phys. Rev.}, D74:085033, 2006.


\bibitem{Karch:2007pd}
Andreas Karch and Andy O'Bannon.
\newblock {Metallic AdS/CFT}.
\newblock {\em JHEP}, 09:024, 2007.


\bibitem{Karch:2008fa}
A.~Karch, D.~T.~Son and A.~O.~Starinets.
\newblock {Zero Sound from Holography}.




\bibitem{Kirsch:2005uy}
Ingo Kirsch and Diana Vaman.
\newblock {The D3/D7 background and flavour dependence of Regge trajectories}.
\newblock {\em Phys. Rev.}, D72:026007, 2005.


\bibitem{Kirsch:2006he}
Ingo Kirsch.
\newblock {Spectroscopy of fermionic operators in AdS/CFT}.
\newblock {\em JHEP}, 09:052, 2006.


\bibitem{Klebanov:1997kc}
Igor~R. Klebanov.
\newblock {World-volume approach to absorption by non-dilatonic branes}.
\newblock {\em Nucl. Phys.}, B496:231--242, 1997.


\bibitem{Klebanov:2000hb}
I.~R.~Klebanov and M.~J.~Strassler.
\newblock {Supergravity and a confining gauge theory: Duality cascades and chiSB-resolution of naked singularities}.
\newblock {\em JHEP}, 0008:052, 2000.


\bibitem{Klebanov:1998hh}
I.~R.~Klebanov and E.~Witten.
\newblock {Superconformal field theory on threebranes at a Calabi-Yau  singularity}.
\newblock {\em Nucl.\ Phys.\  B}, 536:199, 1998.


\bibitem{Klevansky:1992qe}
S.~P.~Klevansky.
\newblock {The Nambu-Jona-Lasinio model of quantum chromodynamics}.
\newblock {\em Rev. Mod. Phys.}, 64, 649, 1992.


\bibitem{Kobayashi:2006sb}
Shinpei Kobayashi, David Mateos, Shunji Matsuura, Robert~C. Myers, and Rowan~M.
  Thomson.
\newblock {Holographic phase transitions at finite baryon density}.
\newblock {\em JHEP}, 02:016, 2007.




\bibitem{Kovtun:2003wp}
P.~Kovtun, D.~T.~Son and A.~O.~Starinets.
\newblock {Holography and hydrodynamics: Diffusion on stretched horizons}.
\newblock {\em JHEP}, 0310:064 (2003).


\bibitem{Kovtun:2004de}
P.~Kovtun, D.~T.~Son and A.~O.~Starinets.
\newblock {Viscosity in strongly interacting quantum field theories from black hole physics}.
\newblock {\em Phys.\ Rev.\ Lett.}, 94:111601 (2005).


\bibitem{Kovtun:2005ev}
P.~K.~Kovtun and A.~O.~Starinets.
\newblock {Quasinormal modes and holography}.
\newblock {\em  Phys.\ Rev.\  D}, 72:086009 (2005). 




\bibitem{Kruczenski:2003be}
Martin Kruczenski, David Mateos, Robert~C. Myers, and David~J. Winters.
\newblock {Meson spectroscopy in AdS/CFT with flavour}.
\newblock {\em JHEP}, 07:049, 2003.


\bibitem{Kruczenski:2003uq}
Martin Kruczenski, David Mateos, Robert~C. Myers, and David~J. Winters.
\newblock {Towards a holographic dual of large-N(c) QCD}.
\newblock {\em JHEP}, 05:041, 2004.






\bibitem{Kuperstein:2008cq}
S.~Kuperstein and J.~Sonnenschein.
\newblock {A New Holographic Model of Chiral Symmetry Breaking}.
\newblock {\em JHEP}, 0809:012, 2008.


\bibitem{Landau:1987fl}
L.~D. Landau and E.~M.~Lifshitz.
\newblock {Fluid Mechanics}.
\newblock Butterworth Heinemann (1987) 552 p.





\bibitem{Leigh:1989jq}
R.~G. Leigh.
\newblock {Dirac-Born-Infeld Action from Dirichlet Sigma Model}.
\newblock {\em Mod. Phys. Lett.}, A4:2767, 1989.




\bibitem{Lifschytz:2009si}
{G.~Lifschytz and M.~Lippert}.
\newblock {Anomalous conductivity in holographic QCD}.
\newblock {\em  Phys.\ Rev.\  D}, 80:066005, 2009.


\bibitem{Lifschytz:2009sz}
{G.~Lifschytz and M.~Lippert}.
\newblock {Holographic Magnetic Phase Transition}.
\newblock {\em  Phys.\ Rev.\  D}, 80:066007, 2009.




\bibitem{Maldacena:1997re}
{Juan M. Maldacena}.
\newblock {The large N limit of superconformal field theories and supergravity}.
\newblock {\em  Adv. Theor. Math. Phys.}, 2 (1998) 231-252.


\bibitem{Marolf:2003ye}
{D.~Marolf, L.~Martucci and P.~J.~Silva}.
\newblock {Fermions, T-duality and effective actions for D-branes in bosonic backgrounds}.
\newblock {\em  JHEP}, 0304:051, 2003.


\bibitem{Marolf:2003vf}
{D.~Marolf, L.~Martucci and P.~J.~Silva}.
\newblock {Actions and fermionic symmetries for D-branes in bosonic backgrounds}.
\newblock {\em  JHEP}, 0307:019, 2003.


\bibitem{Martucci:2005rb}
{L.~Martucci, J.~Rosseel, D.~Van den Bleeken and A.~Van Proeyen}.
\newblock {Dirac actions for D-branes on backgrounds with fluxes}.
\newblock {\em  Class.\ Quant.\ Grav.}, 22:2745, 2005.


\bibitem{Mateos:2007vc}
D.~Mateos, S.~Matsuura, R.~C.~Myers and R.~M.~Thomson.
\newblock {Holographic phase transitions at finite chemical potential}.
\newblock {\em JHEP}, 0711:085, 2007.


\bibitem{Mateos:2006nu}
David Mateos, Robert~C. Myers, and Rowan~M. Thomson.
\newblock {Holographic phase transitions with fundamental matter}.
\newblock {\em Phys. Rev. Lett.}, 97:091601, 2006.


\bibitem{Mateos:2007vn}
David Mateos, Robert~C. Myers, and Rowan~M. Thomson.
\newblock {Thermodynamics of the brane}.
\newblock {\em JHEP}, 05:067, 2007.


\bibitem{McGreevy:2009xe}
J.~McGreevy.
\newblock {Holographic duality with a view toward many-body physics}.





\bibitem{Miransky:2002eb}
V.~A. Miransky.
\newblock {Dynamics of QCD in a strong magnetic field}.
\newblock 2002.


\bibitem{MM}
{I. Montvay, G. Munster}.
\newblock {Quantum Fields on a Lattice}.
\newblock {\em Cambridge, UK: Univ. Press}, 1997.




\bibitem{Myers:2007we}
R.~C.~Myers, A.~O.~Starinets and R.~M.~Thomson.
\newblock {Holographic spectral functions and diffusion constants for fundamental matter}.
\newblock {\em  JHEP}, 0711:091 (2007).


\bibitem{Nambu:1961tp}
Y.~Nambu and G.~Jona-Lasinio.
\newblock {Dynamical model of elementary particles based on an analogy with superconductivity. I}.
\newblock {\em  Phys.\ Rev.}, 122:345 (1961).


\bibitem{O'Bannon:2007in}
Andy O'Bannon.
\newblock {Hall Conductivity of flavour Fields from AdS/CFT}.
\newblock {\em Phys. Rev.}, D76:086007, 2007.


\bibitem{Paredes:2008nf}
A.~Paredes, K.~Peeters and M.~Zamaklar.
\newblock {Mesons versus quasi-normal modes: undercooling and overheating}.
\newblock {\em  JHEP}, 0805:027 (2008). 


\bibitem{Parnachev:2007bc}
 A.~Parnachev.
\newblock {Holographic QCD with Isospin Chemical Potential}.
\newblock {\em  JHEP}, 0802:062 (2008).


\bibitem{Parnachev:2006dn}
A.~Parnachev and D.~A.~Sahakyan.
\newblock {Chiral phase transition from string theory}.
\newblock {\em  Phys. Rev. Lett.}, 97:111601 (2006).


\bibitem{Parnachev:2006ev}
A.~Parnachev and D.~A.~Sahakyan.
\newblock {Photoemission with chemical potential from QCD gravity dual}.
\newblock {\em Nucl.\ Phys.\  B}, 768:177, (2007).


\bibitem{Peeters:2006iu}
K.~Peeters, J.~Sonnenschein and M.~Zamaklar.
\newblock {Holographic melting and related properties of mesons in a quark gluon plasma}.
\newblock {\em  Phys.\ Rev.\  D}, 74:106008 (2006). 




\bibitem{Peeters:2007ti}
K.~Peeters and M.~Zamaklar.
\newblock {Dissociation by acceleration}.
\newblock {\em  JHEP}, 0801:038 (2008).


\bibitem{Peskin:1995ab}
M.~E.~Peskin and D.~V.~Schroeder.
\newblock {An Introduction To Quantum Field Theory}.
\newblock Westview Press (1995) 864 p. 


\bibitem{Polchinski:1995mt}
J.~Polchinski.
\newblock {Dirichlet-Branes and Ramond-Ramond Charges}.
\newblock {\em Phys.\ Rev.\ Lett.}, 75:4724, (1995).


\bibitem{Polchinski:1998rq}
J.~Polchinski.
\newblock {String theory. Vol. 1: An introduction to the bosonic string}.
\newblock Cambridge, UK: Univ. Pr. (1998) 402 p.


\bibitem{Polchinski:1998rq2}
J.~Polchinski.
\newblock {String theory. Vol. 2: Superstring theory and beyond}.
\newblock Cambridge, UK: Univ. Pr. (1998) 531 p.


\bibitem{Polchinski:2000uf}
Joseph Polchinski and Matthew~J. Strassler.
\newblock {The string dual of a confining four-dimensional gauge theory}.
\newblock 2000.


\bibitem{Policastro:2001yc}
G.~Policastro, D.~T.~Son and A.~O.~Starinets.
\newblock {The shear viscosity of strongly coupled N = 4 supersymmetric Yang-Mills plasma}.
\newblock {\em Phys.\ Rev.\ Lett.}, 87:081601 (2001).


\bibitem{Policastro:2002se}
G.~Policastro, D.~T.~Son and A.~O.~Starinets.
\newblock {From AdS/CFT correspondence to hydrodynamics}.
\newblock {\em JHEP}, 0209:043 (2002).


\bibitem{Policastro:2002tn}
G.~Policastro, D.~T.~Son and A.~O.~Starinets.
\newblock {From AdS/CFT correspondence to hydrodynamics. II: Sound waves}.
\newblock {\em  JHEP}, 0212:054 (2002). 


\bibitem{Politzer:1973fx}
{H. D. Politzer}.
\newblock {Reliable perturbative results for strong interactions?}.
\newblock {\em Phys. Rev. Lett.}, 30:1346, 1973.


\bibitem{Rajagopal:2000wf}
{K.~Rajagopal and F.~Wilczek}.
\newblock {The condensed matter physics of QCD}.


\bibitem{Rangamani:2009xk}
M.~Rangamani.
\newblock {Gravity and Hydrodynamics: Lectures on the fluid-gravity correspondence}.
\newblock {\em Class. Quant. Grav.}, 26:224003, 2009.


\bibitem{Romans:1984an}
L.~J.~Romans.
\newblock {New Compactifications Of Chiral N=2 D = 10 Supergravity}.
\newblock {\em Phys.\ Lett.\  B}, 153:392, 1985.


\bibitem{Rozali:2007rx}
M.~Rozali, H.~H.~Shieh, M.~Van Raamsdonk and J.~Wu.
\newblock {Cold Nuclear Matter In Holographic QCD}.
\newblock {\em  JHEP}, 0801:053 (2008).


\bibitem{Sakai:2004cn}
Tadakatsu Sakai and Shigeki Sugimoto.
\newblock {Low energy hadron physics in holographic QCD}.
\newblock {\em Prog. Theor. Phys.}, 113:843--882, 2005.


\bibitem{Sakai:2005yt}
Tadakatsu Sakai and Shigeki Sugimoto.
\newblock {More on a holographic dual of QCD}.
\newblock {\em Prog. Theor. Phys.}, 114:1083--1118, 2006.


\bibitem{Schafer:2005ff}
{T.~Schafer}.
\newblock {Phases of QCD}.


\bibitem{Schwarz:1999dj}
T.~M.~Schwarz, S.~P.~Klevansky and G.~Papp.
\newblock {The phase diagram and bulk thermodynamical quantities in the NJL model  at finite temperature and density}.
\newblock {\em Phys. Rev.  C}, 60, 055205, 1999.


\bibitem{Schwinger:1951nm}
J.~S.~Schwinger.
\newblock {On gauge invariance and vacuum polarization}.
\newblock {\em Phys. Rev.}, 82, 664, 1951.


\bibitem{Semenoff:1999xv}
G.~W. Semenoff, I.~A. Shovkovy, and L.~C.~R. Wijewardhana.
\newblock {Universality and the magnetic catalysis of chiral symmetry
  breaking}.
\newblock {\em Phys. Rev.}, D60:105024, 1999.


\bibitem{Shuryak:2003xe}
E.~Shuryak.
\newblock {Why does the quark gluon plasma at RHIC behave as a nearly ideal fluid?}.
\newblock {\em Prog.\ Part.\ Nucl.\ Phys.}, 53:273, 2004.


\bibitem{Shuryak:2004cy}
E.~V.~Shuryak.
\newblock {What RHIC experiments and theory tell us about properties of  quark-gluon plasma?}.
\newblock {\em  Nucl.\ Phys.\  A}, 750:64, 2005.


\bibitem{Skenderis:2002wp}
K.~Skenderis.
\newblock {Lecture notes on holographic renormalization}.
\newblock {\em Class.\ Quant.\ Grav.}, 19:5849, (2002).


\bibitem{Son:2006em}
D.~T.~Son and A.~O.~Starinets.
\newblock {Hydrodynamics of R-charged black holes}.
\newblock {\em  JHEP}, 0603:052 (2006). 


\bibitem{Son:2009tf}
D.~T.~Son and P.~Surowka.
\newblock {Hydrodynamics with Triangle Anomalies}.
\newblock {\em Phys. Rev. Lett.}, 103:191601, 2009.


\bibitem{Song:2007ux}
{H.~Song and U.~W.~Heinz}.
\newblock {Causal viscous hydrodynamics in 2+1 dimensions for relativistic heavy-ion collisions}.
\newblock {\em Phys.\ Rev.\  C}, 77:064901 (2008).


\bibitem{Starinets:2002br}
Andrei~O. Starinets.
\newblock {Quasinormal modes of near extremal black branes}.
\newblock {\em Phys. Rev.}, D66:124013, 2002.


\bibitem{Starinets:2008fb}
A.~O.~Starinets.
\newblock {Quasinormal spectrum and the black hole membrane paradigm}.
\newblock {\em  Phys.\ Lett.\  B}, 670:442 (2009).   


\bibitem{Susskind:1994vu}
Leonard Susskind.
\newblock {The World as a hologram}.
\newblock {\em J. Math. Phys.}, 36:6377--6396, 1995.


\bibitem{Thompson:2008qw}
E.~G.~Thompson and D.~T.~Son.
\newblock {Magnetized baryonic matter in holographic QCD}.
\newblock {\em  Phys.\ Rev.\  D}, 78:066007 (2008).


\bibitem{'t Hooft:1973jz}
{'t Hooft, Gerard}.
\newblock {A PLANAR DIAGRAM THEORY FOR STRONG INTERACTIONS}.
\newblock {\em Nucl. Phys.}, B72:461, 1974.


\bibitem{'tHooft:1993gx}
G.~'t Hooft.
\newblock {Dimensional reduction in quantum gravity}.


\bibitem{Tseytlin:1996it}
{A.~A.~Tseytlin}.
\newblock {Self-duality of Born-Infeld action and Dirichlet 3-brane of type IIB superstring theory}.
\newblock {\em Nucl.\ Phys.\ B}, 469:51, 1996.


\bibitem{Vuorinen:2003fs}
{A.~Vuorinen}.
\newblock {The pressure of QCD at finite temperatures and chemical potentials}.
\newblock {\em Phys.\ Rev.\  D}, 68:054017 (2003).


\bibitem{Witten:1979vv}
{E.~Witten}.
\newblock {Current Algebra Theorems For The U(1) Goldstone Boson}.
\newblock {\em Nucl.\ Phys.\  B}, 156:269 (1979).


\bibitem{Witten:1998qj}
Edward Witten.
\newblock {Anti-de Sitter space and holography}.
\newblock {\em Adv. Theor. Math. Phys.}, 2:253--291, 1998.


\bibitem{Witten:1998zw}
{E.~Witten}.
\newblock {Anti-de Sitter space, thermal phase transition, and confinement in  gauge theories}.
\newblock {\em Adv.\ Theor.\ Math.\ Phys.}, 2:505 (1998).


\bibitem{Jean:2002}
{J.~Zinn-Justin}.
\newblock {Quantum Field Theory and Critical Phenomena}.
\newblock {\em Oxford Univ. Press, USA}, 4th edition, 1074p, 2002.

  

\end{thebibliography}
\end{document}